\UseRawInputEncoding
\documentclass[aps, prd, superscriptaddress,nofootinbib, preprint]{revtex4-1}

\usepackage{amsmath,amssymb,amsfonts,bm,slashed,graphicx,array,multirow,xspace}
\usepackage{placeins}
\usepackage{setspace}
\usepackage{enumitem}
\usepackage{hhline}
\usepackage[labelfont=bf,font={small,stretch=1}, justification=centerlast]{caption}
\usepackage{subcaption}
\usepackage{tikz}
\usepackage[official]{eurosym}

\usepackage{breakurl}
\usepackage{comment}
\definecolor{red}{rgb}{0.9, 0,0}
\definecolor{cerulean}{rgb}{0., 0.42,0.9}
\definecolor{navy}{rgb}{0.05, 0.05,0.8}

\definecolor{nicered}{rgb}{0.7,0.1,0.1}
\definecolor{nicegreen}{rgb}{0.1,0.5,0.1}
\definecolor{niceblue}{rgb}{0.1,0.1,0.7}
\usepackage[bookmarks=false]{hyperref}
\hypersetup{colorlinks,citecolor= nicegreen,linkcolor= niceblue}

\newcommand{\CODEXb}{\mbox{CODEX-b}\xspace}
\newcommand{\CODEXbeta}{\mbox{CODEX-$\beta$}\xspace}

\newcommand{\nn}{{\nonumber}}
\newcommand{\lp}{\left(}
\newcommand{\rp}{\right)}
\newcommand{\order}[1]{\mathcal{O}\lp#1\rp}
\newcommand{\abs}[1]{\left| #1\right|}
\newcommand{\vev}[1]{\left< #1\right>}

\usepackage{extramarks}

\usepackage{fancyhdr}
\allowdisplaybreaks

\makeatletter
\g@addto@macro\bfseries{\boldmath}
\makeatother

\let\origfootnote\footnote
\renewcommand{\footnote}[1]{%
   \begingroup
   \renewcommand{\baselinestretch}{1}%
   \origfootnote{#1}%
   \endgroup}
\begin{document}

\title{Expression of Interest for the CODEX-b Detector}

\author{Giulio Aielli}
\affiliation{Universit\`a e INFN Sezione di Roma Tor Vergata, Roma, Italy}

\author{Eli Ben-Haim}
\affiliation{LPNHE, Sorbonne Universit{\'e}, Paris Diderot Sorbonne Paris Cit{\'e}, CNRS/IN2P3, Paris, France}

\author{Roberto Cardarelli}
\affiliation{INFN Sezione di Roma Tor Vergata, Roma, Italy}

\author{Matthew John Charles}
\affiliation{LPNHE, Sorbonne Universit{\'e}, Paris Diderot Sorbonne Paris Cit{\'e}, CNRS/IN2P3, Paris, France}

\author{Xabier Cid Vidal}
\affiliation{Instituto Galego de F\'isica de Altas Enerx\'ias (IGFAE), Universidade de Santiago de Compostela, Santiago de Compostela, Spain}

\author{Victor Coco}
\affiliation{European Organization for Nuclear Research (CERN), Geneva, Switzerland}

\author{Biplab Dey}
\affiliation{ELTE E\"otv\"os Lor\'and University, Budapest, Hungary}

\author{Raphael Dumps}
\affiliation{European Organization for Nuclear Research (CERN), Geneva, Switzerland}

\author{Jared A.~Evans}
\affiliation{Department of Physics, University of Cincinnati, Cincinnati, Ohio 45221, USA}

\author{George Gibbons}
\affiliation{University of Birmingham, Birmingham, United Kingdom}

\author{Olivier Le Dortz}
\affiliation{LPNHE, Sorbonne Universit{\'e}, Paris Diderot Sorbonne Paris Cit{\'e}, CNRS/IN2P3, Paris, France}

\author{Vladimir V.~Gligorov}
\affiliation{LPNHE, Sorbonne Universit{\'e}, Paris Diderot Sorbonne Paris Cit{\'e}, CNRS/IN2P3, Paris, France}

\author{Philip Ilten}
\affiliation{University of Birmingham, Birmingham, United Kingdom}

\author{Simon Knapen}
\affiliation{School of Natural Sciences, Institute for Advanced Study, Princeton, NJ 08540, USA}

\author{Jongho Lee}
\affiliation{European Organization for Nuclear Research (CERN), Geneva, Switzerland}
\affiliation{Kyungpook National University (KNU), Daegu, Korea}

\author{Saul L\'{o}pez Soli\~{n}o}
\affiliation{Instituto Galego de F\'isica de Altas Enerx\'ias (IGFAE), Universidade de Santiago de Compostela, Santiago de Compostela, Spain}

\author{Benjamin Nachman}
\affiliation{Ernest Orlando Lawrence Berkeley National Laboratory, University of California, Berkeley, CA 94720, USA}

\author{Michele Papucci}
\affiliation{Ernest Orlando Lawrence Berkeley National Laboratory, University of California, Berkeley, CA 94720, USA}
\affiliation{Walter Burke Institute for Theoretical Physics, California Institute of Technology, Pasadena, CA 91125, USA}

\author{Francesco Polci}
\affiliation{LPNHE, Sorbonne Universit{\'e}, Paris Diderot Sorbonne Paris Cit{\'e}, CNRS/IN2P3, Paris, France}

\author{Robin Quessard}
\affiliation{LPENS, D\'epartement de Physique de l'Ecole Normale Sup\'erieure, Paris, France}

\author{Harikrishnan Ramani}
\affiliation{Ernest Orlando Lawrence Berkeley National Laboratory, University of California, Berkeley, CA 94720, USA}
\affiliation{Department of Physics, University of California, Berkeley, CA 94720, USA}

\author{Dean J.~Robinson}
\affiliation{Ernest Orlando Lawrence Berkeley National Laboratory, University of California, Berkeley, CA 94720, USA}

\author{Heinrich Schindler}
\affiliation{European Organization for Nuclear Research (CERN), Geneva, Switzerland}

\author{Michael D. Sokoloff}
\affiliation{Department of Physics, University of Cincinnati, Cincinnati, Ohio 45221, USA}

\author{Paul Swallow}
\affiliation{University of Birmingham, Birmingham, United Kingdom}

\author{Riccardo Vari}
\affiliation{INFN Sezione di Roma La Sapienza, Roma, Italy}

\author{Nigel Watson}
\affiliation{University of Birmingham, Birmingham, United Kingdom}

\author{Mike Williams}
\affiliation{Laboratory for Nuclear Science, Massachusetts Institute of Technology, Cambridge, MA 02139, USA}

\begin{figure*}[b]
\vspace{-0.75cm}
\includegraphics[width=0.3\textwidth, trim=0 0 0 -1.cm]{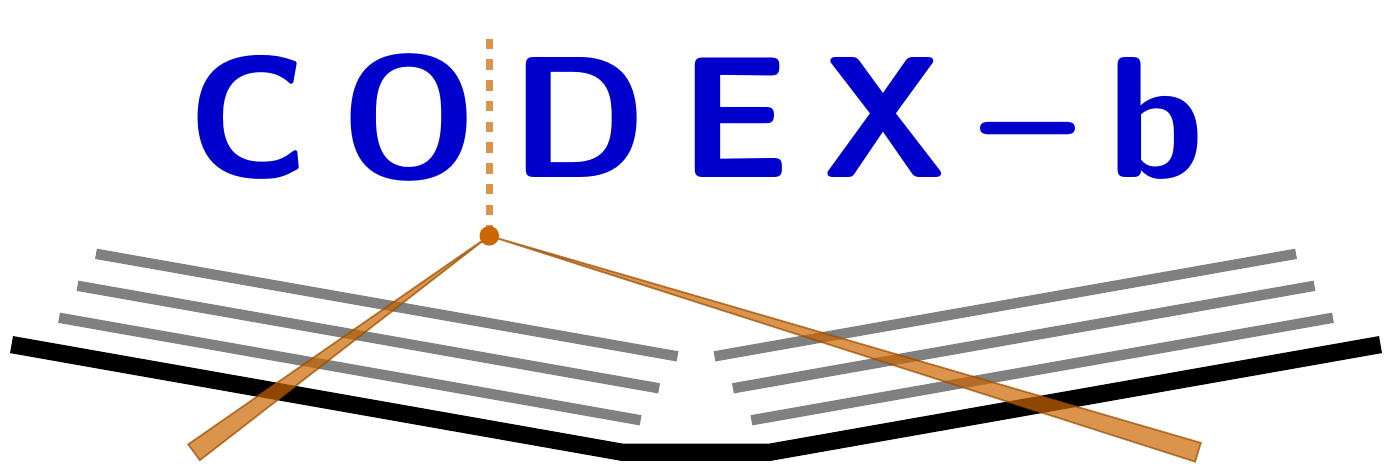}
\end{figure*}

\begin{abstract}
This document presents the physics case and ancillary studies for the proposed CODEX-b long-lived particle (LLP) detector, 
as well as for a smaller proof-of-concept demonstrator detector, \CODEXbeta, to be operated during Run~3 of the LHC.
Our development of the \CODEXb physics case synthesizes `top-down' and `bottom-up' theoretical approaches, providing a detailed survey of both minimal and complete models featuring LLPs.
Several of these models have not been studied previously, and for some others we amend studies from previous literature:
In particular, for gluon and fermion-coupled axion-like particles.
We moreover present updated simulations of expected backgrounds in CODEX-b's actively shielded environment, including the effects of shielding propagation uncertainties, high-energy tails and variation in the shielding design.
Initial results are also included from a background measurement and calibration campaign. 
A design overview is presented for the \CODEXbeta demonstrator detector, which will enable background calibration and detector design studies.
Finally, we lay out brief studies of various design drivers of the \CODEXb experiment and potential extensions of the baseline design, including the physics case for a calorimeter element, precision timing, 
event tagging within LHCb, and precision low-momentum tracking.
\end{abstract}

\maketitle

\tableofcontents

\clearpage
\section*{Executive summary}

The Large Hadron Collider (LHC) provides unprecedented sensitivity to short-distance physics.
Primary achievements of the experimental program include the discovery of the Higgs boson \cite{Aad:2012tfa,Chatrchyan:2012xdj}, 
the ongoing investigation of its interactions \cite{Khachatryan:2016vau},
and remarkable precision Standard Model (SM) measurements. Furthermore,
a multitude of searches for physics beyond the Standard Model (BSM) have been conducted over a tremendous array of channels.
These have resulted in greatly improved BSM limits, with no new particles or force carriers having been found.

The primary LHC experiments (ATLAS, CMS, LHCb, ALICE) have proven to be remarkably versatile and complementary in their
BSM reach. As these experiments are scheduled for upgrades and data collection over at least another 15 years,
it is natural to consider whether they can be further complemented by one or more detectors specialized for well-motivated but currently hard-to-detect BSM signatures.
A compelling category of such signatures are long-lived particles (LLPs), which generally appear in any theory
containing a hierarchy of scales or small parameters, and are therefore ubiquitous in BSM scenarios.

\begin{figure}[b]
	\includegraphics[width = 0.95\linewidth]{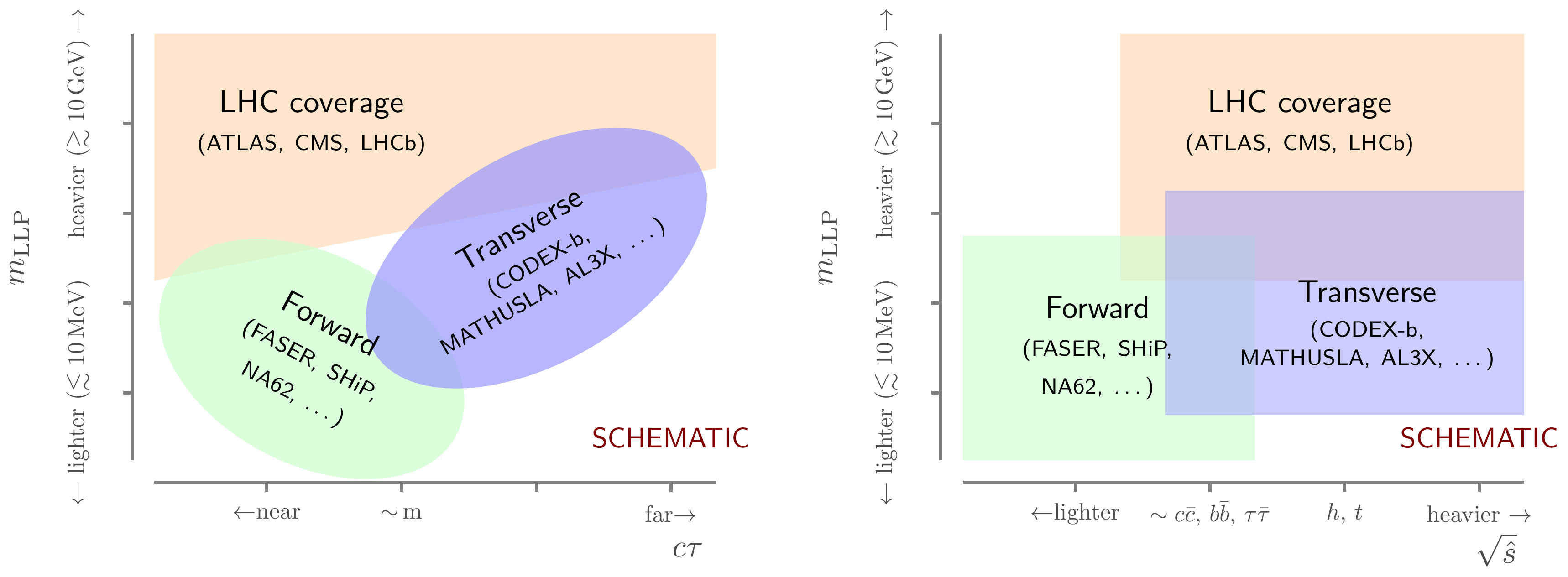}
	\caption{Schematic summary of reach and coverage of current, planned or proposed experiments in terms of the  LLP mass, lifetime and the required parton center-of-mass energy, $\sqrt{\hat{s}}$.}
	\label{fig:llpschexec}
\end{figure}

The central challenge in detecting LLPs is that not only their masses but also their lifetimes may span many orders of magnitude.
This makes it impossible from first principles to construct a single detector which would have the ultimate sensitivity
to all possible LLP signatures; multiple complementary experiments are necessary, as summarized
in Fig.~\ref{fig:llpschexec}.

\begin{figure}
\includegraphics[width=\textwidth]{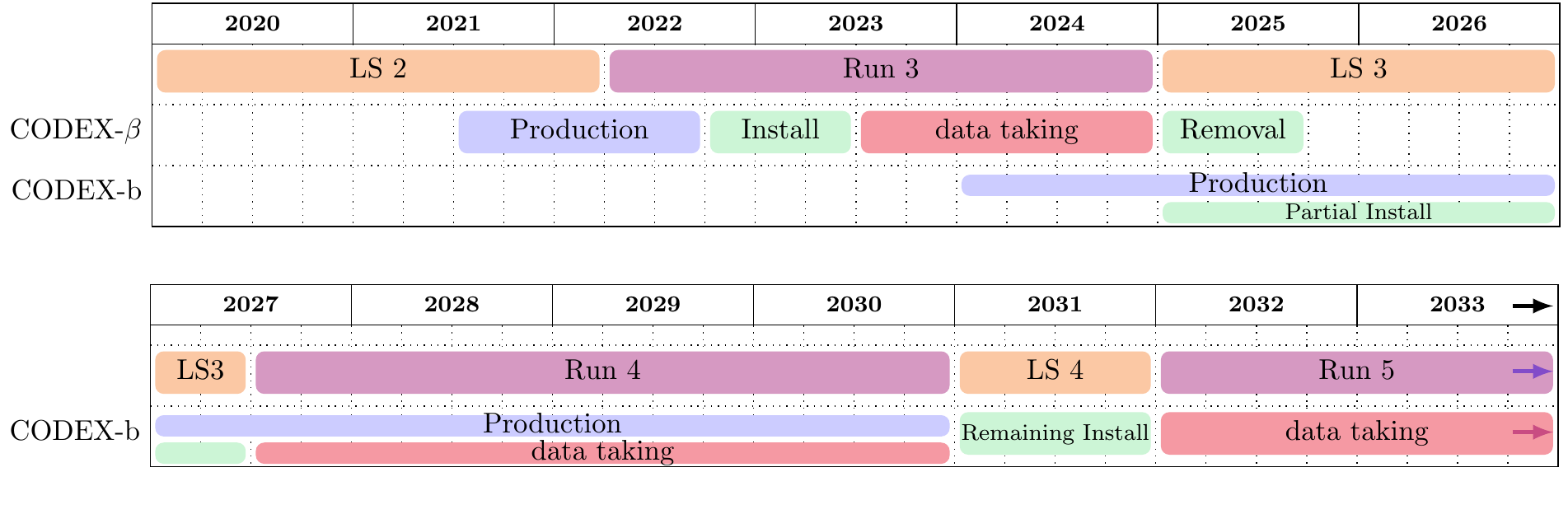}
\caption{Approximate production, installation and data-taking timelines for demonstrator (\CODEXbeta) and full (\CODEXb) detectors. (Updated according to COVID constraints.)\label{timelineexec}}
\end{figure}

In this expression of interest we advocate for CODEX-b (\emph{``COmpact Detector for EXotics at LHCb"}), a LLP detector that would be installed in the  DELPHI/UXA cavern next to LHCb's interaction point (IP8). 
The approximate proposed timeline is given in Fig.~\ref{timelineexec}: Here ``\CODEXbeta'' refers to a smaller proof-of-concept detector
with otherwise the same basic geometry and technology as CODEX-b.

The central advantages of CODEX-b are:
\begin{itemize}
\item Very competitive sensitivity to a wide range of LLP models, either exceeding or complementary to the sensitivity of other existing or proposed detectors;
\item An achievable zero background environment, as well as an accessible experimental location in the DELPHI/UXA cavern with all necessary services already in place;
\item Straightforward integration into LHCb's trigger-less readout and the ability to tag events of interest with the LHCb detector;
\item A compact size and consequently modest cost, with the realistic possibility to extend detector capabilities for neutral particles in the final state.
\end{itemize}
We survey a wide range of BSM
scenarios giving rise to LLPs and demonstrate how these advantages translate
into competitive and complementary reach with respect to other proposals. We furthermore detail the experimental and simulation studies
carried out so far, showing that CODEX-b can be built as planned and operate as a zero background experiment.
We also discuss possible technology options that may further enhance the reach of CODEX-b.
Finally, we discuss the timetable for the construction and data taking of \CODEXbeta, and show that
it may also achieve new reach for certain BSM scenarios.

\clearpage

\section{Introduction}
\subsection{Motivation\label{sec:motivation}}
New Physics (NP) searches at the LHC and other experiments have primarily been motivated by the predictions of various extensions of the SM, 
designed to address long-standing open questions. These include \emph{e.g.}~the origin and nature of dark matter, 
the detailed dynamics of the weak scale, the mechanism of baryogenesis, among many others. 
However, in the absence of clear experimental NP hints, the solutions to these puzzles remain largely mysterious. 
Combined with increasing tensions from current collider data on the most popular BSM extensions, 
it has become increasingly imperative to consider whether the quest for NP requires new and innovative strategies: 
A means to diversify LHC search programs with a minimum of theoretical prejudice, 
and to seek signatures for which the current experiments are trigger and/or background limited.

\enlargethispage{-2\baselineskip}

A central component of this program will be the ability to probe `dark' or `hidden' sectors,
comprised of degrees of freedom that are `sterile' under the SM gauge interactions. Hidden sectors are ubiquitous in many BSM scenarios,
and typically may feature either suppressed renormalizable couplings, heavy mediator
exchanges with SM states, or both.\footnote{This heavy mediator may itself be an exotic state or part of the SM electroweak sector, such as the Higgs. The possibility of exploring
the `Higgs portal' is particularly compelling, because
our theoretical understanding of Higgs interactions is likely incomplete, and new states might interact with it.
In addition, the Higgs itself  may have a sizable branching ratio to exotic states since its SM partial width is suppressed by the $b$-quark Yukawa coupling. 
Experiments capable of leveraging large samples of Higgs bosons are then natural laboratories to search for NP. Understanding the properties of the Higgs sector will be central to ongoing and future particle physics programs.} 
 The sheer breadth of possibilities for these hidden sectors mandates search strategies that are as model-independent as possible.

Suppressed dark--SM couplings or heavy dark--SM mediators may in turn give rise to relatively long lifetimes for light degrees of freedom in the hidden spectrum,
by inducing suppressions of their total widths via either small couplings, the mediator mass, loops and/or phase space.
This scenario is very common in many models featuring \emph{e.g.}~Dark Matter (Sec.~\ref{sec:inelasticDM}, \ref{sec:DMscatt}, \ref{sec:coannih}, \ref{sec:ADM} and \ref{sec:otherDM}), Baryogenesis (Sec.~\ref{sec:baryogenesis}), Supersymmetry (Sec.~\ref{sec:RPV}) or Neutral Naturalness (Sec.~\ref{sec:neutralnaturalness}).
The canonical examples in the SM are the long lifetimes of the $K^0_L$, $\pi^\pm$, neutron and muon, whose widths are suppressed by the weak interaction scale required for flavor changing processes, as well as phase space.
Vestiges of hidden sectors may then manifest in the form of striking morphologies within LHC collisions,
in particular the \emph{displaced decay-in-flight} of these metastable, light particles in the hidden sector, commonly referred to as `long-lived particles' (LLPs). 
Surveying a wide range of benchmark scenarios, we demonstrate in this document that by searching for such LLP decays, \CODEXb would permit substantial improvements in reach for many well-motivated NP scenarios,
well beyond what could be gained by an increase in luminosity at the existing detectors.

\subsection{Experimental requirements\label{sec:experimentalcoverage}}

In any given NP scenario, the decay width of an LLP may exhibit strong power-law dependencies on \emph{a priori} unknown ratios of various physical scales.
As a consequence, theoretical priors for the LLP lifetime are broad, such that LLPs may occupy a wide phenomenological parameter space.
In the context of the LHC, LLP masses from several MeV up to $\mathcal{O}(1)$\,TeV may be contemplated,
and proper lifetimes as long as $\lesssim0.1$ seconds may be permitted before tensions with Big Bang Nucleosynthesis arise \cite{PhysRevD.37.3441,Dimopoulos:1987fz,Dimopoulos:1988ue}.

Broadly speaking, the ability of any given experiment to probe a particular point in this space of LLP mass and lifetimes will depend strongly not only on the 
center-of-mass energy available to the experiment, but also on its fiducial detector volume, distance from the interaction point (IP), triggering limitations, 
and the size of irreducible backgrounds in the detector environment \cite{Alimena:2019zri}. The latter is large for light LLP searches, requiring a \emph{shielded, background-free} detector. 
Further, LLP production channels involving the decay of a heavy parent state -- \emph{e.g.} a Higgs decay -- 
 require sufficient partonic center-of-mass energy, $\sqrt{\hat{s}}$, to produce an abundant sample of heavy parents. 
Such channels are thus probed most effectively \emph{transverse} to an LHC interaction point.
Taken together, these varying requirements prevent any single experimental approach from attaining comprehensive coverage over the full parameter space.

Experimental coverage of LLP searches is also determined by the \emph{morphology} of LLP decays. 
The simplest scenario contemplates a large branching ratio for $2$-body LLP decays to two charged SM particles -- for instance $\ell^+\ell^-$, $\pi^+\pi^-$ or $K^+K^-$.  In many well-motivated benchmark scenarios (See Sec.~\ref{sec:physicscase}),
however, the LLP may decay to various final states involving missing energy, photons, or high multiplicity, softer final states. In any experimental environment, these more complex decay morphologies can be much more 
challenging to detect or reconstruct: Reconstructing missing energy final states requires the ability to measure track momenta; detecting photons requires a calorimeter element or preshower component;  identifying 
high multiplicity final states requires the suppression of soft hadronic backgrounds. The \CODEXb baseline concept, as described below in Sec.~\ref{sec:detector}, is well-suited to reconstruct several of these morphologies, 
in addition to the simple 2-body decays. Extensions of the baseline design may permit some calorimetry or pre-shower capabilities, which would enable the reconstruction of photons and other neutral hadrons.

\subsection{Baseline detector concept}
\label{sec:detector}

\begin{figure*}[bt]
	\begin{subfigure}[b]{0.5\textwidth}
        \centering
      	\includegraphics[width = 1.2\linewidth]{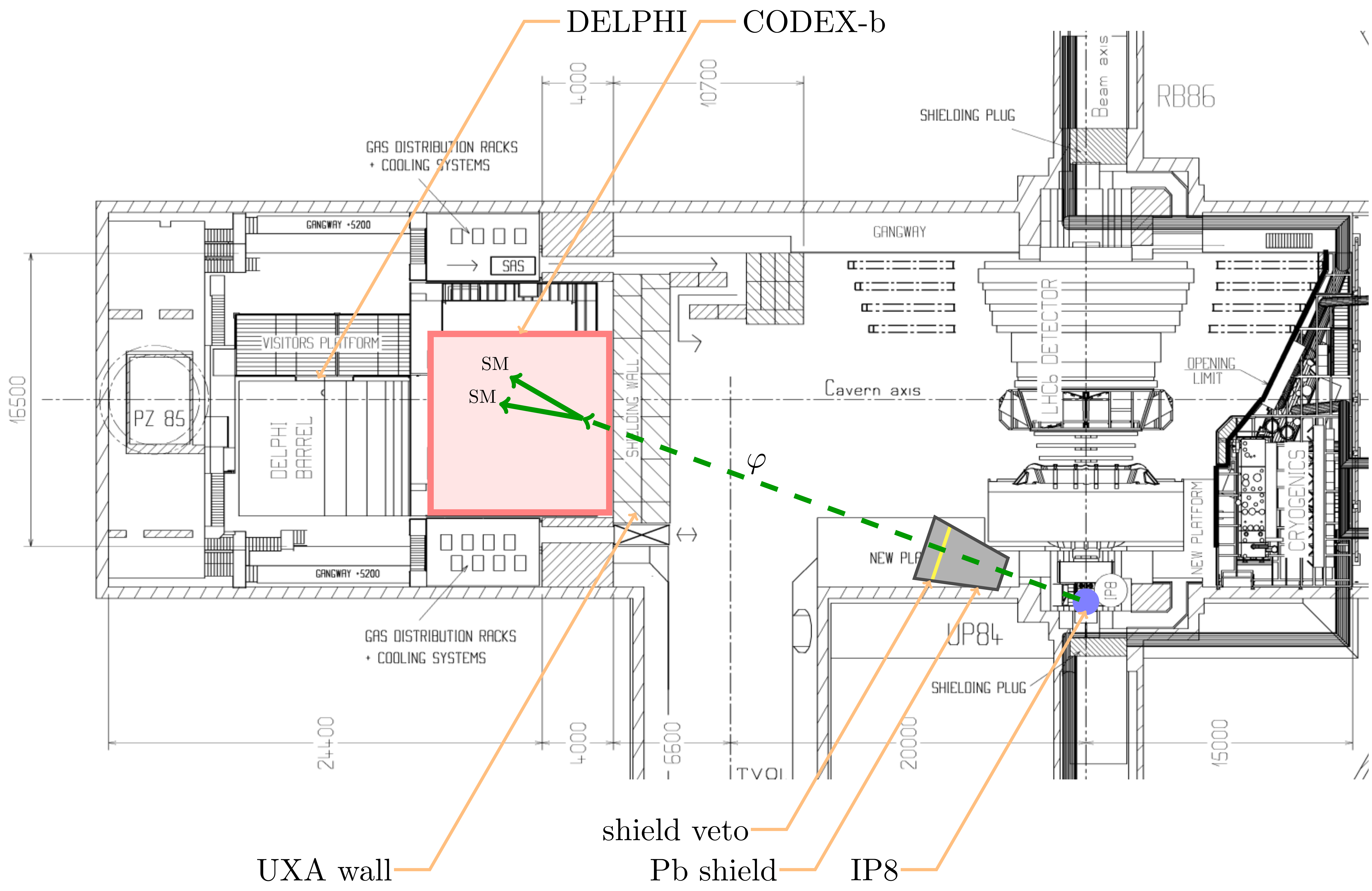}
        \caption{Location in the cavern\label{fig:codexBlocation}}
        \end{subfigure}\hfill
	\begin{subfigure}[b]{0.4\textwidth}
        \centering
       	\includegraphics[width = 0.6\linewidth]{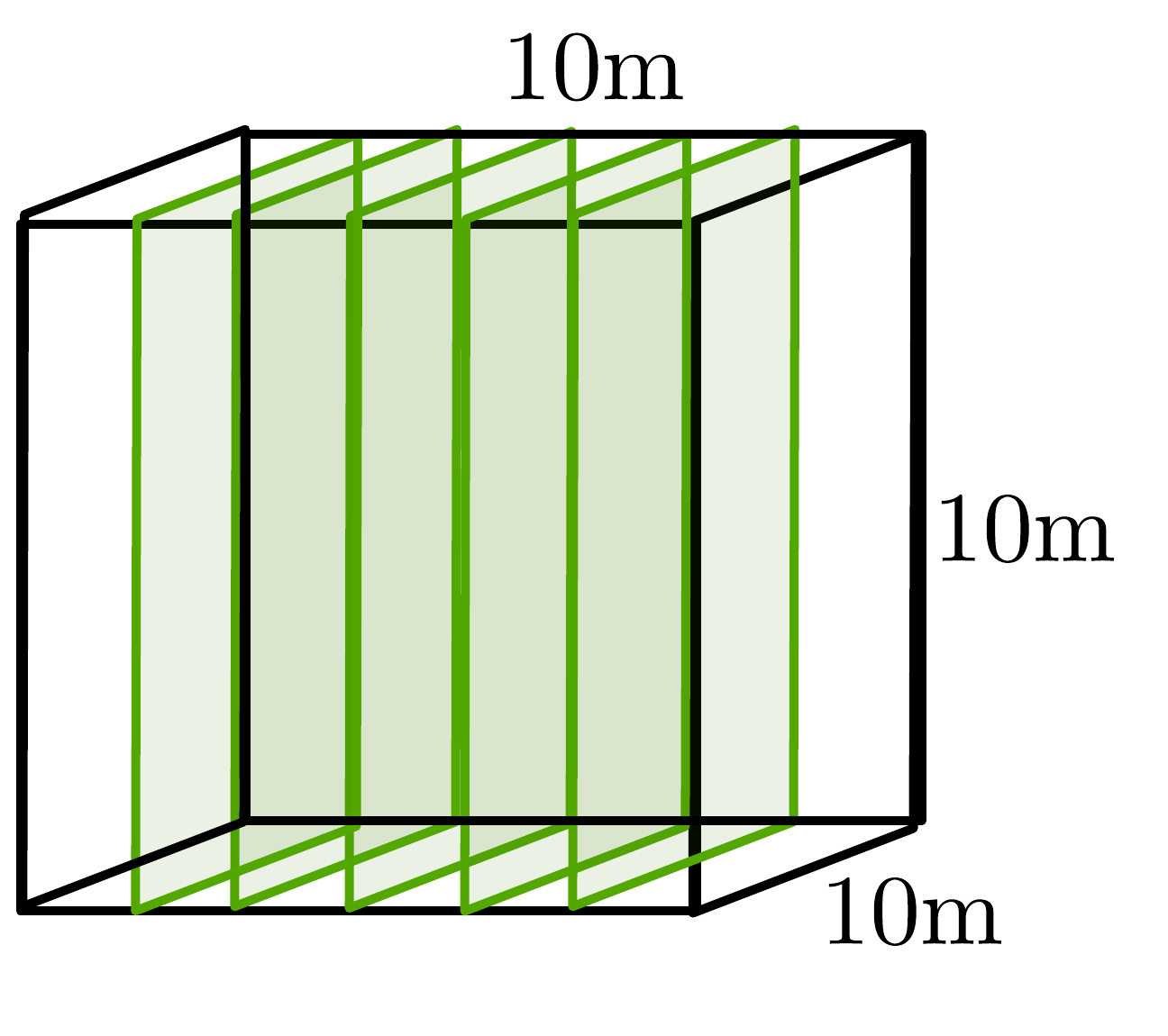}
        \caption{Detector geometry\label{fig:codexBdetector}}
        \end{subfigure}
	\caption{Left: Layout of the LHCb experimental cavern UX85 at point 8 of the LHC~\cite{cavern}, overlaid with the \CODEXb volume, 
	as reproduced from Ref.~\cite{Gligorov:2017nwh}. Right: Schematic representation of the proposed detector geometry.}
	\label{fig:LHCbCav}
\end{figure*}

The proposed \CODEXb location is in the UX85 cavern, roughly 25 meters from the interaction point 8 (IP8), with a nominal fiducial volume of $10\text{ m}\times10\text{ m} \times 10$\,m (see Fig.~\ref{fig:codexBlocation}) \cite{Gligorov:2017nwh}.
Specifically, the fiducial volume is defined by $26\text{ m}<x<36\text{ m}$, $-7\text{ m}<y<3\text{ m}$ and $5\text{ m}<z<15\text{ m}$, where the $z$ direction is aligned along the beam line and the origin of the coordinate
system is taken to be the interaction point. This location roughly corresponds to the pseudorapidity range $0.13<\eta<0.54$. Passive shielding is partially provided by the existing UXA wall, while the remainder is achieved by
a combination of active vetos and passive shielding located nearer to the IP. A detailed description of the backgrounds and the required amount of shielding can be found in Sec.~\ref{sec:backgrounds}.

The actual reach of any LLP detector will be tempered by various efficiencies, 
including efficiencies for tracking and vertex reconstruction. In particular, no magnetic field will be available in the \CODEXb fiducial volume.
To design an LLP \emph{detection} program, rather than only an exclusionary one, it is therefore important to be able to confirm the presence of exotic physics and reject possibly mis-modeled backgrounds. 
This requires capabilities for particle identification, mass reconstruction and/or event reconstruction. 

To address these considerations, several detector concepts are being considered.
The baseline \CODEXb conceptual design makes use of Resistive Plate Chambers (RPC) tracking stations with $\mathcal{O}(100)$\,ps timing resolution.
A hermetic detector, with respect to the LLP decay vertex, is needed to achieve good signal efficiency and background rejection. In the baseline design, 
this is achieved by placing six RPC layers on each surface of the detector.
To ensure good vertex resolution five additional triplets of RPC layers are placed equally spaced along the depth of the detector, as shown in Fig.~\ref{fig:codexBdetector}.
Other, more ambitious options are being considered, that use both RPCs as well as large scale calorimeter technologies such as liquid~\cite{Wagner:2018ajx}
or plastic scintillators, used in accelerator neutrino experiments such as NO$\nu$A~\cite{Ayres:2007tu}, T2K upgrade~\cite{Iwamoto:2018uxt} or Dune~\cite{Yang:2018psz}.
If deemed feasible, implementing one of these options would permit measurement of decay modes involving neutral final states, 
improved particle identification and more efficient background rejection techniques.

Because the baseline \CODEXb concept makes use of proven and well-understood technologies for tracking and precision timing resolution, 
any estimation or simulation of the net reconstruction efficiencies is expected to be reliable.
These estimates must be ultimately validated by data-driven determinations from a demonstrator detector, which we call \CODEXbeta. (See Sec.~\ref{sec:demonstrator}.)
Combined together, the baseline tracking and timing capabilities will permit mass reconstruction and particle identification for some benchmark scenarios.

The transverse location of the detector permits reliable background simulations based on well-measured SM transverse production cross-sections.
The SM particle propagation through matter -- necessary to simulate the response of the UXA radiation wall and the additional passive and active shielding --
is also well understood for the typical particle energies generated in that pseudorapidity range.
The proposed location behind the UXA radiation wall will also permit regular maintenance of the experiment, e.g.~during technical or other stops.
In addition to background simulations, the active veto and the ability to vary the amount of shielding over the detector acceptance permit LLP measurements or
exclusions to be determined with respect to data-driven baseline measurements or calibrations of relevant backgrounds (see Sec.~\ref{sec:backgrounds}).

\subsection{Search power, complementarities and unique features}

Although ATLAS, CMS and LHCb were not explicitly designed with LLP searches in mind, they have been remarkably effective at probing a large region of the LLP parameter space (see \cite{Alimena:2019zri,Lee:2018pag} for recent reviews).
The main variables which provide the necessary discrimination for triggering and off-line background rejection are often the amount of energy deposited and/or the number of tracks connected to the displaced vertex.
In most searches, the signal efficiency therefore drops dramatically for low mass LLPs, especially when they are not highly energetic (e.g.~from Higgs decays.)
For instance, the penetration of hadrons into the ATLAS or CMS muon systems, combined with a reduced trigger efficiency, attenuates the LHC reach for light LLPs, $m_{\text{LLP}} \lesssim 10$\,GeV, decaying in the muon systems.

Beam dump experiments such as SHiP~\cite{Bonivento:2013jag,Alekhin:2015byh,Anelli:2015pba}, NA62~\cite{NA62:2017rwk} in beam-dump mode, 
as well as forward experiments like FASER~\cite{Feng:2017uoz,Ariga:2018zuc,Ariga:2018pin} evade this problem by employing passive and/or active shielding
to fully attenuate the SM backgrounds.
The LLPs are moreover boosted in a relatively narrow cone, and very high beam luminosities can be attained.
This results in excellent reach for light LLPs that are predominantly produced at relatively low center-of-mass energy, such as a kinetically mixed dark photon.
The main trade-off in this approach is, however, the limited partonic center-of-mass energy, 
which severely limits their sensitivity to heavier LLPs or LLPs primarily produced through heavy portals (e.g.~Higgs decays).

Finally, proposals in pursuit of shielded, transverse, background-free detectors such as MATHUSLA~\cite{Chou:2016lxi,Alpigiani:2018fgd}, \CODEXb~\cite{Gligorov:2017nwh} and AL3X~\cite{Gligorov:2018vkc}
aim to operate at relatively low pseudorapidity $\eta$, but with far greater shielding compared to the ATLAS and CMS muon systems.
This removes the background rejection and triggering challenges even for low mass LLPs, $m_{\text{LLP}} \lesssim 10$\,GeV, 
though at the expense of a reduced geometric acceptance and/or reduced luminosity.
Because of their location transverse from the beamline, they can access processes for which a high parton center-of-mass energy is needed, such as Higgs and $Z$ production.

In this light, the regimes for which existing and proposed experiments have the most effective coverage can be roughly summarized as follows:
\begin{enumerate}
	\item ATLAS \& CMS: Heavy LLPs ($m_{\text{LLP}} \gtrsim 10$\,GeV) for all lifetimes ($c\tau \lesssim 10^7$\,m).
	\item LHCb: Short to medium lifetimes ($c\tau \lesssim 1$\,m) for light LLPs \mbox{($0.1\,\text{GeV}\lesssim m_{\text{LLP}} \lesssim 10$\,GeV)}. 
	\item Forward/beam dump detectors (FASER, NA62, SHiP): Medium to long lifetime regime ($0.1 \lesssim c\tau \lesssim 10^7$\,m) for light LLPs ($m_{\text{LLP}} \lesssim $ few GeV), for low $\sqrt{\hat{s}}$ production channels.
	\item Shielded, transversely displaced detectors (MATHUSLA, \CODEXb, AL3X): Relatively light LLPs\footnote{
	The degree to which each of these detectors can compete with ATLAS and CMS in the high mass regime, $m_{\text{LLP}} \gtrsim 10$ GeV, depends on their angular acceptance and integrated luminosity.
	The larger volume MATHUSLA and AL3X proposals therefore typically remain more competitive with the main detectors for higher LLP masses than \CODEXb.}
	($m_{\text{LLP}} \lesssim 10$--$100$\,GeV) in the long lifetime regime ($1 \lesssim c\tau \lesssim 10^7$\,m), and high $\sqrt{\hat{s}}$ production channels. 
\end{enumerate}
 
\begin{figure}[t]
	\includegraphics[width = 0.95\linewidth]{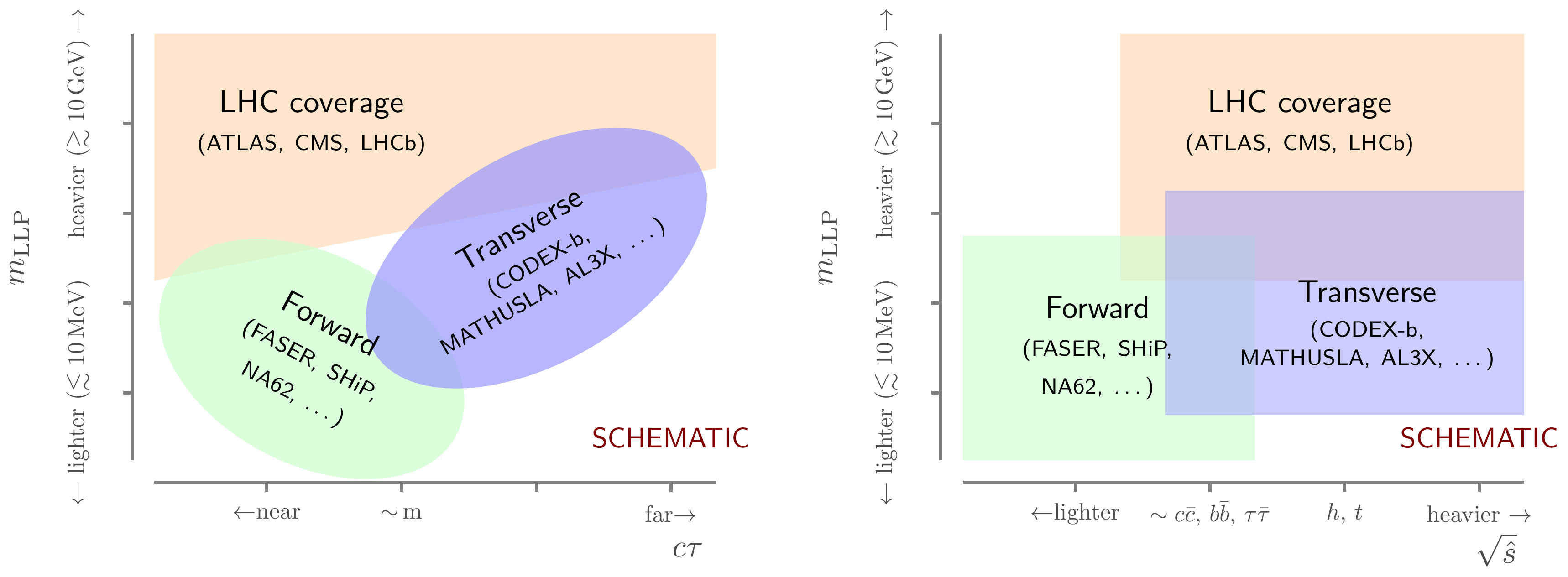}
	\caption{Schematic summary of reach and coverage of various current, planned or proposed experiments in the  LLP mass, lifetime, and $\sqrt{\hat{s}}$ space.}
	\label{fig:llpsch}
\end{figure}

In Fig.~\ref{fig:llpsch} we provide a visual schematic summarizing these LLP coverages, showing slices in the space of LLP mass, lifetime, and $\sqrt{\hat{s}}$,
that provide a sketch of the complementarity and unique features of various LLP search strategies and proposals. 
Relative to the existing LHC detectors, \CODEXb will be able to probe unique regimes of parameter space over a large range of well motivated models and portals, explored further in the Physics Case in Sec.~\ref{sec:physicscase}.
A more extensive discussion and evaluation of the landscape of LLP experimental proposals can be found in Refs.~\cite{Alimena:2019zri} and \cite{Beacham:2019nyx}.

While the ambitiously sized `transverse' detector proposals such as MATHUSLA and AL3X would explore even larger ranges of the parameter space, 
the more manageable and modest size of \CODEXb provides a substantially lower cost alternative with good LLP sensitivity. 
It also allows for the possibility of additional detector subsystems, such as precision tracking and calorimetry. 
Furthermore, the proximity of \CODEXb to the LHCb interaction point (IP8) and LHCb's trigger-less readout (based on standardized and readily available
technologies) makes it straightforward to
integrate the detector into the LHCb readout for triggering and/or partial event reconstruction. This capability is
not available to any other proposed LLP experiment at the LHC interaction points, and may prove crucial to authenticate
any signals seen by \CODEXb.
For a further discussion of the experimental design drivers and preliminary case studies of how different detector capabilities can effect the sensitivity for different models, we refer to Sec.~\ref{sec:design}.

\subsection{Timeline}
\label{sec:time}

\begin{figure}[t]
\includegraphics[width=\textwidth]{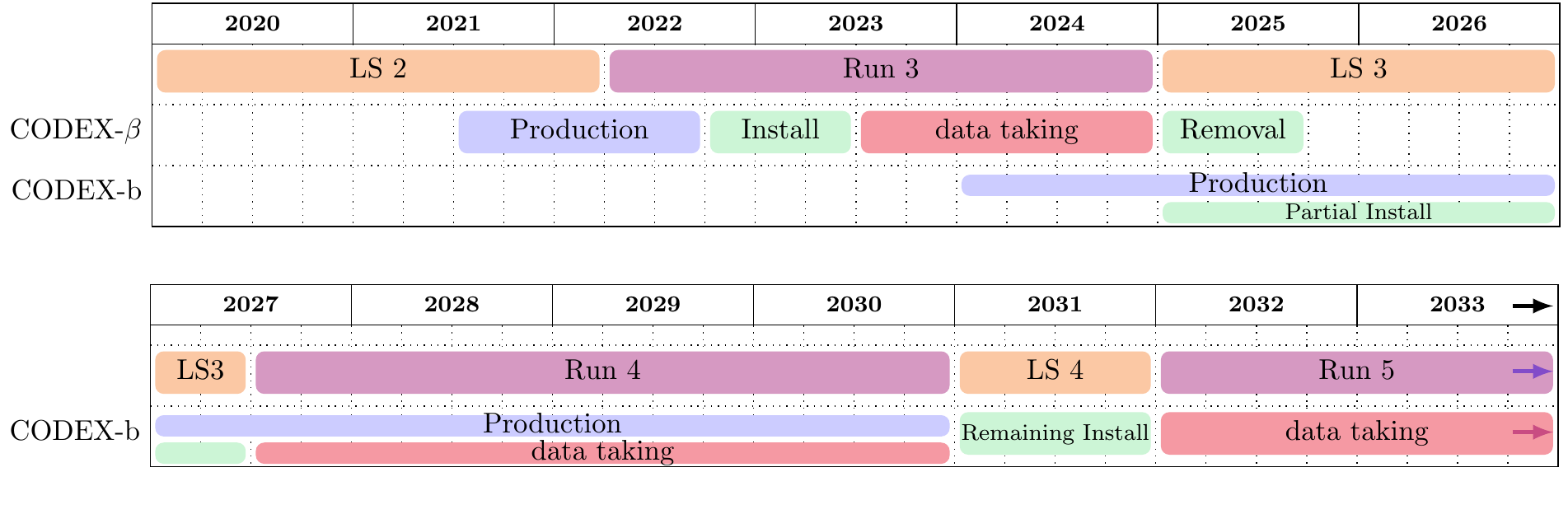}
\caption{Approximate production, installation and data-taking timelines for demonstrator (\CODEXbeta) and full (\CODEXb) detectors. (Updated according COVID to constraints.)\label{fig:timeline}}
\end{figure}

The \CODEXbeta demonstrator detector is proposed for Run~3 and is therefore complementary in time to the other funded proposals such as FASER.
In contrast, the full version of \CODEXb, as well as FASER2, SHiP, MATHUSLA, and AL3X are all proposed to operate in Runs 4 or 5 during the HL-LHC.
We show the nominal timeline for \CODEXbeta and \CODEXb in Fig.~\ref{fig:timeline}.
Results as well as design  and construction lessons from \CODEXbeta are expected to inform the final design choices for the full detector, 
and may also inform the evolution of the schedule shown in Fig.~\ref{fig:timeline}.
The modest size of \CODEXb, the accessibility of the DELPHI cavern, and the use of proven technologies in the baseline design, 
is expected to imply not only lower construction and maintenance costs but also a relatively short construction timescale.
It should be emphasized that \CODEXb may provide complementary data both in reach and in time, at relatively low cost, 
to potential discoveries in other more ambitious proposals, should they be built, as well as to existing LHC experiments.

\FloatBarrier
\clearpage

\section{Physics case}
\label{sec:physicscase}

\subsection{Theory survey strategies}

Long-lived particles occur generically in theories with a hierarchy of mass scales and/or couplings (see Sec.~\ref{sec:motivation}), such as the Standard Model and many of its possible extensions.
This raises the question how best to survey the reach of any new or existing experiment in the theory landscape.
Given the vast range of possibilities, injecting some amount of ``theory prejudice'' cannot be avoided.
We therefore consider two complementary strategies to survey the theory space: 
i)  Studying \emph{minimal models} or ``portals'', where one extends the Standard Model with a single new particle that is inert under all SM gauge interactions. 
The set of minimal modes satisfying this criteria is 
both predictive  and relatively small -- we restrict ourselves to the set of minimal models generating operators of dimension 4 or lower, as well as the well-motivated dimension 5 operators for  axion-like particles. 
It is important to keep in mind, however, that minimal models are merely simplified models,  meant to parametrize different classes of phenomenological features that may arise in more complete models. 
To mitigate this deficiency to some extent, we then also consider: 
ii) Studying a number of \emph{complete models}, which are more elaborate but aim to address one or more of the outstanding problems of the Standard Model, such as the gauge hierarchy problem, 
the mechanism of baryogenesis, or the nature of dark matter.
These complete models feature LLPs as a consequence of the proposed mechanisms introduced to solve these problems.

\subsection{Novel studies}

While many of the models surveyed below have been studied elsewhere and are recapitulated here, several of the studies in this section contain new and novel results,
either correcting previous literature, recasting previous studies for the case of \CODEXb, or introducing new models not studied before. Specifically, we draw the reader's attention to:
\begin{enumerate}
	\item The axion-like particles (ALPs) minimal model (Sec.~\ref{sec:alps}), which includes new contributions to ALP production from parton fragmentation.
	This can be very important in LHC collisions, significantly enhancing production estimates and consequent reaches, but was overlooked in previous literature.
	\item The heavy neutral leptons (HNLs) minimal model (Sec.~\ref{sec:HNL}), which includes modest corrections to the HNL lifetime and $\tau$ branching ratios, compared to prior treatments.
	\item The neutral naturalness complete model (Sec.~\ref{sec:neutralnaturalness}), which is recast for \CODEXb from prior studies.
	\item The coscattering dark matter complete model (Sec.~\ref{sec:DMscatt}), which contains LLPs produced through an exotic $Z$ decay, and has not been studied previously.
\end{enumerate}

\subsection{Minimal models}

The underlying philosophy of the minimal model approach is the fact that the symmetries of the SM already strongly restrict the portals through which a new, neutral state can interact with our sector. 
The minimal models can then be classified via whether the new particle is a scalar ($S$), pseudo-scalar ($a$), a fermion ($N$) or a vector ($A'$). 
In each case there are a only a few operators of dimension 4 or lower (dimension 5 for the pseudo-scalar) which are allowed by gauge invariance. 
The most common nomenclature of the minimal models and their corresponding operators are
\begin{subequations}
\begin{align}
\mathrm{Abelian\; hidden\; sector:\quad}& F_{\mu\nu}F'^{\mu\nu},\quad h A'_{\mu}A'^{\mu}\label{eq:vectorportal}\\
\mathrm{Scalar\text{-}Higgs\; portal:\quad}&S^2H^\dagger H, \quad S H^\dagger H\\
  \mathrm{Heavy\; neutral\; leptons:\quad}&\tilde{H} \bar{L}N\\
 \text{Axion-like\;particles:\quad}&\partial^\mu a\, \bar \psi \gamma_\mu\gamma_5 \psi,\quad a W_{\mu\nu}\tilde W^{\mu\nu}, \quad a B_{\mu\nu}\tilde B^{\mu\nu}, \quad a G_{\mu\nu}\tilde G^{\mu\nu}
\end{align}
\end{subequations}
where $F'^{\mu\nu}$ is the field strength operator corresponding to a $U(1)$ gauge field $A'$, $H$ is the SM Higgs doublet, and $h$ the physical, SM Higgs boson.\footnote{
The second operator in Eq.~\eqref{eq:vectorportal} is strictly speaking not gauge invariant, but can be trivially generated by the kinetic term of a heavy dark scalar charged under the $U(1)$ 
that acquires a vacuum expectation value (VEV) and mixes with the SM Higgs (see e.g.~\cite{Schabinger:2005ei,Gopalakrishna:2008dv,Curtin:2014cca}).}
Where applicable, we consider cases in which a different operator is responsible for the production and decay of the LLP, as summarized in Fig.~\ref{fig:minimalmodels}. 
Note that the $h A'_{\mu}A'^{\mu}$ and $S^2 H^\dagger H$ operators respect a $\mathbb{Z}_2$ symmetry for the new fields and will not induce a decay for the LLP on their own.

For the axion portal, the ALP can couple independently to the $SU(2)$ and $U(1)$ gauge bosons. 
In the infrared, only the linear combination corresponding $a F\tilde F$ survives, though the coupling to the massive electroweak bosons can contribute to certain production modes.
Moreover, the gauge operators mix into the fermionic operators through renormalization group running. 
Classifying the models according to production and decay portals obscures this key point\footnote{An example of when such identification is not as straightforward may be provided by the case of ALP coupled to photons,
where the main production mechanism relevant for CODEX-b is via an effective ALP coupling to quarks.}, and we have therefore chosen to present the model space for the ALPs in Fig.~\ref{fig:minimalmodels} in terms of UV operators. 
Once the UV boundary condition at a scale $\Lambda$ is given, such a choice fully specifies both the ALP production and the decay modes, which often proceed via a combination of the
listed operators. 

\newcolumntype{x}[1]{>{\centering\arraybackslash}m{#1\textwidth}}

\begin{figure}[t]
\def\cellwidth{28}
\def\medcellwidth{24}
\def\cellheight{9}
\def\indentwidth{40}
\def\scale{1}
\tikzset{
every node/.style={shape=rectangle, anchor=north west, outer sep=0mm, draw=black},
x=1mm,%
y=-1mm,%
medcell/.style={fill=black!10, minimum height=\cellheight mm, minimum width=\medcellwidth mm}, 
decaycell/.style={fill=blue!40, minimum height=\scale*\cellheight mm, minimum width=\medcellwidth mm, align = center},
prodcell/.style={fill=green!40, minimum height=\cellheight mm, minimum width=\cellwidth mm},
uvcell/.style={fill=orange!40, minimum height=\cellheight mm, minimum width=\cellwidth mm},
nocell/.style={ fill=black!20, minimum height=\scale*\cellheight mm, minimum width=\medcellwidth mm}, 
figcell/.style={ fill=white, minimum height=\scale*\cellheight mm, minimum width=\cellwidth mm}
}
\def\y{0}
\def\x{0}
\scalebox{0.65}{\parbox{1.5\linewidth}{
\begin{tikzpicture}[line width= 1 pt]
\foreach \mediator/\decayrows/\prods/\dtype/\ptype in {
	{Vector ($A'$)}/{%
	{$F'F$}/{Fig.~\ref{fig:vectorportalHiggsdecay}, no reach}/{1}%
	}/{$hA'A'$, $F'F$ }/{decaycell}/{prodcell},
	{Scalar ($S$)}/{%
	{$S H^\dagger H$}/{Fig.~\ref{fig:limitBtoKS}, Fig.~\ref{fig:limitBtoKS_lambda}}/{1}%
	}/{ $S H^\dagger H$, $S^2 H^\dagger H$ }/{decaycell}/{prodcell},
	{HNL ($N$)}/{ %
	{$\tilde{H} \bar{L}N$}/{Fig.~\ref{fig:Ne}}/{1}%
	}/{ $\tilde{H} \bar{L}N$ }/{decaycell}/{prodcell},
	{ALP ($a$)}/{ %
	{}/{Fig.~\ref{fig:alpBC10}, Fig.~\ref{fig:alpBC11}, pending, pending}/{1}%
	}/{ $\partial_\mu a \bar{q} \gamma^\mu \gamma^5 q$, $a \tilde{G} G$, $a \tilde{F} F$,  $a(W\tilde{W} \!-\! B\tilde{B})$}/{nocell}/{uvcell}%
}{
	\node[medcell] at (\x,\y) {\mediator};	
	\xdef\x{\x+\medcellwidth} 
	\foreach [count=\kx] \prod in \prods{	
		\node[\ptype] at (\x,\y) {\prod};	
		\xdef\x{\x+\cellwidth} 
		\xdef\kxx{\kx}
	}
	\xdef\x{\x-\kxx*\cellwidth-\medcellwidth}
	\xdef\y{\y+\scale*\cellheight} 
	\foreach  \decay/\row/\sc in \decayrows{
		\def\scale{\sc}
		\node[\dtype] at (\x,\y) {\decay};	
		\xdef\x{\x+\medcellwidth}
		\foreach [count=\jx] \rowitem in \row{
			\node[figcell] at (\x,\y) {\rowitem};
			\xdef\x{\x+\cellwidth}	
			\xdef\jxx{\jx}
		}
		\xdef\x{\x-\jxx*\cellwidth-\medcellwidth}	
		\xdef\y{\y+\scale*\cellheight} 
	}
	\xdef\x{\x+\kxx*\cellwidth+\medcellwidth-\indentwidth}
}
\xdef\scale{1}
\node[prodcell, scale = 0.4] (prod) at (0.5*\cellwidth, \y-1.5*\cellheight) {};
\draw (prod) node [prodcell, draw = none, fill= none, minimum width=1.5\medcellwidth mm] at +(0.25*\cellwidth, -0.5*\cellheight)  {Production portal};
\node[decaycell, minimum width=\cellwidth mm, scale = 0.4] (dec) at (0.5*\cellwidth, \y-1.0*\cellheight) {};
\draw (dec) node [prodcell, draw = none, fill= none, minimum width=1.5\medcellwidth mm] at +(0.25*\cellwidth, -0.5*\cellheight)  {Decay portal};
\node[uvcell, minimum width=\cellwidth mm, scale = 0.4] (uv) at (0.5*\cellwidth, \y-0.5*\cellheight) {};
\draw (uv) node [prodcell, draw = none, fill= none, minimum width=1.5\medcellwidth mm] at +(0.25*\cellwidth, -0.5*\cellheight)  {UV operator};
\end{tikzpicture}
}}
\caption{Minimal model tabular space formed from either: production (green) and decay (blue) portals, where well-defined by symmetries or suppressions; or, 
UV operators (orange), where either the production and decay portal may involve linear combinations of operators under RG evolution or field redefinitions.
Each table cell corresponds to a minimal model:
cells for which the \CODEXb reach is known refer to the relevant figure in this document;
cells denoted `pending' indicate a model that may be probed by \CODEXb, but no reach projection is presently available.}
\label{fig:minimalmodels}
\end{figure}

\subsubsection{Abelian hidden sector\label{sec:abelianhiggs}}
The Abelian hidden sector model \cite{Schabinger:2005ei,Gopalakrishna:2008dv,Curtin:2014cca} is a simple extension of the Standard Model, consisting of an additional, massive $U(1)$ gauge boson ($A'$) and its 
corresponding Higgs boson ($H'$).
(See e.g.~\cite{Strassler:2006ri,Barbieri:2005ri,Barger:2007im,Morrissey:2009ur,Curtin:2013fra,Bauer:2018onh,Deppisch:2019ldi} for an incomplete list of References containing other models with similar phenomenology.)
The $A'$ and the $H'$ can mix with respectively the SM photon \cite{Holdom:1985ag,Galison:1983pa} and Higgs boson, each of which provide a portal into this new sector.
In the limit where the $H'$ is heavier than the SM Higgs, it effectively decouples from the phenomenology, such that only the operators in \eqref{eq:vectorportal} remain in the low energy effective theory.

The mixing of the $A'$ with the photon through the $F_{\mu\nu}F'^{\mu\nu}$ operator can be rewritten as a (millicharged) coupling of the $A'$ to all charged SM fermions.
In the limit that the $h A'_{\mu}A'^{\mu}$ coupling is negligible (along with higher dimension operators, such as $h F'_{\mu\nu}F'^{\mu\nu}$), 
the mixing with the photon alone can induce both the production and decay
 of the $A'$ in a correlated manner, which has been studied in great detail (see e.g.~\cite{Beacham:2019nyx} and references therein). 
\CODEXb has no sensitivity to this scenario, because the large couplings required for sufficient production cross-sections imply an $A'$ lifetime that is too short for any $A'$s to reach the detector. 
However, the LHCb VELO and various forward detectors are already expected to greatly improve the reach for this scenario~\cite{Aaij:2017rft,Aaij:2019bvg,Ilten:2016tkc,Ilten:2015hya,Gardner:2015wea,Berlin:2018pwi,CERN-SHiP-NOTE-2016-004,Moreno:2013mja,Feng:2017uoz}.

\begin{figure}[t]
	\includegraphics[width=0.9\textwidth]{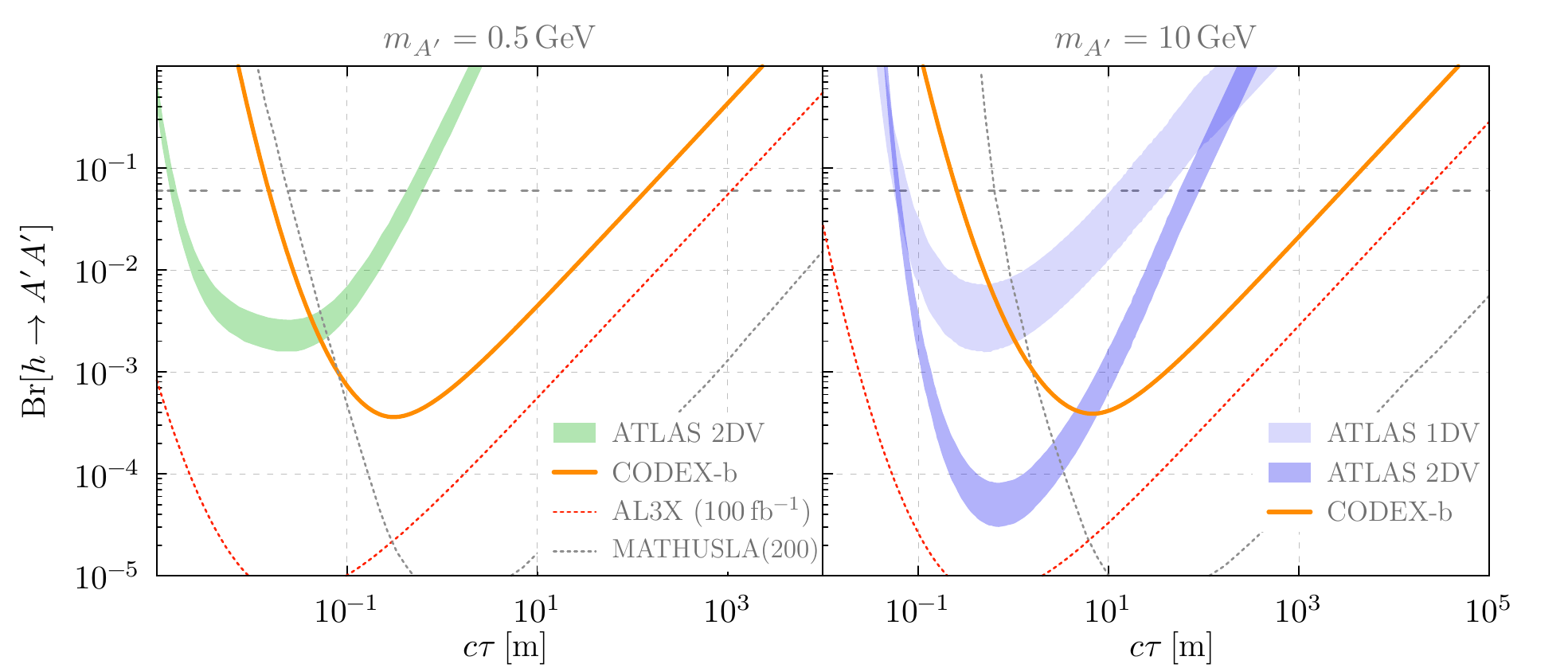}
	\caption{Reach for $h\to A'A'$, as computed in Ref.~\cite{Gligorov:2017nwh}. Shaded bands refer to the optimistic and conservative estimates of the ATLAS sensitivity \cite{ATL-PHYS-PUB-2019-002,Aaboud:2018aqj}
	 for $3\,\text{ab}^{-1}$, as explained in the text. The horizontal dashed line represents the estimated HL-LHC limit on the invisible branching fraction of the Higgs \cite{Cepeda:2019klc}. The MATHUSLA reach is shown for
	 its 200m$\times$200m configuration with 3 ab$^{-1}$; for AL3X $100\text{ fb}^{-1}$ of integrated luminosity was assumed.
	\label{fig:vectorportalHiggsdecay}}
\end{figure}

The $h A'_{\mu}A'^{\mu}$ operator, by contrast, is controlled by the mixing of the $H'$ with the SM Higgs.  This can arise from the kinetic term $\abs{D_\mu H'}^2$, with $\vev{H'}\neq0$ and $H-H'$ mixing. 
This induces the exotic Higgs decay $h\to A'A'$. 
In the limit where the mixing with the photon is small, this becomes the dominant production mode for the $A'$, which then decays through the kinetic mixing portal to SM states. 
\CODEXb would have good sensitivity to this mixing due to its transverse location, with high $\sqrt{\hat{s}}$. 
Importantly, the coupling to the Higgs and the mixing with the photon are independent parameters, so that the lifetime of the $A'$ and the $h\to A'A'$ branching ratio are themselves independent, 
and therefore convenient variables to parameterize the model. 
Fig.~\ref{fig:vectorportalHiggsdecay} shows the reach of \CODEXb for two different values of the $A'$ mass, as done in Ref.~\cite{Gligorov:2017nwh} (see commentary therein), as well as the reach of 
AL3X~\cite{Gligorov:2018vkc} and MATHUSLA~\cite{Chou:2016lxi}.

For ATLAS and CMS, the muon spectrometers have the largest fiducial acceptance as well as the most shielding, thanks to the hadronic calorimeters. 
The projected ATLAS reach for 3 $\text{ab}^{-1}$ was taken from Ref.~\cite{ATL-PHYS-PUB-2019-002} for the low mass benchmark. 
In Ref.~\cite{Aaboud:2018aqj} searches for one displaced vertex (1DV) and two displaced vertices (2DV) were performed with $36.1\text{fb}^{-1}$ of 13 TeV data. 
We use these results to extrapolate the reach of ATLAS for the high mass benchmark to the full HL-LHC dataset, 
where the widths of the bands corresponds to a range between 
a `conservative' and `optimistic' extrapolation for each of the 1DV and 2DV searches. 
Concretely, the 1DV search in Ref.~\cite{Aaboud:2018aqj} is currently background limited, with comparable systematic and statistical uncertainties. 
For our optimistic 1DV extrapolation we assume that the background scales linearly with the luminosity and that the systematic uncertainties can be made negligible with further analysis improvements. 
This corresponds to a rescaling of the current expected limit with $\sqrt{36.1\, \text{fb}^{-1}/3000\,\text{fb}^{-1}}$. 
For our conservative 1DV extrapolation we assume the systematic uncertainties remain the same, with negligible statistical uncertainties. 
This corresponds to an improvement of the current expected limit with roughly a factor of $\sim 2$. 
The 2DV search in Ref.~\cite{Aaboud:2018aqj} currently has an expected background of 0.027 events, which implies $\sim 3$ expected background events, if the background is assumed to scale linearly with the luminosity. 
For our optimistic 2DV extrapolation we assume the search remains background free, which corresponds to a rescaling of the current expected limits with ${36.1\, \text{fb}^{-1}/3000\,\text{fb}^{-1}}$. 
For the conservative 2DV extrapolation we assume 10 expected and observed background events, leading to a slightly weaker limit than with the background free assumption.

Upon rescaling $c\tau$ to account for difference in boost distributions, the maximum \mbox{\CODEXb} reach is largely insensitive to the mass of the $A'$, modulo minor differences in reconstruction 
efficiency for highly boosted particles (See~Sec.~\ref{sec:tracking}).
This is not the case for ATLAS and CMS, where higher masses generate more activity in the muon spectrometer, which helps greatly with reducing the SM backgrounds.

\subsubsection{Scalar-Higgs portal\label{sec:higgsmixing}}
The most minimal extension of the SM consists of adding a single, real scalar degree of freedom ($S$). Gauge invariance restricts the Lagrangian to
\begin{equation}\label{eq:minimalHiggs}
	\mathcal{L}\supset  A_S\, S H^\dagger H + \frac{\lambda}{2}\, S^2 H^\dagger H + \cdots
\end{equation}
where the ellipsis denotes  higher dimensional operators, assumed to be suppressed.
This minimal model is often referred to as simply the ``Higgs portal'' in the literature, though the precise meaning of the latter can vary depending on the context.
LHCb has already been shown to have sensitivity to this model \cite{Aaij:2016qsm,Aaij:2015tna}, and \CODEXb would greatly extend its sensitivity into the small coupling/long lifetime regime.

\begin{figure}[b]
	\begin{subfigure}[b]{0.3\textwidth}
        \centering
       \includegraphics[width=4.5cm]{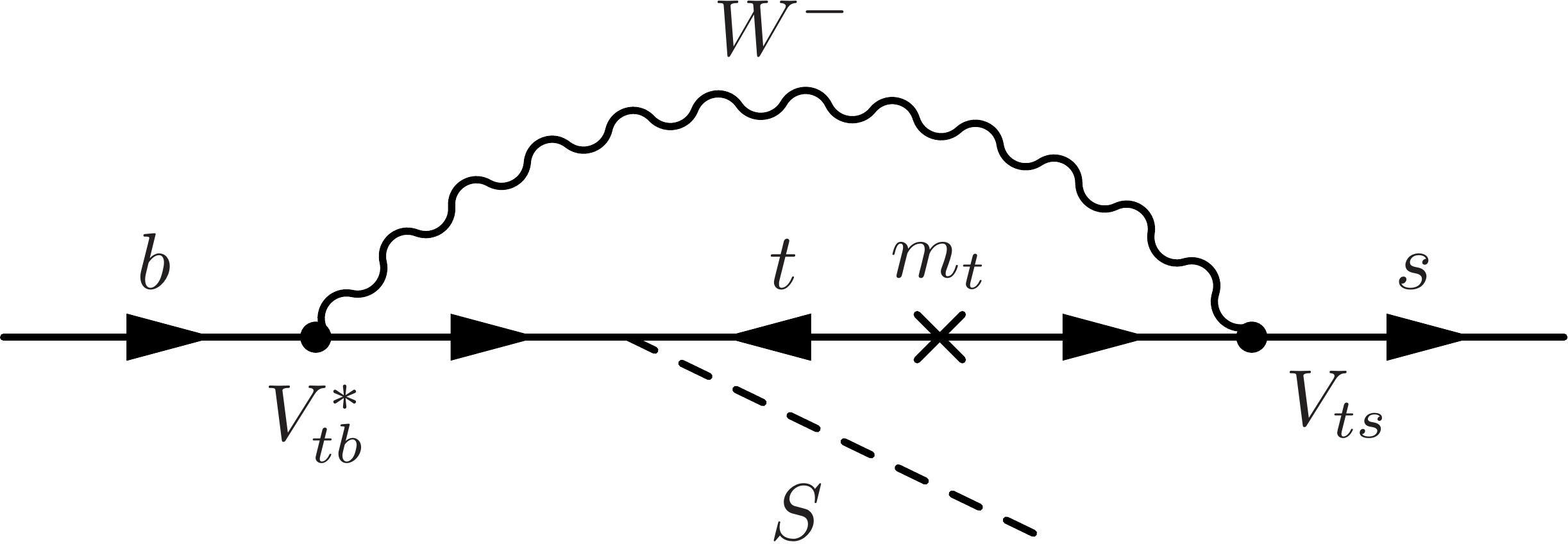}\vspace{0.92cm}
        \caption{\label{fig:diagramBtoKS}}
        \end{subfigure}
	\hfill
	\begin{subfigure}[b]{0.3\textwidth}
        \centering
       \includegraphics[width=4.5cm]{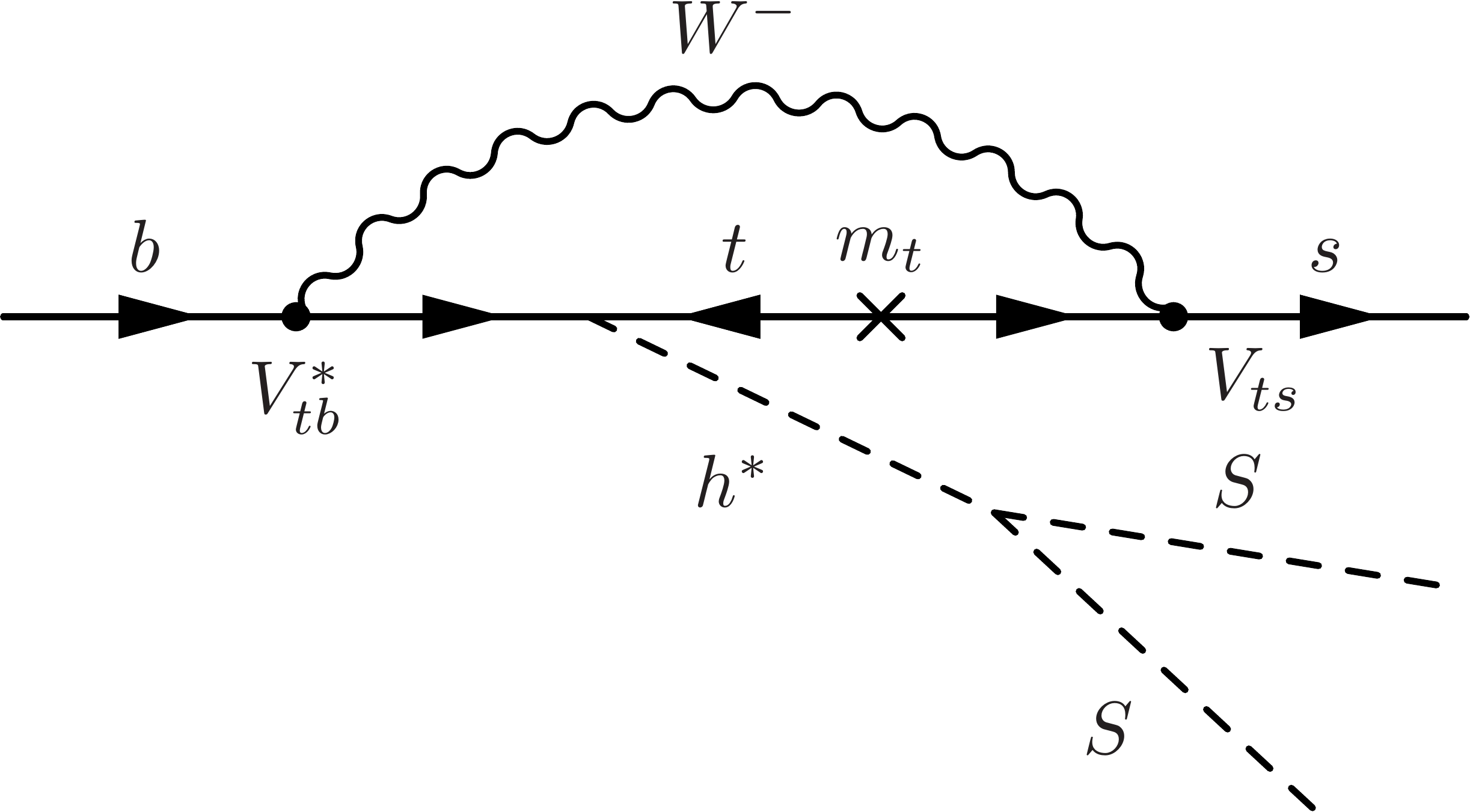}
        \caption{\label{fig:diagramBtoKSS}}
        \end{subfigure}
	\hfill
	\begin{subfigure}[b]{0.3\textwidth}
        \centering
       \includegraphics[height=2.5cm,width=4cm]{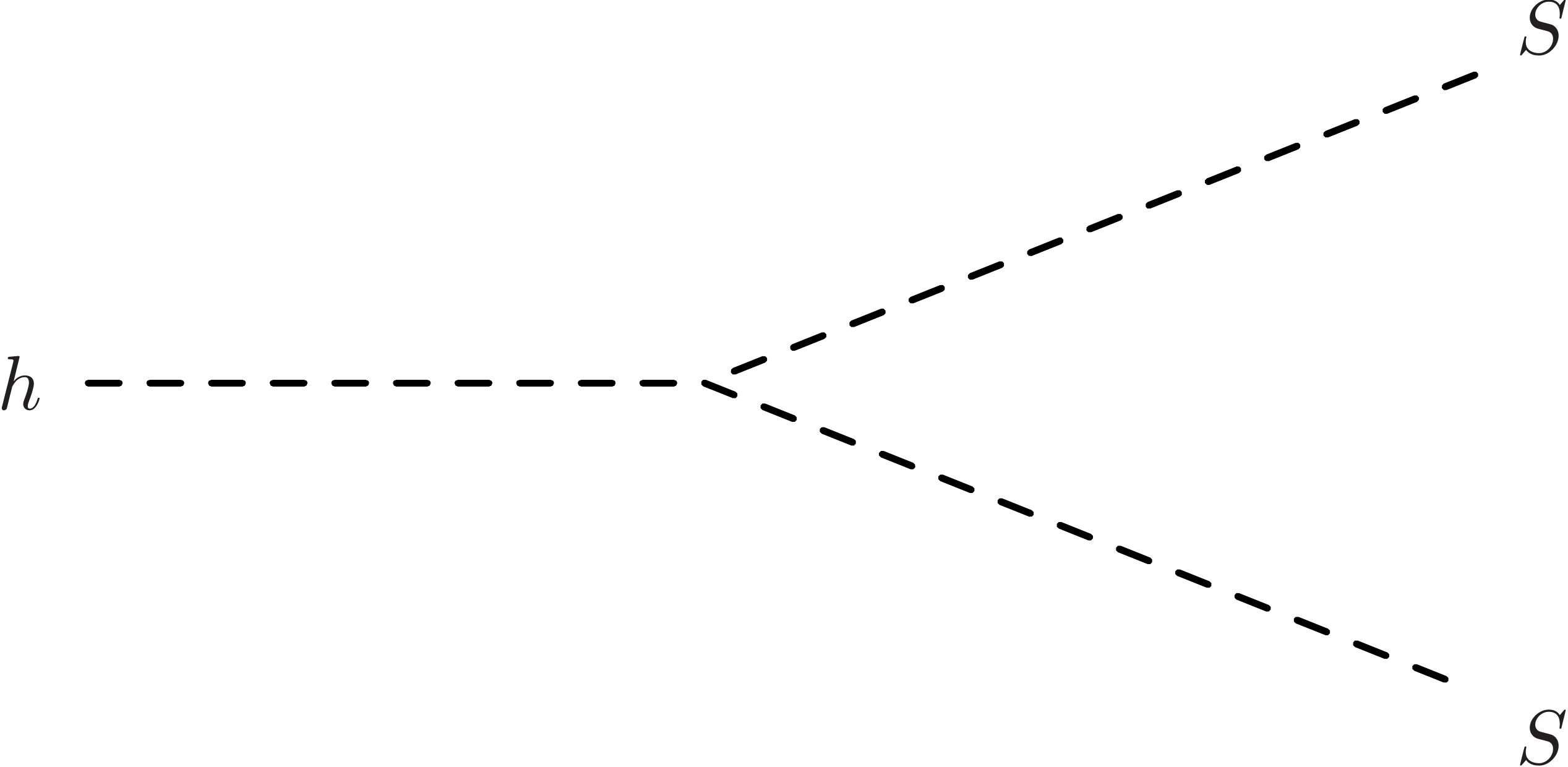}\vspace{0.25cm}
        \caption{\label{fig:diagramhtoSS}}
        \end{subfigure}
	\caption{Diagrams responsible for $S$ production in a minimal extended Higgs sector. (a) is proportional to the mixing between $S$ and Higgs, $\sin^2 \theta$, while (b) and (c) are proportional to the
	square of the quartic coupling, $\lambda^2$.
	\label{fig:Higgsportaldiagrams}}
\end{figure}

The parameter $A_S$ can be exchanged for the mixing angle, $\sin\theta$, of the $S$ with the physical Higgs boson eigenstate.
In the mass eigenbasis, the new light scalar therefore inherits all the couplings of the SM model Higgs: Mass hierarchical couplings with all the SM fermions, as well couplings to photons and gluons at one loop.
All such couplings are suppressed by the small parameter $\sin\theta$.
The couplings induced by Higgs mixing are responsible not only for the decay of $S$~\cite{Donoghue:1990xh,Bezrukov:2009yw,Fradette:2017sdd,Winkler:2018qyg,Bezrukov:2018yvd,Bezrukov:2009yw},
but also contribute to its production cross-section.
Concretely, for $m_K < m_S < m_B$, the dominant production mode is via the $b \to s$ penguin in Fig.~\ref{fig:diagramBtoKS} \cite{Willey:1982dk,Chivukula:1988lo,Grinstein:1988yu},
because $S$ couples most strongly to the virtual top quark in the loop.
If the quartic coupling $\lambda$ is non-zero, the rate is supplemented by a penguin with an off-shell Higgs boson, shown in Fig.~\ref{fig:diagramBtoKSS}~\cite{Batell:2009jf}, as well as direct Higgs decays,
shown in Fig.~\ref{fig:diagramhtoSS}.

\begin{figure}[b]\centering
	\begin{subfigure}[t]{0.49\textwidth}
        \centering
       \includegraphics[width=\textwidth]{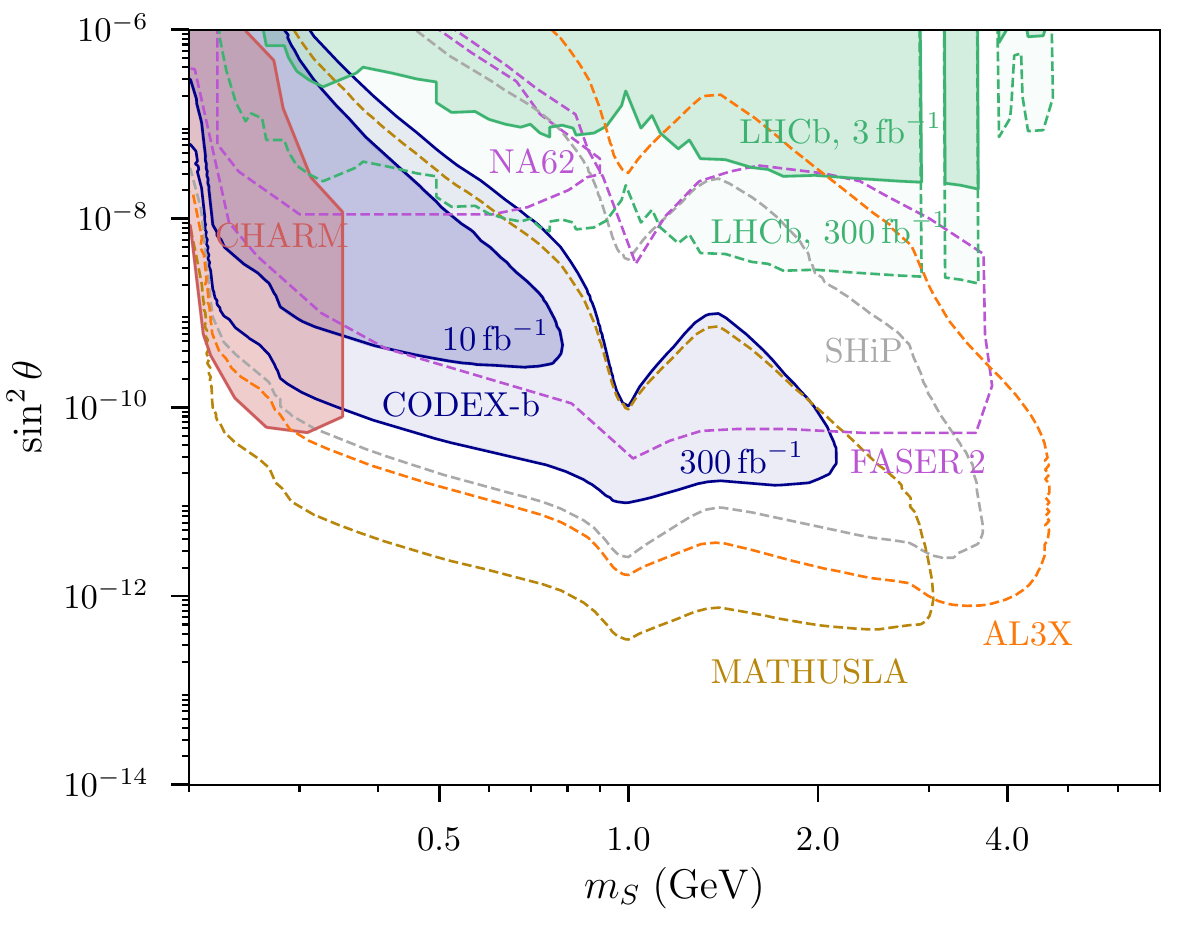}
        \caption{$\lambda=0$\label{fig:limitBtoKS}}
        \end{subfigure}
        \hfill
	\begin{subfigure}[t]{0.49\textwidth}
        \centering
       \includegraphics[width=\textwidth]{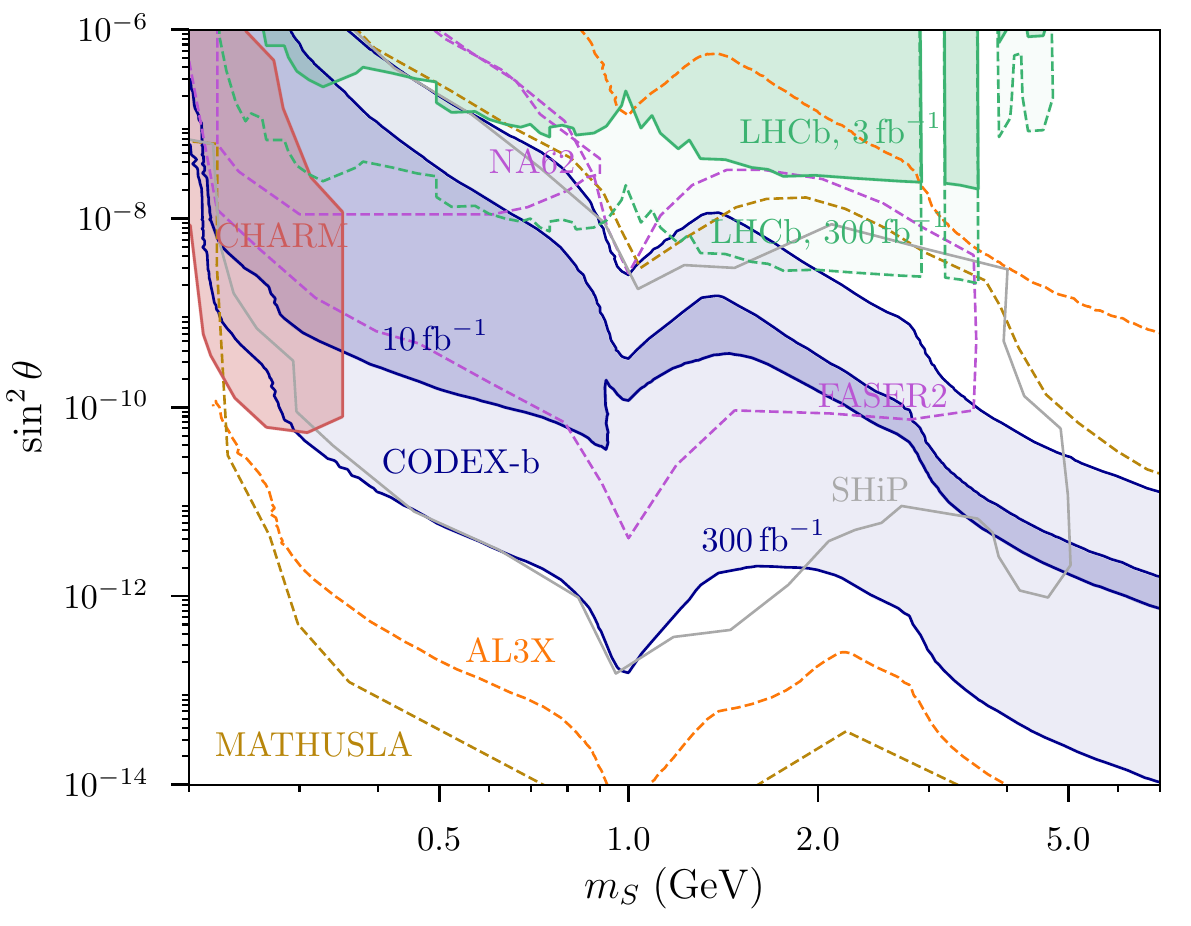}
        \caption{$\lambda=1.6\times 10^{-3}$\label{fig:limitBtoKS_lambda}}
        \end{subfigure}
	\caption{Reach of \CODEXb as a function of $\sin\theta$, for two representative values of $\lambda$. The MATHUSLA reach is shown for its 200m$\times$200m configuration for 3 ab$^{-1}$; 
	for AL3X $100\text{ fb}^{-1}$ of integrated luminosity was assumed.
	\label{fig:Higgsportalreach}}
\end{figure}

In Fig.~\ref{fig:Higgsportalreach} we show the reach of \CODEXb taking two choices of $\lambda$, following \cite{Beacham:2019nyx}:
i) $\lambda=0$, corresponding to the most conservative scenario, in which the production rate is smallest;
ii) $\lambda=1.6\times 10^{-3}$ was chosen such that the $\text{Br}[h\to SS]=0.01$.\footnote{In the specific context of this minimal model, this size of quartic implies that $m_S$ is rather severely fine-tuned for 
$m_S\lesssim10$ GeV.}
The latter roughly corresponds to the future reach for the branching ratio of the Higgs to invisible states.
In this sense it is the most optimistic scenario that would not be probed already by ATLAS and CMS.
The reach for other choices of $\lambda$ therefore interpolates between Fig.~\ref{fig:diagramBtoKS} and Fig.~\ref{fig:diagramBtoKSS}.
Also shown are the limits from LHCb~\cite{Aaij:2016qsm,Aaij:2015tna} and CHARM~\cite{BERGSMA1985458}, and projections for MATHUSLA~\cite{Evans:2017lvd}, FASER2~\cite{Feng:2017vli}, 
SHiP~\cite{Lanfranchi:2243034}, AL3X~\cite{Gligorov:2018vkc} and LHCb, where for the latter we extrapolated the limits from~\cite{Aaij:2016qsm,Aaij:2015tna}, assuming (optimistically) that the large 
lifetime signal region remains background free with the HL-LHC dataset.

\begin{figure}[t]
        \centering
       \includegraphics[width=0.6\textwidth]{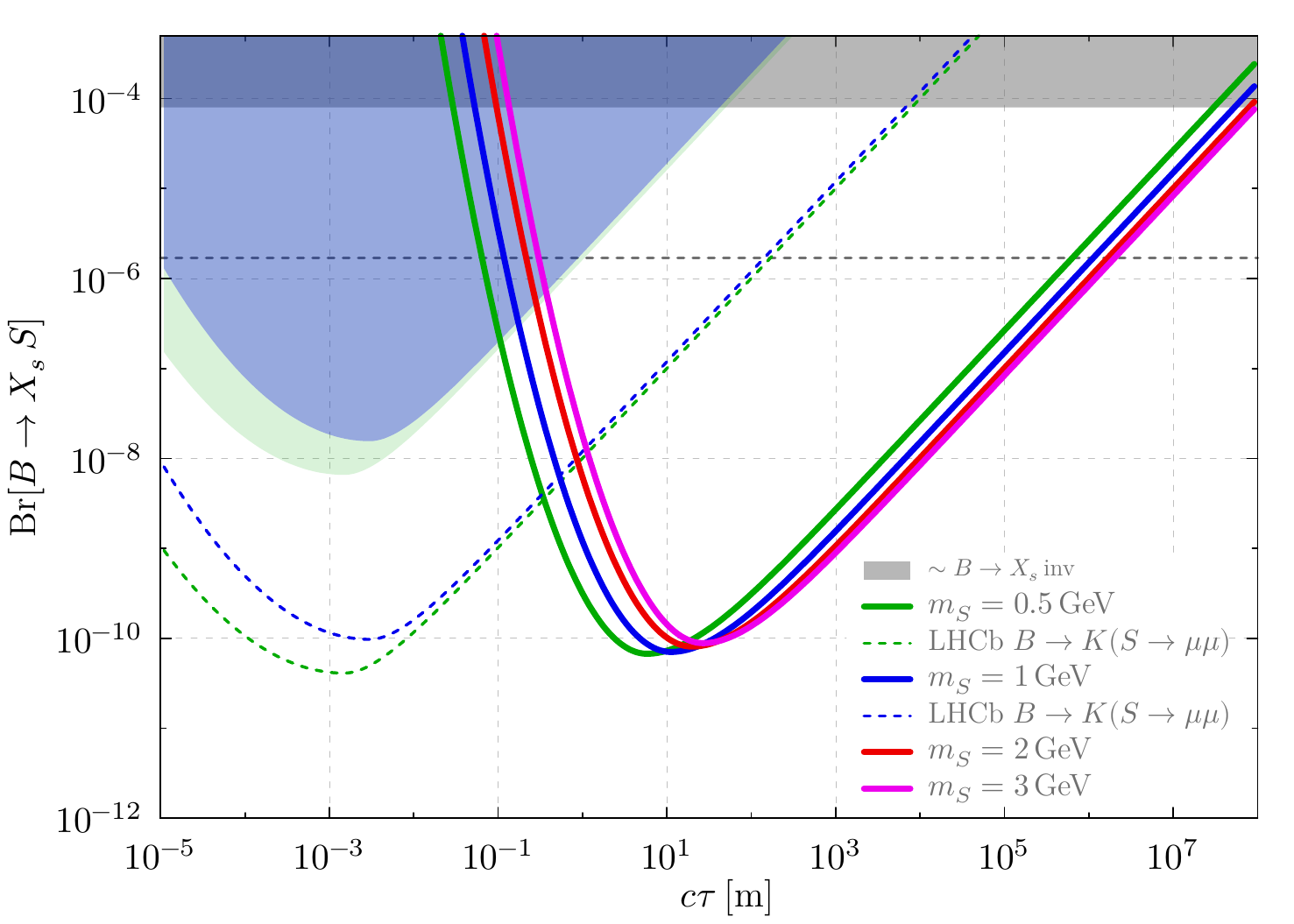}
	\caption{Reach of \CODEXb for LLPs produced in $B$-meson decays, in a non-minimal model. Also shown is the current (shaded) and projected (dashed) reach for: LHCb via $B \to K^{(*)}(S \to \mu\mu)$,
	 for $m_S = 0.5$\,GeV (green) and $m_S = 1$\,GeV (blue), assuming a muon branching ratio of 30\%  and 10\%, respectively (see Ref.~\cite{Gligorov:2017nwh}); and for $B \to K^{(*)}+\text{inv.}$ (gray).
	\label{fig:nonminmalBfull}}
\end{figure}

The scalar-Higgs portal is, by virtue of its minimality, very constraining as a model.
When studying LLPs produced in $B$ decays, it is therefore worthwhile to relax its assumptions, in particular relaxing the full correlation between the lifetime
and the production rate -- the $b \to s S$ branching ratio -- as is the case in a number of non-Minimally Flavor Violating (MFV) models (See e.g.~\cite{Batell:2017kty,Batell:2018fqo,Elor:2018twp,Nelson:2019fln}).
Fig.~\ref{fig:nonminmalBfull} shows the \CODEXb reach in the $b \to s S$ branching ratio for a number of benchmark LLP mass points, as done in Ref.~\cite{Gligorov:2017nwh}.
The LHCb reach and exclusions are taken and extrapolated from Refs.~\cite{Aaij:2016qsm,Aaij:2015tna},
assuming 30\% (10\%) branching ratio of $S\to\mu\mu$ for the 0.5 GeV (1 GeV) benchmark (see Ref.~\cite{Gligorov:2017nwh}).
Also shown are the current and projected limits for $B\to K^{(*)}\nu\nu$ \cite{Tanabashi:2018oca,Kou:2018nap}.
A crucial difference compared to LHCb is that the \CODEXb reach depends only on the total branching ratio to charged tracks, rather than on the branching ratio to muons.

Interestingly, the \CODEXbeta detector proposed for Run~3 (see Sec.~\ref{sec:demonstrator}) may already have novel sensitivity to the $b \to s S$ branching ratio, as shown in Fig.~\ref{fig:nonminmalBdemo}.
This reach is estimated under the requirement that the number of tracks in the final state is at least four, in order to control relevant backgrounds (see Sec.~\ref{sec:backgrounds}).
A more detailed discussion of this reach is reserved for Sec.~\ref{sec:demonstrator}.

\subsubsection{Axion-like particles}
\label{sec:alps}

Axion-like particles (ALPs) are pseudoscalar particles coupled to the SM through dimension-5 operators. 
They arise in a variety of BSM models and when associated with the breaking of approximate Peccei-Quinn-like symmetries they tend to be light. 
Furthermore, their (highly) suppressed dimension-5 couplings naturally renders them excellent candidates for LLP searches. The Lagrangian for an ALP, $a$, can be parameterized 
 as~\cite{Bauer:2017ris}
\begin{align}\label{eq:alp}
{\cal L} &\supset \frac{1}{2}\left(\partial_\mu a\right)^2 -\frac{1}{2}m_a^2 a^2 + \frac{c_q^{ij}}{2\Lambda}(\partial_\mu a)\bar q_i \gamma^\mu \gamma^5 q_j  + \frac{c_\ell^{ij}}{2\Lambda}(\partial_\mu a)\bar \ell_i \gamma^\mu \gamma^5 \ell_j \nn \\ 
&  \quad + \frac{4 \pi \alpha_s c_G}{\Lambda}a G_{\mu\nu}^a\, \tilde G^{a,\mu\nu}+ \frac{4 \pi \alpha_2 c_W}{\Lambda}a \,W_{\mu\nu}^a \tilde W^{a,\mu\nu} + \frac{4 \pi \alpha_1 c_B}{\Lambda}a \,B_{\mu\nu} \tilde B^{\mu\nu} + \dots
\end{align}
where $\tilde{G}_{\mu\nu}=1/2\,\epsilon_{\mu\nu\rho\sigma}G^{\rho\sigma}$. 
The couplings to fermions do not have to be aligned in flavor space with the SM Yukawas, leading to interesting flavor violating effects.
The gauge operators mix into the fermionic ones at 1-loop, and therefore in choosing a benchmark model one needs to specify the values of these couplings as a UV boundary condition at a scale $\Lambda$.
In the following we will focus on the same benchmark models chosen in the Physics Beyond Colliders (PBC) community study~\cite{Beacham:2019nyx} 
based on the ALP coupling to photons (``BC9'', defined as $c_W+c_B\neq 0$),
universally to quarks and leptons (``BC10'', $c_q^{ij}=c\, \delta^{ij}$, $c_\ell^{ij}=c\, \delta^{ij}$, $c\neq0$) and to gluons (``BC11'', $c_G\neq 0$).
 Another interesting benchmark to consider is the so-called photophobic ALP~\cite{Craig:2018kne}, in which the ALP only couples to the $SU(2)\times U(1)$ gauge bosons
such that it is decoupled from the photons in the UV and has highly suppressed photon couplings in the IR. 

\CODEXb is expected to have a potentially interesting reach for all these cases. 
This is true with the nominal design provided the ALP has a sizable branching fraction into visible final states, while for ALPs decaying to
photons one would require a calorimeter element, as discussed below in Section~\ref{sec:calo}. 
In this section, we will present the updated reach plots for BC10 and BC11 and leave the ALP with photon couplings (BC9)
and the photophobic case for future study.

ALPs coupled to quark and gluons can be copiously produced at the LHC even though their couplings are suppressed enough to induce macroscopic decay lengths.
They therefore provide an excellent target for LLP experiments such as \CODEXb.
Based on the fragmentation of partons to hadrons in LHC collisions, we can divide the ALP production into four different mechanisms:
\begin{enumerate}
	\item radiation during partonic shower evolution (using the direct ALP couplings to quarks and/or gluons),
	\item production during hadronization of quarks and gluons via mixing with $(J^{PC} =) 0^{-+}$ $\bar q q$ operators (dominated at low ALP masses via mixing with $\pi^0,\eta,\eta'$),
	\item production in hadron decays via mixing with neutral pseudoscalar mesons, and
	\item production in flavor-changing neutral current bottom and strange hadron decays, via loop-induced flavor-violating penguins. 
\end{enumerate}
The last mechanism has been already considered extensively in the literature. The ALP production probability scales parametrically as $(m_t/\Lambda)^2$ and is proportional to the
number of strange or $b$-hadrons produced. In general, the population of ALPs produced by this mechanism is not very boosted at low pseudorapidities. 
For the PBC study, it was the only production mechanism considered for BC10, and it was included in BC11. 

The second and third mechanism are related as they both incorporate how the ALP couples to low energy QCD degrees of freedom. 
Conventionally the problem is rephrased into ALP mixing with
neutral pseudoscalar mesons. This production is parametrically suppressed by $(f_\pi/\Lambda)^2$ and it quickly dies off for ALP masses much above $1$\,GeV. 
The population of ALPs produced by these mechanisms is not very boosted at low pseudorapidities, while the forward experiments will have access to very energetic ALPs. 
Compared to the PBC study, we treat separately the two cases of hadronization and hadron decays as they give rise to populations of ALPs with different energy distributions, 
and include them both in BC10 and BC11. 

Finally, the first mechanism listed above has been so far overlooked in the literature. 
However, emission in the parton shower can be the most important production mechanism at transverse LHC experiments such as \CODEXb.
Emission of (pseudo)scalars is expected to exhibit neither collinear nor soft enhancements, such that 
ALPs emitted in the shower may then carry an $\mathcal{O}(1)$ fraction of the parent energy and can be emitted at large angles. 

For the case of quark-coupled ALPs (BC10), emission in the parton shower is suppressed by the quark mass  -- a consequence of the soft pion theorem --  i.e. by $m_q^2/\Lambda^2$ 
(or by loop factors to the induced gluon coupling, as below).
The shower contribution may nevertheless still dominate at high ALP masses, where the other production mechanisms are forbidden by phase space or kinematically suppressed.
For gluon-coupled ALPs (BC11), however, no such suppression arises in the shower.
In a parton shower approximation, the ALP emission is attributed to a single parton with a given probability:
While interference terms between ALP emissions from adjacent legs  -- \emph{e.g.} in $g\rightarrow g g a$ -- cannot be neglected,
such an approximation 
still captures the bulk of the production, even when the ALP is not emitted in the soft limit.

The parton shower approximation greatly simplifies the description of ALP emission, allowing the implementation in existing Monte Carlo tools.
In this approximation, the probability for a parton to fragment into an ALP scales parametrically as $Q^2/\Lambda^2$, with $Q$ of the order of the virtuality of the parent parton. 
For example, the $g \to g a$ splitting function
\begin{equation}\label{eq:ps}
	P_{g\rightarrow ag}(t,z) = \pi \alpha_s c_G^2 \frac{t}{\Lambda^2} \left(1 - \frac{m_a^2}{t}\right)^2
\end{equation}
where $t = Q^2$. 
While the population of partons with large energies is much smaller than the final number of hadrons, 
the production rate is enhanced by a large $\mathcal{O}(Q^2/f_\pi^2)$ factor, compared to the second and third mechanisms. 
In LHC collisions this is sufficient to produce a large population of energetic ALPs at low pseudorapidities, with boosts exceeding $10^3$ for ALP masses in the $0.1$--$1$\,GeV range. 
The \CODEXb reach can therefore be extended to higher ALP masses and larger couplings compared to previous estimates if very collimated LLP decays can be detected.

We estimate these production mechanisms using \texttt{Pythia 8}, with the code modified to account for the production of ALPs during hadronization. 
We include ALP production in decays by extending its
decay table in such a way that for each decay mode containing a $\pi^0,\eta,\eta'$ meson in the final state, we add another entry with the meson substituted by the ALP. 
The branching ratio is rescaled by the ALP mixing factor and phase space differences. 

The ALP production from the shower is computed by navigating through the generated QCD shower history
and for each applicable parton an ALP is generated by re-decaying that parton with a weight: The ratio of the ALP branching (integrated) probability over the total (SM+ALP)
(integrated) probabilities. This is correct for time-like showers in the limit that the ALP branching probability is small, because in this limit the branching scale is still controlled by the SM
Sudakov factor. This procedure is not applicable to space-like showers, without also incorporating information from parton distribution functions. 
Such space-like showers, however, provide only a sub-leading contribution to transverse production, 
i.e.~at the low pseudorapidities for the \CODEXb acceptance, and we therefore neglect them.
For forward experiments at the LHC, such as FASER2, we do not include any shower contribution in the reach estimates, 
since a more complete treatment is required in order to fully estimate ALP
production at high pseudorapidities,
and the effect is expected to be at most $\mathcal{O}(1)$.
For the case of a fermion-coupled ALP we include both the emission from heavy quark lines, proportional to $(m_q c_q/\Lambda)^2$, and from loop-induced coupling to gluons, 
taking $c_G= N_f c_q/32\pi^2$ in Eq.~\eqref{eq:ps} above~\cite{Bauer:2017ris}, where $N_f$ is the number of flavors.
Further details will be given in upcoming work.

\begin{figure}[t]
        \centering
       \includegraphics[width=0.75\textwidth]{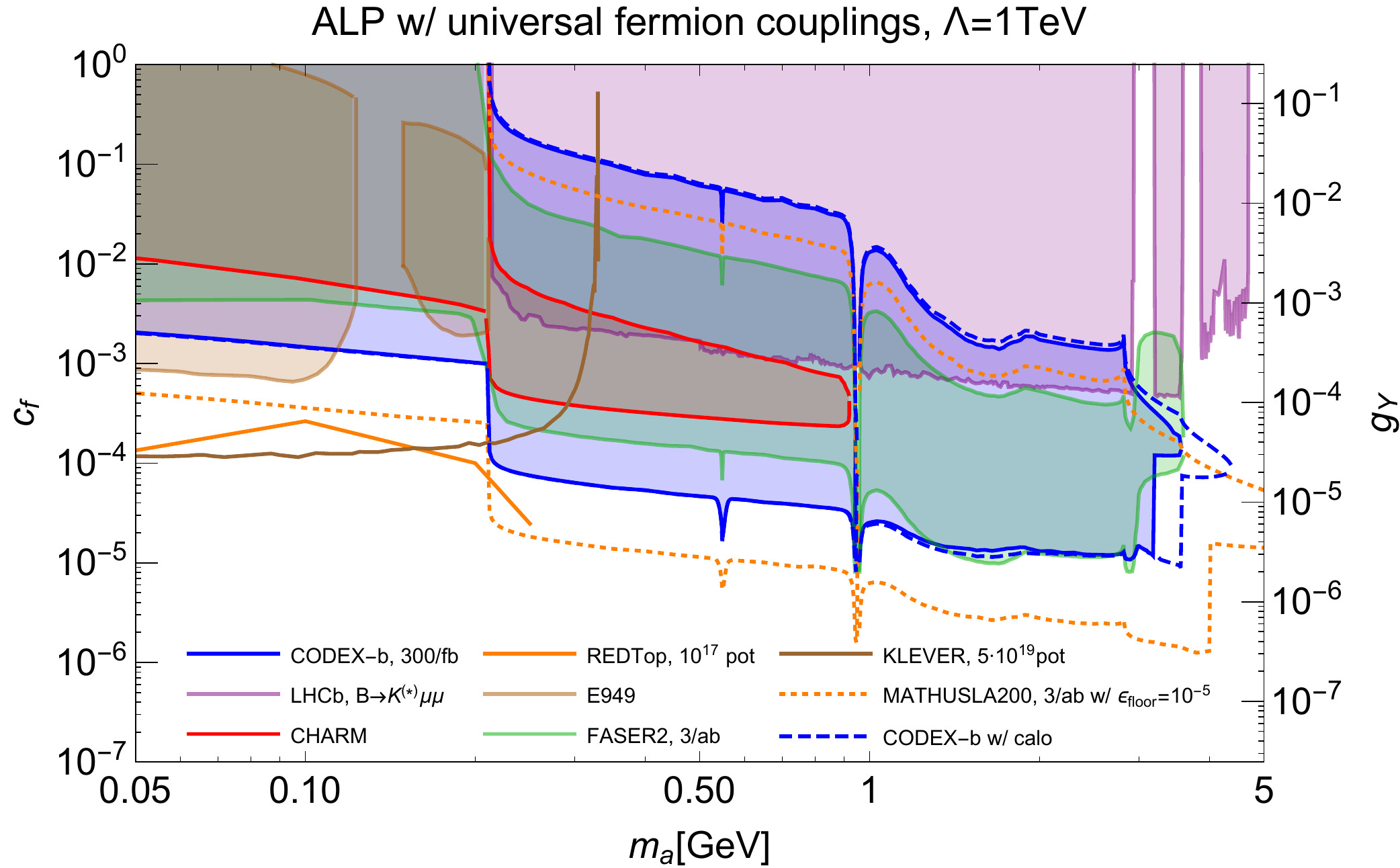}
	\caption{Reach of \CODEXb for fermion-coupled ALPs. The vertical axis on the left corresponds to the couplings defined in Eq.~\eqref{eq:alp}, while the one on the right to the normalization 
	used in the PBC study~\cite{Beacham:2019nyx}.
	 The baseline (tracker only) \CODEXb design is shown as solid, while the gain by a calorimeter option is shown as dashed. 
	 All the curves for the other experiments except MATHUSLA are taken from~\cite{Beacham:2019nyx}, after rescaling with to the 
	 different lifetime/branching ratio calculation used here. The MATHUSLA reach is based on our estimates, see text for details.
	\label{fig:alpBC10}}
\end{figure}

\begin{figure}[t]
        \centering
       \includegraphics[width=0.75\textwidth]{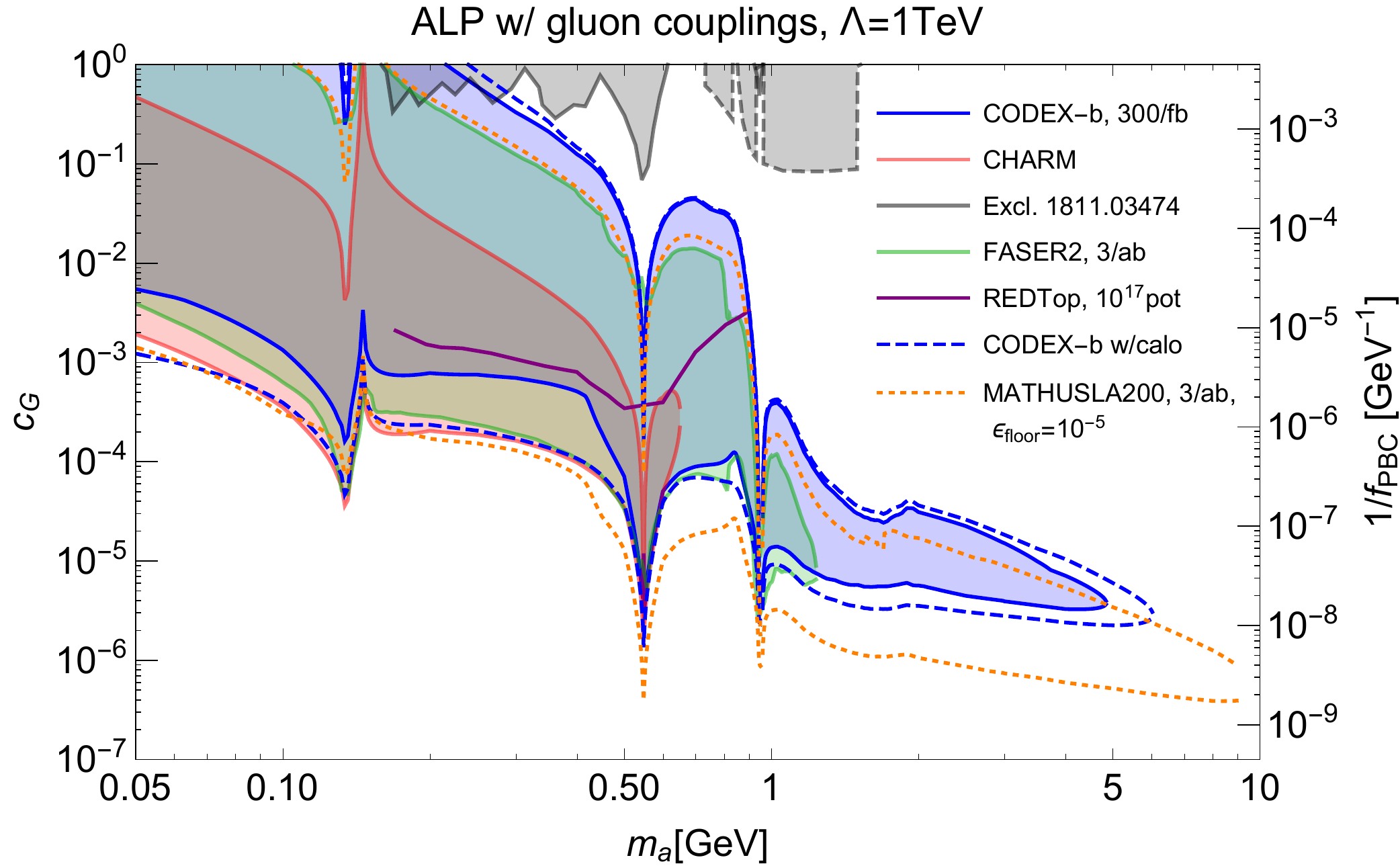}
	\caption{Reach of \CODEXb for gluon-coupled ALPs. The vertical axis on the left corresponds to the couplings defined in Eq.~\eqref{eq:alp}, while the one on the right to the normalization 
	used in the PBC study~\cite{Beacham:2019nyx}. 
	The baseline (tracker only) \CODEXb design is shown as solid, while the gain by a calorimeter option is shown as dashed. 
	See Fig.~\ref{fig:ALPcalo} for further information about how the \CODEXb reach changes with
	 different detector designs. 
	 FASER2 and REDTop curves are taken from~\cite{Beacham:2019nyx}, after rescaling to the different lifetime/branching ratio calculation used here. The CHARM curve has been
	  recomputed with the same assumptions used for the \CODEXb curve. The MATHUSLA reach is based on our estimates, see text for details.
	\label{fig:alpBC11}}
\end{figure}

The reach predictions are shown in Fig.~\ref{fig:alpBC10} for a fermion-coupled ALP (BC10) and in Fig.~\ref{fig:alpBC11} for an ALP coupled to gluons (BC11), 
for the case of the nominal -- i.e. tracker-only -- \CODEXb design. 
In this case, an ALP decaying only to neutral particles such as photons is invisible, and highly boosted ALPs may decay to merged tracks, 
such that the signature resembles more closely a single appearing track inside the detector volume. 
For such a signature, the \CODEXb baseline design is not background-free; we use the background estimates presented in Tab.~\ref{tab:bkg-tracks} (see Sec.~\ref{sec:backgrounds} below),
corresponding to 50 events of background in the entire detector in 300~fb$^{-1}$.
The \CODEXb reach with a calorimeter option (shown here as a dashed line) is further discussed in Sect.~\ref{sec:calo}. 

The MATHUSLA estimates and the CHARM exclusion in Fig.~\ref{fig:alpBC11} have been recomputed, while all the other curves have been taken from~~\cite{Beacham:2019nyx}, after rescaling them to the appropriate 
lifetime and branching ratio expressions used in our plots. For MATHUSLA we used the $200\,{\rm m} \times 200\,{\rm m}$ configuration and assumed that a floor veto for upward going muons entering the decay volume is 
available with a rejection power of $10^5$. Based on the estimates of $10^7$ upward going muons~\cite{Alpigiani:2018fgd, Curtin:2019pc}, we therefore used 100 events as the background for unresolved highly boosted ALPs. 

\begin{figure}[t]
	\begin{subfigure}[t]{0.39\textwidth}
        \centering
       \includegraphics[width=\textwidth]{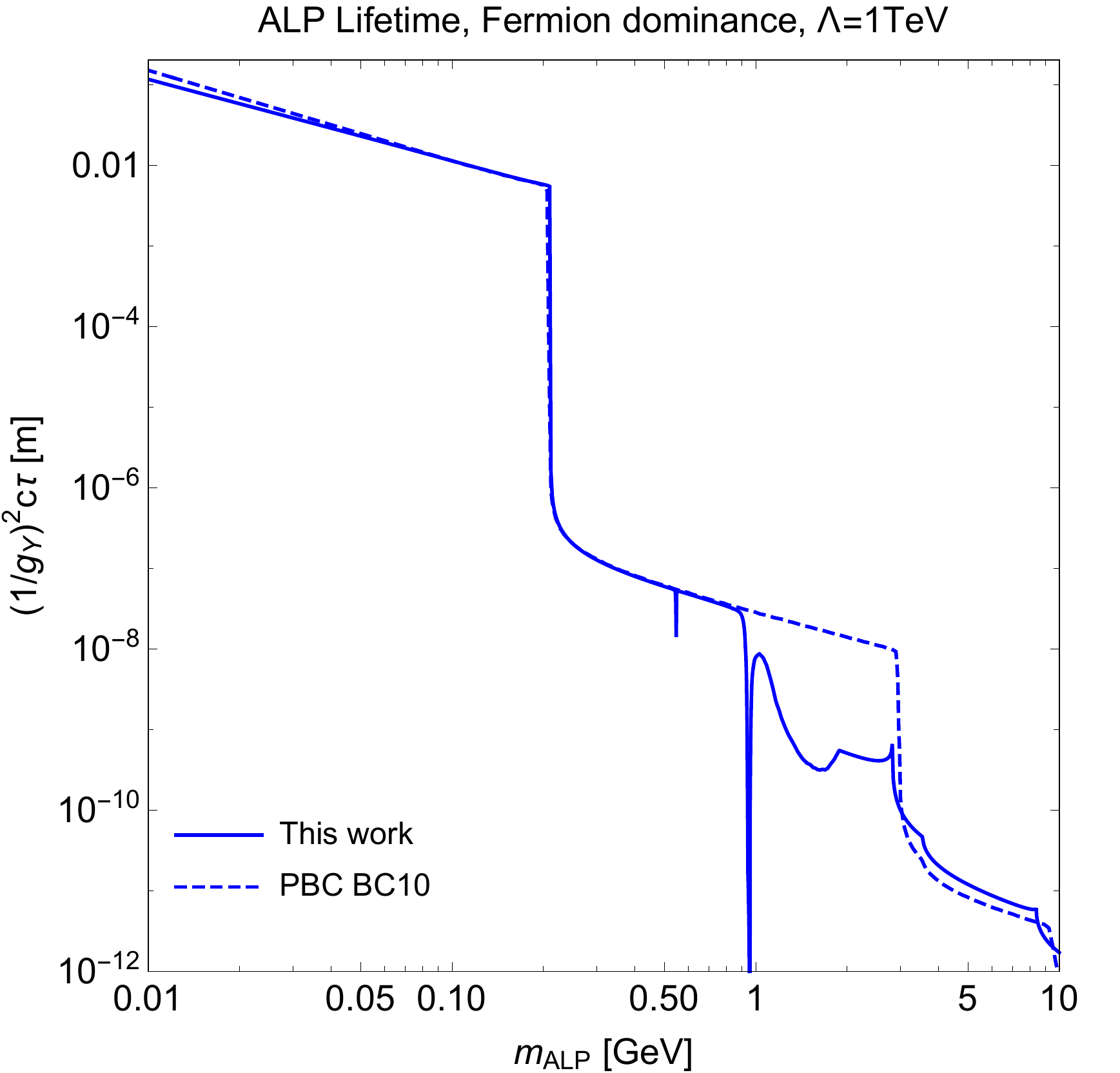}
        \caption{\label{fig:alp-ct}}
        \end{subfigure}
	\hfill
	\begin{subfigure}[t]{0.59\textwidth}
        \centering
       \includegraphics[width=\textwidth]{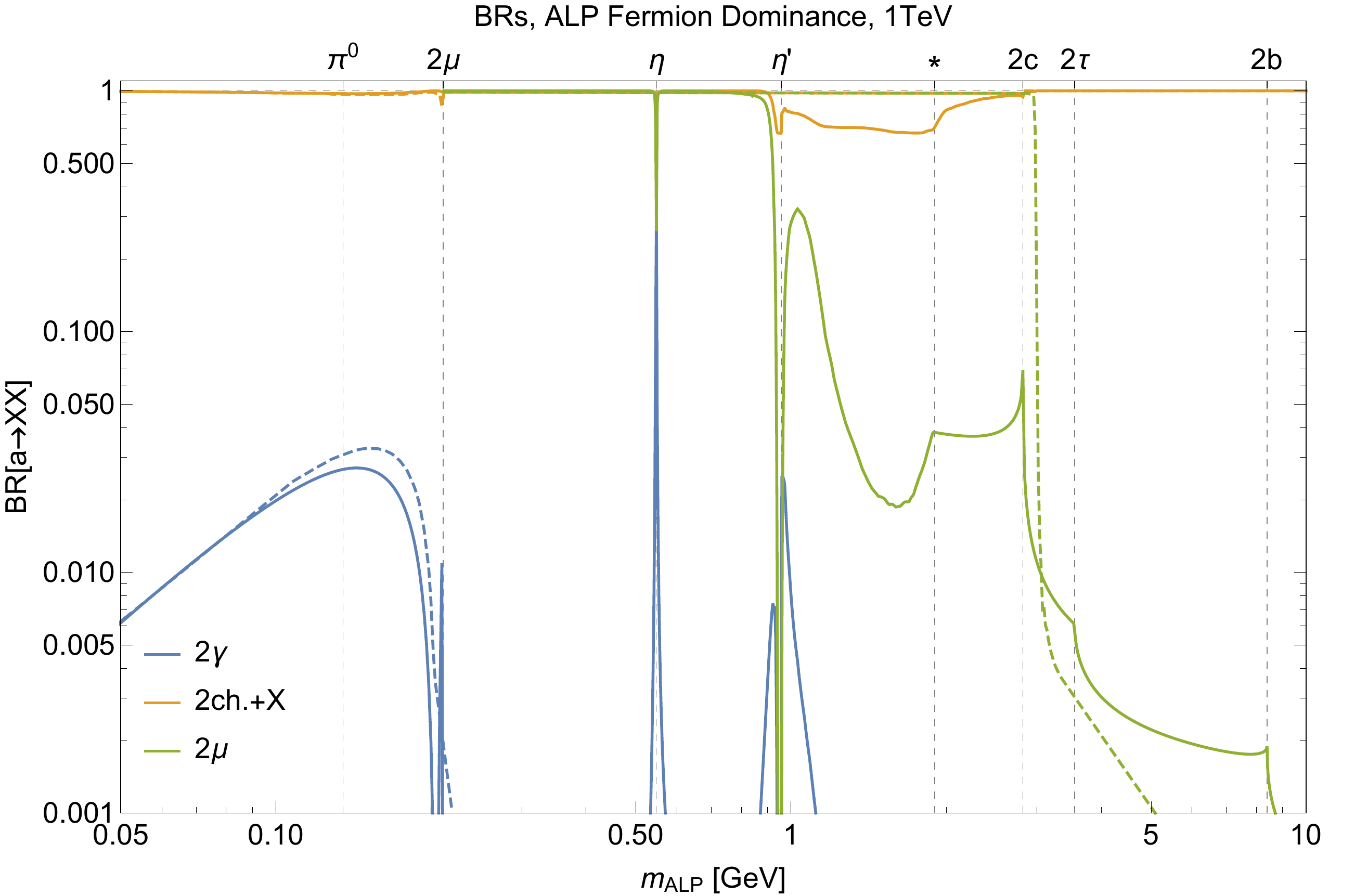}
        \caption{\label{fig:alp-br}}
        \end{subfigure}
        \centering
	\caption{Lifetime (left) and branching ratios (right) for an ALP coupled to fermions, used in Fig.~\ref{fig:alpBC10}. For comparison we plot as dashed lines the corresponding values used for BC10 in the PBC document.
	\label{fig:alp-ctbr}}
\end{figure}

For the case of the fermion-coupled ALP we have further improved the lifetime and branching ratio calculations compared to those used in Refs.~\cite{Beacham:2019nyx,Ariga:2018uku}, 
by including the partial widths of ALP into light QCD degrees of freedom, using the same procedure as in Ref.~\cite{Aloni:2018vki}. 
The result is shown in Fig.~\ref{fig:alp-ctbr}. In particular in the $1\lesssim m_a \lesssim 3$~GeV range, for a given coupling the ALP lifetime is 
$\mathcal{O}(10)$ \emph{smaller} than previously assumed, and the decays are mostly to hadrons instead of muon pairs. 

\subsubsection{Heavy neutral leptons}
\label{sec:HNL}
Heavy neutral leptons (HNLs) may generically interact with the SM sector via the lepton Yukawa portal, mediated by the marginal operator $\bar{L}_i \tilde{H} N$,
or may feature in a range of simplified NP models coupled to the SM via various higher-dimensional operators.
In  the $m_{N} \sim 0.1$--$10$\,GeV regime, that we consider below, these models can be motivated e.g.~by explanations for the neutrino masses \cite{Mohapatra:1986bd}, the $\nu$MSM \cite{Asaka:2005an,Asaka:2005pn}, dark matter theories \cite{Batell:2017cmf}, or by models designed to address various recent semileptonic anomalies \cite{Greljo:2018ogz,Asadi:2018wea,Robinson:2018gza}.

UV completions of SM--HNL operators typically imply an active-sterile mixing $\nu_\ell = U_{\ell j} \nu_j + U_{\ell N} N$, where $\nu_j$ and $N$ are
mass eigenstates, and $U$ is an extension of the PMNS neutrino mixing matrix to incorporate the active-sterile mixings $U_{\ell N}$.
 If $|U_{\ell N}|$ are the dominant couplings of $N$ to the SM and $N$ has negligible partial width to any hidden sector, then the $N$ decay width is electroweak suppressed, 
 scaling as $\Gamma \sim G_{F}^2 |U_{\ell N}|^2 m_N^5$.
Because the mixing $|U_{\ell N}|$ can be very small, $N$ can then become long-lived.
We assume hereafter for the sake of simplicity that $N$ couples predominantly to only a single active neutrino flavor, i.e.
\begin{equation}
	\label{eqn:HNLsingleflavor}
		U_{\ell_i N} \gg U_{\ell_j \not= \ell_iN}\,,
\end{equation}
and refer to $\ell$ as the `valence' lepton.

The width of the HNL can be expressed as
\begin{equation}
	\label{eqn:HNLpw}
	\frac{\Gamma_N}{s \abs{U_{\ell_i N}}^2} = \sum_M (\Gamma_{\nu_i M} +\Gamma_{\ell_i M}) + \sum_j (\Gamma_{\ell_i \ell_j \nu_j} +\Gamma_{\nu_i \ell_j \ell_j}) + \sum_{q,q'} \Gamma_{\ell_i qq'} +\sum_{q} \Gamma_{\nu_i qq} +\Gamma_{\nu_i \nu\nu},
\end{equation}
where $s = 1$ ($s=2$) for a Dirac (Majorana) HNL and the final state $M$ corresponds to a single kinematically allowed (ground-state) meson. Specifically, $M$ considers:
charged pseudoscalars, $\pi^\pm$, $K^\pm$; 
neutral pseudoscalars $\pi^0$, $\eta$, $\eta'$; 
charged vectors, $\rho^\pm$,  $K^{*\pm}$; 
and neutral vectors, $\rho^0$, $\omega$,  $\phi$. 
For $m_N > 1.5$ GeV, we switch from the exclusive meson final states to the inclusive decays widths $\Gamma_{\ell_i qq'}$ and $\Gamma_{\nu_i qq}$, which are disabled below 1.5 GeV. 
Expressions for each of the partial widths may be found in Ref.~\cite{Bondarenko:2018ptm}; 
each is mediated by either the $W$ or $Z$, 
generating long lifetimes for $N$ once one requires $U_{\ell N} \ll 1$. Apart from the $3\nu$, and some fraction of the $\nu M$ and $\nu qq$ (e.g.~$\nu\pi^0\pi^0$) decay modes, 
all the $N$ decays involve two or more tracks, 
so that the decay vertex will be reconstructible in \CODEXb, up to $\mathcal{O}(1)$ reconstruction efficiencies.
We model the branching ratio to multiple tracks by considering the decay products of the particles produced.  Below 1.5 GeV, we consider the decay modes of the meson $M$ to determine the frequency 
of having 2 or more charged tracks; above 1.5 GeV where $\nu qq$ production is considered instead of exclusive single meson modes, we conservatively approximate the frequency 
of having two or more charged tracks as 2/3.

HNLs may be abundantly produced by leveraging the large $b\bar{b}$ and $c\bar{c}$ production cross-section times branching ratios into semileptonic final states. 
In particular, for $0.1\,\text{GeV} \lesssim m_N \lesssim 3\,\text{GeV}$, the dominant production modes are the typically fully inclusive $c \to s \ell N$ and $b \to c \ell N$.
In order to capture mass threshold effects, production from these heavy flavor semileptonic decays is estimated by considering a sum of exclusive modes.
The hadronic form factors are treated as constants: An acceptable estimate for these purposes, as corrections are expected to be small, $\sim \Lambda_{\text{qcd}}/m_{c,b}$.
In certain kinematic regimes, the on-shell (Drell-Yan) $W^{(*)} \to \ell N$ or $Z^{(*)} \to \nu N$ channels can become important, as can the two-body $D_s \to \ell N$ and $B_c \to \ell N$ 
decays (a prior study in Ref.~\cite{Helo:2018qej} for HNLs at \CODEXb neglected the latter contributions).

In our reach projections, we assume the production cross-section
$\sigma(b\bar{b}) \simeq 500\,\mu\text{b}$ and $\sigma(c\bar{c})$ is taken to be $20$ times larger, based on FONLL estimates~\cite{Cacciari:1998it,Cacciari:2001td}.
The EW production cross-sections used are $\sigma(W \to \ell \nu) \simeq 20$\,nb and $\sum_j\sigma(Z \to \nu_j \nu_j) \simeq 12$\,nb~\cite{Aad:2016naf}.
The $\sigma(D_s)/\sigma(D)$ production fraction is taken to be $10\%$~\cite{Abelev:2012tca,Aad:2015zix}, and we assume a production fraction $\sigma(B_c)/\sigma(B) \simeq 2\times 10^{-3}$~\cite{Aaij:2013noa,Aaij:2013cda}.

In the case of the $\tau$ valence lepton, with $m_N < m_\tau$,
the HNL may be produced not only in association with the $\tau$, but also as its daughter. For example, both $b \to c\tau N$ and $b \to c (\tau \to N e\nu_e) \nu$ are comparable production channels.
When kinematically allowed, we approximate this effect by including for the valence $\tau$ case an additional factor of $1+$BR$(\tau\to N+X)/\left|U_\tau\right|^2$, where BR$(\tau\to N+X)$ is the HNL mass 
dependent BR of the tau into a valence $\tau$ HNL plus anything \cite{Bondarenko:2018ptm}.   HNL production from Drell-Yan $\tau$'s is also included, but typically sub-leading: 
The relevant production cross-section is estimated with MadGraph \cite{Alwall:2014hca} to be $\sigma(\tau_{\text{DY}}) \simeq 37$\,nb.

\begin{figure}[p]
\begin{subfigure}[t]{0.45\textwidth}   \centering
\includegraphics[width=\textwidth]{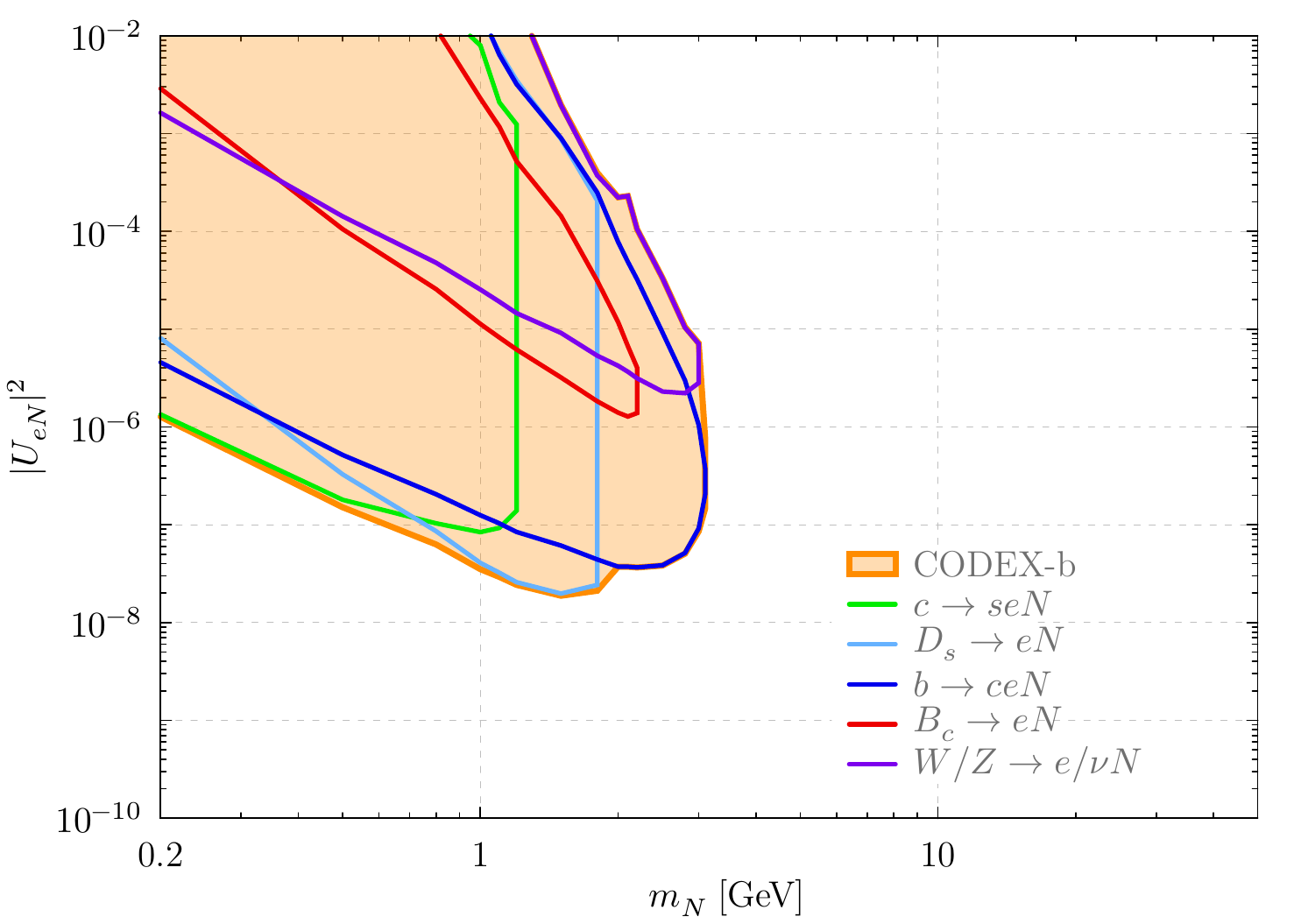}
 \caption{$U_{e N} \gg U_{\mu N},U_{\tau N}$ (Channels)\label{fig:Ne_channels}}
        \end{subfigure}\hfill
\begin{subfigure}[t]{0.45\textwidth}   \centering
\includegraphics[width=\textwidth]{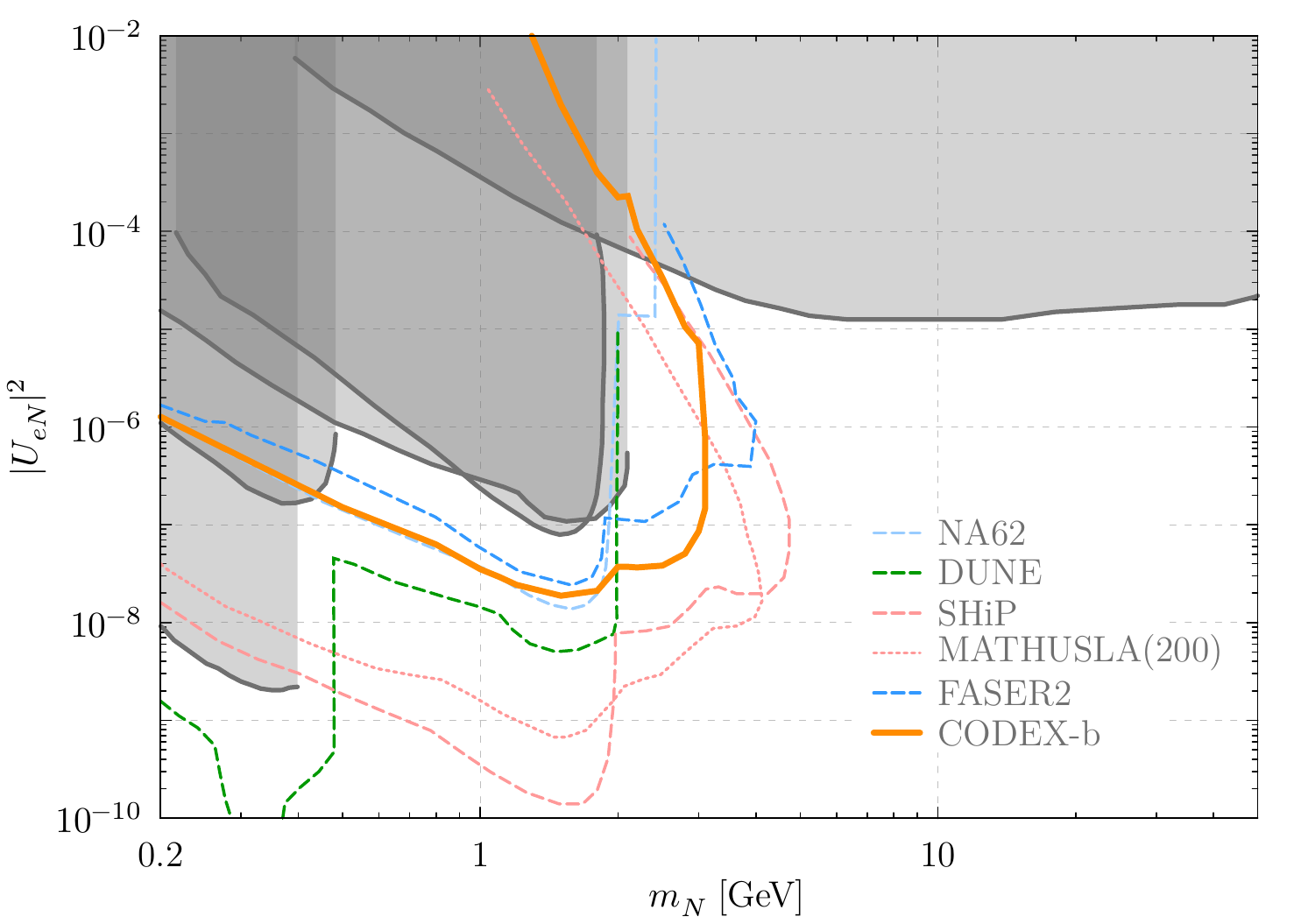}
 \caption{$U_{e N} \gg U_{\mu N},U_{\tau N}$ (Combined)\label{fig:Ne_combined}}
        \end{subfigure}\\\vspace{0.5cm}
%
\begin{subfigure}[t]{0.45\textwidth}   \centering
\includegraphics[width=\textwidth]{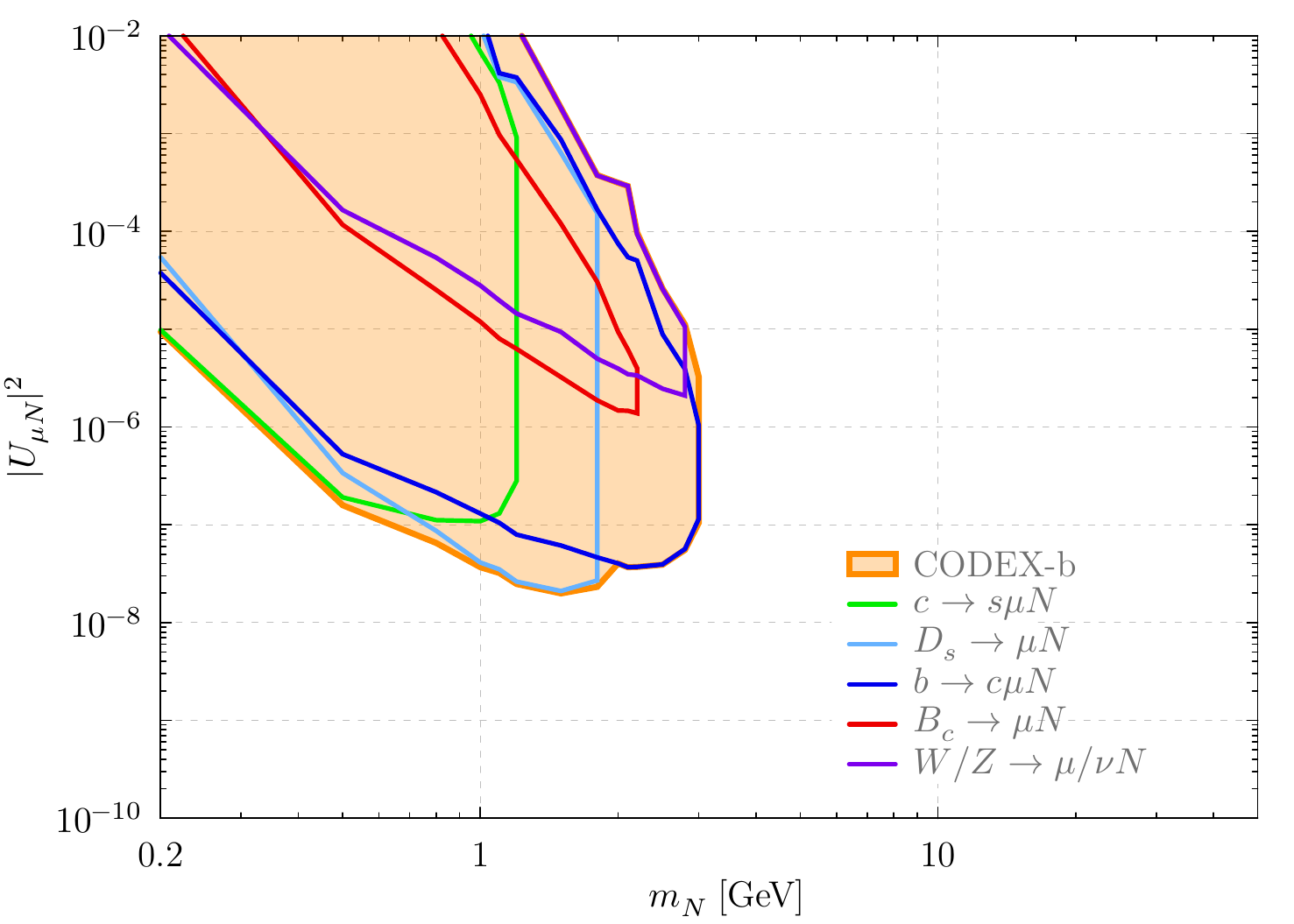}
 \caption{$U_{\mu N} \gg U_{e N},U_{\tau N}$ (Channels)\label{fig:mu_channels}}
        \end{subfigure}\hfill
\begin{subfigure}[t]{0.45\textwidth}   \centering
\includegraphics[width=\textwidth]{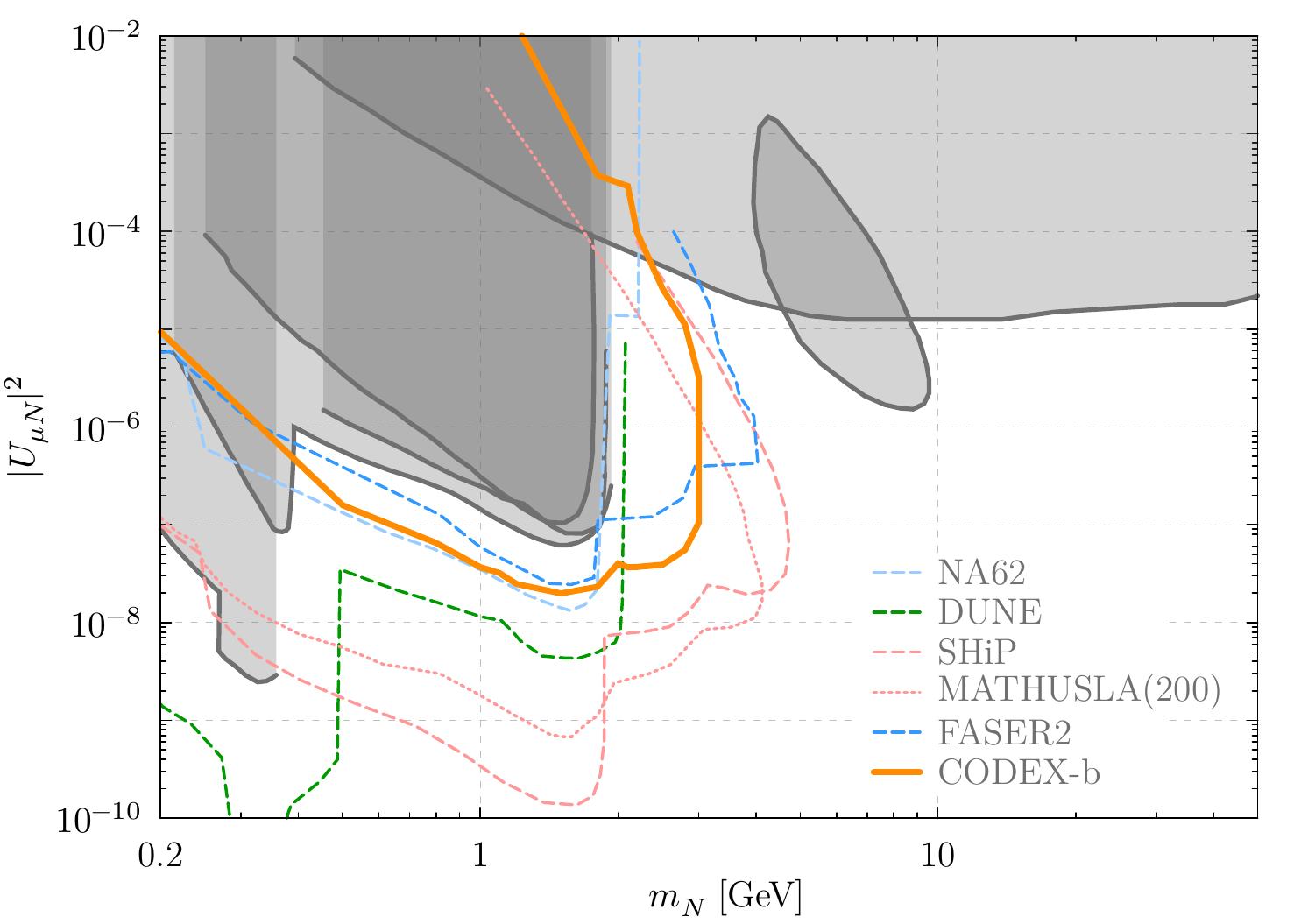}
 \caption{$U_{\mu N} \gg U_{e N},U_{\tau N}$ (Combined)\label{fig:mu_combined}}
        \end{subfigure}\\\vspace{0.5cm}
%
\begin{subfigure}[t]{0.45\textwidth}   \centering
\includegraphics[width=\textwidth]{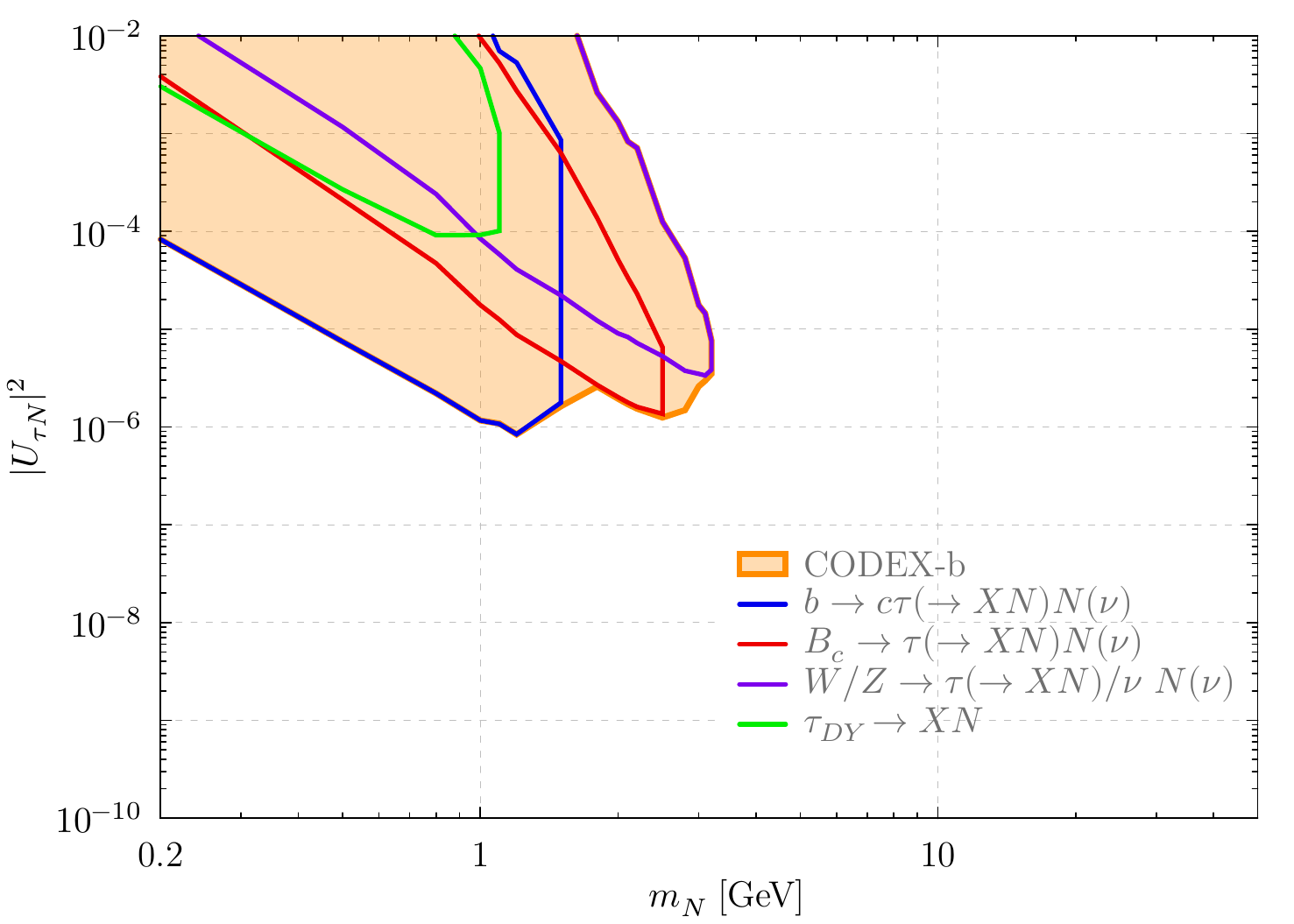}
 \caption{$U_{\tau N} \gg U_{e N},U_{\mu N}$ (Channels)\label{fig:tau_channels}}
        \end{subfigure}\hfill
\begin{subfigure}[t]{0.45\textwidth}   \centering
\includegraphics[width=\textwidth]{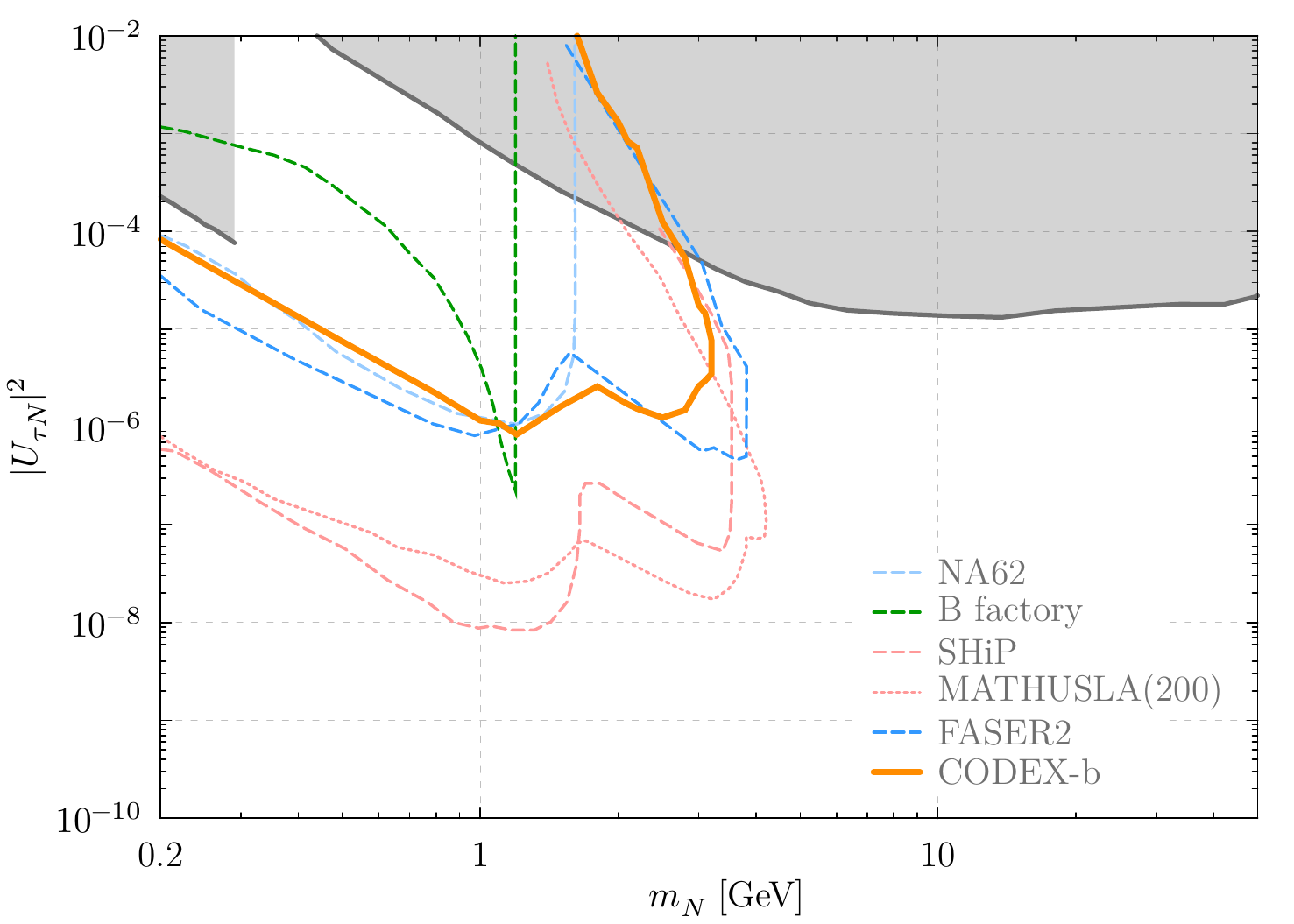}
 \caption{$U_{\tau N} \gg U_{e N},U_{\mu N}$ (Combined)\label{fig:tau_combined}}
        \end{subfigure}
\caption{Projected sensitivity of \CODEXb to Dirac heavy neutral leptons. Left: Contributions from the individual decay channels with the net result (orange).  
Right: Comparison with current constraints (gray) and other proposed experiments, including 
NA62 \cite{Drewes:2018gkc}, DUNE \cite{Adams:2013qkq}, SHiP \cite{Anelli:2015pba}, FASER2 \cite{Kling:2018wct}, and MATHUSLA \cite{Helo:2018qej}, 
which is shown for its 200m$\times$200m configuration. 
ATLAS~\cite{Aad:2019kiz} and CMS~\cite{Sirunyan:2018mtv} constraints on prompt Majorana HNL decays are not shown 
as the sensitivity is currently subdominant to the DELPHI exclusions, and they moreover use a lepton number violating final state only accessible to Majorana HNLs  -- $\mu^{\pm}\mu^\pm e^\mp$ or $e^{\pm}e^\pm\mu^\mp$  -- to place limits.}
\label{fig:Ne}
\end{figure}

The projected sensitivity of \CODEXb to HNLs in the single flavor mixing regime is shown in Fig.~\ref{fig:Ne}.
The breakdown in terms of the individual production modes is shown in the left panels, while the right panels compare \CODEXb sensitivity versus constraints from prior experiments,
including BEBC~\cite{BEBC:1985sf}, PS191~\cite{PS191:1988fl}, CHARM~\cite{CHARMII:1995sf,CHARM:1986as,CHARM:2002de}, JINR~\cite{JINR:1993sf}, and NuTeV~\cite{NuTeV:1999wq}, DELPHI~\cite{DELPHI:1997sf}, and ATLAS~\cite{Aad:2019kiz} (shown collectively by gray regions).
Also included are projected reaches for other current or proposed experiments, including NA62 \cite{Drewes:2018gkc}, DUNE \cite{Adams:2013qkq}, 
SHiP \cite{Anelli:2015pba}, FASER \cite{Kling:2018wct}, and MATHUSLA \cite{Helo:2018qej}. 
We adopt the Dirac convention in all our reach projections; the corresponding reach for the Majorana case is typically almost identical, though relevant exclusions may change.\footnote{With regard to prior measurements,
the PS191~\cite{PS191:1988fl} and CHARM~\cite{CHARM:2002de} measurements
are explicitly quoted for the Dirac HNL case, and the DELPHI measurements~\cite{DELPHI:1997sf} appear also to be implicitly for a Dirac HNL.
Recent ATLAS~\cite{Aad:2019kiz} and CMS~\cite{Sirunyan:2018mtv} measurements are sensitive to prompt Majorana HNLs decays only
via trilepton searches that reject opposite-sign same-flavor lepton pairs that would be produced in the Dirac case;
the lepton number conserving Dirac case is, however, probed via a displaced decay search in Ref.~\cite{Aad:2019kiz}.
The convention of prior CHARM measurements~\cite{CHARMII:1995sf,CHARM:1986as}, as well as for BEBC, JINR and NuTeV \cite{BEBC:1985sf,JINR:1993sf,NuTeV:1999wq} are unclear.}

\FloatBarrier

\subsection{Complete Models}

The LLP search program at the LHC is extensive and rich. In the context of complete models, it has been driven so far primarily by searches for weak scale supersymmetry,
along with searches for dark matter, mechanisms of baryogenesis, and hidden valley models.
In this section, we review the part of the theory space relevant for \CODEXb, which is typically the most difficult to access with the existing experiments.
A comprehensive overview of all known possible signatures is neither feasible nor necessary, the latter thanks to the inclusive setup of \CODEXb.
Instead we restrict ourselves to a few recent and representative examples. For a more comprehensive overview of the theory space we refer to Ref.~\cite{Curtin:2018mvb}.

\subsubsection{R-parity violating supersymmetry}
\label{sec:RPV}
The LHC has placed strong limits on supersymmetric particles in a plethora of different scenarios.
The limits are especially strong if the colored superpartners are within the kinematic range of the collider.
If this is not the case, the limits on the lightest neutralino ($\tilde \chi_1^0$) are remarkably mild, especially if the lightest neutralino is mostly bino-like.
In this case $\tilde \chi_1^0$ can still reside in the $\sim$ GeV mass range, and be arbitrarily separated from the lightest chargino.
Such a light neutralino must be unstable to prevent it from overclosing the universe, which will happen if R-parity is violated~\cite{Bechtle:2015nua}.
The $\tilde \chi^0_1$ then decays through an off-shell sfermion coupling to SM particles through a potentially small R-parity violating coupling.
The combination of these effects typically provide a macroscopic $\tilde \chi^0_1$ proper lifetime.

\begin{figure}
\begin{subfigure}[t]{0.47\textwidth}   \centering
\includegraphics[width=\textwidth]{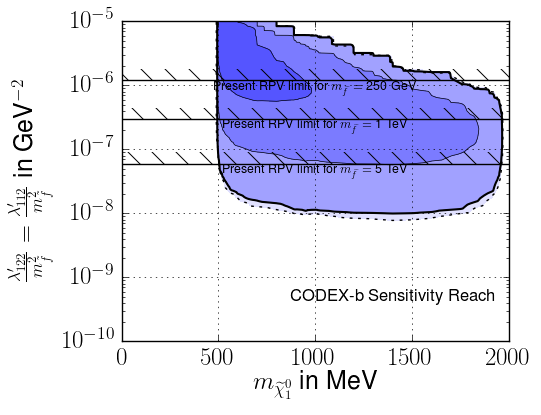}
 \caption{Benchmark 1}
        \end{subfigure}\hfill
\begin{subfigure}[t]{0.45\textwidth}   \centering
\includegraphics[width=\textwidth]{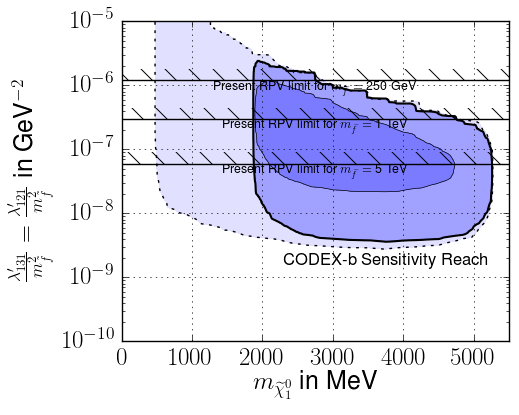}
 \caption{Benchmark 4}
        \end{subfigure}\hfill
\caption{
Projected sensitivity of \CODEXb for light neutralinos with R-parity violating coupling produced in $D$ and $B$ meson decays, reproduced from \cite{Dercks:2018eua} with permission of the authors.
The light blue, blue and dark blue regions enclosed by the solid black lines correspond to $\gtrsim 3$, $3 \times 10^3$ and $3 \times 10^6$ events respectively.
The dashed curve represents the extended sensitivity if one assumes \CODEXb could also detect the neutral decays of the neutralino.
The hashed solid lines indicate the single RPV coupling limit for different values of the sfermion masses. See Ref.~\cite{Dercks:2018eua} for details.}
\label{fig:RPVmesons}
\end{figure}

The sensitivity of  \CODEXb to this scenario was recently studied for $\tilde \chi^0_1$ production through exotic $B$ and $D$ decays~\cite{Dercks:2018eua}, 
as well as from exotic $Z^0$ decays~\cite{Helo:2018qej}.
Dercks~\emph{et~al.}~\cite{Dercks:2018eua} studied the interaction
\begin{equation}
	W_{RPV} = \lambda'_{ijk} L_i Q_j D^c_k
\end{equation}
and considered five benchmarks, corresponding to different choices for the matrix $\lambda'_{ijk}$, each with a different phenomenology.
We reproduce here their results for their benchmarks 1 and 4, and refer the reader to Ref.~\cite{Dercks:2018eua} for the remainder.
The parameter choices, production modes and main decay modes are summarized in Tab.~\ref{tab:RPV}.
The reach of \CODEXb is shown in Fig.~\ref{fig:RPVmesons}. In both benchmarks, \CODEXb would probe more than 2 orders of magnitude in the coupling constants.
For benchmark 4 the reach would be substantially increased if the detector is capable of detecting neutral final states by means of some calorimetry.

\begin{table}[b]
\renewcommand*{\arraystretch}{1.5}
\begin{tabular}{x{0.2}|x{0.2}|x{0.3}|x{0.3}}
\hline
& coupling & production& decay products\\
\hline\hline
benchmark 1 &$ \lambda'_{122},\; \lambda'_{112}$&$D_s^\pm\to\tilde \chi^0_1\; +\;e^\pm $&$\eta,\eta',\phi,K^{0,\pm}$ + $\nu_e,\; e^\mp$\\
benchmark 4 &$ \lambda'_{131},\; \lambda'_{121}$&$B^{0,\pm}\to\tilde \chi^0_1\; + X^{0,\pm} $&$D^\pm, D^{\ast\pm} + e^\mp$\\
\hline
\end{tabular}
\caption{Summary of two of the five benchmark models considered in Ref.~\cite{Dercks:2018eua}.\label{tab:RPV}}
\end{table}
\renewcommand*{\arraystretch}{1}

\begin{figure}\centering
\includegraphics[width=0.45\textwidth]{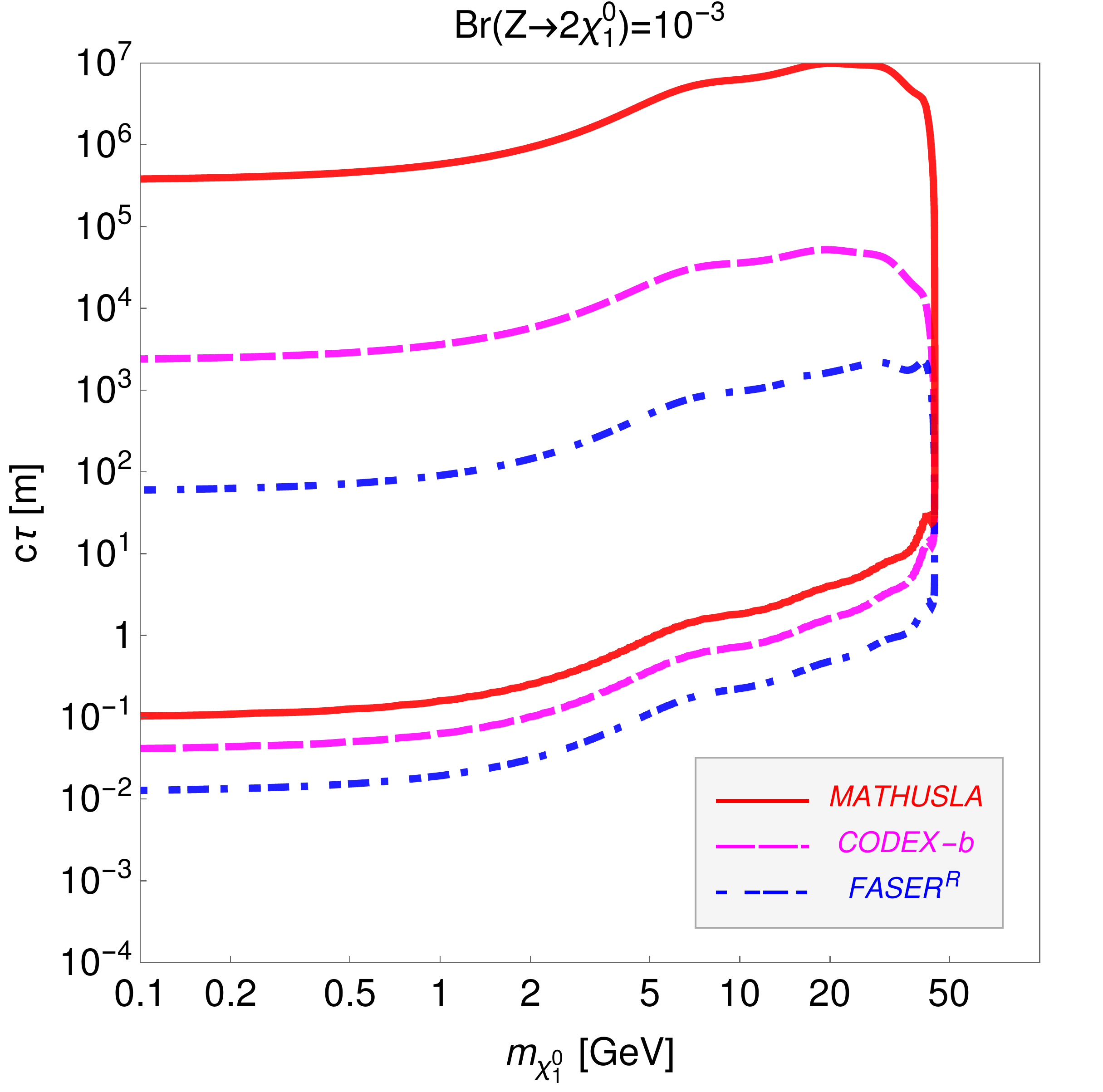}\hfill
\includegraphics[width=0.45\textwidth]{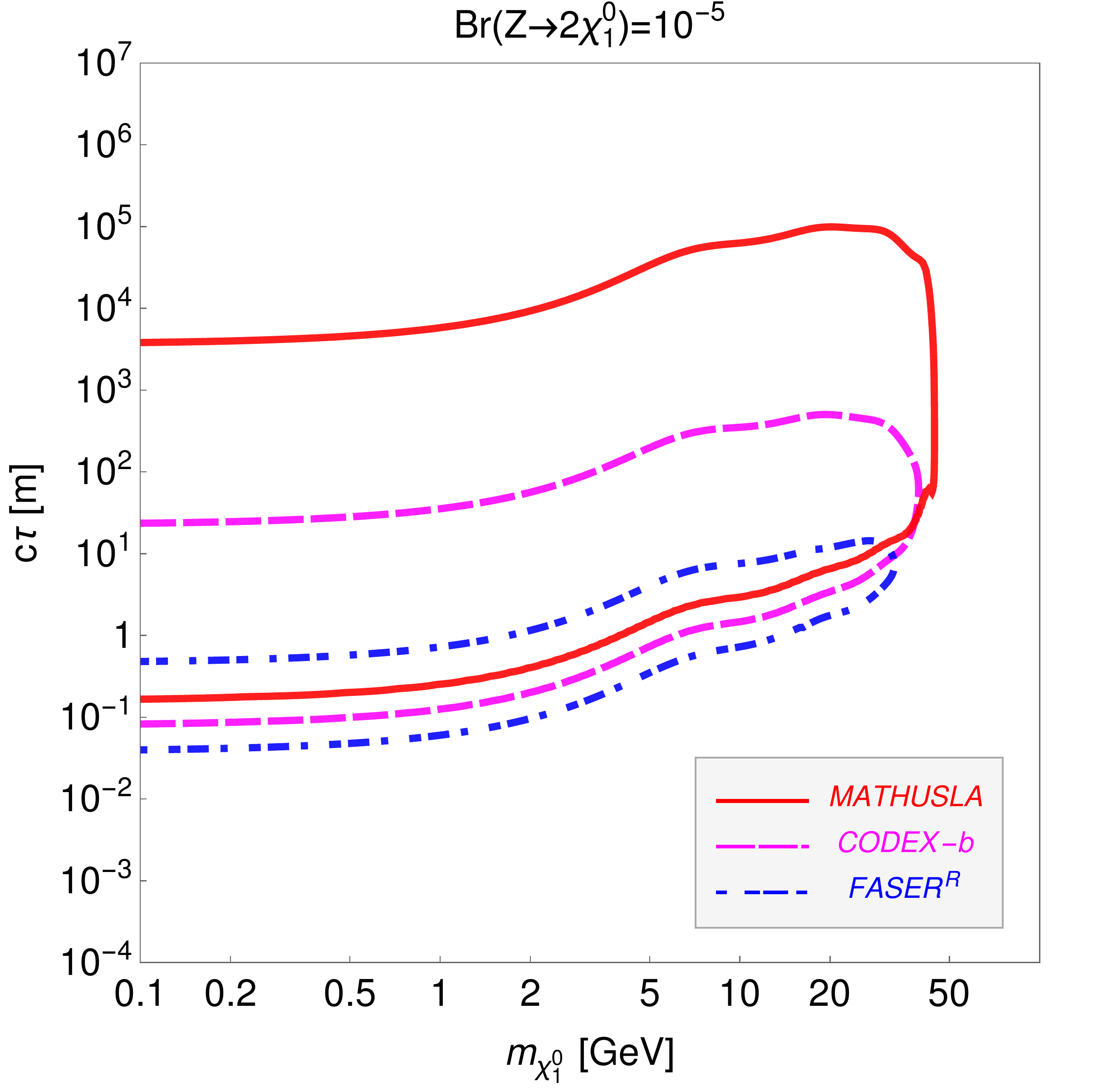}
\caption{Projected sensitivity  of \CODEXb for light neutralinos with R-parity violating coupling, as produced in Z decays, reproduced from Ref.~\cite{Helo:2018qej} with permission of the authors.   
Also shown are projections for the 200m$\times$200m MATHUSLA configuration and FASER$^R$, the 1m radius configuration (referred to as FASER2 elsewhere in this document).}
\label{fig:RPVZdecay}
\end{figure}

The above results assume the wino and higgsino multiplets are heavy enough to be decoupled from the phenomenology. This need not be the case.
For instance, the current LHC bounds allow for a higgsino as light as $\sim 150$ GeV ~\cite{Aaboud:2017leg}, as long as the wino is kinematically inaccessible and the bino decays predominantly outside the detector.
In this case, the mixing of the bino-like $ \tilde\chi^0_1$ can be large enough to induce a substantial branching ratio for the $Z\to  \tilde\chi^0_1 \tilde\chi^0_1$ process.
Helo et.~al.~\cite{Helo:2018qej} showed that the reach of \CODEXb would exceed the $Z\to$ invisible bound for $0.1 \,\mathrm{GeV}< m_{\tilde\chi^0_1}<m_Z/2$
and $10^{-1}\,\mathrm{m}< \;c\tau\; <\; 10^6\, \mathrm{m}$, as shown in Fig.~\ref{fig:RPVZdecay}.
The reach is independent of the flavor structure of the RPV coupling(s), so long as the branching ratio to final states with at least two charged tracks is unsuppressed.
It should be noted that the ATLAS searches in the muon chamber \cite{ATL-PHYS-PUB-2019-002,Aaboud:2018aqj} are expected to have sensitivity to this scenario, although no recasted estimate is currently available.
As with exotic Higgs decays in Sec.~\ref{sec:abelianhiggs}, the expectation is, however, that \CODEXb would substantially improve upon the ATLAS reach for low $m_{\tilde \chi_0}$.

\subsubsection{Relaxion models}
Relaxion models rely on the cosmological evolution of a scalar field -- the relaxion -- to dynamically drive the weak scale towards an otherwise unnaturally low value~\cite{Graham:2015cka}.
The relaxion sector therefore must be in contact with the SM electroweak sector, and the implications of relaxion-Higgs mixing have been studied extensively~\cite{Graham:2015cka,
Espinosa:2015eda,Evans:2016htp,Evans:2017bjs,Batell:2015fma}.
The phenomenological constraints were mapped out in detail in Refs.~\cite{Flacke:2016szy,Choi:2016luu}. (See \cite{Bezrukov:2009yw} for similar phenomenology in a model where the light scalar is identified with the inflaton.)
Following the discussion in Ref.~\cite{Curtin:2018mvb}, the phenomenologically relevant physics of the relaxion, $\phi$, is contained in the term
\begin{equation}
	\label{eq:relaxion}
	\mathcal{L}\supset 2 C \frac{h^2}{\Lambda} \Lambda_N^3 \cos\bigg(\frac{\phi}{f}+\delta\bigg),
\end{equation}
in which $h$ is the real component of the SM Higgs field that obtains a vacuum expectation value $v$, $\Lambda$ is the cut-off scale of the effective theory,
$\Lambda_N$ is the scale of a confining hidden sector, $f$ is the scale at which a UV $U(1)$ symmetry is broken spontaneously, and finally, $C$ and $\delta$ are real constants.
After $\phi$ settles into its vacuum expectation value, $\phi_0$, Eq.~\eqref{eq:relaxion} can be expanded in large $\phi_0/f$, such that
\begin{equation}\label{eq:relaxion2}
	\mathcal{L}\supset 2 \lambda' \sin\bigg(\frac{\phi_0}{f}+\delta\bigg)\frac{v^2}{f}h^2\phi+  \lambda' \cos\bigg(\frac{\phi_0}{f}+\delta\bigg)\frac{v^2}{f^2}h^2\phi^2 + \ldots\,,
\end{equation}
with $\lambda'=C \Lambda_N^3/v^2 \Lambda$. The model in Eq.~\eqref{eq:relaxion2} now directly maps onto the scalar-Higgs portal in Eq.~\eqref{eq:minimalHiggs} of Sec.~\ref{sec:higgsmixing}.
\CODEXb and other intensity and/or lifetime frontier experiments can then probe the model in the regime $\lambda'\sim 1$ and $f\sim \mathrm{TeV}$.
The angle $\phi_0/f+\delta$ controls whether the mixing or quartic term is most important:
On the one hand, if it is small, the lifetime of $\phi$ increases but the quartic in Eq.~\eqref{eq:relaxion2} can be sizable, enhancing the $h\to \phi\phi$ branching ratio (Fig.~\ref{fig:limitBtoKS_lambda}).
On the other hand, for $\phi_0/f+\delta \simeq \pi/2$ the quartic is negligible and the phenomenology is simply that of a scalar field mixing with the Higgs (Fig.~\ref{fig:limitBtoKS}).

\subsubsection{Neutral naturalness}
\label{sec:neutralnaturalness}

The Abelian hidden sector model in Sec.~\ref{sec:abelianhiggs} has enough free parameters to set the mass ($m_{A'}$), the Higgs branching ratio ($\text{Br}(h\to A'A')$) 
and the width ($\Gamma_{A'}$) independently. It therefore allows for a very general parametrization of the reach for exotic Higgs decays
in terms of the lifetime, mass and production rate of the LLP.
The downside of this generality is that the model has too many independent parameters to be very predictive.
In many models, however, the lifetime has a very strong dependence on the mass, favoring long lifetimes for low mass states. 
We therefore provide a second, more constrained example where the lifetime is not a free parameter.

The example we choose is the fraternal twin Higgs~\cite{Craig:2015pha}, which is a recent incarnation of the Twin Higgs paradigm \cite{Chacko:2005pe,Chacko:2005un}, 
which is designed to address the little hierarchy problem. It is itself an example of a hidden valley~\cite{Strassler:2006im,Han:2007ae}.
The model consists of a dark or ``twin'' sector containing an $SU(2)\times SU(3)$ gauge symmetry, that are counterparts of the SM weak and color gauge groups.
It further contains a dark $b$-quark and a number of heavier states which are phenomenologically less relevant.
The most relevant interactions are
\begin{equation}
	\mathcal{L}\supset \lambda\left(\mathcal{H}\mathcal{H}^\dagger -f^2\right)^2 + y_t' H' q'_L t'_R + y_b' H'^c q'_L b'_R\,, \qquad \mathcal{H}\equiv \begin{pmatrix} H \\H' \end{pmatrix}\,,
\end{equation}
with $H$ the SM Higgs doublet and $H'$ the dark sector Higgs doublet. The ``twin quarks'' $q'_L$, $b'_R$ and $t'_R$ are dark sector copies of the 3rd generation quarks.

The Higgs potential of this model has an accidental $SU(4)$ symmetry, which protects the Higgs mass at one loop provided that $y'_t \approx y_t$, with $y_{t}$ the SM top Yukawa coupling.
The corresponding top partner  -- the ``twin top'' -- carries color charge under the twin sector's $SU(3)$ rather than SM color, and is therefore not subject to existing collider constraints from searches for colored top partners.
The accidental symmetry exchanging $H\leftrightarrow H'$ may further be softly broken, such that $\langle H\rangle = v$ and $\langle H'\rangle \approx f$.
The parameter $f$ is typically expressed in terms of the mass of the twin top quark, $m_T$, through the relation $m_T= y_t f/\sqrt{2}$.
The existing constraints on the branching ratio of the SM Higgs already demand $m_T/m_t \gtrsim3$~\cite{Burdman:2014zta}. 

\begin{figure}
\includegraphics[width=0.45\textwidth]{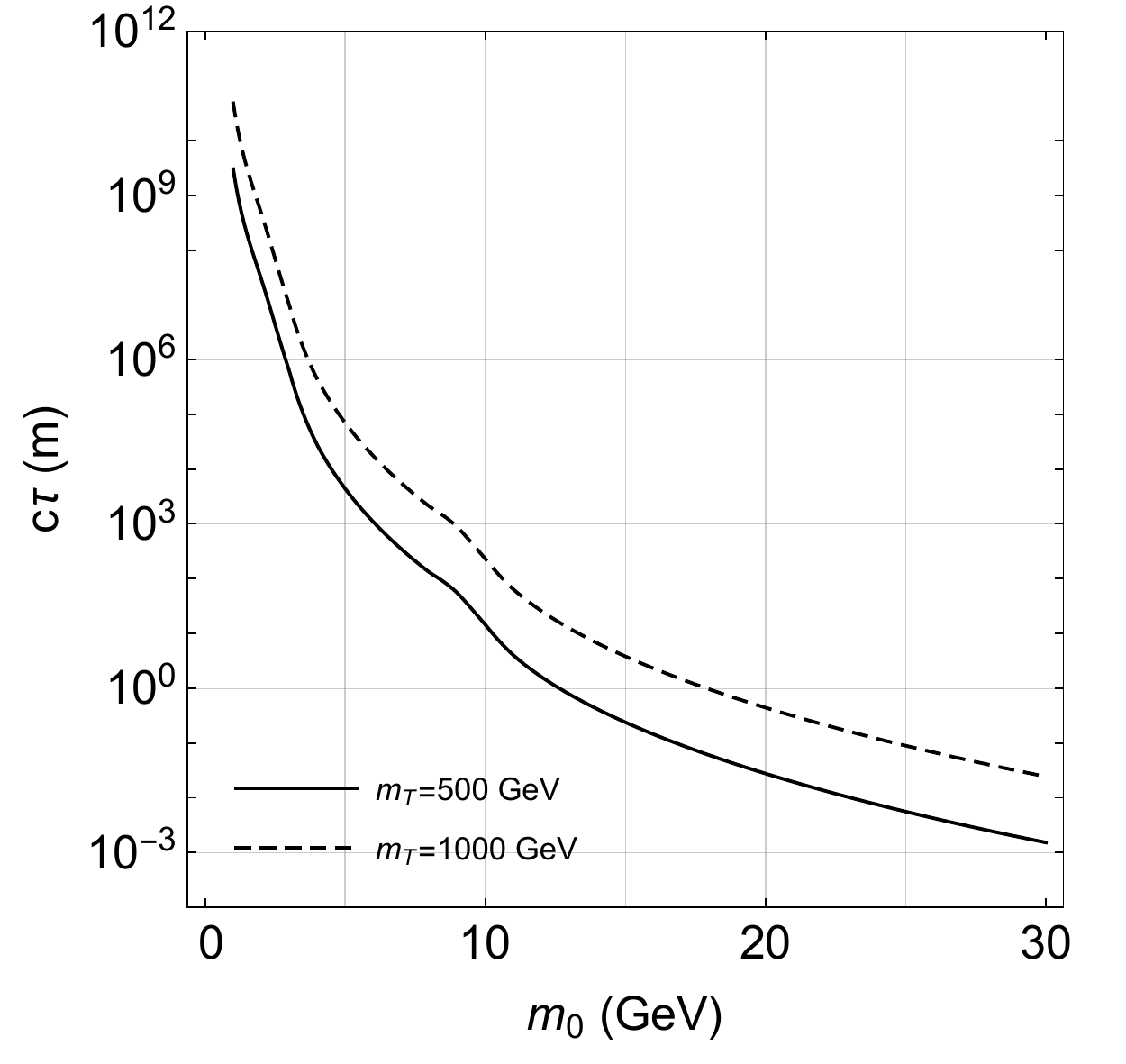}\hfill
\includegraphics[width=0.45\textwidth]{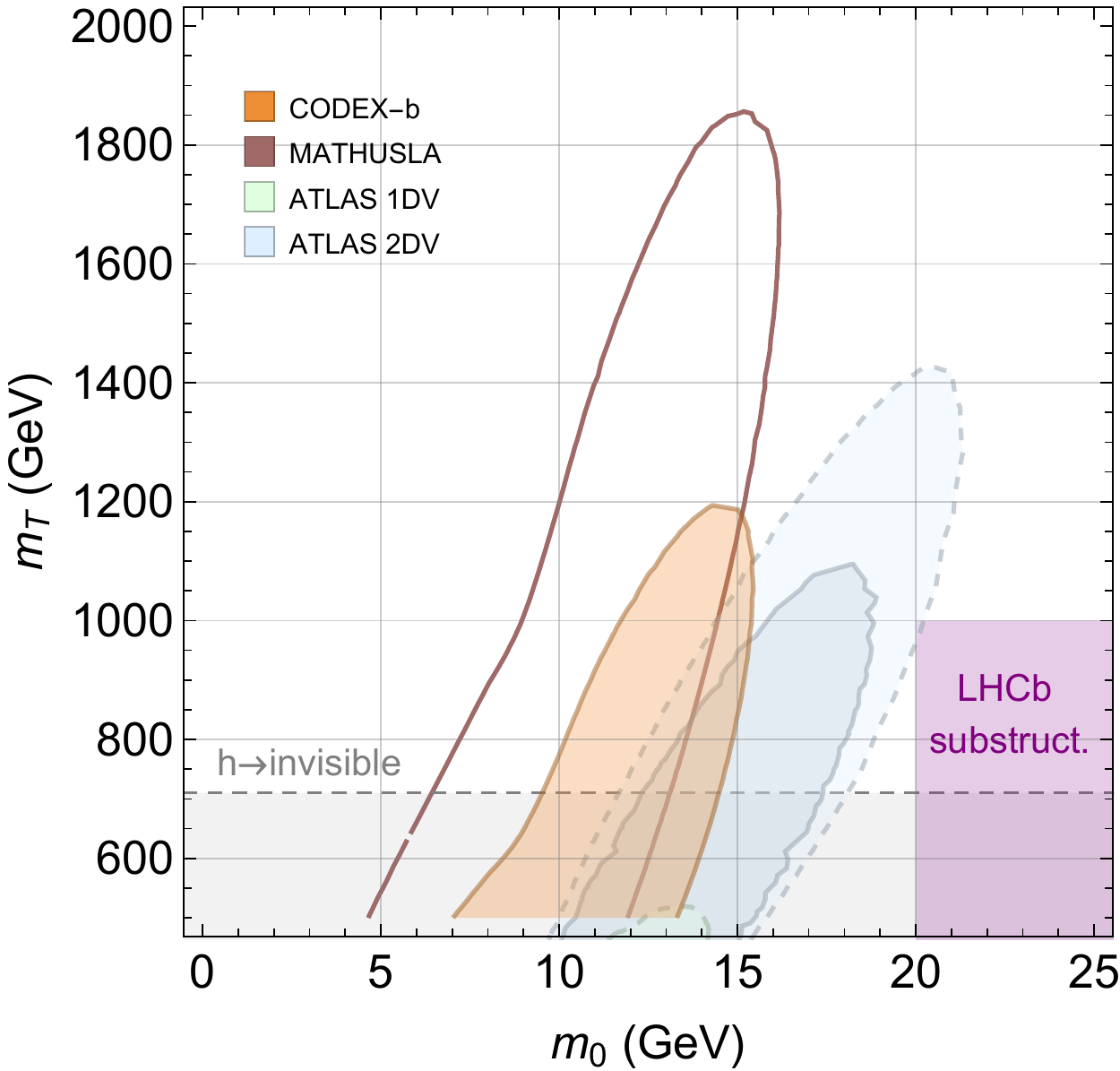}
\caption{Left: Lifetime of the $0^{++}$ glueball as a function of its mass. Right: Projected reach of \CODEXb, MATHUSLA (200m$\times$200m) and ATLAS at the full luminosity of the HL-LHC. 
The solid (dashed) ATLAS contours refer to the optimistic (conservative) extrapolations of the ATLAS reach, as discussed in Sec.~\ref{sec:abelianhiggs}. 
The horizontal dashed line indicates the reach of precision Higgs coupling measurements at the LHC~\cite{Cepeda:2019klc}.\label{fig:glueball}}
\end{figure}

We consider the scenario in which the $b'$ mass is heavier than the dark SU(3) confinement scale, $\Lambda'$,
such that the lightest state in the hadronic spectrum is the $0^{++}$ glueball~\cite{Morningstar:1999rf,Chen:2005mg}, with a mass $m_0\approx 6.8 \Lambda'$.
The $0^{++}$ glueball mixes with the SM Higgs boson through the operator
\begin{equation}
	\mathcal{L}\supset -\frac{\alpha'_3y_t}{6\pi\sqrt{2}}\frac{m_t }{m_T^2}\, h \mathrm{Tr}\Big[G'^{\mu\nu}G'_{\mu\nu}\Big]
\end{equation}
where $h$ is the physical Higgs boson and $\alpha'_3$ the twin QCD gauge coupling.
After mapping the gluon operator to the low energy glueball field, this leads to a very suppressed decay width of the $0^{++}$ state, even for moderate values of $m_t/m_T$.
In particular, the lifetime is a very strong function of the mass, and can be roughly parametrized as
\begin{equation}\label{eq:glueballlifetime}
	c\tau_{0^{++}}\sim18\,m\times\bigg(\frac{10\,\mathrm{GeV}}{m_0}  \bigg)^7\times \bigg(\frac{m_T/m_t}{3}  \bigg)^4.
\end{equation}
This is naturally in the range where displaced detectors like \CODEXb, AL3X and MATHUSLA are sensitive.
The full lifetime curve is shown in the left hand panel of Fig.~\ref{fig:glueball}, where we have accounted for the running of $\alpha'_3$, as in Ref.~\cite{Craig:2015pha,Craig:2016kue}.

For simplicity we assume that the second Higgs is too heavy to be produced in large numbers at the LHC, as is typical in composite UV completions.
However, even in this pessimistic scenario the SM Higgs has a substantial branching ratio to the twin sector.
Specifically, this Higgs has a branching ratio of roughly $\sim m_t^2/m_T^2$ for the $h\to b'b'$ channel.
The $b'$ quarks subsequently form dark quarkonium states, which in turn can decay to lightest hadronic states in the hidden sector.
While this branching ratio is large, the phenomenology of the dark quarkonium depends on the detailed spectrum of twin quarks (see e.g. Ref.~\cite{Craig:2016kue}).
There is however a smaller but more model-independent branching ratio of the SM Higgs directly to twin gluons, given by~\cite{Curtin:2015fna}
\begin{equation}
	\text{Br}[h\to g'g']\approx\text{Br}[h\to gg] \times \bigg(\frac{\alpha_s'(m_h)}{\alpha_s(m_h)}\frac{m_t^2}{m_T^2} \bigg)^2
\end{equation}
with $\text{Br}[h\to gg]=0.086$. $\alpha_s(m_h)$ and $\alpha'_s(m_h)$ are the strong couplings, respectively in the SM and twin sectors, evaluated at $m_h$.
The hidden glueball hadronization dynamics is not known from first principles, and we have assumed that the Higgs decays to the twin sector on average produces two $0^{++}$ glueballs.
Especially at the rather low $m_0$ of interest for \CODEXb, this is likely a conservative approximation.

The projected reach of \CODEXb, MATHUSLA and ATLAS is shown in the right hand panel of Fig.~\ref{fig:glueball}.
The projections for ATLAS were obtained as in Sec.~\ref{sec:abelianhiggs}.
The high mass, short lifetime regime may be covered with new tagging algorithms for the identification of merged jets at LHCb~\cite{Bediaga:2018lhg,Bokan:1753426}.
We find that \CODEXb would significantly extend the reach of ATLAS for models of neutral naturalness.
For hidden glueballs, the factor of $\sim30$ larger geometric acceptance times luminosity for MATHUSLA only results in roughly a factor of $\sim 2$ more reach in $m_0$ for a fixed $m_T$, 
because of the scaling in Eq.~\eqref{eq:glueballlifetime}.
For higher glueball masses, \CODEXb outperforms MATHUSLA due to it shorter baseline. However, this region will likely be covered by ATLAS.

In summary, this hidden glueball model serves to illustrate an important point: For light hidden sector states, the lifetime often grows as a strong power-law of its mass, as illustrated by Fig.~\ref{fig:glueball}.
For ATLAS and CMS, this means that the standard background rejection strategy of requiring two vertices becomes extremely inefficient for such light hidden states.
Instead, displaced detectors like \CODEXb, MATHUSLA and FASER  are needed to cover the low mass part of the parameter space.

\subsubsection{Inelastic dark matter}
\label{sec:inelasticDM}
Berlin and Kling \cite{Berlin:2018jbm} have studied the reach for various (proposed) LLP experiments in the context of a simple model for inelastic dark matter \cite{TuckerSmith:2001hy,Izaguirre:2015zva}.
The ingredients are two Weyl spinors with opposite charges under a dark, higgsed $U(1)$ gauge interaction.
In the low energy limit, the model reduces to
\begin{equation}
	\mathcal{L} \supset -i e_D \bar \chi_2 \slashed{A}' \chi_1 + \frac{\epsilon}{2}F'^{\mu\nu}F_{\mu\nu}+ \ldots\,,
\end{equation}
where the second term indicates the mixing of the dark gauge boson with the SM photon. The ellipsis represents sub-leading terms which do not significantly contribute to the phenomenology.
The pseudo-Dirac fermions $\chi_1$ and $\chi_2$ are naturally close in mass, which leads to a phase space suppression of the width of $\chi_2$. 
The fractional mass difference is parameterized by $\Delta \equiv (m_2 - m_1)/m_1\ll1$.

\begin{figure}
\includegraphics[width=0.65\textwidth]{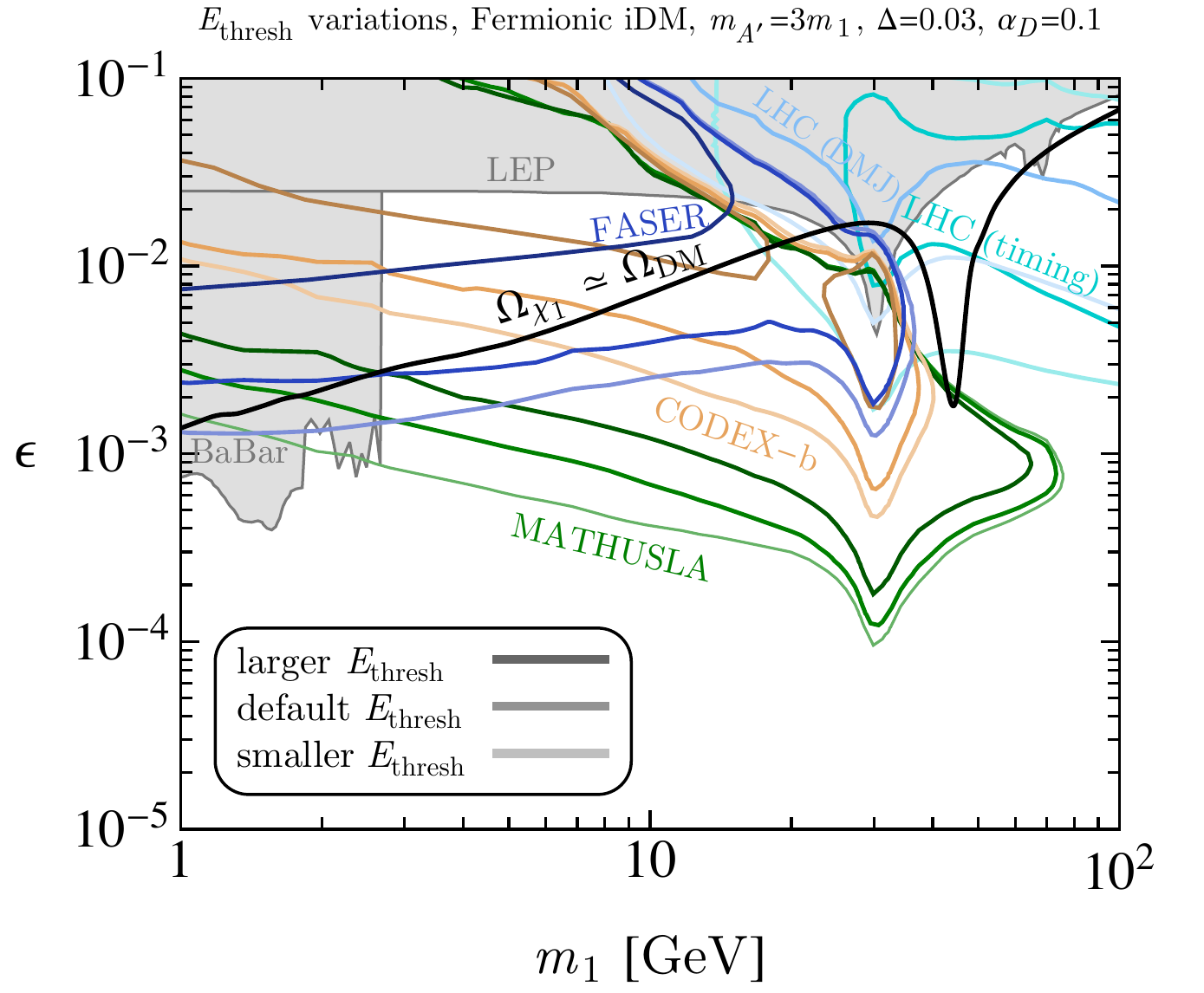}
\caption{Sensitivity estimates for the inelastic dark matter benchmark, reproduced from Ref.~\cite{Berlin:2018jbm} with permission of the authors.
The black line indicates the line on which the correct dark matter relic density is predicted by the model. Darker/lighter shades correspond to larger/smaller
minimum energy thresholds for the decay products of $\chi_2$, with CODEX-b shown in orange shades.
For \CODEXb and MATHUSLA (200m$\times$200m), the minimum energy is taken to $1200$, $600$, or $300$\,MeV per track. For FASER2 (1m radius), 
the total visible energy deposition is taken to be greater than $200$, $100$, or $50$\,GeV.
For a displaced muon-jet search at ATLAS/CMS and a timing analysis at CMS with a conventional monojet trigger, the minimum required transverse 
lepton momentum is $10$, $5$, or $2.5$\,GeV and $6$, $3$, or $1.5$\,GeV, respectively.
}
\label{fig:inelasticDM}
\end{figure}

At the LHC, the production occurs through $q\bar q \rightarrow A' \rightarrow \chi_2 \chi_1$, which is controlled by the mixing parameter $\epsilon$. The decay width of $\chi_2$ is given by
\begin{equation}
\Gamma (\chi_2 \rightarrow \chi_1 \ell^+\ell^-) \simeq\frac{4 \epsilon^2 \alpha_{em}\alpha_D \Delta^5 m_1^5}{15\pi m_{A'}^4}\,,
\end{equation}
where $\alpha_D =e_D^2/4\pi$ is the dark gauge coupling. \CODEXb, MATHUSLA, FASER and the existing LHC experiments can search for the pair of soft, displaced fermions from the $\chi
_2$ decay.
The expected sensitivity of the various experiments  is shown in Fig.~\ref{fig:inelasticDM} for an example slice of the parameter space.
In particular, \CODEXb will be able to probe 
a large fraction of
 the parameter space 
 that produces the observed 
 dark matter relic density, as indicated by the black line in Fig.~\ref{fig:inelasticDM}.
It is worth noting that for this model, the minimum energy threshold per track is an important parameter in determining the reach, which should inform the design of the detector.
For more benchmark points and details regarding the cosmology, we refer to Ref.~\cite{Berlin:2018jbm}.

\subsubsection{Dark matter coscattering}
\label{sec:DMscatt}
The process of coscattering \cite{DAgnolo:2017dbv,DAgnolo:2019zkf} has been studied as a way to generate the correct relic DM abundance.
Coscattering has a similar framework to coannihilating dark matter models:  Both models contain at least one dark matter particle $\chi$, a second state charged under the $\mathbf Z_2$ of the dark sector, $\psi$,
and a third particle $X$ that allows the two particles to transition into one another via an interaction such as a Yukawa, $y X \chi\psi$.
In many coannihilation scenarios $\psi\psi \leftrightarrow XX$ (or SM) is an efficient annihilation mechanism, while $\chi\chi, \chi\psi \leftrightarrow XX$ (or SM) is not.
Throughout the coannihilation, the ``coscattering'' process $\psi X \leftrightarrow \chi X$ (or similar) remains efficient and allows the $\chi$ and $\psi$ species to interchange, without changing the dark particle number.
Eventually, $\psi\psi \leftrightarrow XX$ freezes out, and the total dark particle number is fixed.

By contrast, one may consider coscattering DM \cite{DAgnolo:2017dbv}, in which the $\psi X \leftrightarrow \chi X$ coscattering process drops out of equilibrium before the $\psi\psi \leftrightarrow XX$ coannihilation process.
This requires three ingredients:   $m_X \sim m_\psi \sim m_\chi$; a large $\psi\psi \leftrightarrow XX$ cross-section; and a small $\psi X \leftrightarrow \chi X$ cross-section.
As $\chi$ does not have any sizable interactions other than with $\psi$ by assumption, there are no interactions beyond $\psi X \leftrightarrow \chi X$ that
allow for $\chi$ to maintain a thermal distribution while it is in the process of decoupling from the thermal bath.
This results in important non-thermal corrections that require tracking the full phase space density, rather than just the particle number $n_\chi$, in order to correctly evaluate the relic abundance~\cite{DAgnolo:2017dbv}.

The vector portal model we consider throughout the rest of this subsection is similar to the one in the Sec.~\ref{sec:inelasticDM}.
Here we introduce a new U(1)$_D$ gauge group with fairly strong couplings, a scalar charged under the U(1)$_D$ that obtains a VEV, a Dirac spinor $\chi_2$ charged under the gauge group, 
and a second Dirac spinor $\chi_1$ that is not.
The Lagrangian for the model is
\begin{equation}
	\mathcal{L} \supset -i g_D \bar \chi_2 \slashed{Z}_D \chi_2 + m_2 \bar \chi_2 \chi_2 + m_1 \bar \chi_1 \chi_1 + y_{21} \phi \bar \chi_2 \chi_1+y_{12} \phi^* \bar \chi_1 \chi_2 + \frac{\epsilon}{2\cos\theta_W}Z_D^{\mu\nu}B_{\mu\nu}+\ldots\,.
\end{equation}
The scalar VEV $\vev{\phi}$ gives a mass to the dark vector and generates a small mixing between the U(1)$_D$ active $\chi_2$ and sterile $\chi_1$.
For simplicity, we set  $y\equiv y_{12}=y_{21}$.  When $\Delta m \equiv m_2-m_1 \gg y \vev{\phi}$ a small mixing angle $\theta \approx y \vev{\phi} / \Delta m$ is generated.
We assume that $m_\phi \gtrsim m_{Z_D}$, so that then when $y\ll g_D$, the phenomenology is insensitive to the presence of the scalar.

\begin{figure}
\includegraphics[width=0.55\textwidth]{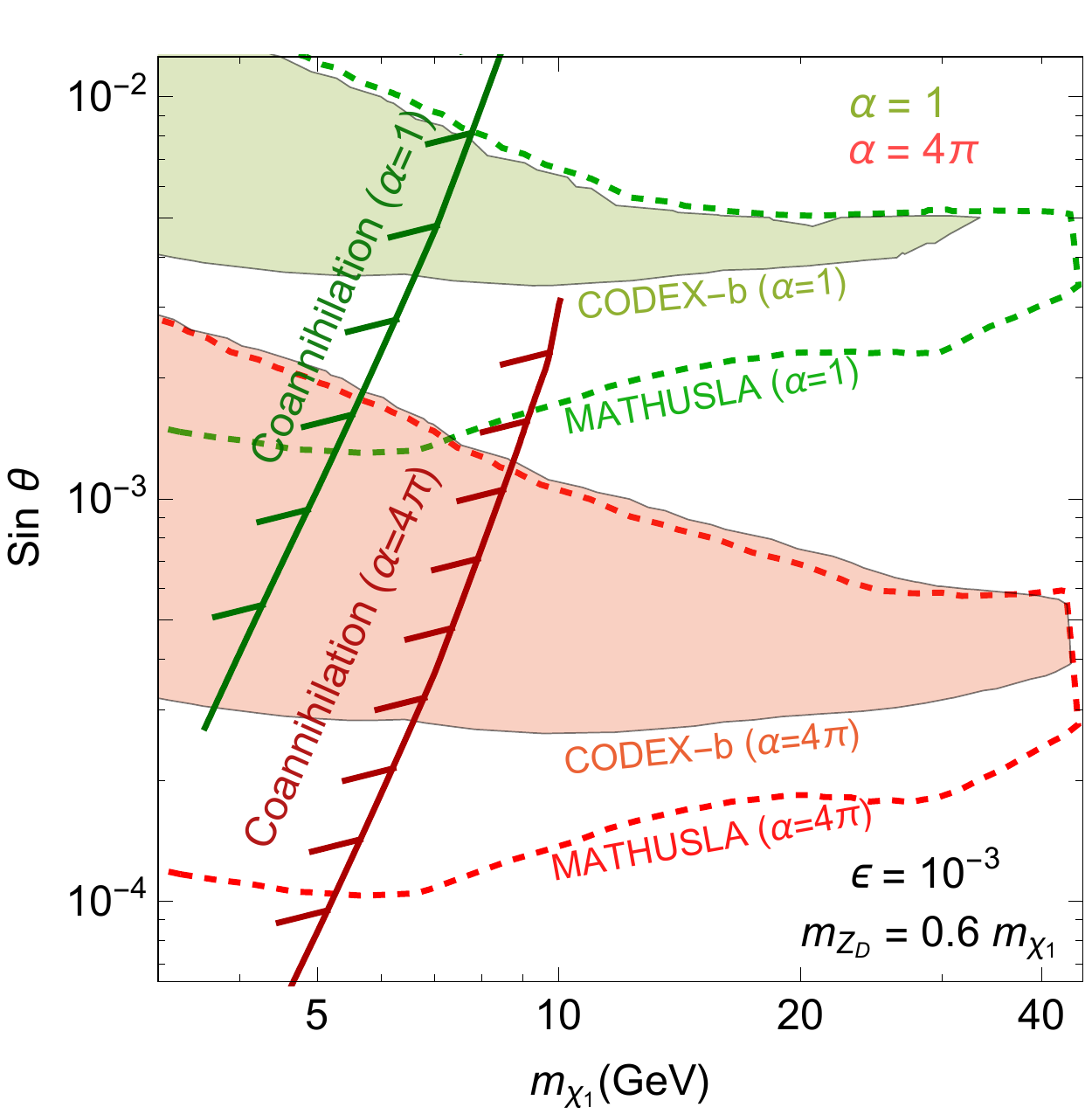}
\caption{Projected sensitivity to the coscattering dark matter benchmark for $\alpha_D=1$ (green) and  $\alpha_D=4\pi$ (red).  The shaded region represents the reach for \CODEXb with 300 fb$^{-1}$.  
The dashed line is the reach for MATHUSLA in the 200m$\times$200m configuration with 3 ab$^{-1}$.  To the left of the dark hatched line, 
the coannihilation process of $\chi_2 \bar \chi_1 \to Z_D Z_D$ remains active long enough to deplete the relic abundance of $\chi_1$ below the observed amount so that the model is inconsistent to the left of these lines.
\label{fig:coscat}}
\end{figure}

 The mixing of $Z_D$ with the $Z$ boson allows for $Z\to \chi_2\bar \chi_2$ with a branching ratio of
 \begin{equation}
	 \text{BR}(Z\to \chi_2\bar \chi_2) = \frac{\alpha_D \epsilon^2 \tan^2\theta_W m_Z^5}{12 (m_Z^2 - m_{Z_D}^2)^2 \Gamma_Z}\lp 1 + 2 \frac{m_{\chi_2}^2}{m_{Z}^2}  \rp \sqrt{1- \frac{m_{\chi_2}^2}{m_{Z}^2} }\,.
\end{equation}
 The daughter $\chi_2$ particles from the $Z$ decay can propagate several meters before decaying to $\chi_1$ through an off-shell dark photon, i.e. $\chi_2 \to \chi_1ff$.  
 The `$ff$' indicates a pair of SM fermions, which CODEX-b can detect. The decay rate is dictated by the splitting between the two states.  For example, the partial width to electrons, neglecting the electron mass
\begin{align}
\Gamma_{ee} =\frac{ \alpha_D \alpha_{em} \epsilon^2 \sin^2\theta}{24 m_{Z_D}^4 \pi m_{\chi_2}^3}
\bigg[  m_{\chi_2}^8
 - 2 m_{\chi_1} m_{\chi_2}^7
 - 8 m_{\chi_1}^2 m_{\chi_2}^6
 - 18 m_{\chi_1}^3 m_{\chi_2}^5
+ 18 m_{\chi_1}^5 m_{\chi_2}^3 \nn \\
+\, 8 m_{\chi_1}^6 m_{\chi_2}^2
+ 2 m_{\chi_1}^7 m_{\chi_2}
-m_{\chi_1}^8
- 24 m_{\chi_1}^3 m_{\chi_2}^3 ( m_{\chi_1}^2 +
     m_{\chi_1}  m_{\chi_2}+ m_{\chi_2}^2  ) \ln\lp\frac{m_{\chi_1}}{
    m_{\chi_2}} \rp
  \bigg].
\end{align}
From this expression we can approximate the lifetime as
\begin{equation}
	c\tau_{\chi_2} = \frac{\text{BR}(Z_D\to ee; \Delta m)}{\Gamma_{ee}} ,
\end{equation}
where $\text{BR}(Z_D\to ee; \Delta m)$ is the branching ratio for a kinetically mixed dark vector of mass $\Delta m$ into $ee$.  This is done to approximate the inclusion of additional accessible final states, 
as splittings in this model are commonly $\order{\text{GeV}}$.
While a more thorough treatment would integrate over phase space for each massive channel separately, this approximation captures the leading effect to well within the precision desired here.
Additionally, $\chi_2$ pairs can be directly produced through an off-shell $Z_D$.  Because the $Z_D$ is off-shell, this does not generate a large contribution unless $m_{\chi_2} \lesssim 10$ GeV.
This model  provides a scenario containing an exotic $Z$ decay into long-lived particles.

In Fig.~\ref{fig:coscat} we show the projected sensitivity for \CODEXb (shaded) and MATHUSLA (dashed) \cite{Curtin:2018mvb} to the model setting $\epsilon = 10^{-3}$, $m_{Z_D} = 0.6 m_{\chi_1}$,
and for two choices of $\alpha_D = 1$ and $4\pi$ (green and red, respectively).
With these parameters fixed, the choice of $\sin\theta$ fixes the mass splitting from the DM relic abundance criteria.
At small masses, the $\bar\chi_2\chi_1 \leftrightarrow Z_DZ_D$ coannihilation process remains in equilibrium long enough to deplete the $\chi_1$ number density below the relic abundance today.
This region is illustrated by the dark hatched lines.

\subsubsection{Dark matter from sterile coannihilation}
\label{sec:coannih}
D'Agnolo et.~al.~\cite{DAgnolo:2018wcn} have explored the mechanism of sterile coannihilation,  for which the number density in the dark sector is set by the annihilation of states 
that are heavier than the dark matter. In this scenario the dark matter remains in chemical equilibrium with these heavy states until after their annihilation process freezes out, 
which naturally allows for much lighter dark matter than in standard thermal freeze-out models.

\begin{figure}
\includegraphics[width=0.9\textwidth]{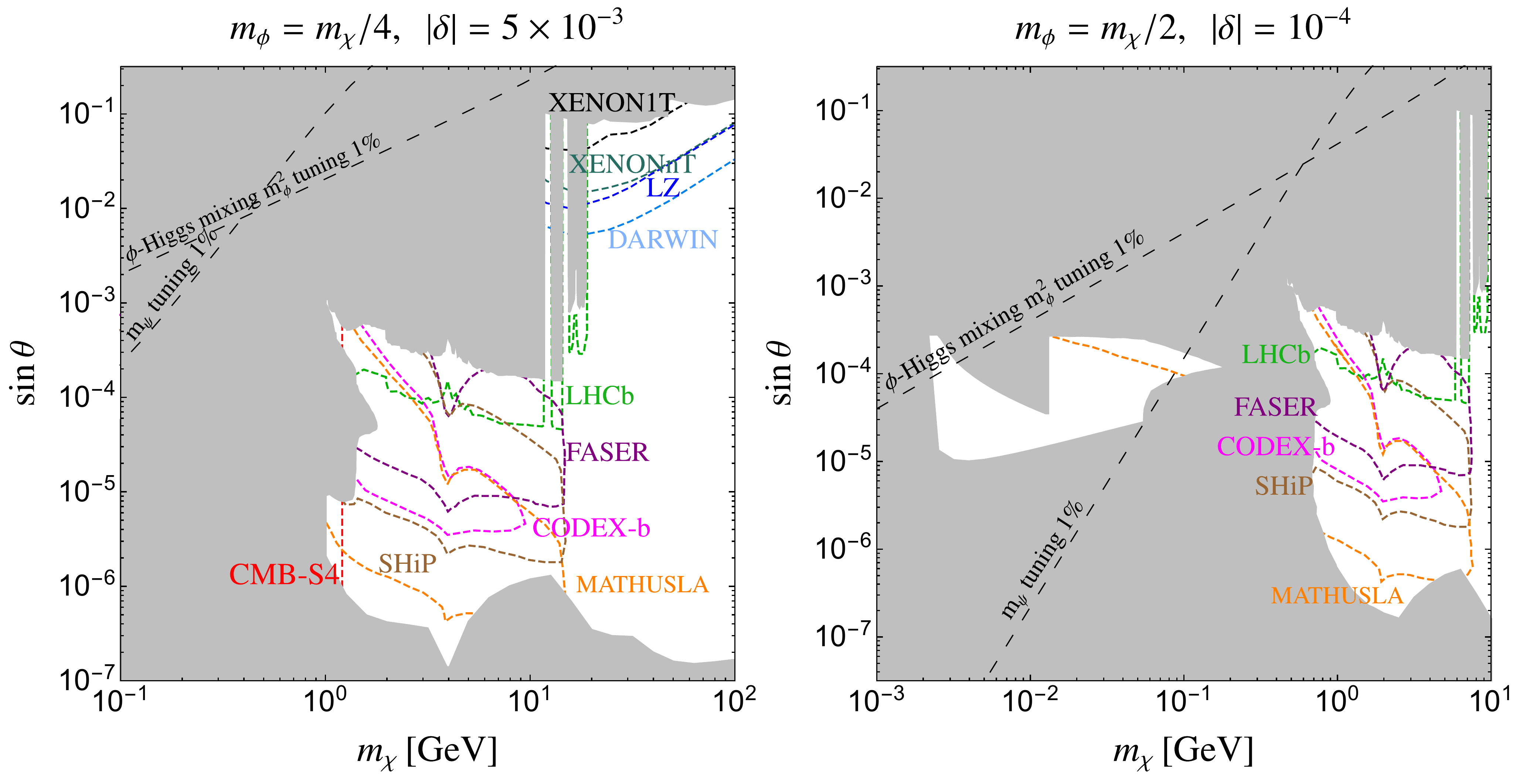}
\caption{Projected sensitivity to the coannihilation dark matter benchmark, reproduced from Ref.~\cite{DAgnolo:2018wcn} with permission of the authors.
The projections are effectively the same as those in Fig.~\ref{fig:limitBtoKS}.
In every point of the plot, $\Delta$ is fixed to reproduce the observed relic density.
Left: The remaining parameters are set as $y = e^{i\pi/4}$, $m_\phi = m_\chi/4$ and $\delta = 5 \times 10^{-3}e^{i\pi/4}$, with $\delta \equiv \delta m/m_\chi$.
Right: The remaining parameters are set instead as $y = e^{i\pi/4}$, $m_\phi = m_\chi/2$ and $\delta =10^{-4}e^{i\pi/4}$.  
MATHUSLA sensitivity is shown for the 200m$\times$200m configuration, while FASER sensitivity is shown for the 1m radius configuration, now referred to as FASER2.
\label{fig:coannihilation} }
\end{figure}

Concretely, the example model that is considered in Ref.~\cite{DAgnolo:2018wcn} is given by
\begin{equation}
-i\mathcal{L}\supset \frac{1}{2} m_\phi^2 \phi^2 + \frac{1}{2}m_\psi \psi^2+ \delta m\, \chi\psi +\frac{1}{2}m_\chi \chi^2+\frac{y}{2}\phi\psi^2
\end{equation}
where the parameter $\delta m \ll m_\psi, m_\chi$ generates a small mixing between $\psi$ and $\chi$. For the choice $m_\psi \gtrsim m_\chi > m_\phi$, 
the relic density of $\chi$ is effectively set by $\psi\psi \rightarrow \phi\phi$ annihilations.
Finally, $\phi$ is assumed to mix with the SM Higgs, and it is this coupling which keeps the dark sector in thermal equilibrium with the SM sector.
For a summary of the direct detection and cosmological probes of this model, we refer to Ref.~\cite{DAgnolo:2018wcn}.
From a collider point of view, the most promising way to probe the model is to search for the scalar $\phi$ through its mixing with the Higgs.
This scenario is identical to the scalar-Higgs portal model with $\lambda=0$, which is discussed in Sec.~\ref{sec:higgsmixing}.
Fig.~\ref{fig:coannihilation} shows the projected reach for \CODEXb, overlaid with the relevant constraints and projections from dark matter direct detection and CMB measurements.

\subsubsection{Asymmetric dark matter}
\label{sec:ADM}

In many asymmetric dark matter models, the DM abundance mass is directly tied to the matter anti-matter asymmetry in the SM sector~\cite{Kaplan:2009ag,Kim:2013ivd,Zurek:2013wia}.
Therefore the generic expectation for the DM is to carry $B-L$ quantum numbers and have  a mass $\simeq$~GeV.
For this mechanism to operate, the DM sector interactions with the SM should be suppressed and both sectors communicate in the early universe through operators of the form
\begin{equation}
	\frac{1}{M^{\Delta_{SM}+\Delta_{X}-4}}\mathcal{O}_X \mathcal{O}_{SM}\,,
\end{equation}
where $\mathcal{O}_X$ and $\mathcal{O}_{SM}$ are operators consisting of dark sector and SM fields respectively, with $\Delta_{X}$ and $\Delta_{SM}$ their respective operator dimensions.
In supersymmetric models of asymmetric dark matter~\cite{Kaplan:2009ag}, the simplest operators in the superpotential are of the form
\begin{equation}
	W= X LH,\quad \frac{1}{M}Xu^c d^c d^c,\quad \frac{1}{M} X QLd^c,\quad \frac{1}{M}X LLe^c\,,
\end{equation}
with $X\equiv \tilde x +\theta x$ the chiral superfield containing the DM, denoted by $x$.
The phenomenology of this scenario is very similar to that of RPV supersymmetry, with decay chains such as $\tilde \chi^{0}\to \tilde x u^c d^c d^c$ (see Sec.~\ref{sec:RPV}).
To accommodate the correct cosmology, macroscopic lifetimes $c\tau\sim 10$\,m are typically required~\cite{Chang:2009sv,Kim:2013ivd}. Moreover, $\tilde x$ itself may or may not be stable, depending on the model.

More generally, if the dark sector has additional symmetries and multiple states in the GeV mass range, as occurs naturally in hidden valley models with asymmetric dark matter 
(see e.g.~Refs.~\cite{Strassler:2006im,Schwaller:2015gea}),
these excited states often must decay to the DM plus some SM states. Such decays must necessarily occur through higher dimensional operators, and macroscopic lifetimes are therefore generic.
As for previous portals, LLP searches in the GeV mass range are best suited to displaced, background-free detectors such as \CODEXb.

\subsubsection{Other Dark Matter models}
\label{sec:otherDM}

There are many other dark matter models that could provide signals observable with CODEX-b.  Presenting projections for all possibilities is beyond the scope of this work, 
but here we briefly summarize many of the existing scenarios that can provide long-lived particles.  Below we detail: SIMPs, ELDERs, co-decaying DM, dynamical DM, and freeze-in DM.

\textcolor{darkgray}{\textbf{Strongly Interacting Massive Particles (SIMPs)}}~\cite{Hochberg:2014dra,Hochberg:2014kqa,Choi:2017zww,Choi:2018iit} obtain their relic density 
through a $3 \to 2$ annihilation process mediated by a strong, hidden sector force.
The preferred mass scale in this scenario is the GeV scale, and the strong nature of the hidden sector implies the presence a whole spectrum of dark pions, dark vector mesons etc.
The $3 \to 2$ annihilation process, however, heats up the dark sector in the early universe, which would drive the dark matter to be exponentially hotter than the SM if it were completely decoupled.
Since the dark matter is known to be cold, this means there must exist a sufficiently strong portal keeping both sectors in thermal equilibrium.
These interactions predict a variety of signatures, in conventional dark matter detection and at colliders.
In this sense SIMP models provide more motivation for exploring the Hidden Valley framework: At the LHC, it is possible to produce the hidden quarks through the aforementioned portal (e.g.~a kinetically mixed dark photon),
which would subsequently shower and fragment to hidden mesons with masses around the GeV scale.
Some of these mesons will be stable and invisible, such as the dark matter, while other will decay back to the standard model, often with macroscopic displacements.
This phenomenology is studied in detail in Refs.~\cite{Hochberg:2015vrg,Berlin:2018tvf}.

\textcolor{darkgray}{\textbf{ELastically DEcoupling Relics (ELDERs)}}~\cite{Kuflik:2015isi} share many of the features of the SIMP models, including the strong $3\to 2$ annihilation process in the hidden sector
and the mandatory portal with the SM to prevent the dark sector from overheating.
In contrast to SIMP models, the elastic scattering processes between the dark sector and the SM freeze out \emph{before} the end of the $3\to 2$ annihilations,
such that the dark matter cannibalizes itself for some time during the evolution of the universe.
ELDER models are also examples of hidden valleys, and the collider phenomenology is therefore qualitatively similar to that of SIMP models.

In \textcolor{darkgray}{\textbf{co-decaying dark matter models}}~\cite{Dror:2016rxc,Okawa:2016wrr,Farina:2016llk}, the dark matter state, $\chi_1$, 
is kept in equilibrium with a slightly heavier dark state, $\chi_2$,
though efficient $\chi_1\chi_1\leftrightarrow \chi_2\chi_2$ processes in the early universe, but the dark sector does not maintain thermal equilibrium with the SM.
The $\chi_2$ state is, however, unstable and decays back to the SM.
Because both states remain in equilibrium, this also depletes the $\chi_1$ number density once the temperature of the dark sector drops below the mass of $\chi_2$.
For this mechanism to operate, $\chi_2$ should have a macroscopic lifetime.
On the one hand, the heavier $\chi_2$ could very well be produced at the LHC through a heavy portal, however this is not strictly required for the co-decaying dark matter framework to operate.
On the other hand, if implemented in the context of e.g. neutral naturalness~\cite{Farina:2015uea,Freytsis:2016dgf}, a production mechanism at the LHC is typically a 
prediction and the phenomenology is once again that of a hidden valley.

In \textcolor{darkgray}{\textbf{dynamical dark matter models}}~\cite{Dienes:2011ja,Dienes:2011sa,Dienes:2012jb}, the dark sector contains a large ensemble of decaying dark states with a wide range of lifetimes.
Their collective abundance makes up the DM abundance we see today, by balancing their share in the universe's energy budget against their lifetime.
Some of the states in the ensemble are expected to have lifetimes that can be resolved on collider length scales.
Just as for co-decaying dark matter, an observable cross-section at the LHC is possible but not required.
If the dynamical dark sector can be accessed, however, the collider phenomenology is rich~\cite{Dienes:2012yz,Dienes:2014bka,doi:10.1063/1.4953280}
and auxiliary, displaced, background-free LLP detectors can play an important role~\cite{Curtin:2018ees}.

Finally, in \textcolor{darkgray}{\textbf{Freeze-in models}}~\cite{Hall:2009bx}, the dark matter is never in equilibrium with the SM sector, but instead the dark sector is slowly populated 
through either scattering or the decay of a heavy state.
This mechanism demands very weak couplings, which in the case of freeze-in through decay predicts a long-lived state decaying to DM plus a number of SM states.
In the models considered so far, the preferred mass range for the decaying state tends to be in the $100$\,GeV to $1$\,TeV regime~\cite{Co:2015pka,Evans:2016zau,DEramo:2017ecx,No:2019gvl},
such that ATLAS and CMS ought to be well equipped to find these decays.
Should the final states however prove to be difficult to resolve at ATLAS and CMS, or should the parent particle be lighter than currently predicted, \CODEXb could provide the means to probe these models.

\subsubsection{Baryogenesis}
\label{sec:baryogenesis}

There exists a wide range of models explaining the baryon asymmetry in the Universe, some of which reside in the deep UV, while others are tied to the weak scale,
such as electroweak baryogenesis (see e.g. Ref.~\cite{Morrissey:2012db}) and WIMP baryogenesis~\cite{Cui:2012jh,Cui:2014twa}.
The latter in particular predicts long-lived particles at LHC, with a phenomenology that is qualitatively similar to displaced decays for RPV supersymmetry (see Sec.~\ref{sec:RPV}).
We refer to Ref.~\cite{Curtin:2018mvb} for a discussion of the discovery potential of WIMP baryogenesis at the lifetime frontier.

Instead we focus here in more depth on a recent idea which generates the baryon asymmetry through the CP-violating oscillations of heavy flavor baryons~\cite{McKeen:2015cuz,Aitken:2017wie}.
(See Refs.~\cite{Elor:2018twp,Nelson:2019fln} for similar ideas involving heavy flavor mesons and Ref.~\cite{Alonso-Alvarez:2019fym} for a supersymmetric realization.)
This enables very low reheating temperatures and is moreover directly testable by experiments such as \CODEXb, as well as Belle II.

The model relies on the presence of the light Majorana fermions $\chi_{1}$ and $\chi_2$ which carry baryon number. (Two generations are needed to allow for CP-violation.) They couple to the SM quarks through
\begin{equation}\label{eq:baryongenesis1}
	\mathcal L \supset \frac{g_{ijk\ell}}{\Lambda^2}\chi_i u_j  d_k d_\ell + \mathrm{h.c.}
\end{equation}
The out-of-equilibrium condition necessary for baryogenesis can be satisfied, for instance, by a late decay of a third dark fermion to the SM heavy flavor quarks.
The operator in Eq.~\eqref{eq:baryongenesis1} generates a dimension-9 operator with $\Delta B=2$, of the form $(u_j  d_k d_\ell )^2$, that is responsible for the baryon oscillations.
For these oscillations to be sufficiently large to generate the observed baryon asymmetry, one needs $m_{\chi_{1,2}}\lesssim m_B$, which has intriguing phenomenological consequences.
Moreover, stringent constraints from the dinucleon decay of $O^{16}$ imply that
\begin{equation}
	c\tau_\chi \gtrsim 100\, \text{m}\, \bigg(\frac{5\,\text{GeV}}{m_{\chi_{1,2}}}\bigg)^5 \bigg(\frac{\Lambda /\sqrt{g_{uss}}}{20\,\text{TeV}}\bigg)^4\,.
\end{equation}
These low masses and long lifetimes are precisely where \CODEXb would have a substantial advantage over ATLAS, CMS and LHCb.
The branching ratios for $B$ baryons and mesons into $\chi_{1,2}$ can even be as large as $10^{-3}$, which means that the rate of $\chi_{1,2}$ production at IP8 could be very large.
Part of the parameter space of this model might therefore be probed already by the \CODEXbeta during Run~3 (see Sec.~\ref{sec:demophysicsreach}).

\subsubsection{Hidden valleys}

Hidden Valley models \cite{Strassler:2006im} are hidden sectors with non-trivial dynamics, which can lead to a relatively large multiplicity of final states in decays of hidden particles.
Confining hidden sectors provide a canonical example, because of their non-trivial spectrum of hidden sector hadrons and the ``dark shower'' that may arise when energy is injected 
in the hidden sector through a high energy portal.
(Fully perturbative examples can also be constructed easily.)
Some hidden valleys naturally arise in models which aim to address various shortcomings of the SM:
Examples discussed in the preceding sections are neutral naturalness (Sec.~\ref{sec:neutralnaturalness}), asymmetric dark matter (Sec.~\ref{sec:ADM}) and SIMP dark matter (Sec.~\ref{sec:otherDM}).

 The phenomenology of hidden valleys can vary widely~\cite{Strassler:2006qa,Han:2007ae,Strassler:2008bv,Strassler:2008fv,Schwaller:2015gea,Knapen:2016hky,Pierce:2017taw},
 both in terms of the energy and angular distributions of the final states, as well as the lifetime of the dark sector particles.  A handful of initial searches have already been performed at 
 ATLAS, CMS, and LHCb (see e.g. Refs.~\cite{Sirunyan:2018njd,Aaij:2017mic,Aaij:2020ikh,CERN-EP-2019-140}).
The various opportunities afforded by the experiments, as well as the challenges involved in constructing a comprehensive search plan were recently summarized in Ref.~\cite{Alimena:2019zri}.   
In particular,
 \begin{itemize}
 	\item While the LLPs are typically in the $\sim$ GeV range, 
	their lifetimes can easily take phenomenologically relevant values spanning many orders of magnitude. 
	In the short lifetime regime, backgrounds can be suppressed by demanding multiple displaced vertices in the same event, provided that a suitable trigger can be found, 
	but this strategy is much less effective in the long-lifetime regime.
 	\item There are generically multiple species of LLPs, with vastly different lifetimes, and some may decay (quasi-)promptly or to a high multiplicity of soft final states.
	 In practice, this means that a displaced decay to SM final states from a dark shower is likely to fail traditional isolation criteria or $p_T$ thresholds, further complicating searches at ATLAS, CMS and LHCb.
 	\item The energy flow in the event may be non-standard, and is poorly understood theoretically. This means that standard jet-clustering algorithms are expected to fail for a large subclass of models.
  \end{itemize}
Because of both its ability to search inclusively for LLP decays and its background-free setup, \CODEXb is not limited by many of these challenges,
and would be sensitive to any hidden valley model which has at least one LLP species in the spectrum with both a sizable branching fraction to charged final states and a moderately large lifetime, i.e $c\tau\gtrsim 1$\,m.
The latter requirement in particular makes \CODEXb highly complementary to ATLAS, CMS and LHCb in the context of these models, 
as the short lifetime regime can be probed with a multi-vertex strategy in the main detectors,
provided that the putative trigger challenges can be addressed.

\FloatBarrier
\clearpage

\section{Backgrounds\label{sec:backgrounds}}
Crucial to the \CODEXb programme is the creation and maintenance of a background-free environment.
An in-depth discussion of relevant primary and secondary backgrounds may be found in Ref.~\cite{Gligorov:2017nwh} as well as Ref.~\cite{Gligorov:2018vkc}.

In this section we re-examine the core features of the relevant backgrounds, and the required active and passive shielding required to ensure a background-free environment in the detector.
This study includes an updated and more realistic \texttt{Geant4} simulation of the shielding response that incorporates uncertainties, charged-neutral particle correlations,
as well as an updated simulation of the high energy tails of the primary backgrounds, and simulation of multitrack production in the detector volume.
Further, the details and results of a measurement campaign, conducted in the LHCb cavern in 2018 as a preliminary data-driven validation of these simulations, are presented.

\subsection{Overview}

An LHC interaction point produces a large flux of primary hadrons and leptons. Many of these may be fatal to a background-free environment either because they are themselves neutral long-lived particles,
e.g.~(anti)neutrons and $K_L^0$'s, that can enter the detector and then decay or scatter into tracks, or because they may generate such neutral LLP secondaries 
by scattering in material, e.g. muons, pions or even neutrinos.
In the baseline \mbox{\CODEXb} design, LLP-like events are comprised of tracks originating within the detector volume, with the track momentum as low as $400$\,MeV.
This threshold is conservative with respect to likely minimum tracking requirements for a signal (cf. Ref.~\cite{LHCbtwiki:2015}).

\begin{figure}[t]
\centering{
\includegraphics[width = 0.75\linewidth]{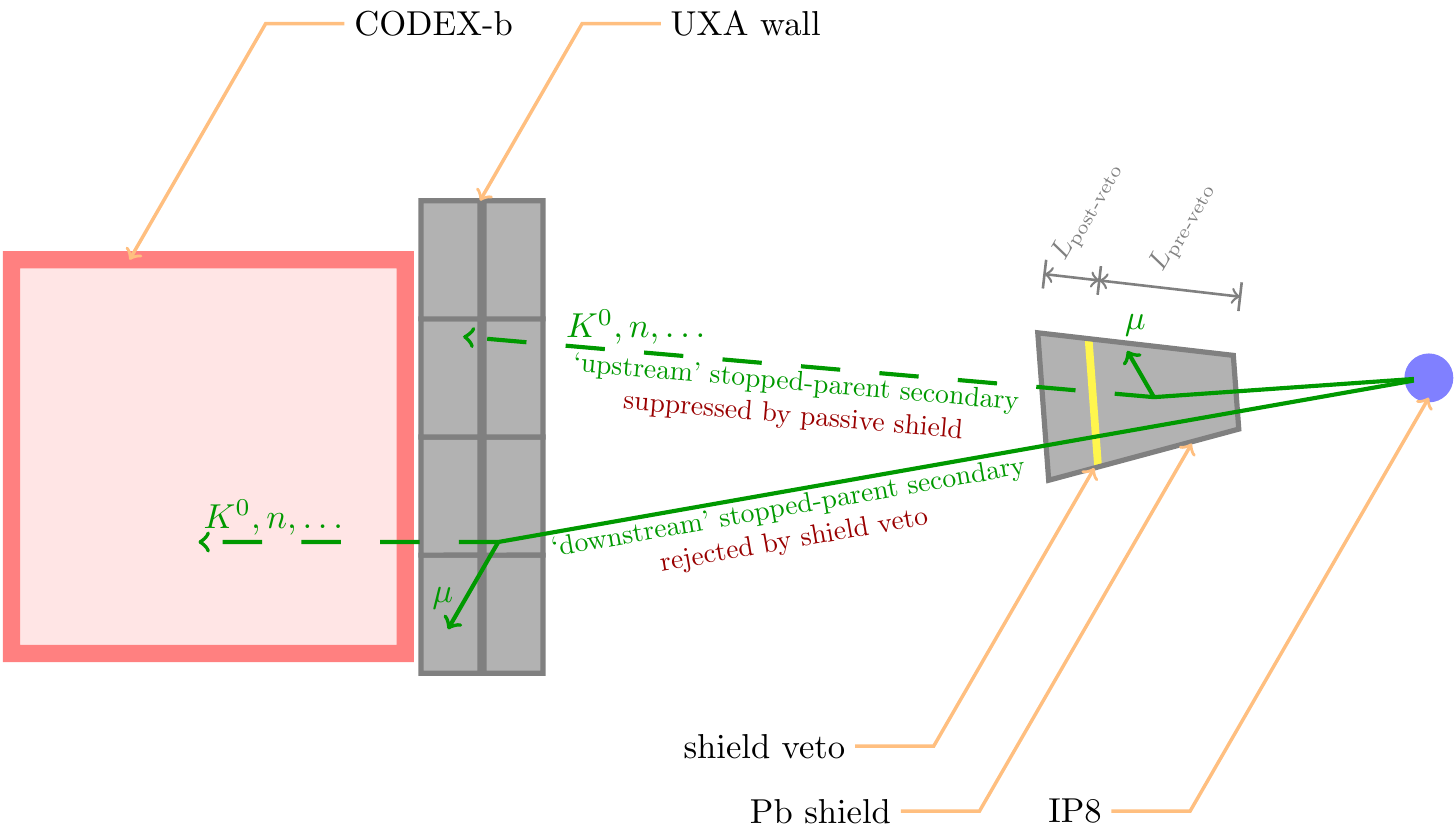}
}
\caption{Cross-section of the shielding configuration of the Pb shield (gray), active shield veto (gold), and concrete UXA wall with respect to IP8 and the detector volume. 
Also shown are typical topologies for production of upstream and downstream stopped-parent secondaries, which are suppressed by passive shielding or rejected by the active shield veto, respectively.}
\label{fig:shld_cnfg}
\end{figure}

Suppression of primary hadron fluxes can be achieved with a sufficient amount of shielding material: roughly $10^{14}$ neutrons and $K_L^0$'s are produced per $300$\,fb$^{-1}$ at IP8,
requiring $\log(10^{14}) \simeq 32\lambda$ of shield for full attenuation, where $\lambda$ is a nuclear interaction length.
In the nominal \CODEXb design, the 3\,m of concrete in the UXA radiation wall, corresponding to $7\lambda$,
\footnote{The UXA wall is $3.2$m in depth, corresponding to $\simeq 7.5\lambda$ for `standard' concrete~\cite{Tanabashi:2018oca}. 
Since the precise composition of the concrete used in the wall is not available we treat the wall as $3$m of standard concrete as a conservative estimation.}
is supplemented with an additional $25\lambda$ Pb shield, 
corresponding to about $4.5$\,m, as shown in Fig.~\ref{fig:shld_cnfg}.
(We focus here on a shield comprised of lead, though composite shielding making use of e.g. tungsten might also be considered, with similar performance~\cite{Gligorov:2018vkc}.)
However, this large amount of shielding material in turn may act as a \emph{source} of neutral LLP secondaries, produced by muons 
(or neutrinos, see Sec.~\ref{sec:bkg-neutrinos}) that stream through the shielding material.
The most concerning neutral secondaries are those produced in the last few $\lambda$ by high energy muons that themselves slow down and stop before reaching the detector veto layers.
Such parent muons are not visible to the detector, while the daughter neutral secondaries, because they pass through only a few $\lambda$ of shield, 
may themselves escape the leeward side of the shield and enter the detector volume:
We call these `stopped-parent secondaries'; a typical topology is shown in Fig.~\ref{fig:shld_cnfg}.

As a rough example, a $10$\,GeV muon has a `CSDA' (Continuous Slowing-Down Approximation) range in lead of approximately $6$\,m~\cite{Tanabashi:2018oca}, corresponding to $32\lambda$ ($\lambda_{\text{Pb}} \simeq0.18$\,m).
Approximately $10^9$ such muons are produced per $300$\,fb$^{-1}$ in the \CODEXb acceptance (see Sec.~\ref{sec:bkg_pf}), and by the last few $\lambda$ they have slowed to $\lesssim$\,GeV kinetic energy.
The strange muoproduction cross-section for a GeV muon is $\sim 0.01\,\mu$b per nucleon, so that in the last  $\lambda$ approximately $\text{few}\times 10^3$ $K_L^0$'s are produced by these muons.
The kaon absorption cross-section on a Pb atom is $\sim2$\,b, so the reabsorption probability in the last $\lambda$ is $\sim 30\%$, 
with the result that $\sim 10^3$ stopped-parent secondary $K_L^0$'s can still escape into the detector.
This behavior is more properly modelled by a system of linear differential equations, that capture the interplay of the muon $dE/dx$ with the energy-dependence of the secondary muoproduction cross-section
and their (re)absorption cross-sections; in practice we simulate this with \texttt{Geant4}, as described below in Sec.~\ref{sec:bkg_sim}.

The \CODEXb proposal resolves this secondary background problem by the addition of a veto layer placed deep inside the shield itself: an active shield element, shown in gold in Fig.~\ref{fig:shld_cnfg}.
This veto layer may then trigger on the parent muons before they produce neutral secondaries and stop.
The veto layer must be placed deep enough in the shield -- shielded sufficiently from the IP -- so that the efficiency required to veto the stopped-parent secondaries produced downstream is not too high.
At the same time there must be sufficient shielding downstream from the veto to attenuate the stopped-parent secondaries with respect to the shield veto itself:
That is, neutrals produced upstream before the veto layer that could still reach the detector (see Fig.~\ref{fig:shld_cnfg}).
An additional consideration is that the veto rejection rate itself should be much smaller than the overall event rate, in order not to degrade the LLP signal detection efficiency.
The nominal shield we consider has a so-called `$(20 + 5)\lambda$' configuration, with $20\lambda$ of Pb before the shield veto and $5\lambda$ afterwards, 
plus the additional $3$\,m of concrete ($7\lambda$) from the UXA wall.

\subsection{Simulation}
\label{sec:bkg_sim}

\subsubsection{Primary fluxes}
\label{sec:bkg_pf}
\begin{figure}[p]
\centering{\hfill
\includegraphics[width = 0.4\linewidth]{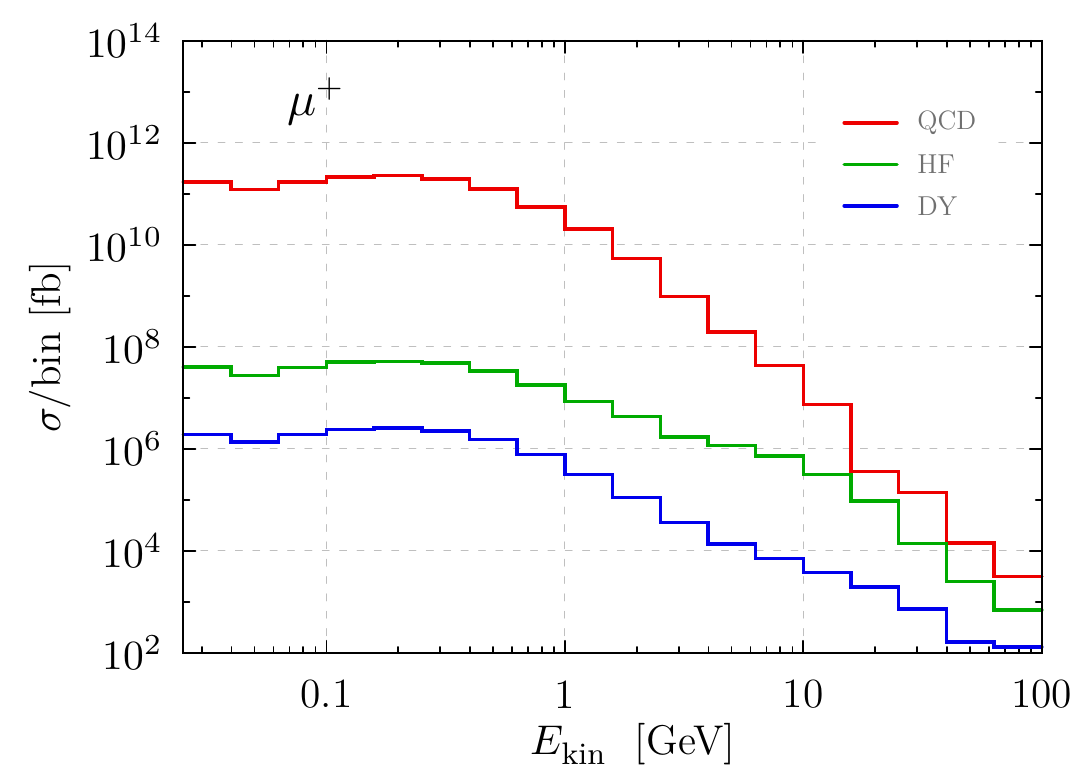}\hfill
\includegraphics[width = 0.4\linewidth]{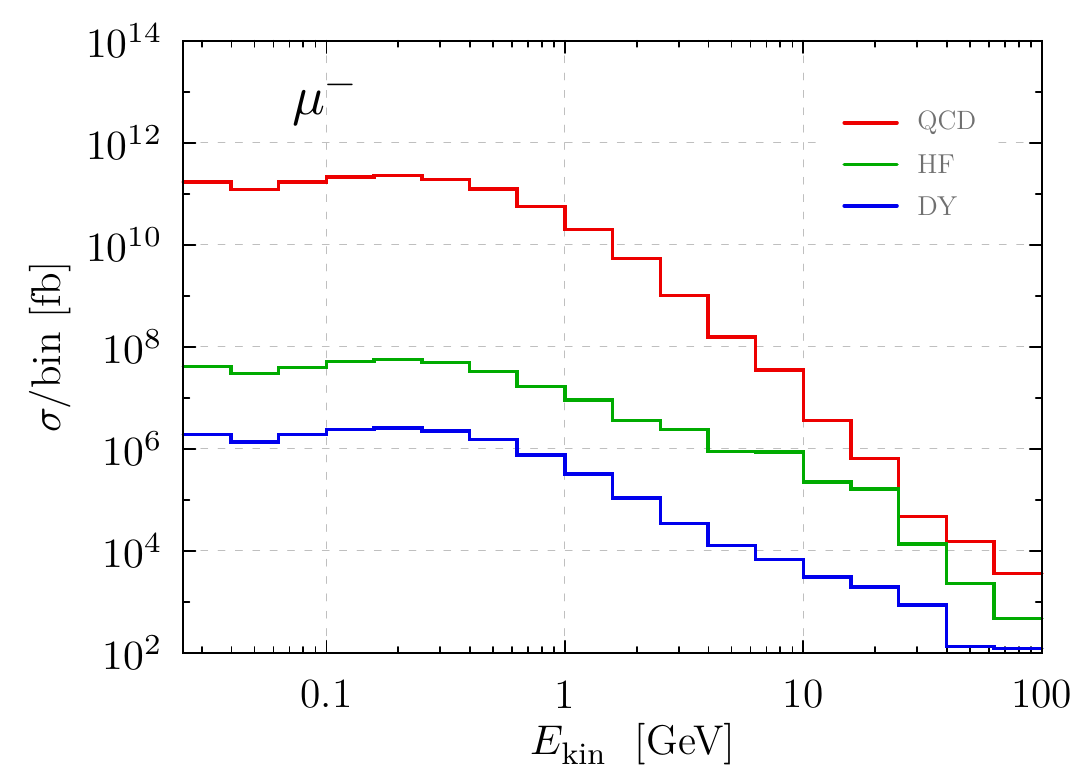}\hfill \\
\hfill
\includegraphics[width = 0.4\linewidth]{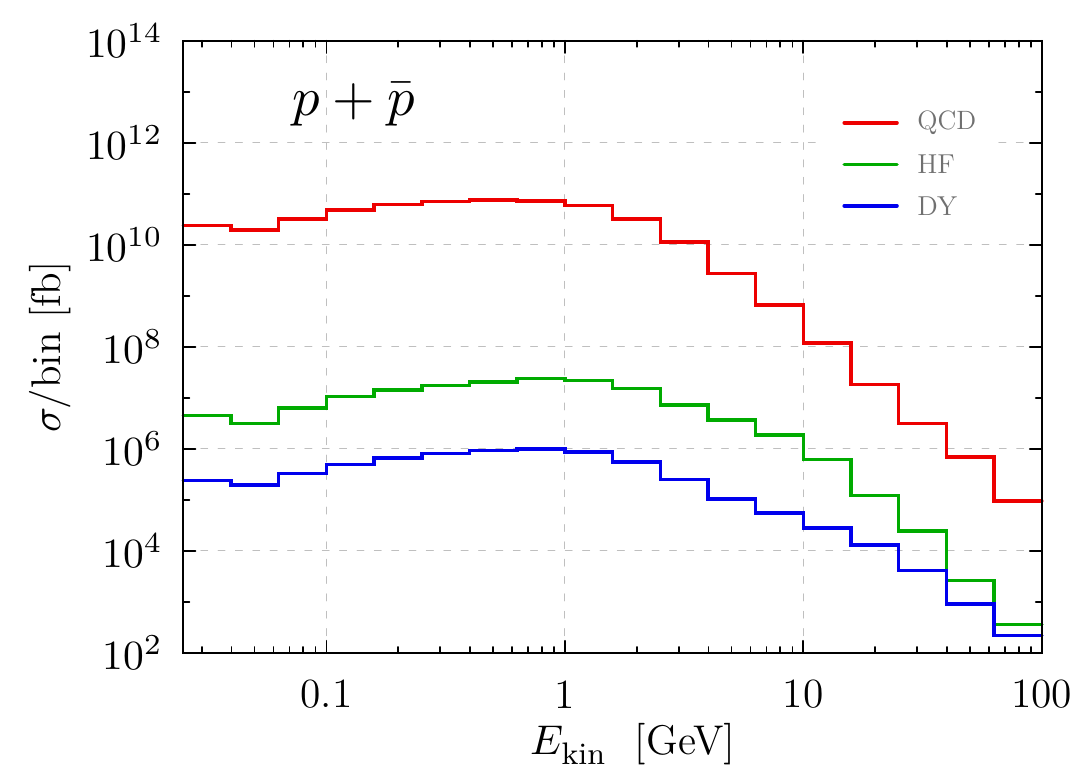} \hfill
\includegraphics[width = 0.4\linewidth]{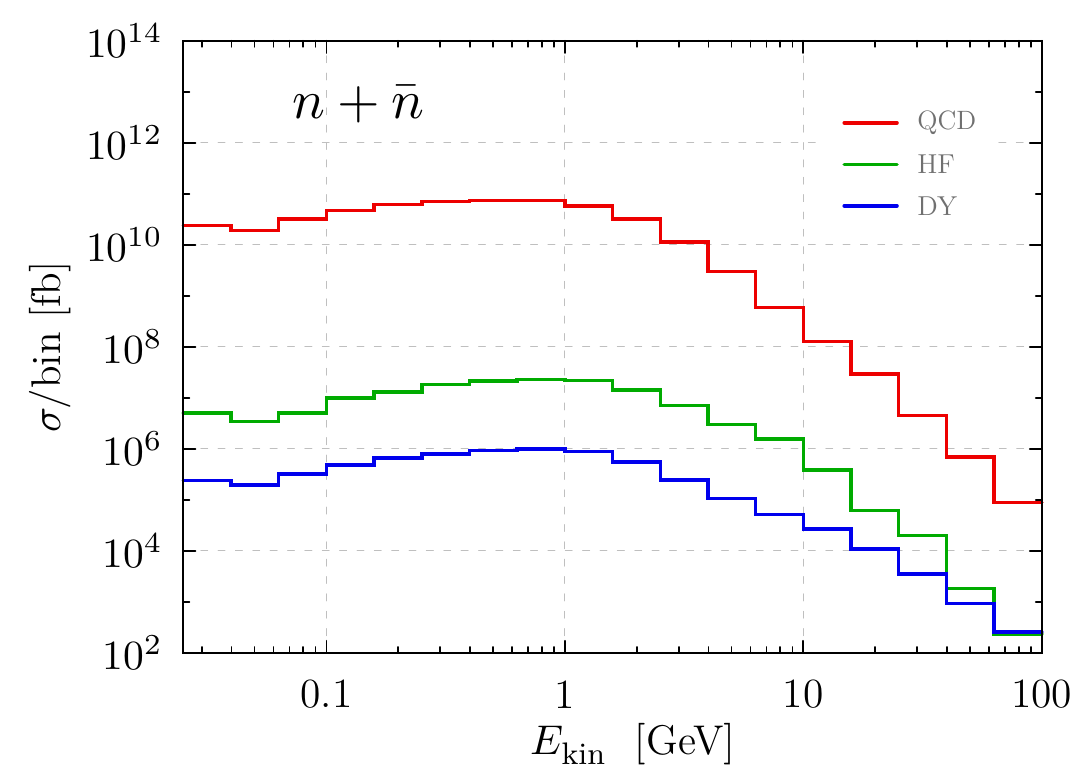} \hfill \\
\hfill
\includegraphics[width = 0.4\linewidth]{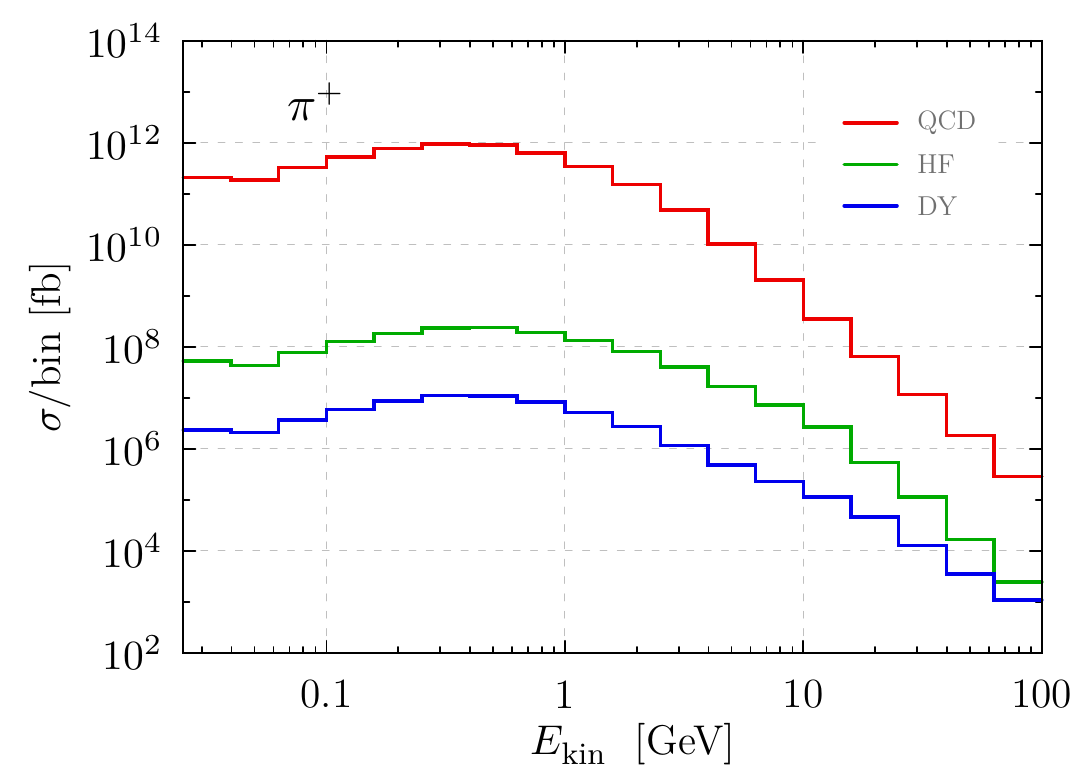}\hfill
\includegraphics[width = 0.4\linewidth]{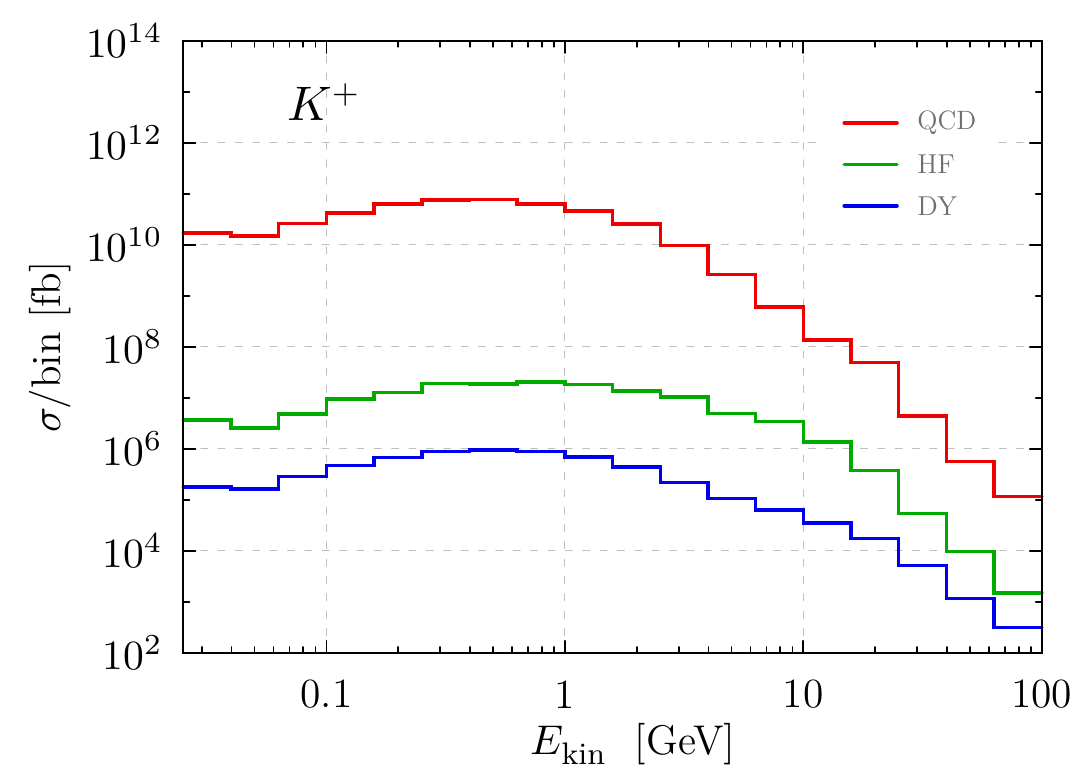} \hfill \\
\hfill
\includegraphics[width = 0.4\linewidth]{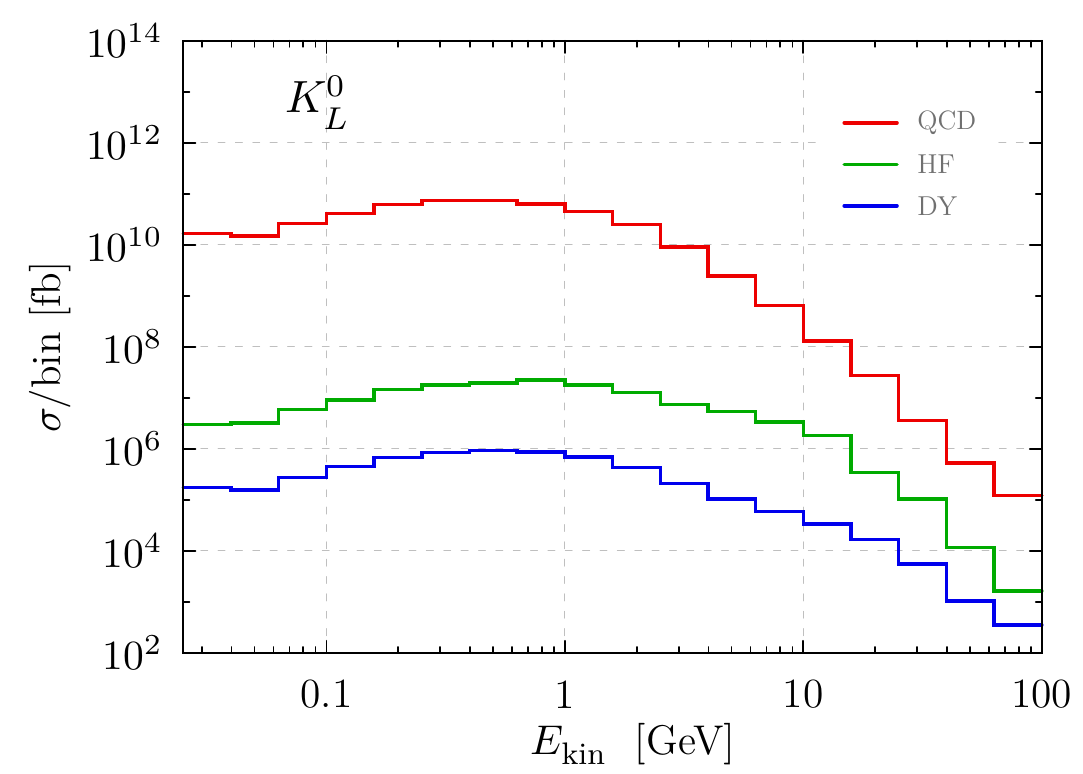} \hfill
\includegraphics[width = 0.4\linewidth]{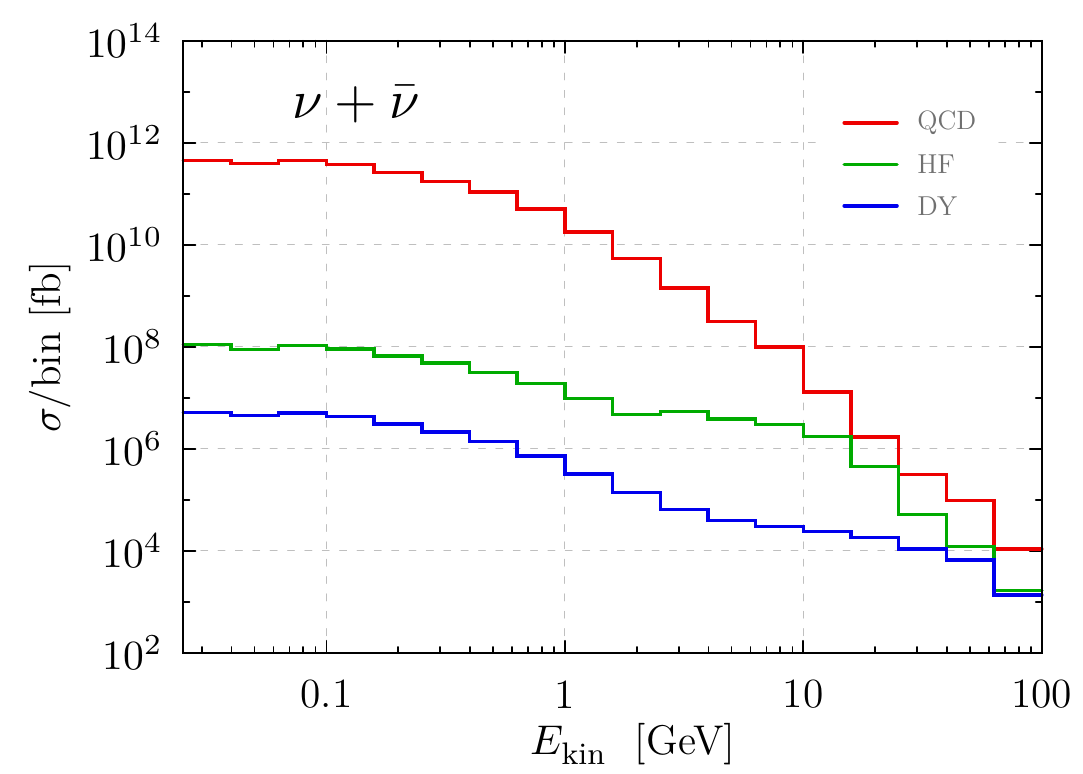} \hfill \\
}
\caption{IP production cross-section per kinetic energy bin, for minimum bias (QCD), Heavy Flavor (HF), and Drell-Yan (DY) production channels.}
\label{fig:IPfluxes}
\end{figure}

Generation of the primary IP fluxes is achieved via simulation of the production of pions, neutral and charged kaons, (anti)muons, (anti)neutrons, (anti)protons and neutrino fluxes
with \texttt{Pythia~8}~\cite{Sjostrand:2006za,Sjostrand:2014zea}. Included production channels span minimum bias (QCD), heavy flavor decays (HF), as well as Drell-Yan production (DY).
Leptons produced from pion decay vertices inside a cylindrical radius $r<5\,$m and  $z < 2$\,m are included.
We simulate weighted \texttt{Pythia~8} events biasing  the primary collisions in $\hat p_T$ in order to achieve approximately flat statistical errors in $\log(\hat s)$ up to $\sqrt{\hat s}$ of a few TeV. 
Under the same procedure we also combine soft and hard QCD processes with a $\hat p_T$ cut of 20~GeV. A similar cut is used to define the HF sample. 
For the DY case we both standard $2\rightarrow2$ Drell-Yan processes and $V+j$, suitably combined to avoid double counting.
In Fig.~\ref{fig:IPfluxes} we show all the relevant generated fluxes, broken down by production channel. In most cases, QCD production dominates, however, 
the HF and DY production can be important for high energy muon tails.

\subsubsection{Simulated shield propagation}
\label{sec:bkg_prop}
Particle propagation and production of secondary backgrounds inside the shield is simulated with \texttt{Geant4} 10.3 using the \texttt{Shielding}~2.1 physics list.
The \texttt{FTFP\_BERT} physics list is used to model high-energy interactions, based on the Fritiof~\cite{ANDERSSON1987289,Andersson1996,NilssonAlmqvist:1986rx,Ganhuyag:1997gz}
and Bertini intra-nuclear cascade~\cite{Guthrie:1968ue,Bertini:1971xb,Karmanov:1979if} models and the standard electromagnetic physics package~\cite{1462617}.

Propagating $\sim 10^{14}$--$10^{17}$ particles though the full shield is obviously computationally prohibitive.
Instead, as in Refs.~\cite{Gligorov:2017nwh} and~\cite{Gligorov:2018vkc} we use a ``particle-gun'' on a shield subelement, typically either $5$ or $2\lambda$ deep for Pb, and $7\lambda$ for concrete.
The subelement geometry is chosen to be a conical section with the same opening angle as the CODEX-b geometric acceptance -- approximately $23^\circ$ -- 
in order to conservatively capture forward-propagating backgrounds after mild angular rescattering.
The particle-gun input and output is binned logarithmically in kinetic energy, in 20 bins from $E_{\text{kin}} = 10^{-1.6}$\,GeV to $100$\,GeV and by particle species, including:
$\gamma$, $e^{\pm}$, $p^{\pm}$, $n^{\pm}$, $\pi^{\pm,0}$, $K^{\pm}$, $K^{0}_{S,L}$, $\mu^{\pm}$, $\nu$. 
Propagation of charged and neutral particles and anti-particles are treated separately for kaons, pions, neutrons and muons.
For each particle-gun energy bin and species, $10^5$ events are simulated; $10^7$ events are simulated for muons and anti-muons to properly capture strange muoproduction of secondary $K^0_L$s.
To also properly capture the `CSDA' or slowing-down behavior of high energy muons when transiting through a large number of shield subelements,
the particle-gun energy for muons was distributed uniformly in kinetic energy, within each bin.

Combining these results together one generates a ``transfer matrix'' between all incoming and outgoing backgrounds fluxes in the shield subelement for each chosen depth and material type.
These transfer matrices may then be further composed together with the primary IP fluxes to obtain the attenuation and response of the full shield.
Neutrino production of neutral hadrons occurs at a prohibitively small rate and is included separately.
As muons may often generate problematic secondaries via forward scattering $\mu \to X\mu$, an additional handle on the capability to veto neutral secondaries is obtained
by keeping track of the presence of any associated charged particles in the particle-gun event, that may trigger relevant veto layers: a charged-neutral correlation.
This `correlation veto' is implemented by an additional binning in outgoing particle kinetic energy \emph{and} the kinetic energy of the hardest charged particle in the event.
This information is then used to generate an additional transfer matrix in which the outgoing particles are produced in association with a charged particle above a chosen kinetic energy threshold.
We conservatively set the correlation veto threshold to be $E_{\text{kin}} > 0.6$\,GeV.
At both the shield veto and detector we apply an additional suppression of neutral secondaries according to their charged-neutral correlation.

In order to incorporate statistical uncertainties in the \texttt{Geant4} simulation, an array of $50$ pseudo-datasets are generated by Poisson-distributing the statistics of each simulated particle-gun event.
Thus in practice one obtains $50$ separate transfer matrix compositions for the shield simulation, from which the statistics of overall shield performance and uncertainties may be extracted.

In Tab.~\ref{tab:bkg-fluxes} we show the results of this simulation for $(20+5)\lambda$ shield configuration made of Pb, plus the $3$\,m concrete UXA wall, 
with a shield veto efficiency of $1-\varepsilon_{\text{veto}} = 10^{-4}$.
For outgoing neutral particle fluxes in Tab.~\ref{tab:bkg-fluxes}, a kinetic energy cut $E_{\text{kin}} > 0.4$\,GeV was applied, as required by minimum tracking requirements to produce at least one track.
Tab.~\ref{tab:bkg-fluxes} includes the background fluxes rejected by the shield veto, both with and without application of the charged-neutral correlation veto in the detector.
The $\sim 60$ neutrons may each produce a single track scattering event along the $10$\,m depth of detector (see Sec.~\ref{sec:tracksBG}).
The neutron incoherent scattering cross-section on air is $\sim 1$\,b, so that the probability of a neutron scattering on air into two tracks along the $10$\,m depth of detector is at most $\sim 5\%$.
Requiring neutrons with $E_{\text{kin}} > 0.8$\,GeV for at least two tracks, results in a total neutron flux of $\sim 3$ per $300$\,fb$^{-1}$, so that the net scattering rate to two or more tracks is $< 1$.
The shield veto rejection rate for the $(20+5)\lambda$ configuration is $\simeq 2.2$~kHz, assuming the projected instantaneous luminosity of $10^{-34}$\,cm$^{-2}$\,s$^{-1}$ at IP8.
(By comparison, for only $15\lambda$ of Pb before the veto, this would increase to $\simeq 6.0$~kHz.)
This is dominated mainly by the incoming muon flux, and is far smaller than the total event rate, and therefore has a negligible effect on the detector efficiency.

One sees in Tab.~\ref{tab:bkg-fluxes} that the shield veto is crucial to achieve a zero background environment.
Moreover, for a low, rather than zero, background environment, the shield veto data provides a data-driven means to calibrate the background simulation,
from which any residual backgrounds in the detector can then be more reliably estimated and characterized.

\begin{table*}[t]
\renewcommand*{\arraystretch}{1.5}
\newcolumntype{C}{ >{\centering\arraybackslash $} m{5cm} <{$}}
\newcolumntype{D}{ >{\centering\arraybackslash $} l <{$}}
\newcolumntype{E}{ >{\centering\arraybackslash $} c <{$}}
\scalebox{0.75}{\parbox{1.25\linewidth}{
\begin{tabular*}{\linewidth}{D|CCC|E}
\hline
 & \multicolumn{3}{c|}{Particle yields} & \\\cline{2-4}
\multirow{2}{*}{ \text{BG species\quad} } & \multirow{2}{*}{ \text{Net ($E^{\text{neutral}}_{\text{kin}} > 0.4$\,GeV)} }  & \text{Shield veto rejection}  & \text{Shield veto rejection} &  \multirow{2}{*}{ \text{Net yield} } \\
& & \text{(total)} & \text{($\pm$/$0$~correlation)} & \\
 \hline \hline
 \gamma	  & 	0.54 \pm 0.12	  & 	(8.06 \pm 0.60) \times 10^{4}	  & 	(2.62 \pm 1.03) \times 10^{3}	& 	\text{--}	\\
n	  & 	58.10 \pm 4.63	  & 	(4.59 \pm 0.15) \times 10^{5}	  & 	(3.44 \pm 0.51) \times 10^{4}	& 	\text{--}	\\
n~\text{($ > 0.8$\,GeV)}	  & 	2.78 \pm 0.25	  & 	(1.03 \pm 0.06) \times 10^{5}	  & 	(7.45 \pm 1.92) \times 10^{3}	& 	\lesssim 1	\\
\bar{n}~\text{(no cut)}	  & 	(3.24 \pm 0.72) \times 10^{-3}	  & 	34.40 \pm 25.80	  & 	(7.12 \pm 2.19) \times 10^{-2}	& 	\ll 1	\\
K^0_L	  & 	0.49 \pm 0.05	  & 	(1.94 \pm 0.74) \times 10^{3}	  & 	54.40 \pm 19.20	& 	\lesssim 0.1	\\
K^0_S	  & 	(6.33 \pm 1.39) \times 10^{-3}	  & 	93.90 \pm 45.80	  & 	0.74 \pm 0.19	& 	\ll 1	\\
\nu + \bar{\nu}	  & 	(5.69 \pm 0.00) \times 10^{13}	  & 	(7.35 \pm 0.12) \times 10^{6}	  & 	(7.31 \pm 0.11) \times 10^{6}	& 	\text{--}	\\
\hline
p^{\pm}	  & 	(2.07 \pm 0.26) \times 10^{2}	  & 	(9.24 \pm 0.36) \times 10^{5}	  & 	(9.24 \pm 0.36) \times 10^{5}	& \text{--} \\
e^{\pm}	  & 	(4.53 \pm 0.02) \times 10^{3}	  & 	(4.38 \pm 0.02) \times 10^{7}	  & 	(4.38 \pm 0.02) \times 10^{7}	& \text{--} \\
\pi^+	  & 	34.70 \pm 2.27	  & 	(2.96 \pm 0.20) \times 10^{5}	  & 	(2.96 \pm 0.20) \times 10^{5}	& \text{--} \\
\pi^-	  & 	31.40 \pm 2.12	  & 	(2.68 \pm 0.19) \times 10^{5}	  & 	(2.68 \pm 0.19) \times 10^{5}	& \text{--} \\
K^+	  & 	0.83 \pm 0.30	  & 	(3.08 \pm 1.24) \times 10^{3}	  & 	(3.08 \pm 1.24) \times 10^{3}	& \text{--} \\
K^-	  & 	0.23 \pm 0.12	  & 	(1.12 \pm 0.63) \times 10^{3}	  & 	(1.12 \pm 0.63) \times 10^{3}	& \text{--} \\
\mu^+	  & 	(1.04 \pm 0.00) \times 10^{6}	  & 	(1.04 \pm 0.00) \times 10^{10}	  & 	(1.04 \pm 0.00) \times 10^{10}	& \text{--} \\
\mu^-	  & 	(8.07 \pm 0.01) \times 10^{5}	  & 	(8.07 \pm 0.01) \times 10^{9}	  & 	(8.07 \pm 0.01) \times 10^{9}	& \text{--} \\
\hline
\end{tabular*}
}}
\caption{
Results from the \texttt{Geant4} background simulation for $(20 + 5)\lambda$\,Pb shield, i.e. with an active shield veto at $20\lambda$, applying a veto efficiency of $1-\varepsilon_{\text{veto}} = 10^{-4}$.
For outgoing neutral particles a kinetic energy cut $E_{\text{kin}} > 0.4$\,GeV was applied as required by minimum tracking requirements, 
except for anti-neutrons in order to exclude $\bar{n} + N$ annihilation processes.
We also show the rate for neutrons with $E_{\text{kin}} > 0.8$\,GeV, required for production of at least two tracks via scattering.
For total luminosity $\mathcal{L} = 300$fb$^{-1}$, the column ``Net ($E^{\text{neutral}}_{\text{kin}} > 0.4$\,GeV)'' shows the net background particle yields after traversing the shield plus veto rejection,
including veto correlations (denoted `$\pm/0$') between charged particles with $E_{\text{kin}} > 0.6$\,GeV and neutral particles.
The column ``Shield veto rejection (total)'' shows the corresponding background particle yields entering the detector subject to the shield veto rejection alone, 
without applying the charged-neutral correlation veto.
The column  ``Shield veto rejection ($\pm$/$0$~correlation)'' shows the corresponding background particle yields entering the detector  subject to the shield veto rejection, 
after application of the charged-neutral correlation veto on the detector front face. 
The final column lists the net background yield including detector rejection, scattering or decay probabilities.}
\label{tab:bkg-fluxes}
\end{table*}

\begin{figure}[p]
\centering{\hfill
\includegraphics[width = 0.4\linewidth]{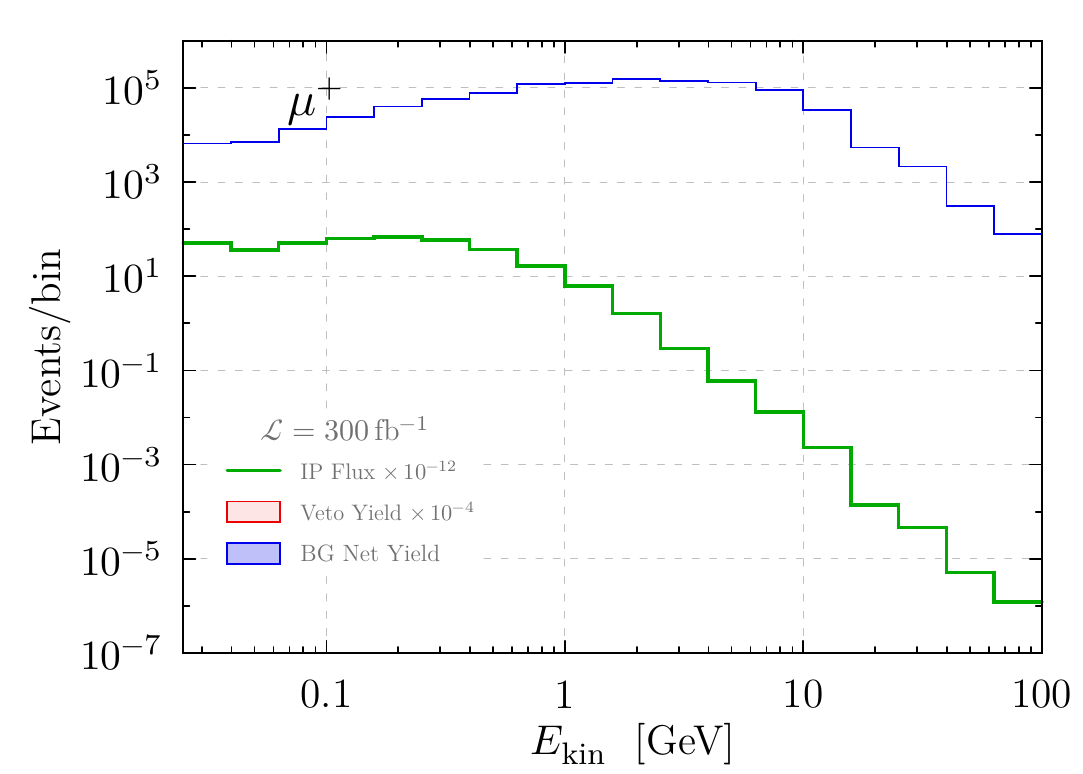}\hfill
\includegraphics[width = 0.4\linewidth]{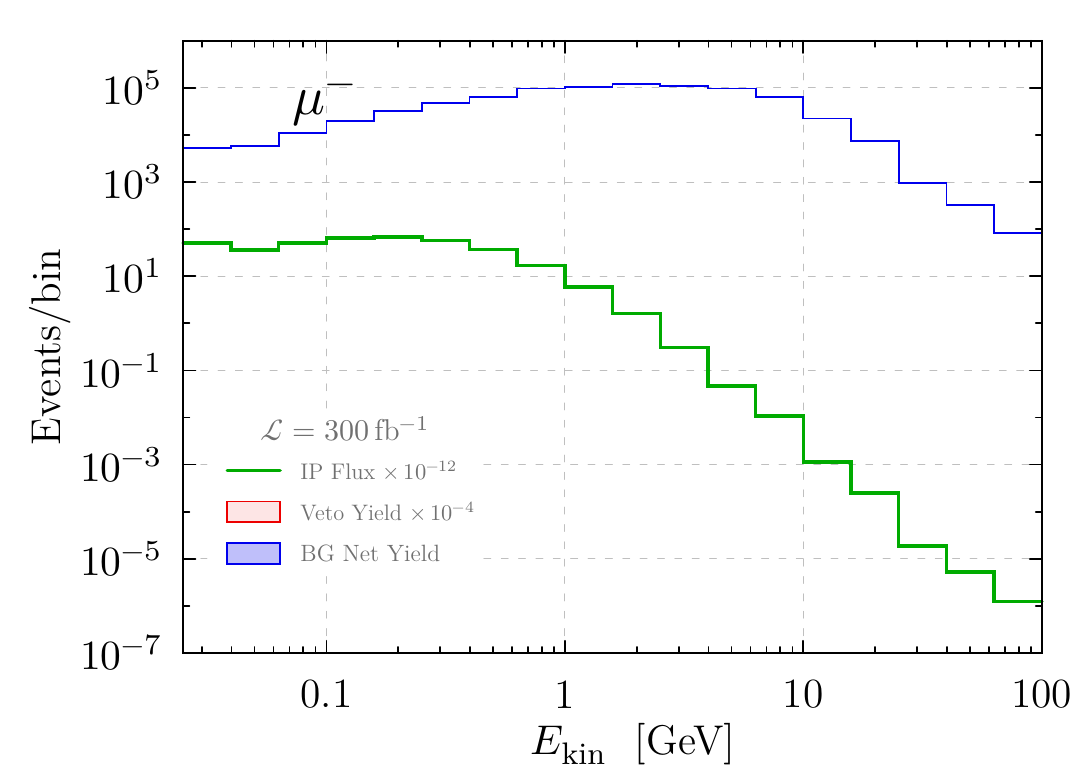}\hfill \\
\hfill
\includegraphics[width = 0.4\linewidth]{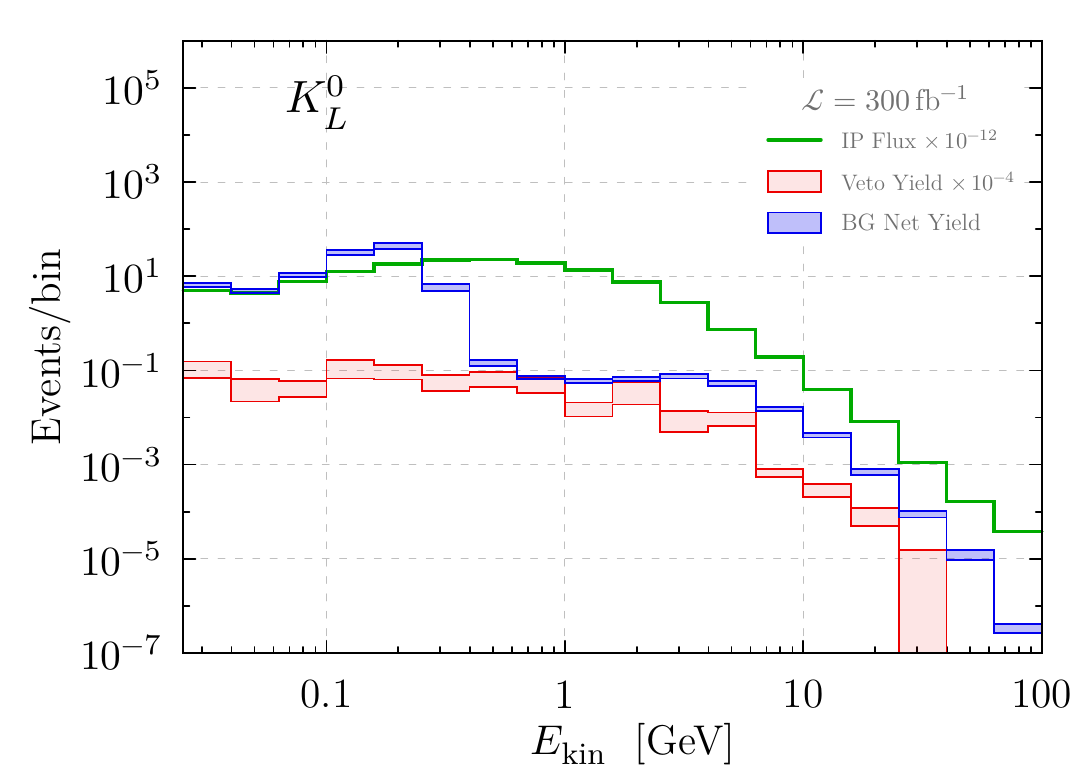} \hfill
\includegraphics[width = 0.4\linewidth]{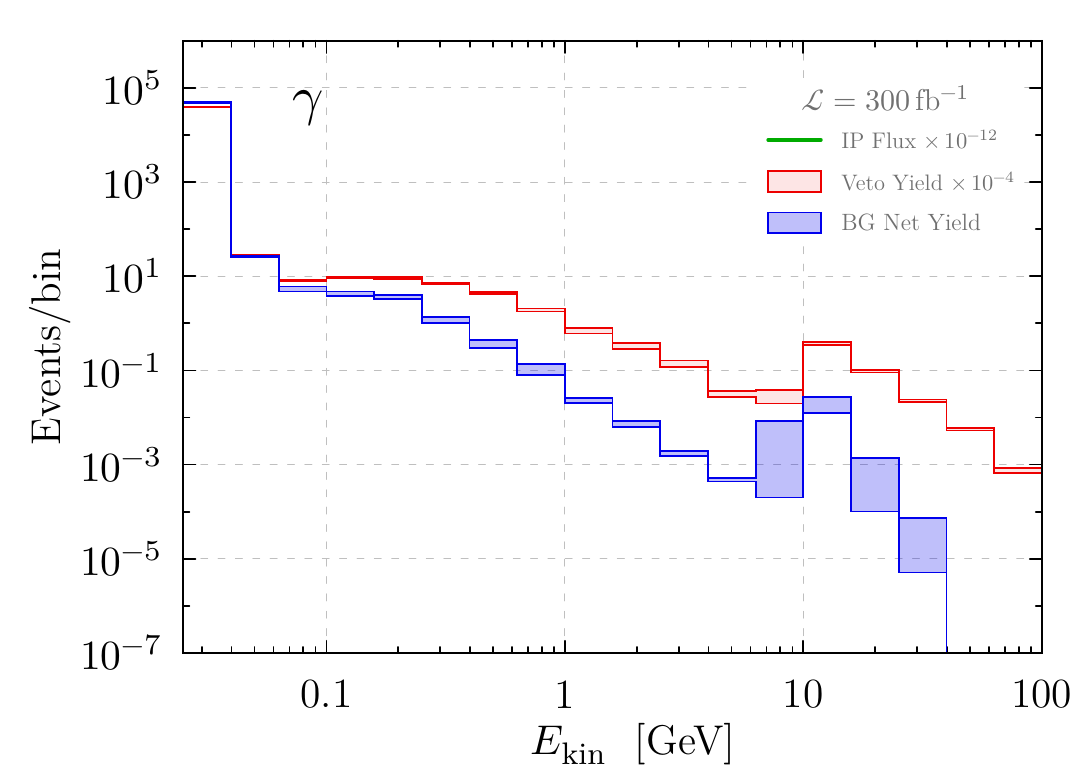} \hfill \\
\hfill
\includegraphics[width = 0.4\linewidth]{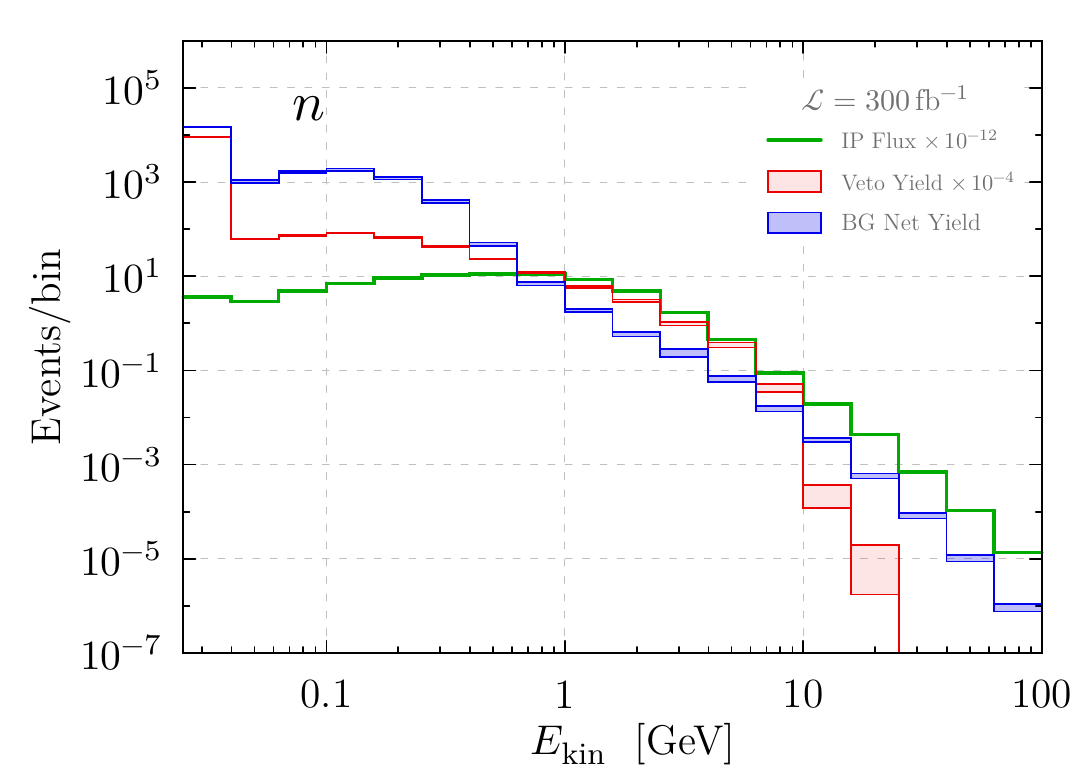} \hfill
\includegraphics[width = 0.4\linewidth]{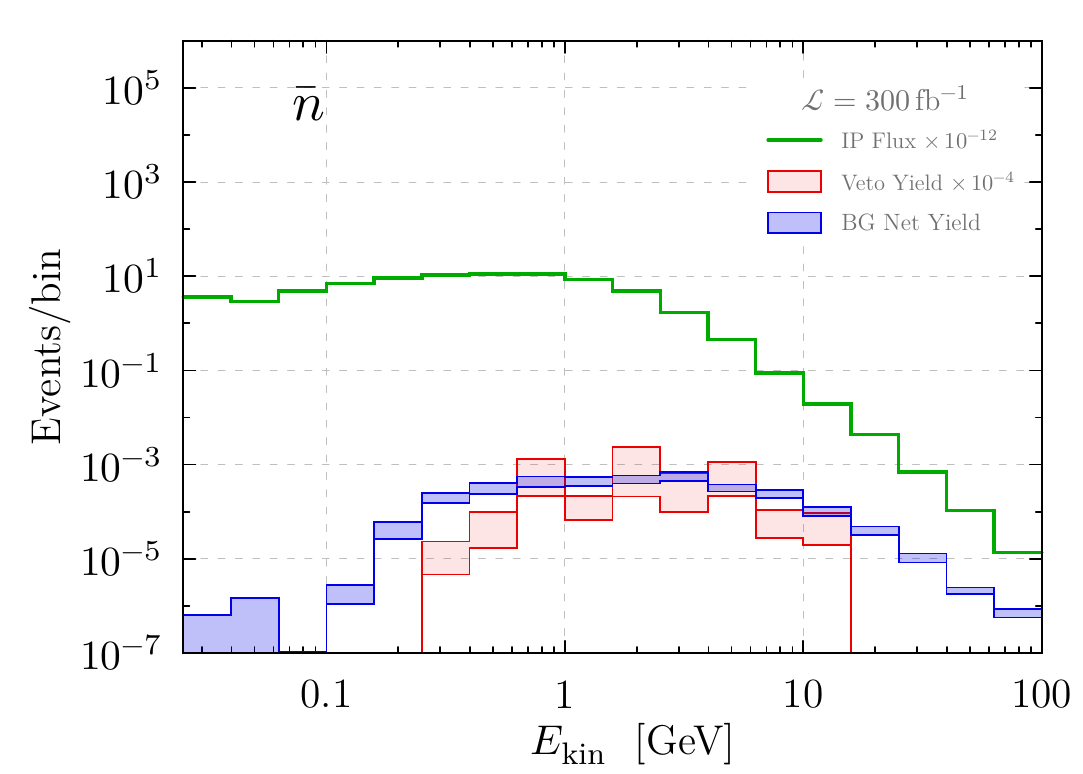} \hfill \\
\hfill
\includegraphics[width = 0.4\linewidth]{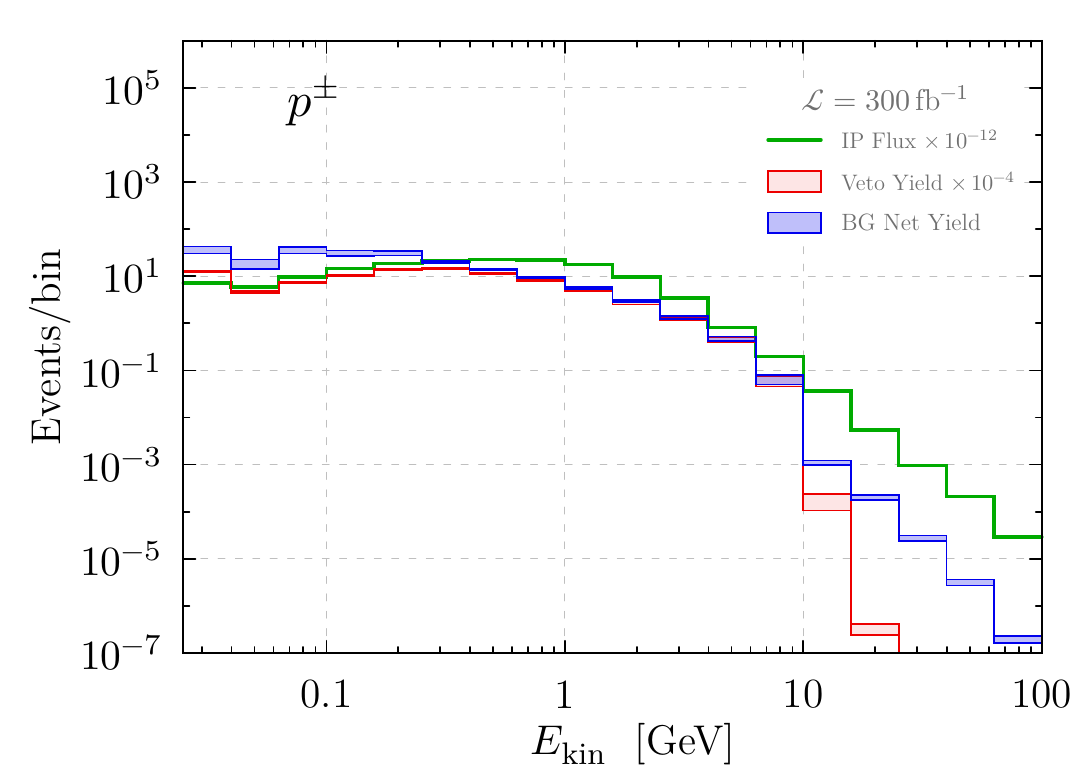} \hfill
\includegraphics[width = 0.4\linewidth]{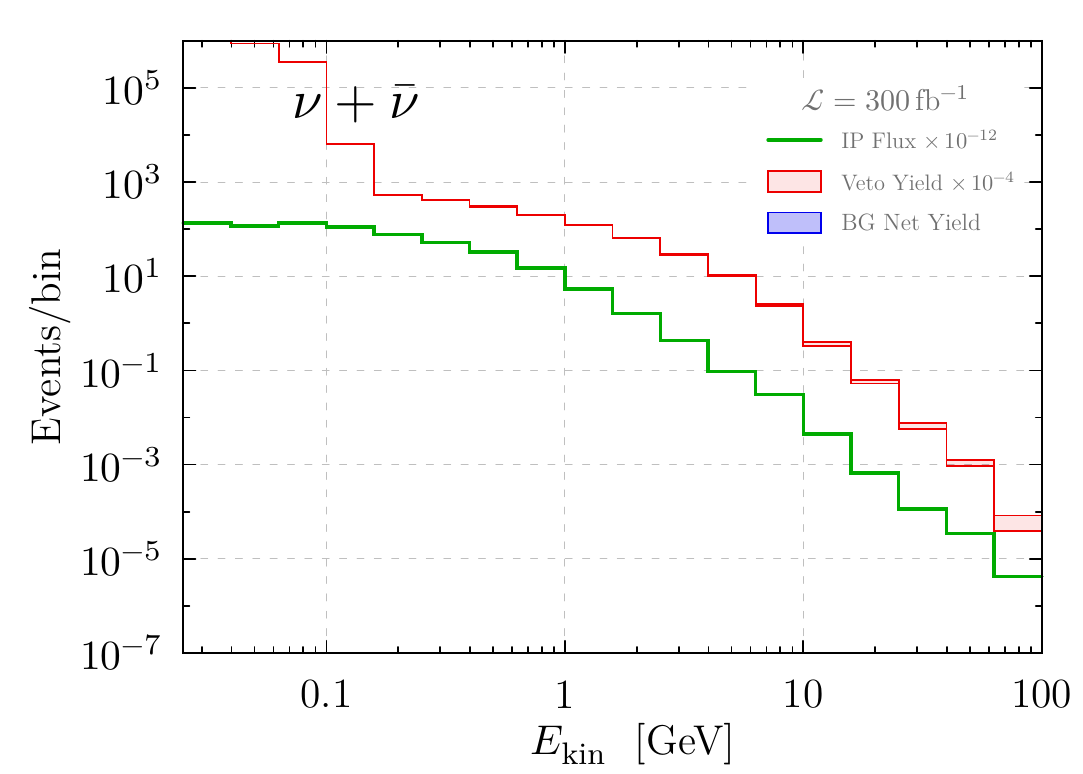} \hfill \\
}
\caption{Background fluxes per kinetic energy bin, comparing primary IP fluxes $\times 10^{-12}$ (green) with the irreducible background flux (blue) after the $(20+5)\lambda$ passive and active shield. 
Also shown are background  fluxes $\times 10^{-4}$ entering the detector that are rejected by the shield veto (red). }
\label{fig:BGfluxes}
\end{figure}

In Fig.~\ref{fig:BGfluxes} we show the net background flux distributions in kinetic energy (blue) for a variety of neutral and charged species, including uncertainties, and without any $E_{\text{kin}}$ cuts.
These may be compared to the background flux distributions of particles reaching the detector that are rejected by the shield veto (red, with $10^{-4}$ scaling) and the IP fluxes (green, with $10^{-12}$ scaling).

\FloatBarrier

\subsubsection{Neutrinos}
\label{sec:bkg-neutrinos}
An additional background may arise through production of neutral secondaries from neutrinos, that stream through the shield unimpeded.
In particular, with a sufficiently high neutrino flux, $\bar\nu p \to \ell n$ quasi-elastic scattering may produce a non-negligible amount of neutrons in the last few $\lambda$ that reach the detector volume,
while the charged lepton is too soft or misses the acceptance.
(The cross-section for the neutral current scattering $\nu n \to \nu n$ or $\bar\nu n \to \bar\nu n$ is approximately $10$ times smaller than for the charged current process~\cite{Formaggio:2013kya}.)
From Tab.~\ref{tab:bkg-fluxes} and Fig.~\ref{fig:BGfluxes}, approximately $5\times10^{13}$ neutrinos are produced per $300$\,fb$^{-1}$ at IP8 in the \CODEXb acceptance, with $E_\nu > 0.4$\,GeV.
The neutrino flux is approximately power-law suppressed by a quartic above $E_\nu \sim 1$\,GeV.
Hence, although the charged current cross-section for $\bar\nu p \to \ell n$ is only $\sim 0.01(E_\nu/\text{GeV})$\,pb~\cite{Formaggio:2013kya},
the large flux of $\mathcal{O}(\text{GeV})$ neutrinos streaming through the shield implies as many as $\sim 10$ neutrons might be generated per $\lambda$ of concrete in the UXA wall with $E_{\text{kin}} > 0.4$\,GeV.
Neutral kaon production, such as $\nu n \to \nu K_L^0 \Lambda$, has a cross-section $\sim 0.1$\,fb for $E_\nu \sim 3.5$\,GeV~\cite{Marshall:2016yho}, scaling approximately linearly with neutrino energy, 
and therefore may be safely neglected.

Composing the IP flux of anti-neutrinos in Fig.~\ref{fig:IPfluxes} with $E_\nu > 0.4$\,GeV against the measured energy-dependent $\bar\nu p \to \ell n$ cross-section~\cite{Formaggio:2013kya},
and including subsequent attenuation as characterized by the nuclear interaction length $\lambda$, one may estimate a conservative upper bound on the number of neutrons
that might reach the detector with $E_{\text{kin}} > 0.4$\,GeV.
From this procedure one finds that there are at most approximately $5$ neutrino-produced neutrons of this type. Since this estimate is extremely conservative,
neutron production from neutrinos is expected to be negligible compared to secondary neutrons from other primary fluxes.

\subsubsection{Shielding marginal performance}
\begin{figure}[t]
\centering{
\includegraphics[width = 0.65\linewidth]{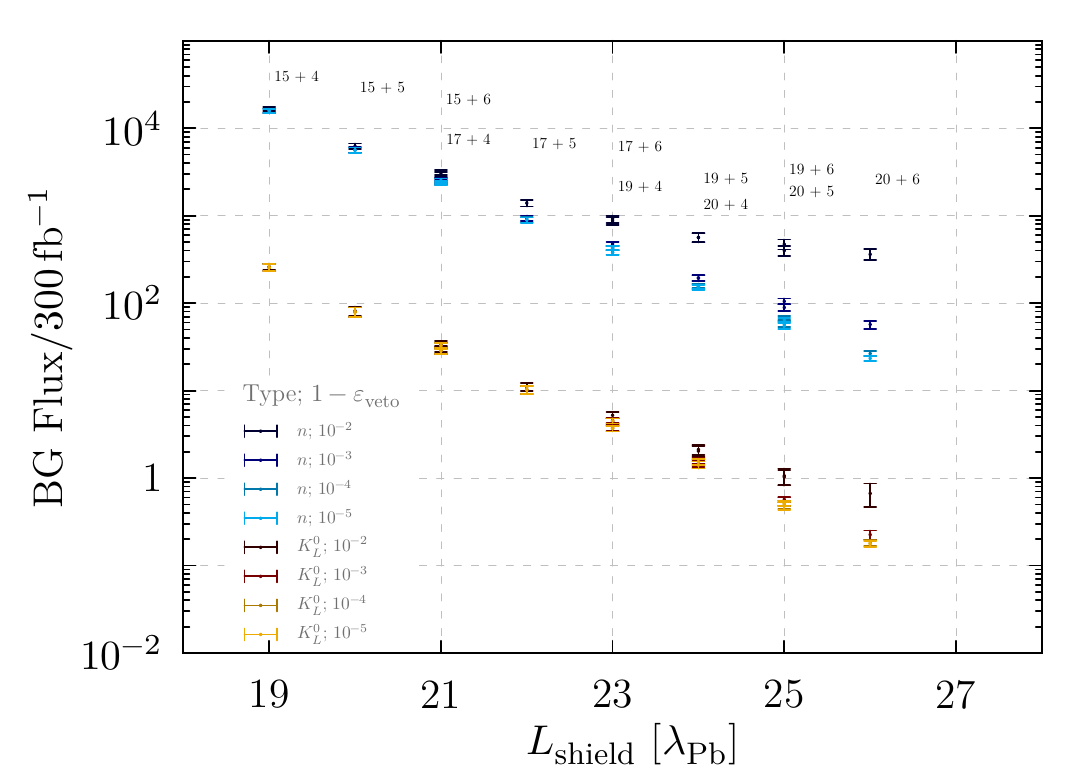}
\includegraphics[width = 0.65\linewidth]{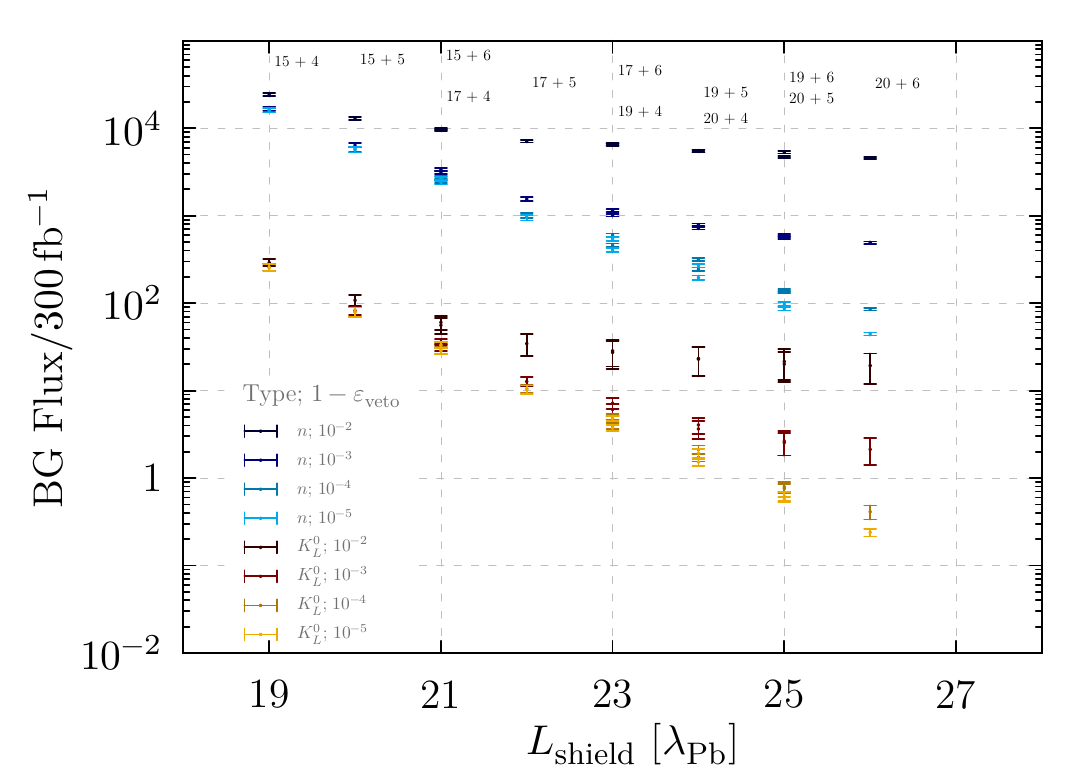}
}
\caption{Neutron (black-blue palette) and $K_L^0$ (red-yellow palette) background fluxes with kinetic energy $E_{\text{kin}} > 0.4$\,GeV versus total Pb shield depth in $\lambda_{\text{Pb}}$, 
under variation of the shield configuration and veto efficiency, including the charged-neutral correlation veto (top) and without the charged-neutral correlation veto (bottom). 
For each value of the total shield depth, the (possibly multiple) corresponding configurations ``$L_{\text{pre-veto}} + L_{\text{post-veto}}$'' are shown by the adjacent labels in units of $\lambda_{\text{Pb}}$.}
\label{fig:shld_mrg}
\end{figure}

As the detector tolerance of backgrounds may vary depending on the ultimately implemented detector technologies, 
it is instructive to assess the performance of the shield under variation of the shielding configuration,
including variation of the total shield depth, $L_{\text{shield}}$, and the placement and efficiency of the shield veto.
We illustrate the marginal changes in shielding performance in terms of the total neutron and $K_L^0$ fluxes for kinetic energy $E_{\text{kin}} > 0.4$\,GeV,
by varying the shield configuration as combinations of the $5$ or $2\lambda$ transfer matrices permit, 
and by permitting the veto efficiency to range from $1-\varepsilon_{\text{veto}} =10^{-5}$ up to $10^{-2}$.
In the top panel of Fig.~\ref{fig:shld_mrg}, the corresponding variation of the neutron (black-blue palette) and $K_L^0$ (red-yellow palette) background fluxes are shown, taking all combinations of
$L_{\text{pre-veto}} \in \{15,17,19,20\}\lambda$ and $L_{\text{post-veto}} \in \{4,5,6\}\lambda$, as defined in Fig.~\ref{fig:shld_cnfg}, and $1-\varepsilon_{\text{veto}} \in \{10^{-2}, 10^{-3}, 10^{-4}, 10^{-5}\}$.

The simulated background fluxes are generally insensitive to marginal variation in the location of the shield veto, e.g.~one sees that $(19+4)\lambda$ performs similarly to the nominal $(20+5)\lambda$ configuration.
At very high efficiencies, i.e. $1-\varepsilon_{\text{veto}} < 10^{-4}$, both background fluxes are roughly exponentially distributed in shield depth.
In this case the backgrounds are either unsuppressed primaries or stopped-parent secondaries produced upstream from the shield.

As the shield veto efficiency is reduced, however, one sees a departure from the exponential suppression: 
Contributions from stopped-parent secondaries produced downstream from the shield veto begin to dominate.
For the $K_L^0$ flux, this departure occurs only at $1-\varepsilon_{\text{veto}} >10^{-2}$ and at a larger $L_{\text{shield}}$, compared to the neutrons.
This arises because of a somewhat larger charged-neutral correlation for production of $K_L^0$'s: Their parent muons are typically somewhat hard and may reach the veto or detector.

\begin{figure}[t]
\centering{
\includegraphics[width = 0.65\linewidth]{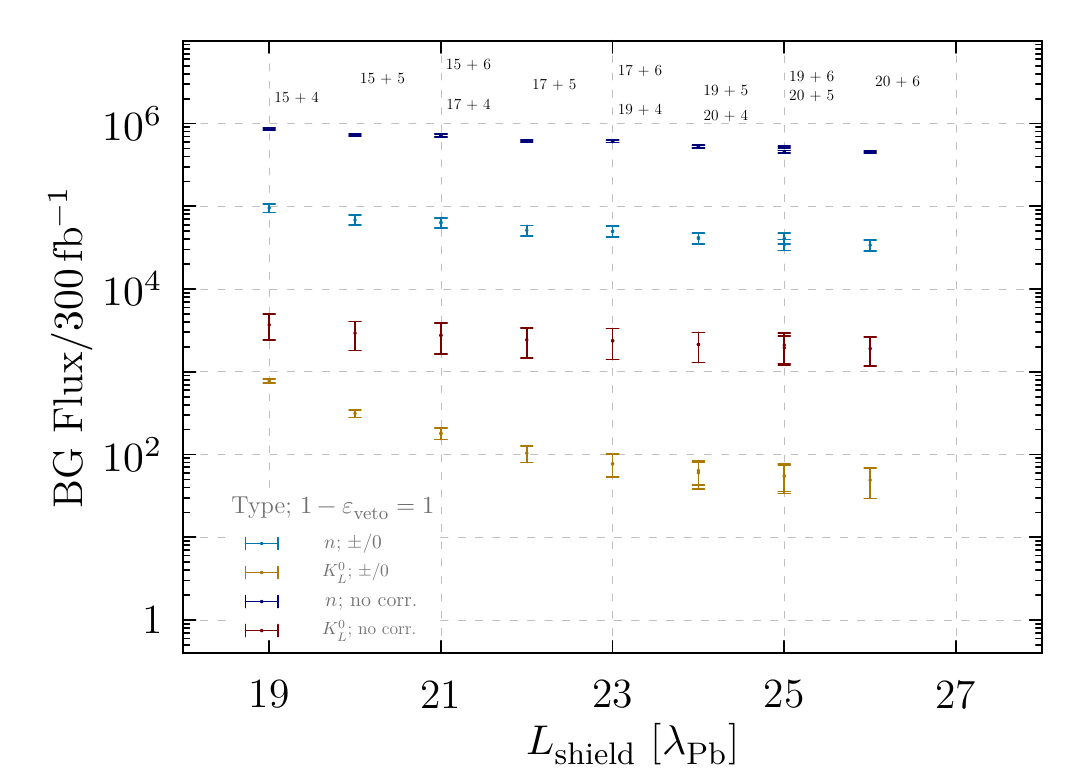}
}
\caption{Variation of neutron and $K_L^0$ background fluxes with kinetic energy $E_{\text{kin}} > 0.4$\,GeV versus total Pb shield depth in $\lambda_{\text{Pb}}$ under variation of the shield configuration 
with no shield veto ($1-\varepsilon_{\text{veto}} = 1$), including the charged-neutral correlation veto in the detector (light palette) and without the charged-neutral correlation veto in the detector (dark palette). 
For each value of the total shield depth, the (possibly multiple) corresponding configurations ``$L_{\text{pre-veto}} + L_{\text{post-veto}}$'' are shown by the adjacent labels in units of $\lambda_{\text{Pb}}$.}
\label{fig:shld_mrg_sv}
\end{figure}

One may assess the degree of this effect -- effectively, the amount of non-stopped parent secondaries -- by considering the case that the charged-neutral correlation veto is not applied
(in practice, it may happen that the associated charged particles do not always trigger the veto). We show the corresponding shielding performance in the bottom panel of Fig.~\ref{fig:shld_mrg}.
For $1-\varepsilon_{\text{veto}} >10^{-3}$, the background fluxes become substantially larger.
One deduces that, especially for $K_L^0$s, charged parents that produce secondaries downstream of the shield may typically reach the detector.
In Fig.~\ref{fig:shld_mrg_sv} we show the neutron and $K_L^0$ fluxes in the absence of a shield veto ($1-\varepsilon_{\text{veto}} = 1$),
both with (light palette) and without (dark palette) the charged-neutral correlation veto in the detector (denoted `$\pm/0$').
One sees that while the charged-neutral correlation veto can significantly reduce the total fluxes, substantial net backgrounds for $K_L^0$s and neutrons remain.

\subsubsection{Simulated track production\label{sec:tracksBG}}
The simulated neutral background fluxes entering the detector may be folded against the probability of scattering into one or more tracks on material inside the detector,
or against the probability of decays into one or more tracks in the detector interior:
Nominally the number of two or more tracks should be $<1$ to ensure a background-free environment.
In Tab.~\ref{tab:bkg-tracks} we show the rates of multitrack production for $1$--$9$ tracks from scattering of the neutral fluxes on air in the $10\times 10 \times 10$\,m$^3$ detector volume,
for the nominal $(20+5)\lambda$ Pb shield configuration with $1-\varepsilon_{\text{veto}} = 10^{-4}$.
This production is simulated with \texttt{Geant4} as in Sec.~\ref{sec:bkg_prop}, requiring each track to have kinetic energy $E_{\text{kin}} > 0.4$\,GeV.

For total luminosity $\mathcal{L} = 300$fb$^{-1}$, one sees that the total number of scatterings or decays into two or more tracks is $\simeq 0.22 \pm 0.03$.
This comports with our simulation and estimation of the background effective yields in Tab.~\ref{tab:bkg-fluxes}.

\begin{table*}[t]
\renewcommand*{\arraystretch}{1.5}
\newcolumntype{C}{ >{\centering\arraybackslash $} m{4.25cm} <{$}}
\newcolumntype{E}{ >{\centering\arraybackslash $} c <{$}}
\begin{tabular*}{\linewidth}{@{\extracolsep{\fill}}E|C|CC}
\hline
 \multirow{2}{*}{Tracks} & (20 +5)\lambda~\text{Pb shield}  & \text{Run~3 (\CODEXbeta)} & \text{Run~3 (\CODEXbeta)}\\
& 1- \varepsilon_{\text{veto}} = 10^{-4} & & K_L^0~\text{contribution}\\
\hline\hline
1	&	53.90 \pm 5.51	&	(3.87 \pm 0.11) \times 10^{8}	&	(2.94 \pm 0.07) \times 10^{8}	 \\
2	&	0.21 \pm 0.02	&	(4.09 \pm 0.13) \times 10^{7}	&	(3.74 \pm 0.13) \times 10^{7}	 \\
3	&	(1.36 \pm 0.34) \times 10^{-2}	&	(5.96 \pm 1.01) \times 10^{5}	&	(2.92 \pm 0.45) \times 10^{5}	 \\
4	&	(1.51 \pm 0.30) \times 10^{-3}	&	(6.78 \pm 1.22) \times 10^{4}	&	(5.12 \pm 1.19) \times 10^{4}	 \\
5	&	(3.80 \pm 0.87) \times 10^{-4}	&	(1.69 \pm 0.50) \times 10^{4}	&	(1.42 \pm 0.50) \times 10^{4}	 \\
6	&	(1.09 \pm 0.27) \times 10^{-4}	&	(3.23 \pm 0.79) \times 10^{3}	&	(2.21 \pm 0.79) \times 10^{3}	 \\
7	&	(1.84 \pm 1.41) \times 10^{-4}	&	(4.23 \pm 2.30) \times 10^{3}	&	(1.75 \pm 0.77) \times 10^{3}	 \\
8	&	(2.98 \pm 1.31) \times 10^{-5}	&	(1.04 \pm 0.63) \times 10^{3}	&	(8.45 \pm 6.11) \times 10^{2}	 \\
9	&	(1.07 \pm 0.33) \times 10^{-5}	&	(2.41 \pm 0.43) \times 10^{2}	&	(1.37 \pm 0.35) \times 10^{2}	 \\
\hline
\end{tabular*}
\caption{Multitrack production on air from the \texttt{Geant4} background simulation for the $(20 + 5)\lambda$\,Pb shield,
in the $10\times 10 \times 10$\,m$^3$ detector volume for total luminosity $\mathcal{L} = 300$fb$^{-1}$, requiring $E_{\text{kin}} > 0.4$\,GeV per track.
Also shown are corresponding rates for total neutral and $K_L^0$ multitrack production during Run~3 in the \CODEXbeta volume for total luminosity $\mathcal{L} = 15$fb$^{-1}$ (see Sec.~\ref{sec:demonstrator}). }
\label{tab:bkg-tracks}
\end{table*}

\subsection{Measurement campaign\label{sec:measurementcampaign}}

To verify the background simulation, a data-driven calibration is needed, using data taken during collisions at IP8.
This section serves as a brief summary of an initial measurement campaign that was undertaken during Run~2 operations at various locations in the UXA cavern,
shielded only by the UXA wall, in August 2018. 
For a detailed description we refer to Ref.~\cite{Dey:2019vyo}, which also features additional detailed background simulations, 
including the effects of infrastructure in the LHCb cavern.

The detector setup used scintillators, light-guides and photomultiplier tubes (PMT) from the HeRSCheL experiment~\cite{CarvalhoAkiba:2018paq} at LHCb. 
The detector itself consisted of two parallel scintillator plates with surface area $300\times300$~mm$^2$: Details can be found in Ref.~\cite{Dey:2019vyo}. 
Before transporting the setup to IP8 it was tested with cosmic rays, indicating an efficiency $>95\%$ for minimum ionizing particles (MIPs), 
i.e. for those particles with kinetic energy $\gtrsim 100$\,MeV. 
In the simulation of Ref.~\cite{Dey:2019vyo}, it was further verified that no collision event produced more than one hit on the scintillator acceptance, 
such that pile up of hits within the trigger window can be ruled out.




A two-fold coincidence between both scintillator plates was required to trigger the detector, 
and the trigger was not synchronized with the collisions. 
However, the coincidence time-window was $5$\,ns, much shorter than the $25$\,ns collision frequency at IP8, so that spill-over effects can be neglected. 
Two waveforms were recorded from each scintillator and the timestamp for all MIP hits.
This timestamp was used to correlate the events with the beam status during data-taking.

\begin{figure}
\centering
    \includegraphics[width=11.5cm]{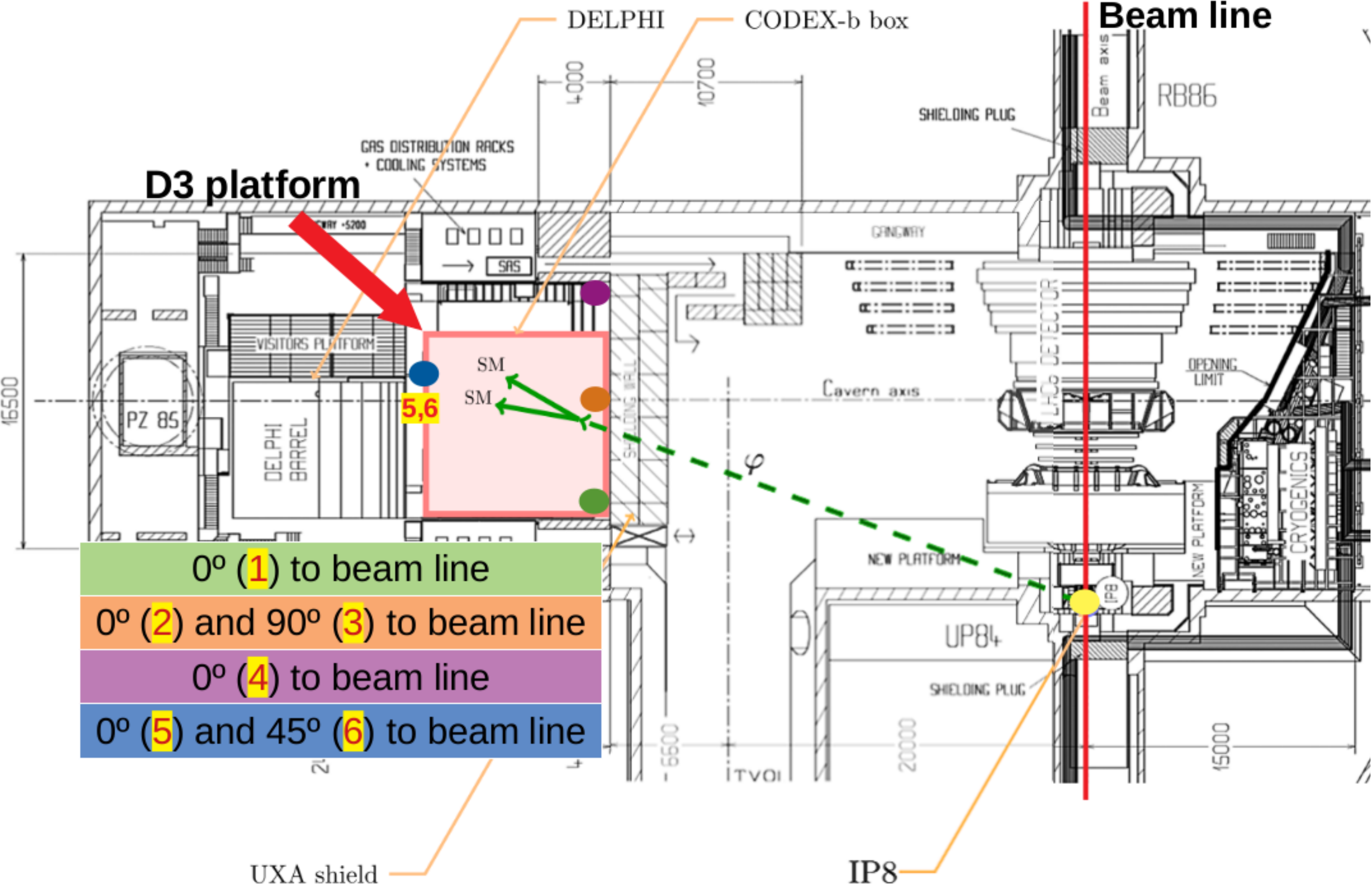}
\caption{\label{fig:posconfig}
    The four measurement locations on the D3 level in the LHCb cavern, shown by red, orange, green and blue dots. The configurations are labelled P1--P6. Figure reproduced and modified from Ref.~\cite{Dey:2019vyo}, as adapted from Ref.~\cite{Gligorov:2017nwh}.
}
\end{figure}

The measurements were taken on the ``D3 platform'' level in the UXA hall, behind the concrete UXA wall.
Fig.~\ref{fig:posconfig} shows this platform, and the different locations and configurations used for the data-taking.
The detector was deployed at three different positions on the passerelle between Data Acquisition (DAQ) server racks and the UXA wall, 
as well as at one location between the DELPHI barrel exhibit and the DAQ racks.
The scintillator stand was oriented either parallel (`$\parallel$'), rotated $45^{\circ}$ or perpendicular (`$\perp$') to the beam line.


\begin{figure}
\centering
    \includegraphics[width=\textwidth]{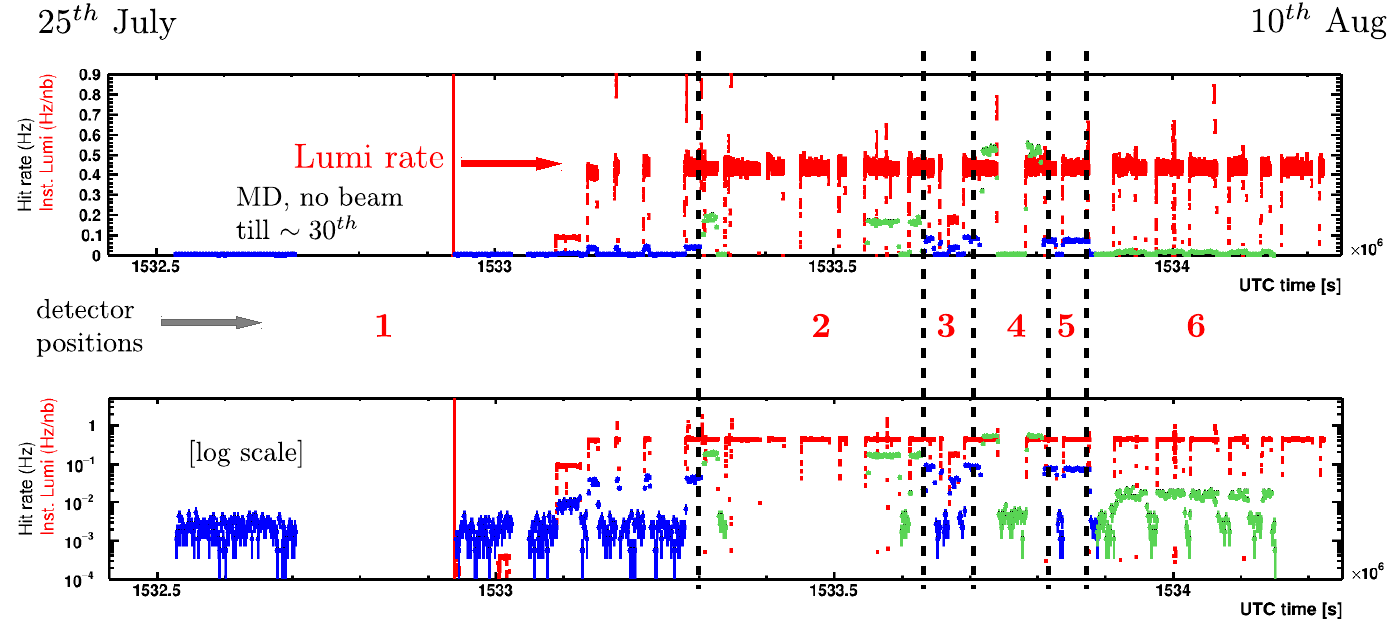}
\caption{\label{fig:meas_results} Hit rates during the run based on the six P1--P6 positions/configurations on a linear (top) and log (bottom) scale.
Red data points denote the luminosity rate of LHCb, blue and green data points denote hit rates. Figure reproduced from Ref.~\cite{Dey:2019vyo}.
}
\end{figure}

The measurement period spanned 17 days between $25$th July and $10$th August, 2018 with 52,036 recorded triggers observed during the run. 
The instantaneous luminosity at IP8 was stable during the measurement period.
There was no beam until July 30th due to machine development and an inadvertent power loss.
Fig.~\ref{fig:meas_results} show the main results from the measurement campaign. 
The red data points represent the instantaneous luminosity measured by LHCb in Hz/nb.
The green and blue data points indicate the hit rate in Hz, where the setup was alternated between the six different configurations/positions. 
The plots are shown in both linear and logarithmic scales.

\begin{table}[h]
\renewcommand*{\arraystretch}{1.5}
\newcolumntype{C}{ >{\centering\arraybackslash } m{6cm} <{}}
\begin{center}
\begin{tabular*}{0.7\linewidth}{@{\extracolsep{\fill}}c|C|c}
  \hline
  Position & Description & Hit rate [mHz] \\
  \hline \hline
   P1 & UXA wall, right corner, $\parallel$ to beam& $1.99\pm0.07$ \\ \hline
   P2 & UXA wall, center, $\parallel$ to beam&  $2.76\pm 0.03$ \\ \hline
   P3 & UXA wall, center, $\perp$ to beam& $ 2.26\pm 0.03$ \\ \hline
   P4 & UXA wall, left corner, $\parallel$ to beam& $ 3.11\pm 0.03$ \\ \hline
   P5 & UXA wall + DAQ racks, center, $\parallel$ to beam& $ 1.95\pm 0.03$ \\ \hline
   P6 & UXA wall + DAQ racks, center, $45^\circ$ to beam& $ 2.22\pm $ 0.02\\ \hline
\end{tabular*}
\caption{\label{table:rate_no_beam}
    Background hit rates based on each position/configuration with no beam. Table reproduced from Ref.~\cite{Dey:2019vyo}.
}
\end{center}
\end{table}

Table~\ref{table:rate_no_beam} contains the hit rate from ambient background without beam, with an average hit rate at each position and configuration of $2$\,mHz.  
The ambient background can therefore be considered negligible for this measurement. 
Table~\ref{table:rate_stable_beam} shows the rate during stable beam. 
This rate is non-negligible, even for the small $300\times300$~mm$^2$ area of the scintillators. 
The rate increases from location P1 to P2 to P4, which, from Fig.~\ref{fig:posconfig}, implies that there is more activity in the downstream region. 
This dependence on the $\eta$ arises from additional concrete near IP8, which screens part of the \CODEXb acceptance, see Ref.~\cite{Dey:2019vyo}. 
Moreover, by comparing the rate at P2 with P5, behind the DAQ server racks, one can see that the racks are adding some amount of shielding material. 
As expected, the flux also depends on the orientation with respect to the beam direction, as indicated by the difference in rate between P5 and P6.  
In absolute numbers, the rate just behind the concrete wall is roughly $0.5$\,Hz over the $900$\,cm$^{2}$ scintillator area.

\begin{table}[h]
\renewcommand*{\arraystretch}{1.5}
\newcolumntype{C}{ >{\centering\arraybackslash } m{6cm} <{}}
\begin{center}
\begin{tabular*}{0.7\linewidth}{@{\extracolsep{\fill}}c|C|c}
  \hline
  Position & Description & Hit rate [mHz] \\
  \hline \hline
   P1 & UXA wall, right corner, $\parallel$ to beam & $ 38.99 \pm 0.99 $\\ \hline
   P2 & UXA wall, center, $\parallel$ to beam& $ 167.10 \pm 1.43$ \\ \hline
   P3 & UXA wall, center, $\perp$ to beam& $ 82.81 \pm 1.55 $ \\ \hline
   P4 & UXA wall, left corner, $\parallel$ to beam& $ 517.45 \pm 3.52 $ \\ \hline
   P5 & UXA wall + DAQ racks, center, $\parallel$ to beam& $ 73.58 \pm 1.18 $ \\ \hline
   P6 & UXA wall + DAQ racks, center, $45^\circ$ to beam& $ 15.71 \pm 0.33 $ \\ \hline
\end{tabular*}
\caption{\label{table:rate_stable_beam}
    Average hit rates measured during stable beam, at various configurations. Table reproduced from Ref.~\cite{Dey:2019vyo}.
}
\end{center}
\end{table}

The predicted charged particle flux from the \texttt{Geant4} simulation of Sec.~\ref{sec:bkg_sim} with just the concrete UXA wall acting as a shield, 
predicts a hit rate $\sim 10$\,Hz at position P2,
assuming an instantaneous luminosity $\sim 0.4$\,Hz\,nb$^{-1}$, as in Fig.~\ref{fig:meas_results}. 
This reduces to $\sim 5$\,Hz treating the full width of UXA wall as $7.5\lambda$ of standard concrete.
This prediction will likely further receive $\mathcal{O}(1)$ reductions from:
relaxing our conservative treatment of forward-propagating backgrounds under angular rescattering; 
accounting for the longer propagation path length through the wall at higher angles of incidence;
variations or uncertainties in the simulation of the primary muon fluxes;
and accounting for possibly additional material in the line-of-sight, such as concrete nearby the IP and the platform adjacent to the LHCb magnet.
A more complete estimate requires a full simulation of the LHCb cavern itself, such as described in Ref.~\cite{Dey:2019vyo}.
Nonetheless comparison to the measured $0.2$\,Hz rate demonstrates that the \texttt{Geant4} simulation of Sec.~\ref{sec:bkg_sim} provides 
conservative estimates of the expected backgrounds.

\clearpage

\section{\CODEXbeta}
\label{sec:demonstrator}
To validate the \CODEXb concept, a proposal has been developed for a small, $2 \times 2 \times 2$\,$\text{m}^3$ demonstrator detector -- ``\CODEXbeta'' -- which will be operational during Run~3. 
This detector will be placed in the proposed location for $\CODEXb$ (UXA hall, sometimes referred to as the `DELPHI cavern') shielded only by the existing, concrete UXA radiation wall.

\subsection{Motivation}
The main goals of the \CODEXbeta setup are enumerated as follows:
\begin{enumerate}[label=\alph*)]
    \item \emph{Demonstrate the ability to detect and reconstruct charged particles which penetrate into the UXA hall, as well as the decay products of neutral particles decaying within the UXA hall.}
\end{enumerate}

This is desirable to provide an accurate and fully data-driven estimate of the backgrounds, so that the design of the eventual shield (both passive and active, instrumented) 
needed by the full experiment can be optimized to be as small as possible. 
We have already made preliminary background measurements in the UXA hall using a pair of scintillators during Run 2 (see Sec.~\ref{sec:measurementcampaign}). 
However these measurements were simply of hit counts; we could not reconstruct particle trajectories. 
\CODEXbeta will allow us to track particles within a volume similar to the \CODEXb fiducial volume, 
and in particular separate charged particles produced outside the decay volume from backgrounds induced by particle scattering inside the decay volume itself.

As shown in Sec.~\ref{sec:backgrounds}, the residual neutron scattering inside the decay volume is one of the most important backgrounds identified in the original \CODEXb proposal~\cite{Gligorov:2017nwh}. 
The tracking capability of \CODEXbeta will also allow us to measure the origin of charged particle backgrounds, 
and in particular potential soft charged particles which could be swept towards \CODEXb by LHCb's magnet ``focusing'' and thereby evade the Pb shield.

\begin{enumerate}[label=\alph*), resume]
    \item \emph{Detect and reconstruct a significant sample of neutral particles decaying inside the hermetic detector volume.}
\end{enumerate}

\begin{figure}[t]
\centering{\hfill
\includegraphics[width = 0.4\linewidth]{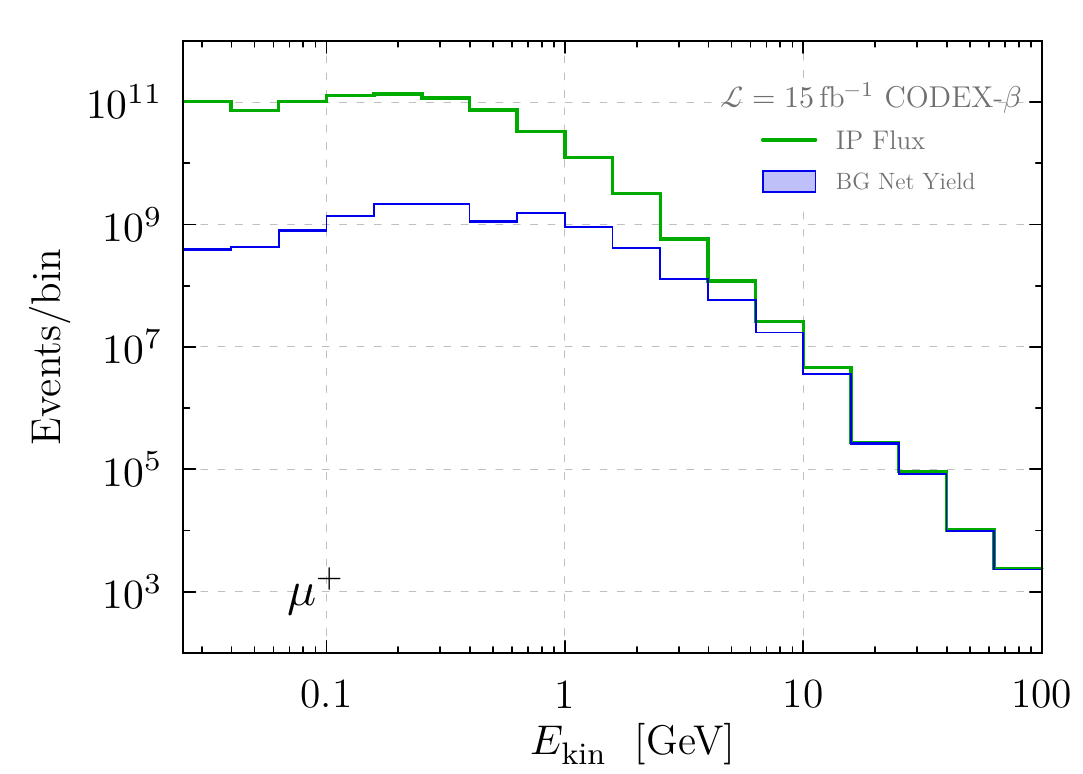}\hfill
\includegraphics[width = 0.4\linewidth]{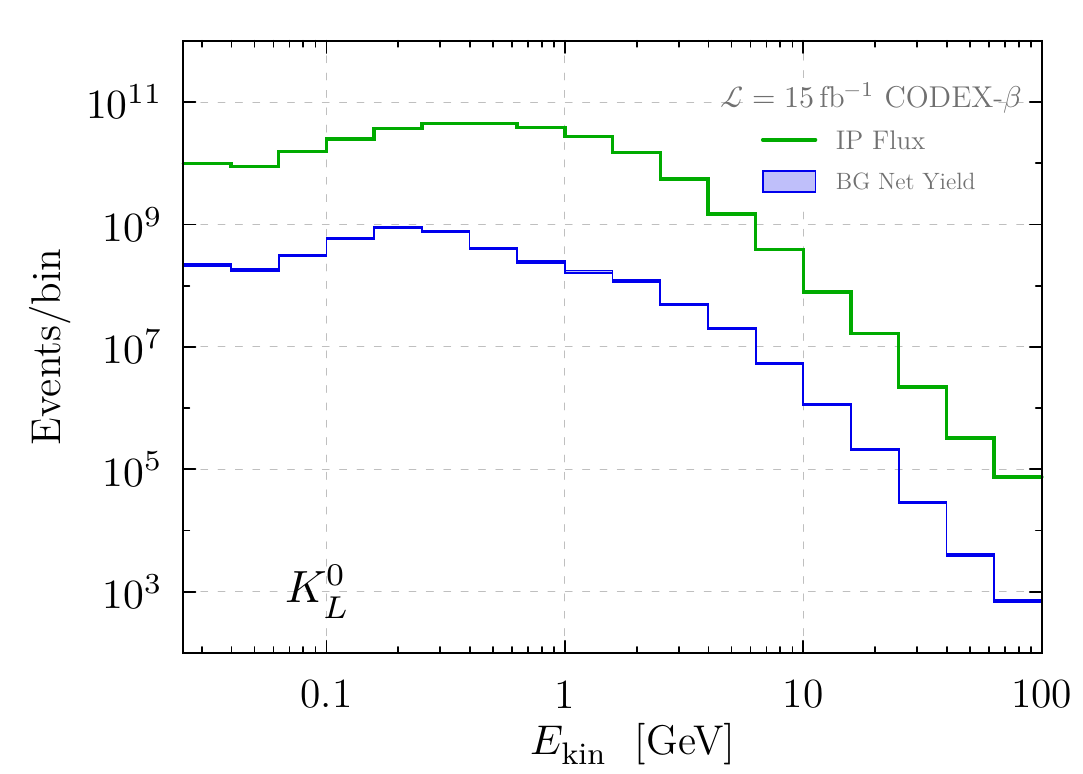} \hfill \\
\hfill
\includegraphics[width = 0.4\linewidth]{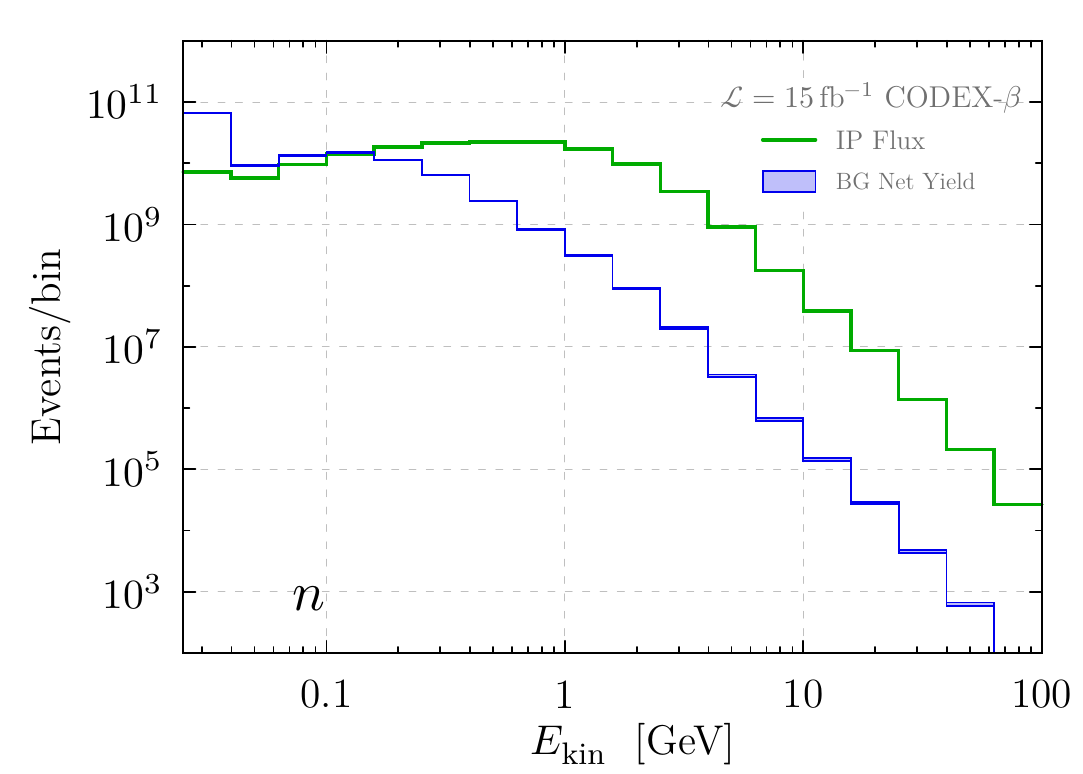} \hfill
\includegraphics[width = 0.4\linewidth]{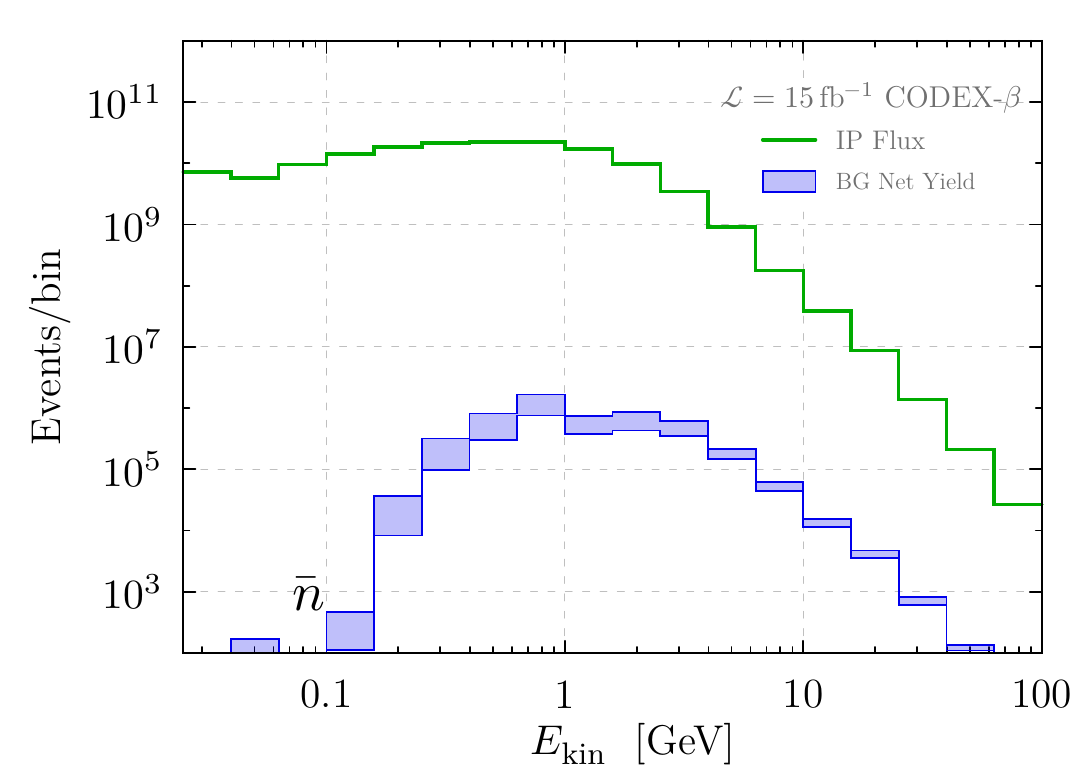} \hfill \\
\hfill
}
\caption{Background fluxes per kinetic energy bin, comparing primary IP fluxes (green) with the background flux (blue) in the  \CODEXbeta acceptance after passage through the $3$\,m UXA concrete wall.}
\label{fig:demo-BGfluxes}
\end{figure}

This will allow us to observe e.g. $K^0_L$ decays and use this data to calibrate our detector simulation.  
Aside from measuring background levels, observing long-lived SM particles decaying inside the detector acceptance
will allow us to calibrate the detector reconstruction and the RPC timing resolution.
The most natural candidates are $K^0_L$ mesons: In Fig.~\ref{fig:demo-BGfluxes} we show the expected differential fluxes of neutrons, antineutrons, $K^0_L$s and (anti)muons, with respect to their kinetic energy, 
for an integrated luminosity of $15$\,fb$^{-1}$ after propagation through the UXA wall. Also shown are the primary fluxes of the same species. 
In Table. \ref{tab:bkg-tracks} we show the expected multitrack production from decay or scattering on air by neutral fluxes entering \CODEXbeta,
requiring $E_{\text{kin}} > 0.4$\,GeV per track. We also show the multitrack contribution just from $K^0_L$s entering \CODEXbeta.

One sees that approximately a $\text{few} \times 10^7$ $K^0_L$ decays to two tracks are expected in the \CODEXbeta volume per nominal year of data taking in Run~3.
The results of the background simulation show that we will be able to reconstruct a variety of $K^0_L$ decays in the \CODEXb demonstrator volume. 
The decay vertex and decay product trajectories moreover allow the boost to be reconstructed independently 
of the time-of-flight information. Comparing the boost distribution of $K^0_L$ mesons observed in \CODEXb, as well as the $K^0_L$ mean decay time which can be inferred from this 
distribution, will allow us to calibrate and validate our detector simulation and reconstruction.

\begin{enumerate}[label=\alph*), resume]
  \item \emph{Show that \CODEXb can be integrated into the LHCb DAQ and demonstrate an ability to give a trigger to LHCb to retain an event that looks interesting in \CODEXb.}
\end{enumerate}

The RPC readout is compatible with LHCb's data acquisition hardware.
Some relatively straightforward firmware 
development will be required to enable LHCb's usual FPGA backend readout boards to receive the \CODEXb data. 
Based on expected data rates, we estimate that a single FPGA backend readout board will be comfortably able to read out the full \CODEXbeta 
detector. From an LHCb point of view the simplest solution would be that this board also clusters the RPCs and performs a basic track reconstruction, so that events which look 
interesting for \CODEXbeta can be kept for further inspection by LHCb's High Level Trigger simply by reading the \CODEXb data raw bank. Given that 
\CODEXb is about the same distance from the interaction point as LHCb's muon system, latency should not be an issue. Our background measurements in the cavern during Run 2 
indicated hit (not track) rates of maximum 500~mHz across a scintillator area of order $10^{-1}$ $\text{m}^{2}$. 
Therefore, even a simple track reconstruction should allow all interesting events in \CODEXb to be kept for offline inspection. It will be desirable to have a possibility to read \CODEXb 
out during beam-off periods, for cosmic-ray data taking and calibration. \CODEXb will therefore ideally appear as a sub-detector within LHCb, though one whose presence/readiness is 
not required for nominal LHCb data taking.

\begin{enumerate}[label=\alph*), resume]
    \item \emph{Integrate the detector into the gas and electricity services in the UXA hall, and demonstrate stable operational behavior.}
    \item \emph{Demonstrate a viable and modular support structure for the RPC layers, to form a basis for the eventual support structure for the full detector.}
    \item \emph{Search for multi-track signatures from BSM physics: Despite its limited acceptance and large backgrounds, \CODEXbeta is expected to have some new reach for LLPs produced in exotic B-meson decays. 
    This is discussed below in Sec.~\ref{sec:demophysicsreach}.}
\end{enumerate}

\subsection{Technical description and timetable}
The high-level requirements listed above drive the design of \CODEXbeta to be a $2 \times 2 \times 2$\,$\text{m}^{3}$ cube. Each side of the cube
will consist of 2 RPC panels, each of which is $2\times 1$~m$^2$ in area. Each such panel block will contain a triplet of RPC layers. In addition there will be two panels of 
the same $2\times 1$~m$^2$ area placed in the middle of the cube, for a total of $(6+1)\times 2\times3=42$ such $2\times 1$~m$^2$ RPC layers.
\CODEXbeta is proposed for installation in the barrack which housed LHCb's Run~1 and Run~2 High Level Trigger farm, and which will be empty
in Run~3 as the High Level Trigger will be housed in a dedicated data processing center on the surface. As a result \CODEXbeta will have ample
space and straightforward access to all required detector services.
The proposed detector technology for \CODEXbeta is that of the ATLAS RPCs for phase I upgrade, while the full \CODEXb detector would follow the phase II design (see \cite{Collaboration:2285580} for technical descriptions). 

The timetable for installation is driven by the primary consideration to not interfere with the building or commissioning of the LHCb upgrade. 
For this reason, we originally proposed installation in winter 2021/2022, integration in the LHCb DAQ during spring 2022 and first data taking in summer 2022. Given the COVID
pandemic and the subsequent delay of LHC Run~3 datataking to spring 2022, we have chosen to push back installation and all subsequent steps by one year. Although this will
mean missing out on 2022 datataking, that is not so crucial for a demonstrator and avoids having to simultaneously commission \CODEXbeta and the upgraded LHCb detector.
Given the modest size of \CODEXbeta and the use of well-understood detector components, we estimate around six months are needed to produce and qualify the RPCs.
Therefore, if approved, it is realistic to complete the bulk of the construction during the first half of 2021. The mechanical support structure will build on existing
structures used in ATLAS but modified to be modular. It will provide the required stability for a cubic arrangement of the detector layers; the design and construction of this structure
is expected to take place in 2020-2021.
The total cost of the detector components is expected to be roughly 150k \euro{}.

\subsection{New physics reach\label{sec:demophysicsreach}}

The acceptance of \CODEXbeta is roughly only $8\times 10^{-3}$ times that of the full \CODEXb detector, and no shielding beyond the existing concrete wall will be in place.
Its reach for BSM physics is therefore limited due to its reduced acceptance and high background environment.
However, roughly $10^{13}$ $b$-hadrons will be produced at IP8 during Run~3. This enables \CODEXbeta to probe some new regions of parameter space for those cases in which
the LLP production branching ratio from e.g. $B$ decays is independent from its lifetime (cf. Fig.~\ref{fig:nonminmalBfull}).

This scenario can arise e.g. in models that address the baryogenesis puzzle~\cite{McKeen:2015cuz,Aitken:2017wie,Elor:2018twp,Nelson:2019fln,Alonso-Alvarez:2019fym}, as described in Sec.~\ref{sec:baryogenesis}.
We take this model as a representative example: In our simplified phenomenological setup we consider a new particle $\chi$, with a coupling
\begin{equation}
	\mathcal{L}\supset\lambda_{ijk} \chi u_i d_j d_k\,,
\end{equation}
where we assume for simplicity that the $\lambda_{bsu}$ and $\lambda_{udd}$ couplings are independent.
The former is responsible for the production via e.g. $B \to X_s \chi$ decays, where $X_s$ here is a SM (multi)hadronic state with baryon number $\pm1$;
the latter induces the decay of $\chi$ to an (anti-)baryon plus a number of light mesons.
$\text{Br}[B\to X_s \chi]$ and $c\tau$ are then independent parameters.
The $\lambda_{udd}$ coupling moreover must be parametrically small, to avoid exotic dinucleon decays \cite{PhysRevD.91.072006,Friedman:2008es}, implying that $\chi$ must be long-lived.

In Run~3, as shown in Fig.~\ref{fig:demo-BGfluxes} we expect roughly $10^9$ $K_L$ and $4\times10^{9}$ neutrons to enter the \CODEXbeta fiducial volume for an integrated luminosity of $15\text{fb}^{-1}$, 
requiring $E_{\text{kin}} > 0.4$\,GeV. This is desirable for calibration purposes, as explained above.
An additional $4\times10^6$ antineutrons also enter, with no kinetic energy cut.
We show the corresponding multitrack production rate for \CODEXbeta in Tab.~\ref{tab:bkg-tracks}.
The multitrack background falls relatively fast with the number of tracks, partly because of the relative softness of the fluxes emanating from the shield and partly 
because the $K^0_L$s mainly decay to no more than two tracks. 
We therefore define the LLP signal region as those events with $4$ or more reconstructed tracks, also requiring $E_{\text{kin}}>0.4$~GeV from expected minimum tracking requirements.
The expected number of background events in the signal region is then roughly $8.5\times 10^4$ per $15\text{fb}^{-1}$.
In the actual experimental setup this number can be calibrated from a control sample with less than $4$ tracks, if the ratio of both regions is taken from Monte Carlo.

\begin{figure}[t]
        \centering
       \includegraphics[width=0.55\textwidth]{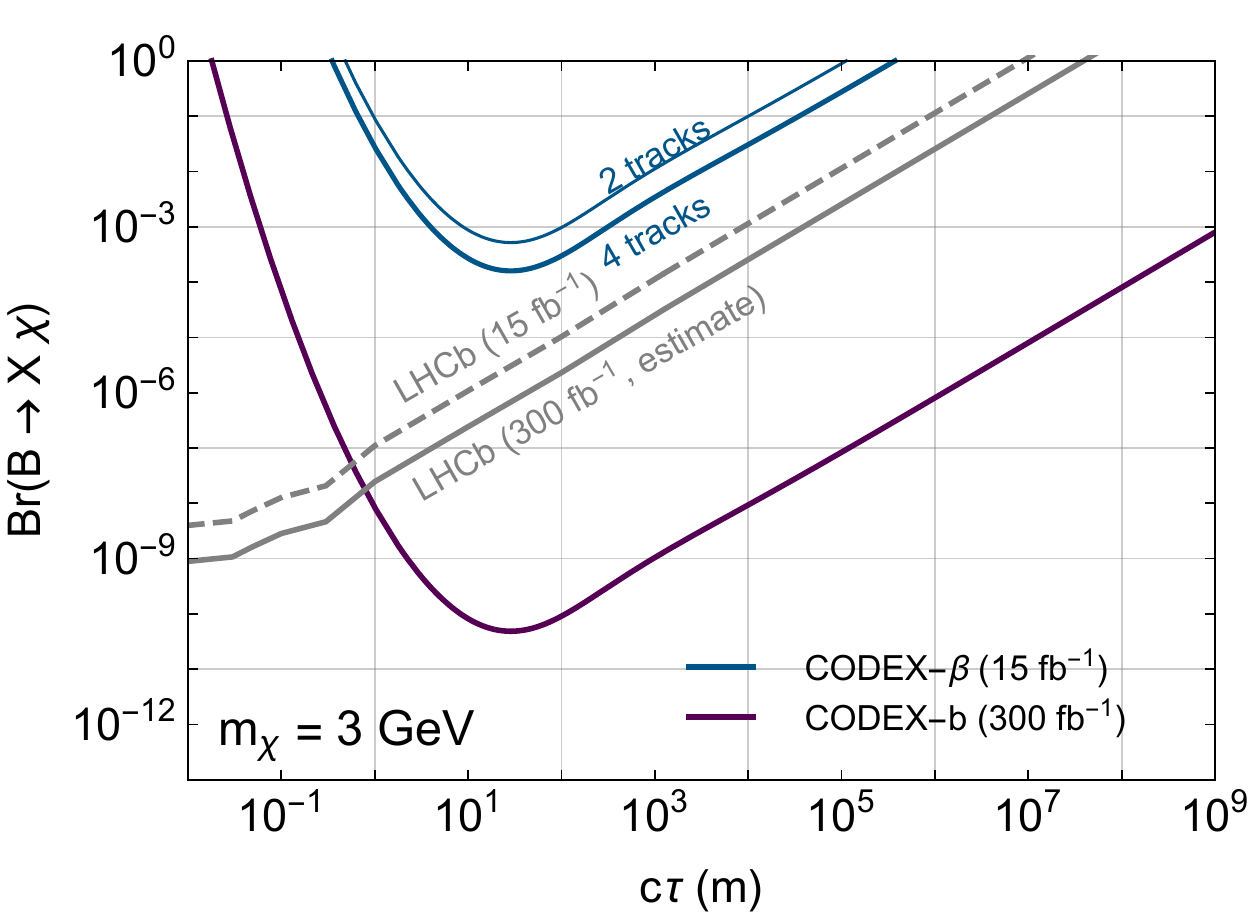}
        \caption{\CODEXbeta reach for a long-lived $\chi$ decaying hadronically. 
        \label{fig:nonminmalBdemo}}
\end{figure}

For a signal benchmark with $m_\chi=3$~GeV, the probability of decaying to 4 tracks with $E_{\text{kin}}>0.4$~GeV is roughly $15\%$, as estimated with \texttt{Pythia 8}. 
In Fig.~\ref{fig:nonminmalBdemo} we show the estimated $2\sigma$ limit reach under these assumptions. 
For comparison, we also show the reach of the full \CODEXb detector and the reach using a $\geq 2$ tracks selection, for which the background is roughly three orders of magnitude higher. 
(This is partially compensated for by a higher signal efficiency.) 
At this stage, no attempt has been made to discriminate signal from background by making use of angular variables, in particular pointing to the interaction point. In this sense the estimated reach is therefore conservative. 

For completeness, we also include a preliminary estimate of the reach of LHCb itself for this signature with 15\,fb$^{-1}$ of data, 
analyzing all decay products that can be reconstructed as a track at LHCb: $e$, $\mu$, $p$, $K^\pm$ and $\pi^\pm$.
For reconstructing a $\chi$ vertex, we first require all pairs of tracks to be vertexed not more than 1 mm away from each other. 
We build the position of these vertices by finding the point that minimizes the distance to each pair of tracks.  
We then average all the resulting vertices to generate the $\chi$ decay vertex. 
In order to build the $B^+$ decays, the $\chi$ vertex is required not to be more than 1 mm away from a $K^+$. 
For the background, SM $B^+$ decays are considered, subject to the same reconstruction criteria. All other backgrounds are neglected.

The analysis cuts are included in Tab.~\ref{tab:cuts}. For the rest of the experimental efficiencies, we estimate a 97\% efficiency per track~\cite{LHCb-DP-2014-002}, and take the remaining efficiencies to be 100\%.
Signal and background were binned according to $4$ or more tracks and $6$ or more tracks in the secondary vertex. 
We estimate the mass resolution, $\sigma$, for 4 and 6 body-decays of the $\chi$ particle to be $\sim$12 and 21 MeV, respectively. 
This estimate is based on a study of 3 and 4-body $B$ and $D$ meson decays at LHCb~\cite{LHCb-PAPER-2018-020,LHCb-PAPER-2016-023,Aaij:2013lla,Aaij:2018hik}, 
 interpolating or extrapolating to the appropriate track multiplicity.
Following a similar procedure, we estimate the $B^+$ meson mass resolution to be  $\sim 24$ and $\sim 36$\,MeV for 5 and 7 body decays, respectively.
To determine the background yields, we cut in $\pm 2\sigma$ windows around the $B^+$ and $\chi$ invariant masses. 
To determine the limits, we take $\sigma_{bb}$ at $\sqrt{s}=14$ TeV to be 500\,$\mu b$~\cite{LHCb-PAPER-2016-031},
and the fraction of $b$ quarks hadronising to a $B^+$ is taken to be $40$\%~\cite{LHCb-PAPER-2012-037}. 
Combining this, we compute the limits on the branching ratio of the $B^+$ decay for both the $4+$ and $6+$ track bins. 
The projected limit shown in Fig.~\ref{fig:nonminmalBdemo} is the strongest of both limits, for each $c\tau$ point. In addition, we show a rough estimate of the reach for the HL LHC, by rescaling the limits with the square root of the ratio of the luminosities.

One sees that \CODEXbeta and the main LHCb detector will have complementary sensitivity to this benchmark scenario, with likely better sensitivity from the LHCb search. 
However, it is conceivable that \CODEXbeta may set an earlier limit than an LHCb analysis on Run~3 data, especially given the comparatively simpler analysis required for the former.
In both estimates no attempt was made to further reduce the backgrounds by means of kinematic cuts, so both projections are conservative.

\begin{table}[th]
\renewcommand*{\arraystretch}{1.5}
\newcolumntype{C}{ >{\centering\arraybackslash } m{4cm} <{}}
    \centering
\begin{tabular*}{0.5\linewidth}{@{\extracolsep{\fill}}C|C}
    	Cut &  Value \\ \hline\hline
         $p_T$ of $\chi$ daughters & $>500$ MeV/$c$  \\
        $\theta$ of $\chi$ daughters  & $<400 $ mrad \\
         $SV_z$   & $<400 $ mm \\
         $SV_{R}$   & $\in [14,25] $ mm \\
         $p_T$ of $K^+$  & $>500$ MeV/$c$  \\
         $\theta$ of $K^+$  & $<400 $ mrad \\
         \hline
    \end{tabular*}
    \caption{List of analysis cuts for the LHCb reach estimate for a hadronically decaying, long-lived particle ($\chi$). $SV_z$ and $SV_R$ respectively stand for the longitudinal and transverse position of the secondary vertex, 
    and $\theta$ is the angle of the track with respect to the beam axis.}
    \label{tab:cuts}
\end{table}

\FloatBarrier
\clearpage

\section{Detector Case Studies}
\label{sec:design}
In this section we discuss various detector studies, as well as possible extensions of the baseline \CODEXb detector configuration. 
As \CODEXbeta is based on the same underlying technology as the full detector, most results apply directly to it as well. 
One caveat, however, is that \CODEXbeta will use a simplified front-end readout based on FPGA cards and therefore have a significantly poorer timing resolution
of around $800/\sqrt{12}$~ps per gas gap. This resolution will nevertheless be comfortably sufficient to integrate \CODEXbeta into the LHCb readout and to validate the detector concept.

\subsection{Tracking\label{sec:tracking}}

\subsubsection{Design drivers}
The geometry and required capabilities of the tracking stations are informed by the signal benchmarks in Sec.~\ref{sec:physicscase}. The main design drivers are:
\begin{enumerate}[label=\alph*)]
	\item Hermeticity:
\end{enumerate}

\begin{figure}
\includegraphics[width=0.3\textwidth]{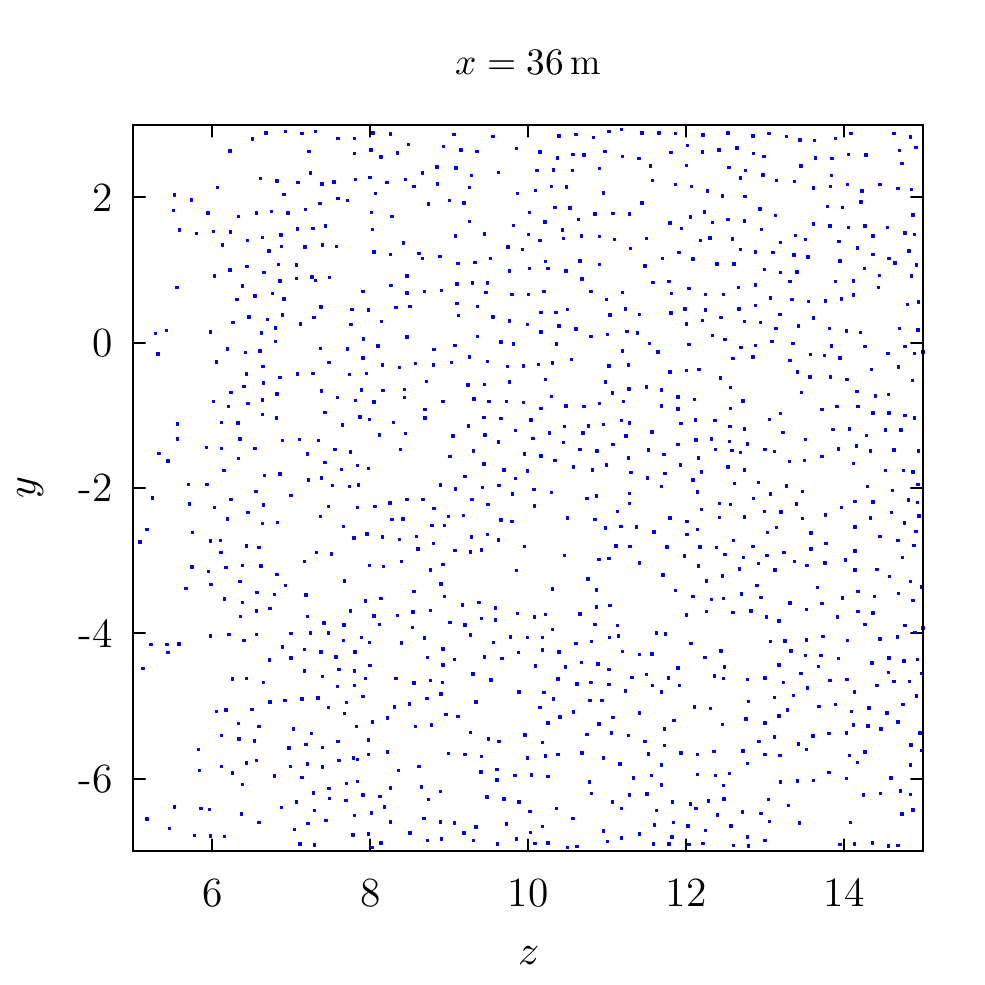}\hfill
\includegraphics[width=0.3\textwidth]{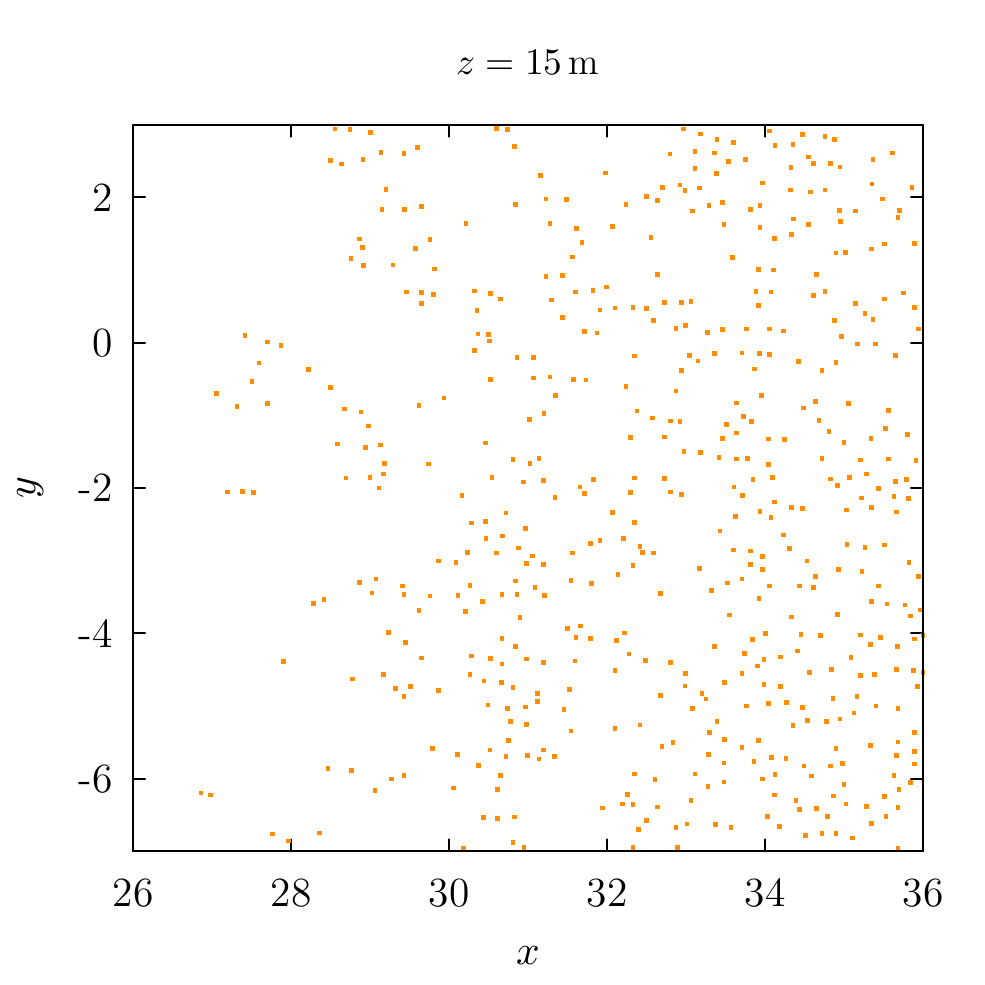}\hfill
\includegraphics[width=0.3\textwidth]{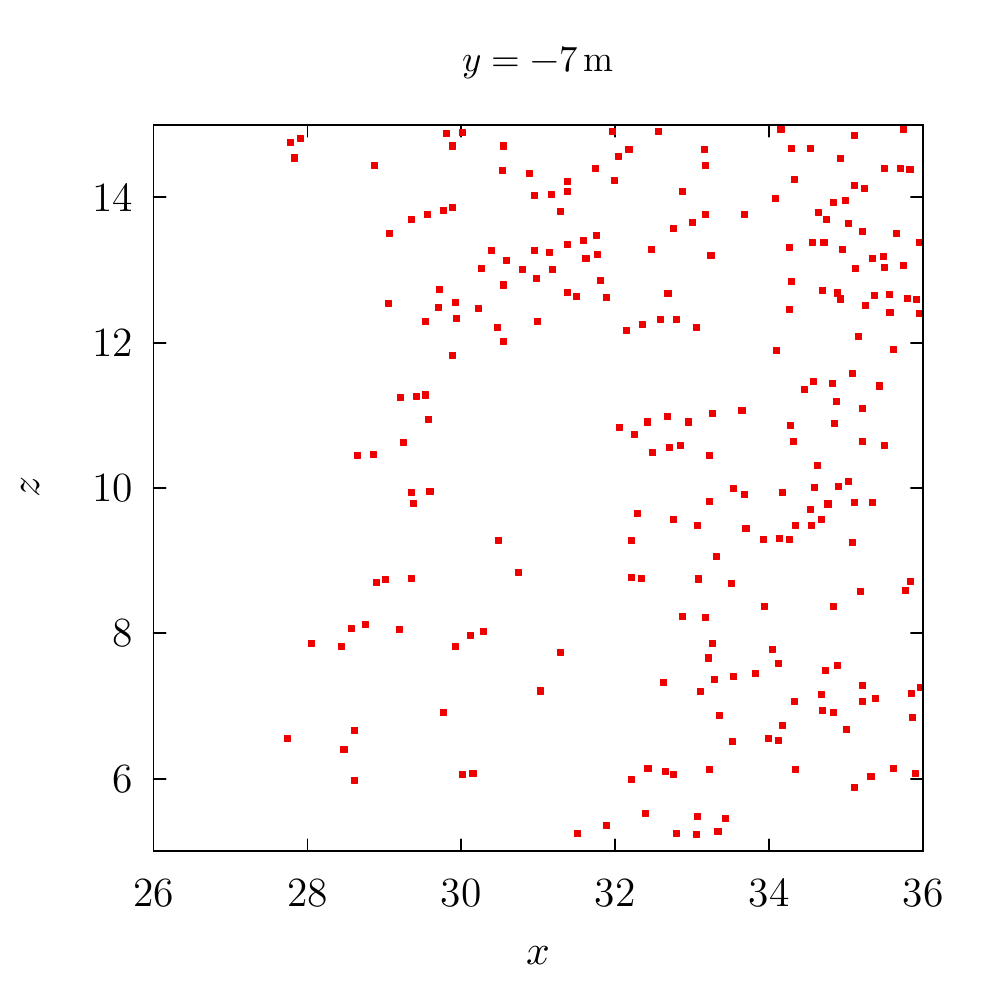}\\
\includegraphics[width=0.3\textwidth]{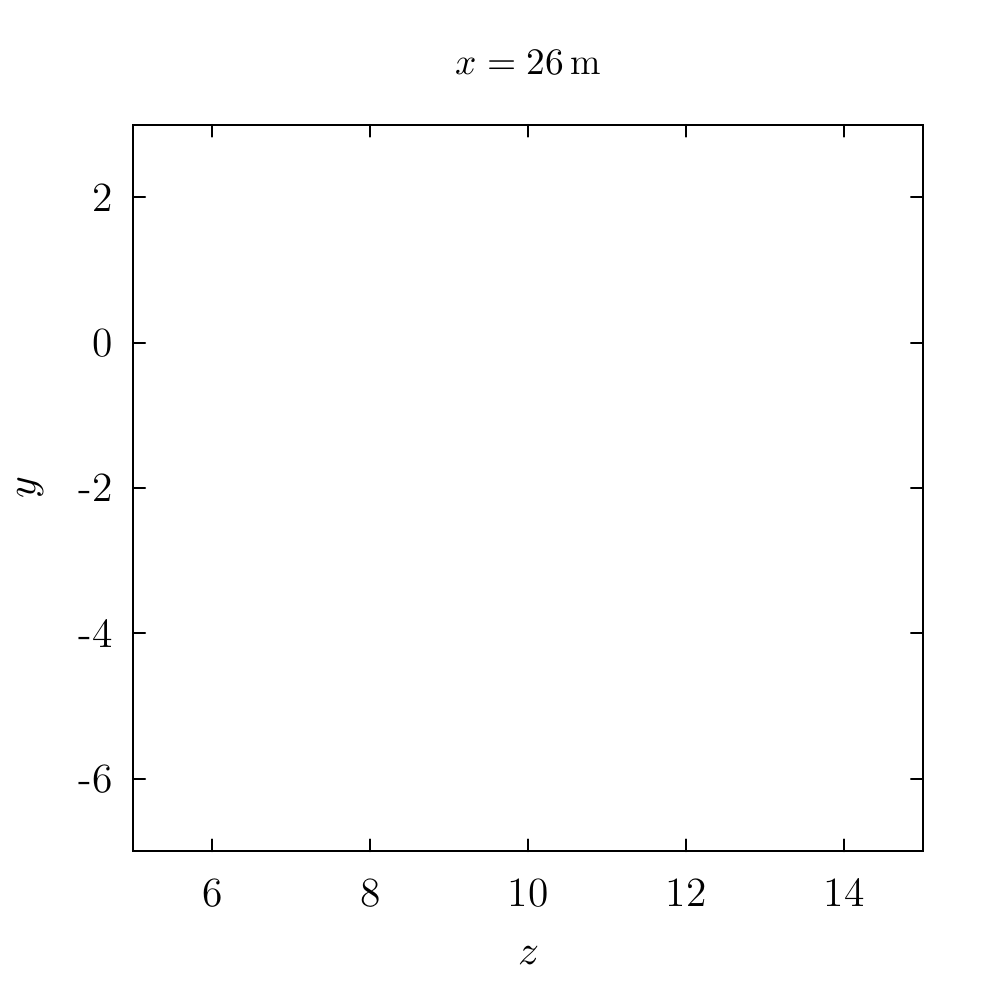}\hfill
\includegraphics[width=0.3\textwidth]{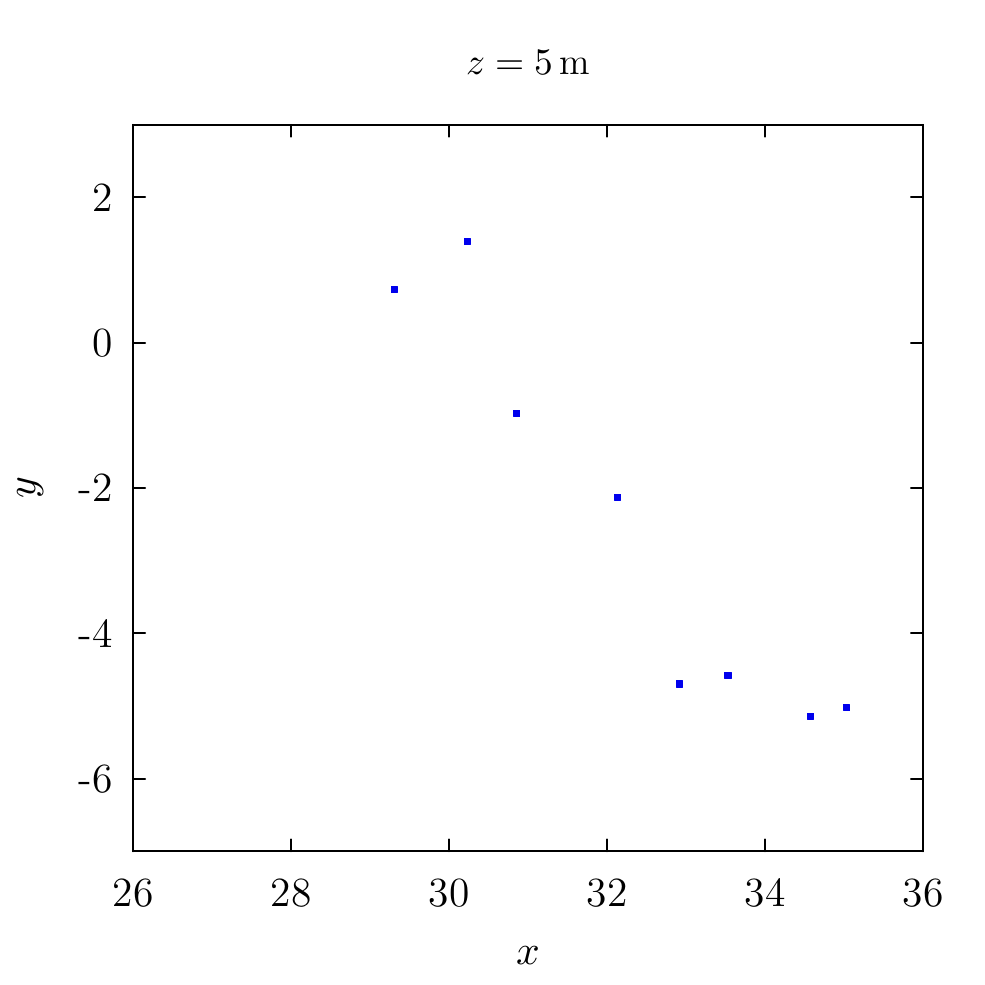}\hfill
\includegraphics[width=0.3\textwidth]{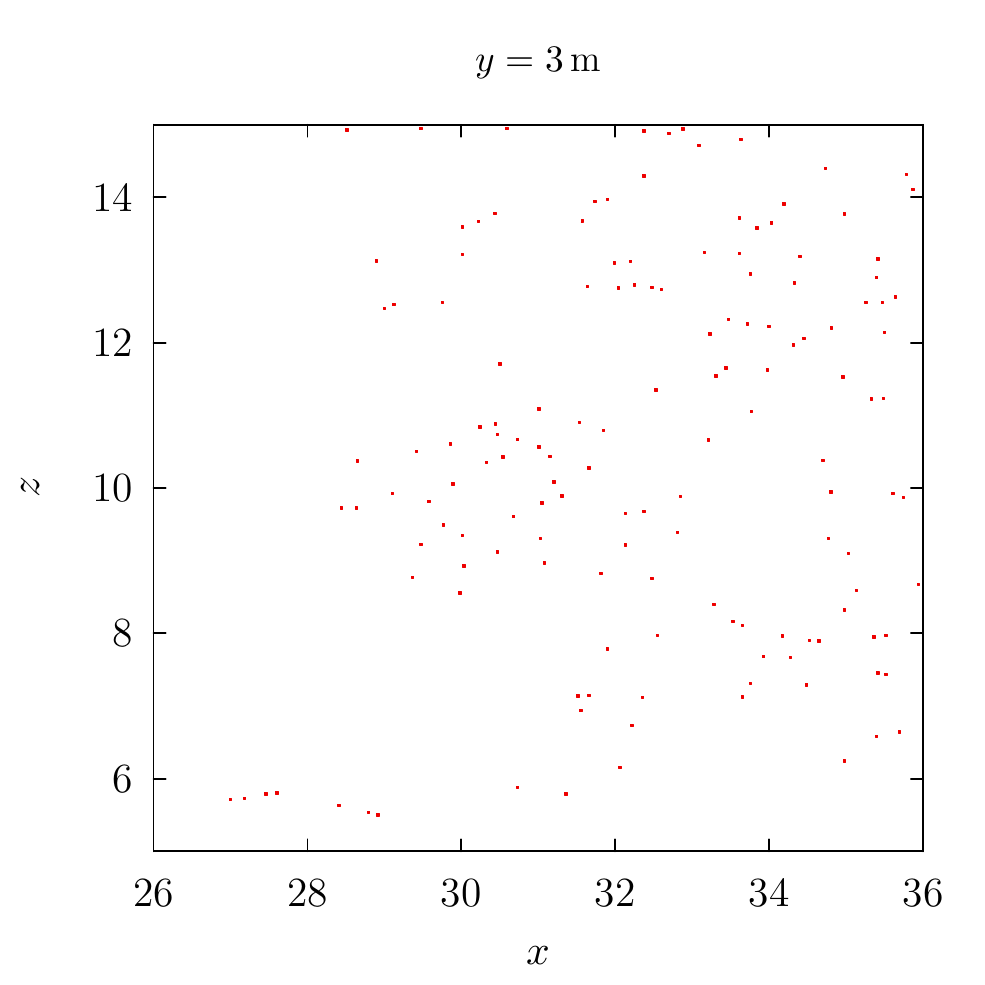}\\
\includegraphics[width=0.4\textwidth]{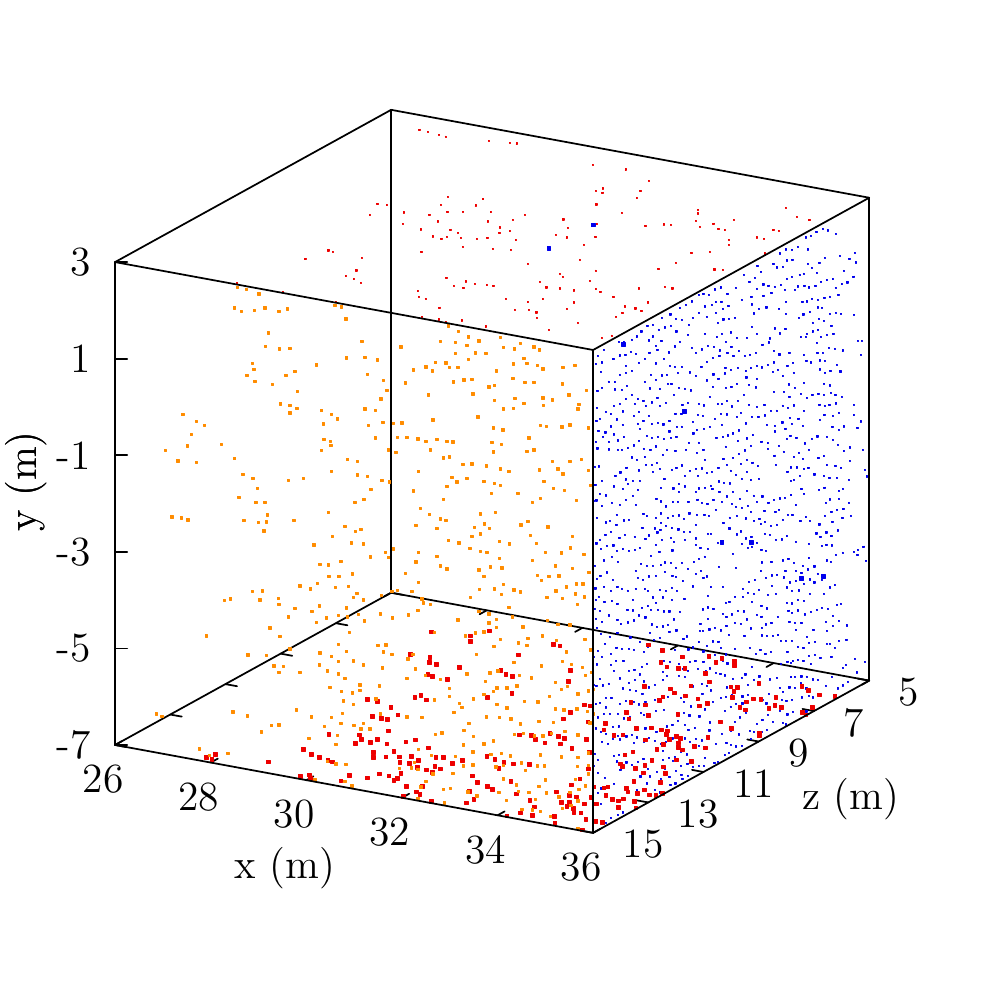}
\caption{
Above: Scatter plot of hits on the faces of the box for 496 decays inside the box, from a $h\to A'A'\to 4 \tau$ benchmark. (From left to right) Top row: $x=36$\,m, $z=15$\,m, $y=-7$\,m. 
Bottom row: $x=26$\,m, $z=5$\,m, $y=3$\,m. 
Below: The corresponding scatter plots shown in projection on the surfaces of the detector, with the same color scheme.
\label{fig:distributionofhits} }
\end{figure}

As discussed in Sec.~\ref{sec:experimentalcoverage}, the primary motivation for \CODEXb is to cover relatively low energy signals, as compared to e.g.~SUSY 
signatures. In many benchmark models (see in particular Secs.~\ref{sec:higgsmixing} and \ref{sec:HNL}) the LLPs are therefore only moderately boosted and large 
opening angles are common. To achieve optimal signal efficiency, it is therefore desirable to place tracking stations on the back-end, top, bottom and sides of the fiducial volume. This is 
illustrated in Fig.~\ref{fig:distributionofhits} which shows the distribution of hits on the various faces of the detector for an example $h\to A'A'$ model (see Sec.~\ref{sec:abelianhiggs}) with $m_{A'}=5$ GeV and 
$A' \to \tau\tau$, and the $\tau$'s decaying in the 3-prong mode.
The majority of the hits land on the back-face of the fiducial volume, but the sides, top and bottom cannot be neglected. This is despite the relatively high boost 
of this benchmark model, as compared to models in which, e.g., the LLP is produced in a heavy flavor decay. It may be feasible to instrument those faces that see typically fewer hits  
more sparsely that the nominal design outlined in Sec.~\ref{sec:detector}. Studies to this effect are under way for a wider range of models and will inform the final design. 

Finally, some 
tracking stations on the front face are needed to reject backgrounds from charged particles emanating from the shield, primarily muons (see Sec.~\ref{sec:backgrounds}). For these stations, resolution is less 
important than efficiency, and alternative technologies (e.g.~scintillator planes) may be considered.

\begin{enumerate}[label=\alph*), resume]
	\item Vertex resolution:
\end{enumerate}

Assuming that no magnetic field will be available in the \CODEXb decay volume, a good vertex resolution is essential to convincingly demonstrate a signal. The baseline design calls for 6 RPC layers in each tracking station covering the walls, while the five internal tracking stations would have 3 RPC layers each. This is however subject to further optimization, as less RPC layers may be needed on some of the less important wall of the cube, as indicated by Fig.~\ref{fig:distributionofhits}.
The most important parameter is the distance to the first tracker plane, which motivates five additional tracking stations spread throughout the fiducial volume, in order to achieve vertex 
resolutions on the order of millimeters rather than centimeters. 
For signals characterized by a high boost (e.g. Higgs decays, Sec.~\ref{sec:abelianhiggs}) the vertex resolution also impacts the signal reconstruction efficiency, as tracks tend to merge. 
The reconstruction efficiency corresponding to the nominal design  for tracks from such an exotic Higgs decay, $h \to A'A'$, is shown in the right-hand panel of Tab.~\ref{tab:effsummary},
under the requirements:
\begin{itemize}
\item the track momentum $>$ 600 MeV (trivially satisfied for this benchmark);
\item each track has at least 6 hits; and,
\item the first hit of each track is unique.
\end{itemize}
It is the last requirement which can fail for a highly boosted LLP. 
 
Once the LLP is required to decay in the fiducial volume, its proper lifetime, $c\tau$, is inversely correlated with its boost, 
so that a larger $c\tau$ typically implies a better reconstruction efficiency, as shown in the right-hand panel of Tab.~\ref{tab:effsummary}.
For particles with a boost factor of $\mathcal{O}(100)$ or more -- roughly, $c\tau < 0.1$\,m -- the nominal design nonetheless achieves $\mathcal{O}(1)$ reconstruction efficiencies.

\begin{table}[t]
\centering
\renewcommand*{\arraystretch}{1.25}
\newcolumntype{C}{ >{\centering\arraybackslash} m{0.7cm} <{}}
\begin{tabular*}{0.9\linewidth}{@{\extracolsep{\fill}}l|CCC|CCCCC}
\hline
$c\tau$ (m) & \multicolumn{3}{c|}{$m_S~\text{(GeV)} ~[B \to X_sS]$} &  \multicolumn{5}{c}{$m_{A'}~\text{(GeV)}~[h \to A'A']$} \\
     & $0.5$ &  $1.0$ & $2.0$ & $0.5$ & $1.2$ & $5.0$ & $10.0$ & $20.0$ \\ \hline\hline
0.05  & --   & --   & --   & 0.39 & 0.48 & 0.50 & --   & --     \\
0.1   & --   & --   & --   & 0.48 & 0.63 & 0.73 & 0.14 & --     \\
1.0   & 0.71 & 0.74 & 0.83 & 0.59 & 0.75 & 0.82 & 0.84 & 0.86   \\
5.0   & 0.55 & 0.64 & 0.75 & 0.60 & 0.76 & 0.83 & 0.86 & 0.88   \\
10.0  & 0.49 & 0.58 & 0.74 & 0.59 & 0.75 & 0.84 & 0.86 & 0.88   \\
50.0  & 0.38 & 0.48 & 0.74 & 0.57 & 0.75 & 0.82 & 0.87 & 0.88   \\
100.0 & 0.39 & 0.45 & 0.73 & 0.62 & 0.77 & 0.83 & 0.87 & 0.89   \\
500.0 & 0.33 & 0.40 & 0.75 & --   & --   & --   & --   & --     \\
\hline
\end{tabular*}
\caption{Efficiency of reconstructing at least two tracks with the nominal design for both $B\to  X_sS$ (Sec.~\ref{sec:higgsmixing}) and $h\to A'A'$ (Sec.~\ref{sec:abelianhiggs}) scenarios, 
for various lifetimes \cite{Gligorov:2017nwh}.}
\label{tab:effsummary}
\end{table}

\begin{enumerate}[label=\alph*), resume]
	\item Track momentum threshold:
\end{enumerate}

The momentum threshold that can be achieved is especially relevant for two types of scenarios:
\begin{itemize}
	\item LLPs \emph{produced in hadron decays} are typically relatively soft, and in order to maintain an $\mathcal{O}(1)$ reconstruction efficiency, 
	the track momentum threshold should be kept roughly around $600$\,MeV or lower. 
	This is illustrated by the efficiency numbers in the left-hand panel of Tab.~\ref{tab:effsummary} for the $B \to X_sS$ portal of Sec.~\ref{sec:higgsmixing}. 
	Here, all losses in efficiency are because of the $600$\,MeV threshold that was assumed in the simulation.
	\item \emph{Inelastic dark matter models} (see Sec.~\ref{sec:inelasticDM}) are characterized by an LLP decaying to a nearby invisible state  -- the dark matter -- and a number of soft SM tracks. 
	Given the low amount of phase space available to the SM decay products, the reach of \CODEXb for this class of models is very sensitive to the threshold that can be achieved, as shown 
	in Fig.~\ref{fig:inelasticDM}.
\end{itemize}

\subsubsection{Studies performed}
A number of initial tracking studies have been performed to validate the design requirement outlined above. They further explore a number of
\CODEXb design configurations, a variety of signals, and novel methods for particle boost reconstruction~\cite{Gibbons:2019xx,Quessard:2019xx}. 
Work is ongoing to integrate the \CODEXb detector into the LHCb simulation framework. 
This will facilitate the optimization of different reconstruction algorithms, as well as the study of how the information from both detectors could be integrated.

For these initial studies, a simplified \texttt{Geant4}~\cite{Agostinelli:2002hh} description of \CODEXb was implemented, 
following the nominal design specifications~\cite{Gligorov:2017nwh}, but without RPC faces on the top 
or bottom of the detector. The active detector material was modeled using silicon planes with 2~$\mathrm{cm}^2$ granularity and the same radiation length as the RPCs proposed in 
the nominal design. Signal events of di-electron and di-muon candidates were then passed through this simulation to determine the detector response. Within this preliminary study, no 
attempt was made at modeling detector noise.

An initial clustering algorithm, based on a CALICE hadronic shower clustering algorithm~\cite{Eigen:2019ccp}, was designed to combine nearest neighbor energy deposits within the 
RPC layers, passing a minimum threshold, into clusters. As expected, this clustering was found to be necessary for electrons, but had little impact on reconstructing charged pion and 
muon signals. After clustering, an iterative linear track-finding algorithm was run. Both back-to-forward and forward-to-back algorithms were implemented, as well as various iterative 
approaches. For the expected signals within \CODEXb the back-to-forward tracking algorithm was 
found to provide the best performance. The tracking also performed well for more complex $n$-body signals without a common decay vertex, \textit{e.g.} emerging jets.

\begin{figure}[t]
	\includegraphics[width = 0.5\linewidth]{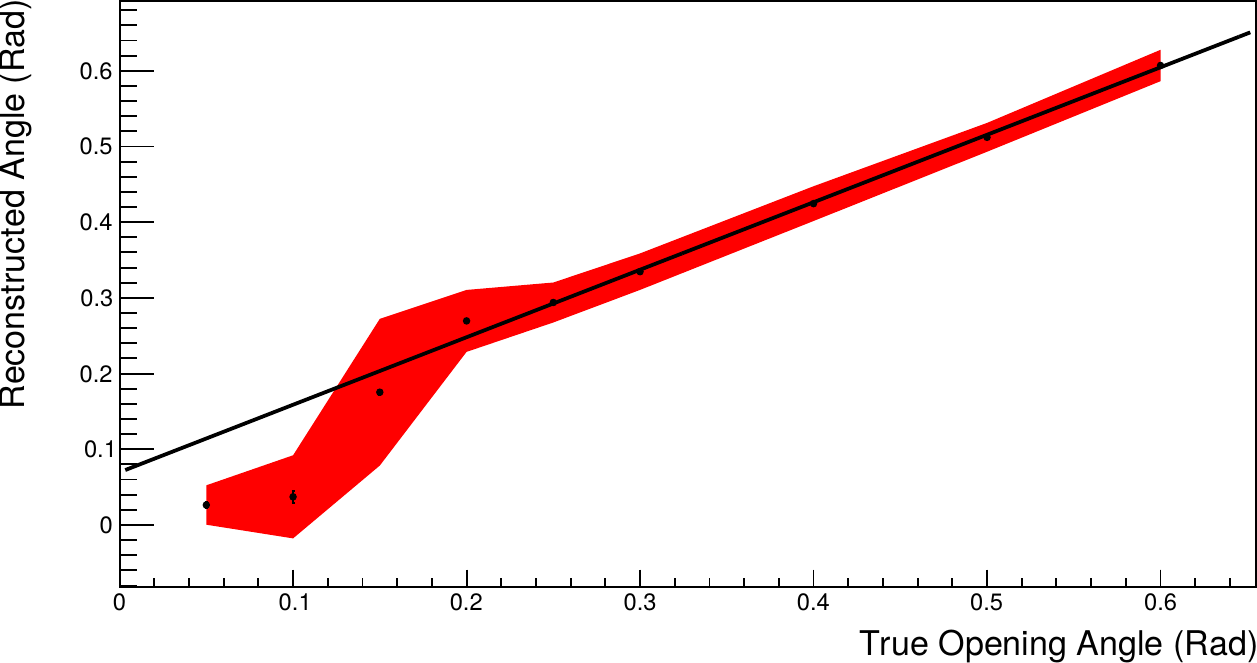}
\caption{Example opening angle reconstruction and resolution for an initial cluster and track building algorithm using $1$\,GeV electrons produced from a two-body decay at the front of the detector.~\cite{Gibbons:2019xx}}
\label{fig:rec_angle}
\end{figure}

The opening angle reconstruction as a function of the true opening angle for electrons with momenta of $1$\,GeV, produced from a two-body decay at the front of the detector, is shown in Fig.~\ref{fig:rec_angle}.
For opening angles above $0.2$\,rad the algorithm provides a flat resolution of $\simeq 20\%$ and a ratio close to unity between the true and reconstructed opening angle.
The tracking efficiency for single electrons with momentum greater than $0.5$\,GeV is $\simeq 0.95$.
This efficiency is also dependent upon the local detector occupancy: For two-body decays at the front of the detector with an opening angle less than $0.05$ radians the efficiency rapidly drops off,
as the individual tracks of the two electron candidates can no longer be resolved. For muons these efficiencies are closer to $\simeq 1.0$, even down to momenta of $0.5$\,GeV.

\begin{figure}[t]
	\includegraphics[width = 0.5\linewidth]{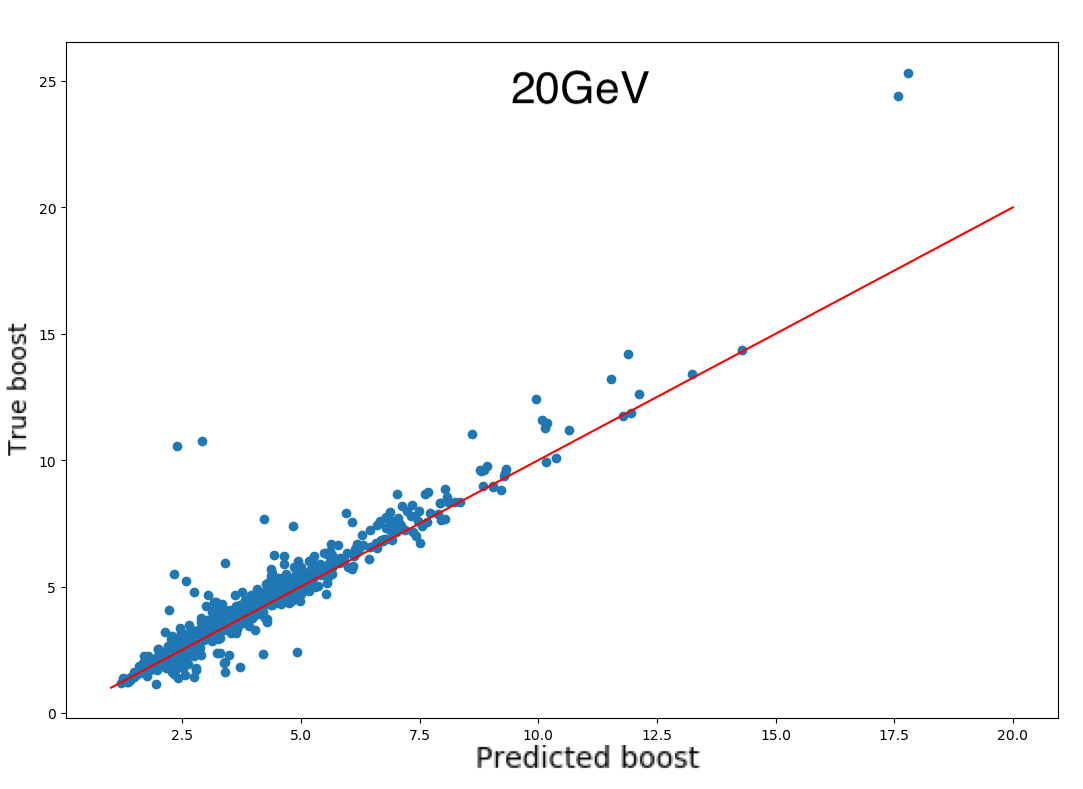}
\caption{Six-body boost reconstruction for a narrow $20~\mathrm{GeV}$
  resonance decay into $\tau \tau$ with $\tau^- \to \pi^- \pi^- \pi^+$.~\cite{Quessard:2019xx}}
\label{fig:rec_boost}
\end{figure}

While momentum information is not available for individual tracks, it is still possible to estimate the boost of an $n$-body signal decay.
For a two-body decay with small masses, the parent signal boost can be analytically approximated by assuming relativistic decay products~\cite{Curtin:2017izq}.
A study was performed looking at a six-body decay of the form $X \to \tau \tau$ with the simplified final state decay $\tau^- \to \pi^- \pi^- \pi^+\nu$,
where the missing energy of the neutrino and the resonance structure of decay was ignored~\cite{Quessard:2019xx}.
A neural net was trained on eight decay topology features: the six opening angles of the pions and the angles of the two three-pion combinations most consistent with a $\tau$-decay topology.
The true boost versus the reconstructed boost for a narrow $20$\,GeV resonance is shown in Fig.~\ref{fig:rec_boost}.
The boost resolution approaches $\simeq 4\%$ for resonance masses greater than $30$\,GeV. While this is a preliminary study, it demonstrates promise for reconstructing complex final states.

\subsection{Timing}
The baseline design employs RPC tracking detectors, which are expected to have a timing resolution of 350 ps per single gas gap of 1 mm \cite{Pizzimento}, which corresponds to $ 350 \text{ps}/\sqrt{6}\simeq
142$ ps resolution per station of 6 layers. The primary function of the timing capability is to synchronize the detector with the main LHCb detector.
This enables one to match LHCb events with \CODEXb events, and to characterize and reject possible backgrounds. 
In particular, backgrounds induced by cosmic muons will be out of time with the collisions.
As explained in Sec.~\ref{sec:backgrounds}, there is a sizable flux of relatively soft, neutral hadrons emanating from the shield. These hadrons can scatter or decay in the detector, respectively in the 
case of neutrons and $K_L$'s, leading to a number of slow moving tracks. For example, over a distance of $2$\,m between two stations, a timing resolution of $\sim 150$ ps would allow 
one to reliably identify particles traveling at $\beta\lesssim0.975$. For the example of a $\pi^{\pm}$, this corresponds to a momentum $\lesssim 0.6$\,GeV.

\begin{figure}[b]
	\includegraphics[width = 0.32\linewidth]{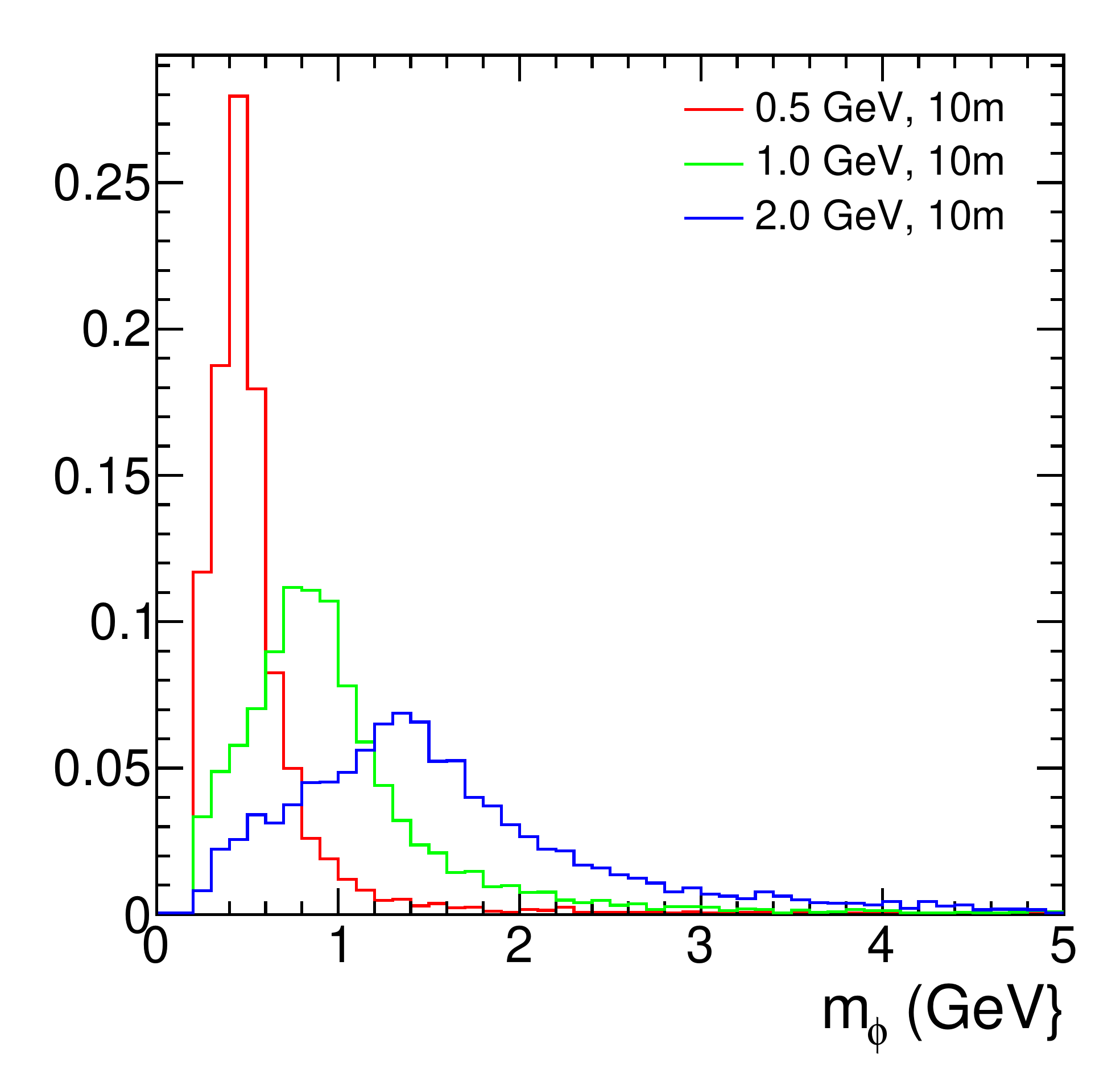}\hfill
	\includegraphics[width = 0.32\linewidth]{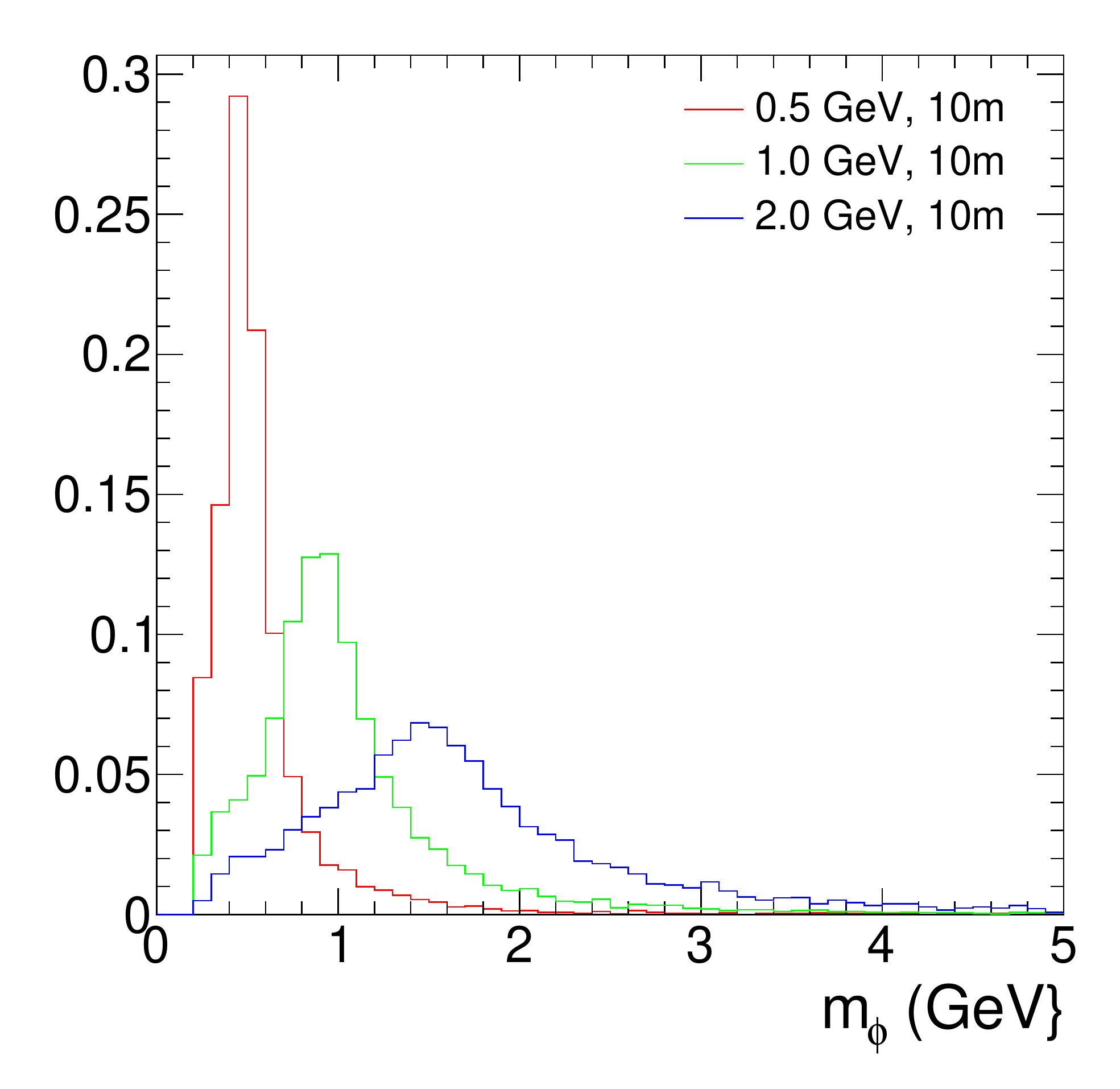}\hfill
	\includegraphics[width = 0.32\linewidth]{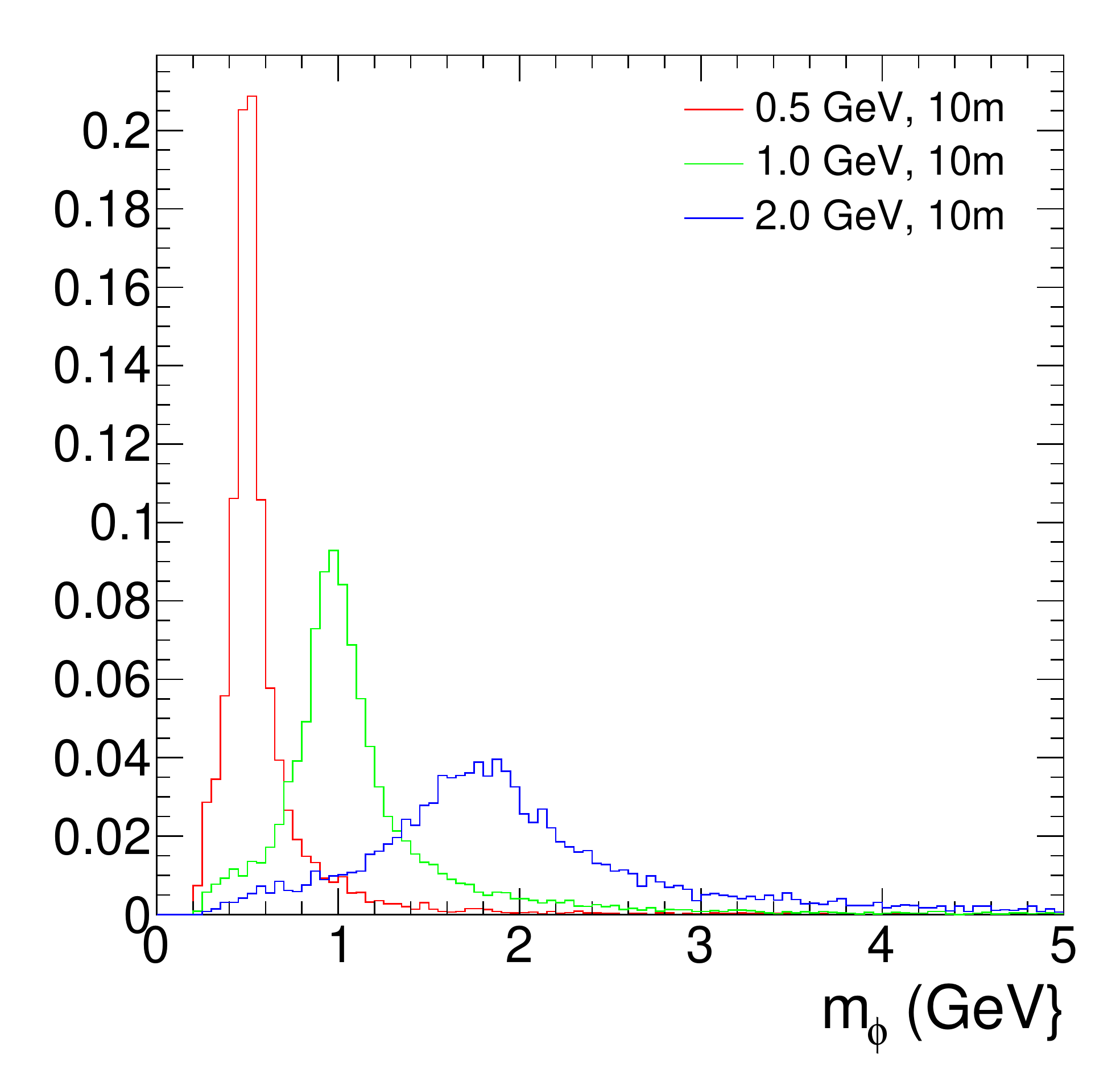}
\caption{Reconstructed LLP mass for different $B\to X_s \varphi$ benchmarks with $c\tau$ = 10 m, for 150 ps (left), 100 ps (middle), and 50 ps (right) timing resolution.~\cite{Gligorov:2017nwh}}
\label{fig:rec_timing}
\end{figure}

The timing capabilities of the RPCs is driven by the fluctuations of the primary ionization in the gas gap, and not to the readout electronics which can be designed 
to achieve resolutions of the order of $10$\,ps. It is possible that further development of the RPC technology may allow us to push this intrinsic timing resolution below 150~ps. 
Fig.~\ref{fig:rec_timing} shows the degree of separation which could be achieved for the $B\to  X_s S$ portal (see Sec.~\ref{sec:higgsmixing}) with some more optimistic assumptions for the timing resolution.

\subsection{Calorimetry}
\label{sec:calo}
Calorimetry would provide several important capabilities, notably particle identification (PID) via energy measurement and mass reconstruction, 
and the ability to expand the visible final states to include neutral hadrons and photons.

PID itself permits determination of the LLP decay modes, which could be crucial to identifying the quantum numbers and physics of the LLP itself.
For instance, reconstructing a $\mu^+ \pi^- \pi^0$ final state might suggest a leptonic LLP coupling to a charged current,
while even measuring the relative $e^+ e^-$ versus $\mu^+ \mu^-$ branching ratios could distinguish a vector from a scalar state.
Moreover, the ability to reliably reconstruct the LLP mass would provide an additional crucial property of the new particle, while also providing an additional handle to reject SM LLP 
backgrounds. Detection and reconstruction of neutral hadrons, especially the $\pi^0$, may permit rejection of $K_L^0$ backgrounds from the $\pi^+\pi^-\pi^0$ final state.

Calorimeter elements may also improve characterization of highly-boosted LLPs, especially when their decay products start to become so collimated that it becomes difficult to separate them from single tracks given the finite track-to-track separation capabilities of the detector. Concretely, the track-to-track separation equals $2\times$pitch$/\sqrt{12}$, for which we take 1 cm as a benchmark, similar to the expected performance ATLAS phase II RPCs. This can however be lowered if needed.
With a tracker-only option, merged tracks will reconstruct as a single `appearing' track in the tracking volume. 
However, for highly boosted LLPs, such as the ALPs of Sec.~\ref{sec:alps}, that may have hadronic final states, such hadrons would develop energetic showers inside the calorimeter.
This renders a signature strikingly different compared to e.g. low-energy neutron scattering. 
(Assuming each of the $\sim 10^6$ background muons passes though at least six tracking layers that are each $95\%$ efficient, 
then the expected number of appearing tracks induced from the muon background is $\sim 10^{-2}$ per $300$fb$^{-1}$.)

Energy measurement and PID may also help in the rejection of backgrounds, 
because they permit comparison of signal and background differential rates (in kinetic energy), rather than just the overall fluxes.
Further, a calorimeter element placed on the front face of the detector, i.e. closest to the IP, may detect and absorb the flux of incoming neutrons (see 
Sec.~\ref{sec:backgrounds}), that might otherwise scatter and produce signal-like tracks:
As seen in Tab.~\ref{tab:bkg-tracks}, single track production from neutron scattering inside the detector is non-negligible, with $\sim 50$ such events expected. 

\begin{figure}[t]
	\includegraphics[width = 0.6\linewidth]{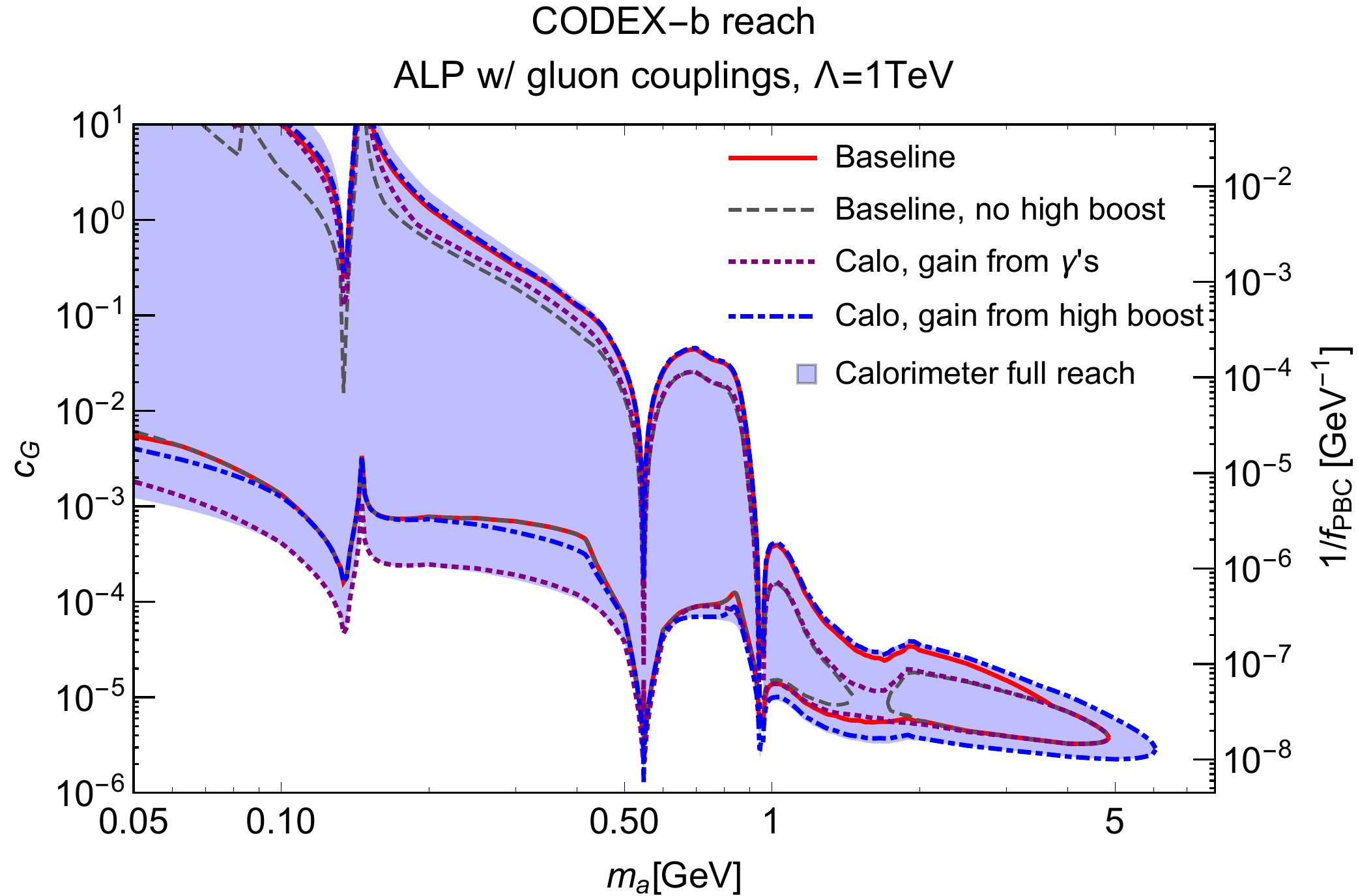}
\caption{Projected sensitivity of \CODEXb to gluon-coupled ALPs with a calorimeter element (blue, shaded) compared to the baseline, i.e. tracking only, detector (solid red). Also shown are the gains in reach 
coming from the ability to detect highly boosted LLPs (blue, dot-dashed) and from the ability to reconstruct photon final states (purple, dotted) separately. The gray dashed line corresponds to the baseline detector reach in which highly boosted ALPs are discarded instead of being considered with 50 events of background.}
\label{fig:ALPcalo}
\end{figure}

Diphoton final states may dominate the branching ratio of (pseudo)scalar LLPs, in particular the ALPs in the sub-GeV mass regime (see Sec.~\ref{sec:alps}), 
such that reaches may be greatly improved with the ability to detect photons.
Deposition of merged photons into the calorimeter will appear as a single highly energetic photon, 
to be compared with the relevant background photon fluxes shown in Fig.~\ref{fig:BGfluxes}. Above $1$\,GeV, these fluxes are $\lesssim 10^{-1}$ per $300$fb$^{-1}$. 
In Fig.~\ref{fig:ALPcalo} we show the ALP coupled to gluon reach for a \CODEXb setup, assuming the that the baseline design can be extended to detect diphoton final states efficiently (shaded area).
For comparison we also show the tracker-only baseline design (solid red line), and the gains obtained by detecting the highly-boosted ALPs with zero background (blue dot-dashed line) or detect the purely photonic decay modes (purple dotted line). For illustration we also include the baseline reach if one completely discards the highly-boosted ALPs (gray dashed line).
The \CODEXb reach attainable with a calorimeter addition is striking, both at high and low ALP masses.

Even more striking improvements are attainable in models where the ALP decays to photon pairs most of the time (as in the Physics Beyond Colliders benchmark BC9), as the first detectable final state with the baseline detector, the Dalitz mode $a \rightarrow e^+e^-\gamma$, has a branching ratio of $\mathcal{O}(10^{-2})$.

\begin{figure}[t]
	\includegraphics[width = 0.5\linewidth]{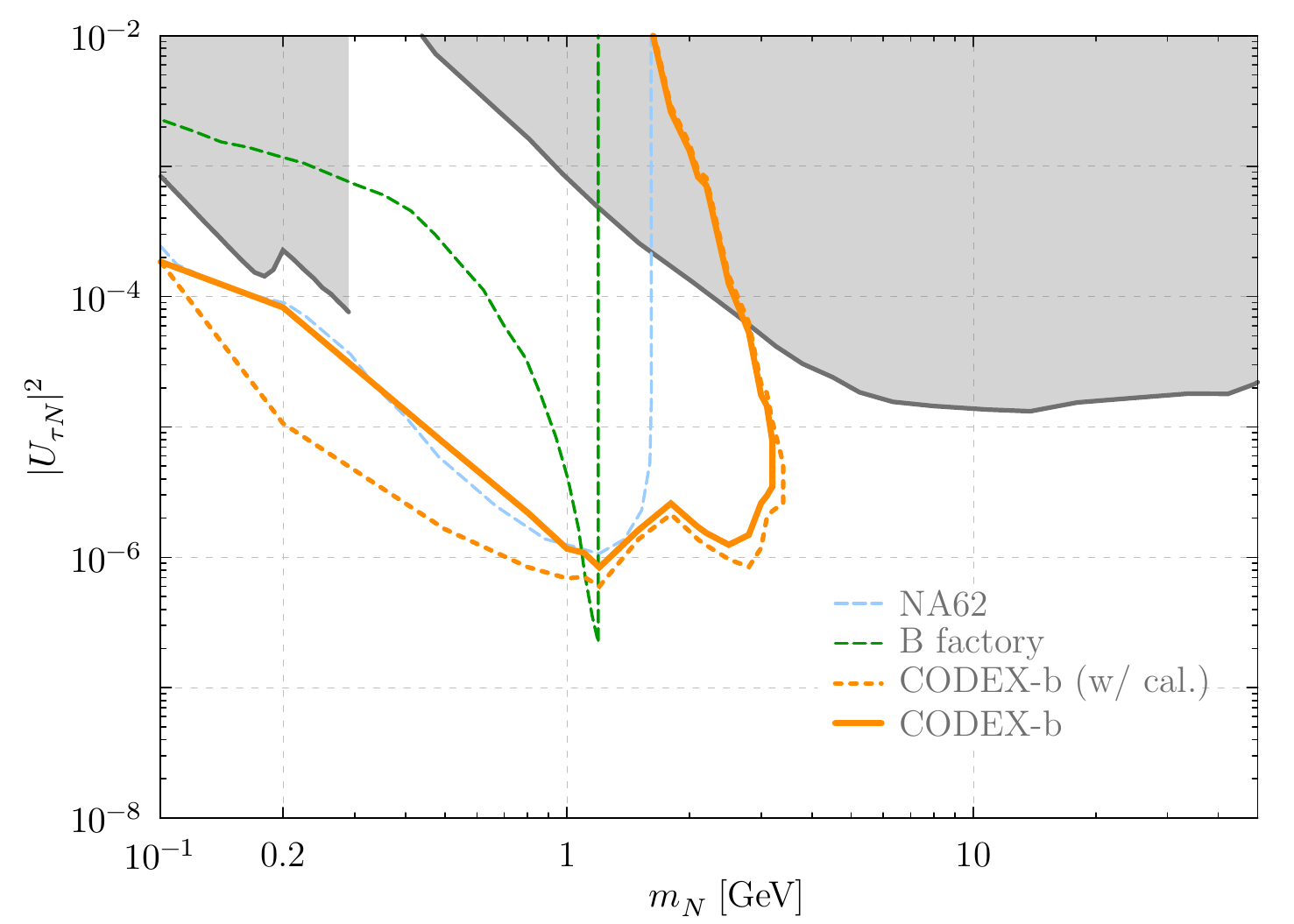}
\caption{Projected sensitivity of \CODEXb to Dirac heavy neutral leptons for $U_{\tau N} \gg U_{e N},U_{\mu N}$ for a tracking detector only (solid) compared to an optimistic case (dashed) with a calorimeter 
capable of reconstructing the $N \to \nu\pi^0$ final state, assuming the background differential flux of single $\pi^0$'s is negligible or reducible.}
\label{fig:hnl_tau_full}
\end{figure}

Similarly, detection and reconstruction of neutral hadrons such as the $\pi^0$ may be important in capturing dominant branching ratios in certain heavy neutral lepton mass regimes.
For example, for that case that the HNL is predominantly coupled to the $\tau$, with $m_N < m_\tau$, the dominant decay mode is $N \to \nu_\tau \pi^0$. 
In Fig.~\ref{fig:hnl_tau_full} we show the improvement in reach assuming this final state is reconstructible, compared to requiring at least two tracks.
In practice, measurement of this final state requires an understanding of the background differential flux of single $\pi^0$'s. While the nominal flux of $\pi^0$'s is vanishing small,
some might be produced from e.g. neutron scattering on air.

\subsection{Tagging of events at LHCb}
The LLPs detected at \CODEXb will be produced in events arising from $pp$ collisions at Interaction Point 8.
Therefore, these events could have information detectable at LHCb that is relevant to further help \CODEXb distinguish interesting phenomena from background. 
In this section we briefly review how  -- and how well -- information from LHCb could be used to tag events at \CODEXb.

To study this tagging, we use as a benchmark a Higgs boson decaying to a pair of long-lived dark photons (see Sec.~\ref{sec:abelianhiggs}), 
which in turn decay to a pair of muons: $h\rightarrow A' (\mu^+\mu^-) A'(\mu^+\mu^-)$.
The $A'$ were assumed to have a mass of $1$\,GeV and a proper lifetime of $c\tau = 1$\,m. The decay was generated using \texttt{Pythia}~\cite{Sjostrand:2014zea} at a center-of-mass energy of $\sqrt{s} = 14$\,TeV.

The first aspect studied was the probability to detect an LLP decay \emph{both} at \CODEXb and LHCb.
For events in which one $A'$ falls in the \CODEXb angular acceptance, $\sim18$\% have the other one in the LHCb acceptance.
However, for the lifetimes of greatest interest with respect to the \CODEXb reach, hardly any of these produce any detectable decay object at LHCb. 
In particular, for the $c\tau =1$\,m benchmark, the probability for such a decay is only $\sim 10^{-5}$. 
In more complicated hidden sectors however, a high multiplicity of LLPs may be produced in the same event, so that one could be detected at LHCb and the other in \CODEXb.

\begin{figure}[t]
	\includegraphics[width = 0.95\linewidth]{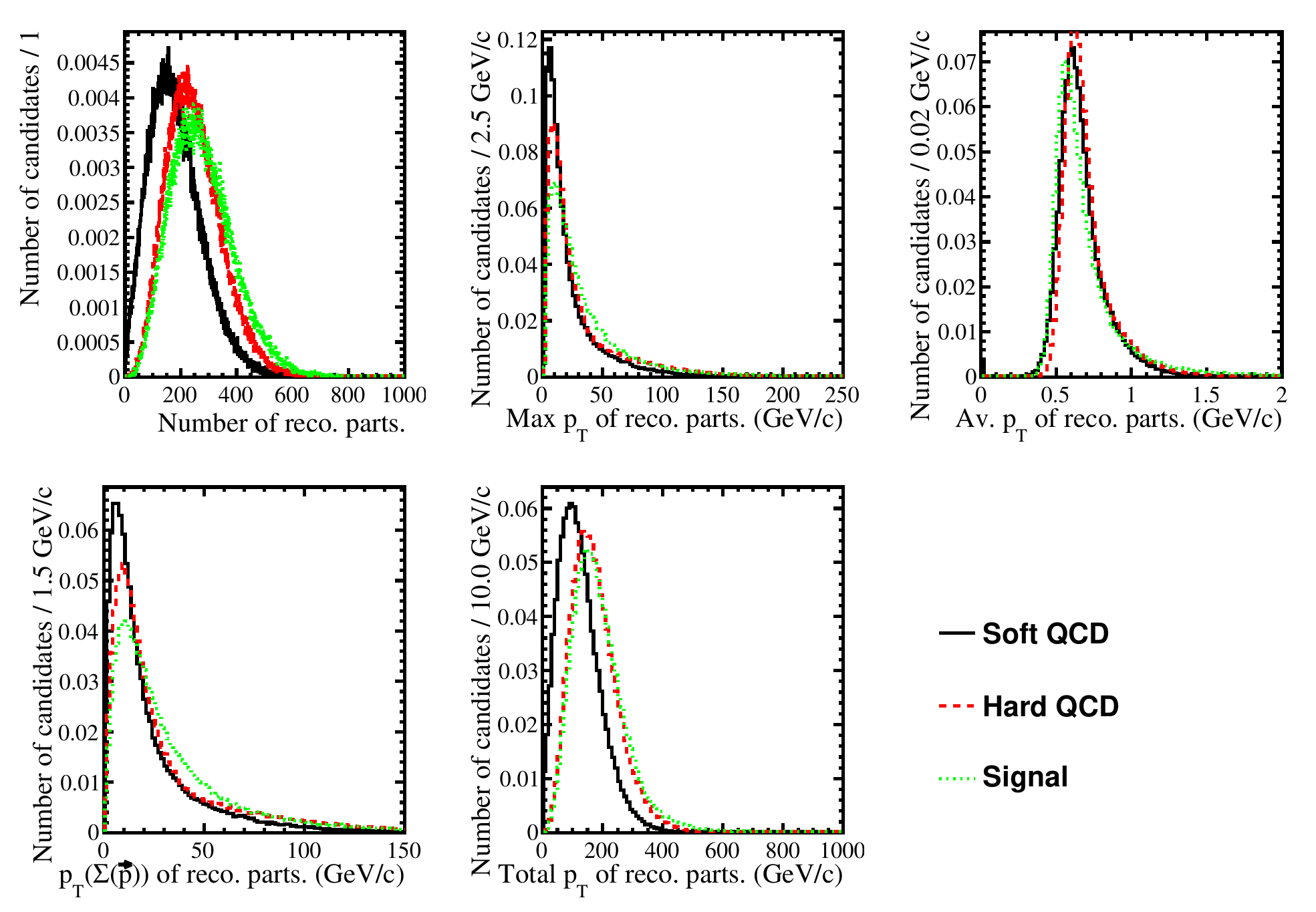}
\caption{Distribution of different general event variables at LHCb in events containing a $h\rightarrow A'(\mu^+\mu^-) A'(\mu^+\mu^-)$ decay (labelled as signal), soft and hard QCD processes.
The processes were all generated with \texttt{Pythia}. The variables displayed are all generated using particles \emph{reconstructible} at LHCb.
They correspond, from left to right and from top to bottom, to: the total number of particles in the event, the maximum $p_T$ of all these particles, their average $p_T$, the $p_T$ of the vectorial sum 
of all particles and the sum of their $p_T$.}
\label{fig:tagging_lhcb}
\end{figure}

The second possibility under study was how the LLP production mechanism can affect the underlying event seen at LHCb.
This should be specifically relevant whenever the LLP is produced through the Higgs portal, such as in our benchmark example.
We performed a general comparison of how events \textit{look} at LHCb at truth level, with no reconstruction involved.
To compare to the signal, we generated softQCD (minimum bias) and hardQCD ($b\bar{b}$) samples with \texttt{Pythia}, under the same conditions as the signal.
For this comparison, we defined reconstructible particles at LHCb as those stable, charged particles that are produced in the LHCb acceptance. 
In Fig.~\ref{fig:tagging_lhcb} we show the distribution of different global variables of interest.
While the figure shows a certain degree of discrimination between the different processes, more detailed studies will be needed.
In particular, for this study $gg$ fusion was chosen as Higgs production mode. 
Production via vector boson fusion, though having a smaller cross section, might provide more power to tag CODEX-b events at LHCb, by searching for a hard jet in the LHCb acceptance.

\clearpage
\section{Outlook}
The immediate priority for \CODEXb is the finalization of the design for the \CODEXbeta demonstrator and approval for its installation.
A Letter of Intent for the full \CODEXb detector will follow this Expression of Interest in the near-term, including further developments of the detector design concept,
although results from the \CODEXbeta demonstrator are expected to inform the final design choices for the detector. In particular, we intend to 
investigate in detail a realistic option for incorporating calorimetry in the CODEX-b design.

\acknowledgments
We thank Wolfgang Altmannshofer, Brian Batell, James Beacham, Florian Bernlochner, David Curtin, Francesco D'Eramo, Jeff Dror, Gilly Elor, Iftah Galon, Stefania Gori, Felix Kling, 
Gaia Lanfranchi, Zoltan Ligeti, Simone Pagan Griso, 
Hiren Patel, Gilad Perez, Maxim Pospelov, Josh Ruderman, Bibhushan Shakya, Jessie Shelton, Brian Shuve, Yotam Soreq, Yuhsin Tsai, Jure Zupan for many helpful discussions. 
We similarly thank all the members of the BSM working group of the Physics Beyond Colliders report.
In addition, we thank Asher Berlin, Raffaele Tito D'Agnolo, Daniel Dercks, Jordy De Vries,  Herbi K.~Dreiner, Juan Carlos Helo, Martin Hirsch, Felix Kling, Dave McKeen and Zeren Simon Wang 
for generously providing plots for reproduction, with permission, as well as for comments on parts of the manuscript. 

We thank the computing and simulation teams of the LHCb collaboration for their generous help in performing the simulation studies presented here. We thank the LHCb technical coordination for making the background studies in the DELPHI cavern possible. We thank Giovanni Passaleva for his support and encouragement,
and for helpful advice regarding the baseline RPC technology and the relationship between CODEX-b and LHCb.

The work of XCV is supported by MINECO (Spain) through the Ramon y Cajal program RYC-2016-20073 and by XuntaGal under the ED431F 2018/01 project.
JAE is supported by U.S. Department of Energy (DOE) grant DE-SC0011784. 
VVG is partially supported by ERC CoG ``RECEPT'' GA number 724777 within the H2020 framework programme. 
SK is supported by U.S. DOE grant DE-SC0009988 and the Paul Dirac fund at the Institute for Advanced Study.  
MP, DR and BN are supported by the U.S. DOE under contract DE-AC02-05CH11231.
HR is supported in part by the U.S. DOE under contract DE-AC02-05CH11231.
MW is supported by NSF grant PHY-1912836.

Significant parts of this work were performed at: 
the Aspen Center for Physics, supported by National Science Foundation grant PHY-1607611; 
the Munich Institute for Astro- and Particle Physics (MIAPP);
and the Galileo Galilei Institute for Theoretical Physics. We thank all these institutions for their support and hospitality.

 This research used resources of the National Energy Research Scientific Computing Center (NERSC), a U.S. Department of Energy Office of Science User Facility 
 operated under Contract No. DE-AC02-05CH11231.
\vspace{-14pt}

\fancyhf{}
\fancyhead[L]{\slshape REFERENCES}

\bibliographystyle{JHEP}
\bibliography{CodexEoI}

\end{document}